\documentclass[preprint,aps, showpacs]{revtex4}
\usepackage{graphicx}
\usepackage[bf,SL,BF]{subfigure}
\usepackage{psfrag}
\usepackage{color}
\usepackage{amssymb}
\usepackage{amsmath}
\usepackage{epstopdf}
\usepackage{bbold}

\newcommand{\be}{\begin{eqnarray}}
\newcommand{\ee}{\end{eqnarray}}

\newcommand{\bfb}{{\bf b}_{\perp}}

\newcommand{\bfp}{{\bf p}_{\perp}}

\newcommand{\Dp}{{\bf \Delta}_{\perp}}

\begin{document}
%%------------------------------------------

\title{Quark Wigner distributions and spin-spin  correlations}
\author{D. Chakrabarti$^1$}
\author{T. Maji$^1$}
\author{C. Mondal$^{1,2}$}
\author{ A. Mukherjee$^3$}
\affiliation{$^1$Department of Physics, Indian Institute of Technology Kanpur, Kanpur 208016, India}
\affiliation{$^2$Institute of Modern Physics, Chinese Academy of Sciences, Lanzhou 730000, China}
\affiliation{$^3$Department of Physics, 
Indian Institute of Technology Bombay,
Mumbai 400076, India}

\date{\today}

\begin{abstract}
We investigate the Wigner distributions for $u$ and $d$ quarks in a light front quark-diquark model of a proton to unravel the spatial and spin structure. The light-front wave functions are modeled from the soft-wall AdS/QCD prediction. We consider the contributions from both the scalar and the axial vector diquarks. The Wigner distributions for unpolarized, longitudinally polarized and transversely polarized proton are presented in transverse momentum plane as well as in transverse impact parameter plane. The Wigner distributions satisfy a Soffer bound type inequality. We also evaluate all the leading twist GTMDs  and show their scale evolution. The spin-spin correlations between the quark and the proton are investigated.  
\end{abstract}

\maketitle

%=========================================
%====================================================
%%%%%%%%%%%%%%%%%%%%%%%%%%%%%%%%%%%%%%
\section{Introduction\label{intro}}
%%%%%%%%%%%%%%%%%%%%%%%%%%%%%%%%%%%%%%
The quark and gluon Wigner distribution  introduced by Ji\cite{xji1, xji2} have been  studied extensively in recent times to understand the three dimensional structure of  proton.  The Wigner distributions encode spatial  as well as  partonic spin and orbital angular momentum structures.  The Wigner distributions are six dimensional phase-space distributions and are not directly measurable. But after some phase-space reductions, they reduce to  the generalized parton distributions(GPDs) and  transverse momentum dependent distributions(TMDs).  The Wigner distributions integrated over transverse momenta reduce to the GPDs at zero skewness and on integration over the transverse impact parameters with zero momentum transfer, they reduce to the TMDs. It is well known that GPDs and TMDs 
encode informations about the three dimensional partonic structure of hadrons.  Recently, generalized transverse momentum dependent PDFs or GTMDs are introduced\cite{meissner08,meissner09,Eche}. Gluon GTMDs have been discussed in \cite{Lorce13}. GTMDs are related to the Wigner distributions and they again in certain kinematical limits reduce to GPDs and TMDs. 
% The Wigner distributions are related  to the fully unintegrated off-diagonal quark-quark correlator in the proton.
The Wigner distributions after integrating over the light cone energy  of the parton are interpreted as a Fourier transform of corresponding generalized transverse momentum dependent distributions (GTMDs) which are functions of the light cone three-momentum of the parton as well as  the momentum transfer to the nucleon.  
 The spin-spin and spin-orbital angular momentum(OAM) correlations between a nucleon and a quark inside the nucleon can  be described from the  phase space average of Wigner distributions. 
 Angular momentum of a quark is extracted from Wigner distributions taking the phase space average. The Wigner distributions have been studied in different models e.g., in lightcone constituent quark model\cite{Lorce12,Lorce11,LCCQM1,LCCQM2}, in  chiral soliton model\cite{chi_QSM, Lorce11, QSM}, light front dressed quark model\cite{MNO1, MNO2,AM}, lightcone spectator model\cite{liu_ma_WD}, light-cone quark-scalar-diquark model\cite{WD_SD}.
 
 In this work, we  study the quark Wigner distributions in  a light front quark-diquark model\cite{MC}  for the proton where the diquark be both scalar or vector. We have studied the distributions for unpolarized as well as longitudinally and transversely polarized proton. The leading twist GTMDs are evaluated from the Wigner distributions. The we study the spin and OAM correlations between the quarks and proton. 
 We find that the quark OAM tend to  anti-align with the quark spin  for $u$ and $d$ quark but  to align with the proton spin.
 
 The paper is organized as follows: We first introduce the light-front quark-diquark model in Sect.\ref{model}. The Wigner distributions and GTMDs are introduced  in Sect.\ref{Wigner} and how the OAM spin  can be extracted from Wigner distributions and GTMDs are discussed in Sect.\ref{oam}. Then, the results in our model are discussed in Sect.\ref{results} for unpolarized, longitudinally polarized and transversely polarized proton in subsections  \ref{unpol}, \ref{longpol} and \ref{transpol} respectively. The spin-spin and spin-OAM correlations are discussed in \ref{spincor}.  A very brief description of the scale evolution of GTMDs is given in Sect.\ref{sec_GTMD}. Some inequalities satisfied among the GTMDs and also among the Wigner distributions in our model are shown in Sect. \ref{inequals}. Finally, we conclude the paper with a summary and discussion in Sect.\ref{con}.
 
 % In this work, we investigate the Wigner distributions for unpolarized and polarized proton and the orbital angular momentum(OAM) and spin-spin and spin-OAM correlations in a scalar diquark model of proton \cite{Gut} with the light front wavefunctions modeled  from AdS/QCD prediction.
 
%%%%%%%%%%%%%%%%%%%%%%%%%%%%%%%%%%%%%%%%%%%%%%%%%%%%%%%
\section{Light-front quark-diquark model for nucleon\label{model}}
%%%%%%%%%%%%%%%%%%%%%%%%%%%%%%%%%%%%%%%%%%%%%%%%%%%%%%
%In the diquark model, we assume that the virtual incoming photon is interacting with a valence and other two valence quark form a diquark of definite mass with spin-0, called scalar diquark, or with spin-1, called vector diquark. The spin-0 diquarks are in in a flavor singlet state and spin-1 diquarks are in flavor triplet state. 

In this model\cite{MC}, the proton state is written as superposition of the quark-diquark states allowed under $SU(4)$ spin-flavor symmetry. Thus the proton can be written 
as a sum of isoscalar-scalar diquark singlet state $|u~ S^0\rangle$, isoscalar-vector diquark state $|u~ A^0\rangle$ and isovector-vector diquark $|d~ A^1\rangle$ state\cite{Jakob97,Bacc08} and the state is written as
% in the spin-flavor $SU(4)$ structure as
\be 
|P; \pm\rangle = C_S|u~ S^0\rangle^\pm + C_V|u~ A^0\rangle^\pm + C_{VV}|d~ A^1\rangle^\pm. \label{PS_state}
\ee
Where $S$ and $A$ represent the scalar and axial-vector diquark having isospin at their superscript. % Under the isospin symmetry, the neutron  state is given by the above formula with $u\leftrightarrow d$.
%The coefficients $C_i$'s are absorbed in the light front wave function $\psi_{\lambda_q \lambda_D}^{\lambda_N(\nu)}$,Eq.(\ref{LFWF_S}-\ref{LFWF_Vm}) corresponding to the nucleon helicity $\lambda_N$, quark helicity $\lambda_q$ and diquark helicity $\lambda_D$ with flavour index $\nu$.

We use the light-cone convention $x^\pm=x^0 \pm x^3$. We choose a frame where the transverse momentum of proton vanishes i,e. $P \equiv \big(P^+,\frac{M^2}{P^+},\textbf{0}_\perp\big)$. In this symmetric frame, the momentum of struck quark  
$p\equiv (xP^+, \frac{p^2+|\bfp|^2}{xP^+},\bfp)$
 and  that of diquark   $P_X\equiv ((1-x)P^+,P^-_X,-\bfp)$. Here $x=p^+/P^+$ is the longitudinal momentum fraction carried by the struck quark. 
The two particle Fock-state expansion for $J^z =\pm1/2$ with spin-0 diquark  is given by
\be
|u~ S\rangle^\pm & =& \int \frac{dx~ d^2\bfp}{2(2\pi)^3\sqrt{x(1-x)}} \bigg[ \psi^{\pm(u)}_{+}(x,\bfp)|+\frac{1}{2}~s; xP^+,\bfp\rangle \nonumber \\
 &+& \psi^{\pm(u)}_{-}(x,\bfp)|-\frac{1}{2}~s; xP^+,\bfp\rangle\bigg],\label{fock_PS}
\ee
and the LF wave functions with spin-0 diquark, for $J=\pm1/2$, are given by\cite{Lepa80}
\be 
\psi^{+(u)}_+(x,\bfp)&=& N_S~ \varphi^{(u)}_{1}(x,\bfp),\nonumber \\
\psi^{+(u)}_-(x,\bfp)&=& N_S\bigg(- \frac{p^1+ip^2}{xM} \bigg)\varphi^{(u)}_{2}(x,\bfp), \label{LFWF_S}\\
\psi^{-(u)}_+(x,\bfp)&=& N_S \bigg(\frac{p^1-ip^2}{xM}\bigg) \varphi^{(u)}_{2}(x,\bfp),\nonumber \\
\psi^{-(u)}_-(x,\bfp)&=&  N_S~ \varphi^{(u)}_{1}(x,\bfp),\nonumber
\ee
where $|\lambda_q~\lambda_S; xP^+,\bfp\rangle$ is the two particle state having struck quark of helicity $\lambda_q$ and a scalar diquark having helicity $\lambda_S=s$(spin-0 singlet diquark helicity is denoted by s to distinguish from triplet diquark). The state with spin-1 diquark is given as \cite{Ellis08}
\be
|\nu~ A \rangle^\pm & =& \int \frac{dx~ d^2\bfp}{2(2\pi)^3\sqrt{x(1-x)}} \bigg[ \psi^{\pm(\nu)}_{++}(x,\bfp)|+\frac{1}{2}~+1; xP^+,\bfp\rangle \nonumber\\
 &+& \psi^{\pm(\nu)}_{-+}(x,\bfp)|-\frac{1}{2}~+1; xP^+,\bfp\rangle +\psi^{\pm(\nu)}_{+0}(x,\bfp)|+\frac{1}{2}~0; xP^+,\bfp\rangle \nonumber \\
 &+& \psi^{\pm(\nu)}_{-0}(x,\bfp)|-\frac{1}{2}~0; xP^+,\bfp\rangle + \psi^{\pm(\nu)}_{+-}(x,\bfp)|+\frac{1}{2}~-1; xP^+,\bfp\rangle \nonumber\\
 &+& \psi^{\pm(\nu)}_{--}(x,\bfp)|-\frac{1}{2}~-1; xP^+,\bfp\rangle  \bigg].\label{fock_PS}
\ee
Where $|\lambda_q~\lambda_D; xP^+,\bfp\rangle$ represents a two-particle state with a quark of helicity $\lambda_q=\pm\frac{1}{2}$ and a axial-vector diquark of helicity $\lambda_D=\pm 1,0(triplet)$.
The LFWFs are, for $J=+1/2$ 
\be 
\psi^{+(\nu)}_{+~+}(x,\bfp)&=& N^{(\nu)}_1 \sqrt{\frac{2}{3}} \bigg(\frac{p^1-ip^2}{xM}\bigg) \varphi^{(\nu)}_{2}(x,\bfp),\nonumber \\
\psi^{+(\nu)}_{-~+}(x,\bfp)&=& N^{(\nu)}_1 \sqrt{\frac{2}{3}} \varphi^{(\nu)}_{1}(x,\bfp),\nonumber \\
\psi^{+(\nu)}_{+~0}(x,\bfp)&=& - N^{(\nu)}_0 \sqrt{\frac{1}{3}} \varphi^{(\nu)}_{1}(x,\bfp),\label{LFWF_Vp}\\
\psi^{+(\nu)}_{-~0}(x,\bfp)&=& N^{(\nu)}_0 \sqrt{\frac{1}{3}} \bigg(\frac{p^1+ip^2}{xM} \bigg)\varphi^{(\nu)}_{2}(x,\bfp),\nonumber \\
\psi^{+(\nu)}_{+~-}(x,\bfp)&=& 0,\nonumber \\
\psi^{+(\nu)}_{-~-}(x,\bfp)&=&  0, \nonumber 
\ee
and for $J=-1/2$
\be 
\psi^{-(\nu)}_{+~+}(x,\bfp)&=& 0,\nonumber \\
\psi^{-(\nu)}_{-~+}(x,\bfp)&=& 0,\nonumber \\
\psi^{-(\nu)}_{+~0}(x,\bfp)&=& N^{(\nu)}_0 \sqrt{\frac{1}{3}} \bigg( \frac{p^1-ip^2}{xM} \bigg) \varphi^{(\nu)}_{2}(x,\bfp),\label{LFWF_Vm}\\
\psi^{-(\nu)}_{-~0}(x,\bfp)&=& N^{(\nu)}_0\sqrt{\frac{1}{3}} \varphi^{(\nu)}_{1}(x,\bfp),\nonumber \\
\psi^{-(\nu)}_{+~-}(x,\bfp)&=& - N^{(\nu)}_1 \sqrt{\frac{2}{3}} \varphi^{(\nu)}_{1}(x,\bfp),\nonumber \\
\psi^{-(\nu)}_{-~-}(x,\bfp)&=& N^{(\nu)}_1 \sqrt{\frac{2}{3}} \bigg(\frac{p^1+ip^2}{xM}\bigg) \varphi^{(\nu)}_{2}(x,\bfp),\nonumber
\ee
having flavor index $\nu=u,d$.
The LFWFs $\varphi^{(\nu)}_i(x,\bfp)$ are  a modified form of the  soft-wall AdS/QCD prediction
\be
\varphi_i^{(\nu)}(x,\bfp)=\frac{4\pi}{\kappa}\sqrt{\frac{\log(1/x)}{1-x}}x^{a_i^\nu}(1-x)^{b_i^\nu}\exp\bigg[-\delta^\nu\frac{\bfp^2}{2\kappa^2}\frac{\log(1/x)}{(1-x)^2}\bigg].
\label{LFWF_phi}
\ee
The wave functions $\varphi_i^\nu ~(i=1,2)$ reduce to the AdS/QCD prediction\cite{BT1, BT2} for the parameters $a_i^\nu=b_i^\nu=0$  and $\delta^\nu=1.0$. We use the AdS/QCD scale parameter $\kappa =0.4~GeV$ as determined in \cite{CM1} and the quarks are  assumed  to be  massless. 

%%%%%%%%%%%%%%%%%%%%%%%%%%%%%%%%%%%%%%%%%%%%%%
\section{Wigner distributions\label{Wigner}}
%%%%%%%%%%%%%%%%%%%%%%%%%%%%%%%%%%%%%%%%%%%%%%
In the same way as the impact-parameter-dependent parton distributions (IPDs) which are obtained by the two-dimensional Fourier transforms
of the generalized parton distributions (GPDs), one can map out the Wigner distributions as the two-dimensional Fourier transforms of the
so-called generalized transverse-momentum-dependent parton distributions (GTMDs). In light-front framework, one defines the 
5-dimensional quark Wigner distributions as \cite{Lorce11,Lorce12}
\be
\rho^{\nu [\Gamma]}(\bfb,\bfp,x;S)=\int \frac{d^2\Dp}{(2\pi)^2} e^{-i\Dp.b_\perp} W^{\nu [\Gamma]}(\Dp,\bfp,x;S).
\label{wig_rho}
\ee
The correlator $W^{[\Gamma]}$ relates the GTMDs \cite{meissner08,meissner09} and in the Drell-Yan-West frame ($\Delta^+=0$) and fixed light-cone time $ z^+=0$ is given by
\be
W^{\nu [\Gamma]}(\Dp,\bfp,x;S)=\frac{1}{2}\int \frac{dz^-}{(2\pi)} \frac{d^2z_T}{(2\pi)^2} e^{ip.z} 
\langle P^{\prime\prime}; S|\bar{\psi}^\nu _i(-z/2)\Gamma \mathcal{W}_{[-z/2,z/2]} \psi^\nu _j(z/2) |P^\prime;S\rangle \bigg|_{z^+=0}.
\label{wigner_W}
\ee
The $\Gamma$ denotes the twist-two Dirac $\gamma$-matrix, $\gamma^+$, $\gamma^+\gamma_5$ or $i\sigma^{j+}\gamma_5$ with $j=1$ or $2$
corresponding to unpolarized, longitudinal polarized or $j$-direction transverse polarized quark respectively. The gauge link Wilson line,  $\mathcal{W}_{[-z/2,z/2]}$ ensures the $SU(3)$ color gauge invariance of the Wigner operator. $P'(P'')$ and $S$ represent the proton momenta 
of the initial (final) state of proton, respectively and the spin of proton state. 
We use the light-front coordinates $v^\mu=[v^+,v^-,\vec v_\perp]$, where $v^\pm=(v^0\pm v^3)$ and $\vec v_\perp=(v^1,v^2)$. 
 The kinematical variables are defined as 
\be
P^\mu=\frac{(P'+P'')^\mu}{2}, \quad\quad \Delta^\mu=(P''-P')^\mu.
\ee
Depending on the various polarization configurations of the proton and the quark, there are 16 independent twist-2 quark Wigner distributions.
In an unpolarized proton, the quark Wigner distributions for unpolarized, longitudinally polarized and transversely polarized quark are defined as \cite{Lorce11,liu2015}
\be 
\rho^{\nu }_{UU}(\bfb,\bfp,x)&=&\frac{1}{2}[\rho^{\nu [\gamma^+]}(\bfb,\bfp,x; +\hat{S}_z) + \rho^{\nu [\gamma^+]}(\bfb,\bfp,x; -\hat{S}_z)],\label{rho_UU_def}\\
\rho^{\nu }_{UL}(\bfb,\bfp,x)&=&\frac{1}{2}[\rho^{\nu [\gamma^+\gamma^5]}(\bfb,\bfp,x; +\hat{S}_z) + \rho^{\nu [\gamma^+\gamma^5]}(\bfb,\bfp,x; -\hat{S}_z)],\label{rho_UL_def}\\
\rho^{\nu j}_{UT}(\bfb,\bfp,x)&=&\frac{1}{2}[\rho^{\nu [i\sigma^{j+}\gamma^5]}(\bfb,\bfp,x; +\hat{S}_z) +\rho^{\nu [i\sigma^{j+}\gamma^5]}(\bfb,\bfp,x; -\hat{S}_z)]\label{rho_UT_def}.
\ee 
For a longitudinally polarized proton and different polarizations of quark the distributions are given by
\be 
\rho^{\nu }_{LU}(\bfb,\bfp,x)&=&\frac{1}{2}[\rho^{\nu [\gamma^+]}(\bfb,\bfp,x; +\hat{S}_z) - \rho^{\nu [\gamma^+]}(\bfb,\bfp,x; -\hat{S}_z)],\label{rho_LU_def}\\
\rho^{\nu }_{LL}(\bfb,\bfp,x)&=&\frac{1}{2}[\rho^{\nu [\gamma^+\gamma^5]}(\bfb,\bfp,x; +\hat{S}_z) - \rho^{\nu [\gamma^+\gamma^5]}(\bfb,\bfp,x; -\hat{S}_z)],\label{rho_LL_def}\\
\rho^{\nu j}_{LT}(\bfb,\bfp,x)&=&\frac{1}{2}[\rho^{\nu [i\sigma^{j+}\gamma^5]}(\bfb,\bfp,x; +\hat{S}_z) -\rho^{\nu [i\sigma^{j+}\gamma^5]}(\bfb,\bfp,x; -\hat{S}_z)]\label{rho_LT_def}.
\ee 
Again the Wigner distributions for a transversely polarized proton with various quark polarization are identified as 
\be 
\rho^{i \nu }_{TU}(\bfb,\bfp,x)&=&\frac{1}{2}[\rho^{\nu [\gamma^+]}(\bfb,\bfp,x; +\hat{S}_i) - \rho^{\nu [\gamma^+]}(\bfb,\bfp,x; -\hat{S}_i)],\label{rho_TU_def}\\
\rho^{i \nu }_{TL}(\bfb,\bfp,x)&=&\frac{1}{2}[\rho^{\nu [\gamma^+\gamma^5]}(\bfb,\bfp,x; +\hat{S}_i) - \rho^{\nu [\gamma^+\gamma^5]}(\bfb,\bfp,x; -\hat{S}_i)],\label{rho_TL_def}\\
\rho^{\nu }_{TT}(\bfb,\bfp,x)&=&\frac{1}{2}\delta_{ij}[\rho^{\nu [i\sigma^{j+}\gamma^5]}(\bfb,\bfp,x; +\hat{S}_i) -\rho^{\nu [i\sigma^{j+}\gamma^5]}(\bfb,\bfp,x; -\hat{S}_i)]\label{rho_TT_def},
\ee 
and the pretzelous Wigner distribution is defined as 
\be 
\rho^{\nu \perp}_{TT}(\bfb,\bfp,x)&=&\frac{1}{2}\epsilon_{ij}[\rho^{\nu [i\sigma^{j+}\gamma^5]}(\bfb,\bfp,x; +\hat{S}_i) -\rho^{\nu [i\sigma^{j+}\gamma^5]}(\bfb,\bfp,x; -\hat{S}_i)]\label{rho_TTp_def}.
\ee 
Here the first subscript denotes the proton polarization, and the second one represents the quark polarization. The Wigner distributions are directly connected to the generalized parton correlation functions \cite{meissner08,Lorce11}. At $z_{\perp=0}$, integrating over momentum $\bfp$ the Wigner distributions reduce to impact-parameter-dependent parton distributions (IPDs) which can be interpreted as quark densities
in the transverse position space. Again the $\bfb$ integration of the distributions give the transverse momentum dependent parton distributions (TMDs) which can be interpreted as quark densities in transverse momentum space. In the Drell-Yan-West frame, the Wigner distributions  may have a quasiprobability interpretation \cite{Lorce11} but the interpretation is lost when one defines a six-dimensional Wigner distribution, $\rho^{\nu [\Gamma]}(\bfb,\xi,\bfp,x; \hat{S})$ by including a longitudinal momentum transfer ($\xi=-\Delta^+/(2P^+)$). One can also obtain the three dimensional quark densities by integrating over two mutually orthogonal components of transverse position and momentum, e,g. $b_y$ and $p_x$ ($b_x$ and $p_y$), which are not constraint by Heisenberg uncertainty principle as \cite{Lorce11}
\be
\int db_y dp_x \rho^{\nu [\Gamma]}(\bfb,\bfp,x;S)=\tilde{\rho}^{\nu [\Gamma]}(b_x,p_y,x;S),
%\int db_x dp_y \rho^{\nu [\Gamma]}(\bfb,\bfp,x;S)&=&\tilde{\rho}^{\nu [\Gamma]}(b_y,p_x,x;S),
\label{rho_bxpy}
\ee
with $\Delta_y=z_x=0$ and
\be
%\int db_y dp_x \rho^{\nu [\Gamma]}(\bfb,\bfp,x;S)&=&\tilde{\rho}^{\nu [\Gamma]}(b_x,p_y,x;S),\nonumber\\
\int db_x dp_y \rho^{\nu [\Gamma]}(\bfb,\bfp,x;S)=\bar{\rho}^{\nu [\Gamma]}(b_y,p_x,x;S),
\label{rho_bypx}
\ee
with $\Delta_x=z_y=0$. For unpolarized and longitudinally polarized proton the distributions $\tilde{\rho}^{\nu [\Gamma]}(b_y,p_x,x;S)$ and  $\bar{\rho}^{\nu [\Gamma]}(b_x,-p_y,x;S)$ are same. 
The Wigner distribution of quarks with longitudinal polarization $\lambda$ in a longitudinally polarised proton $\Lambda$ is defined  for $\Gamma=\gamma^+ \frac{\mathbb{1}+\lambda\gamma^5}{2} $ and $\vec{S}=\Lambda \hat{S}_z$ as \cite{Lorce11}
\be 
\rho^\nu_{\Lambda \lambda}(\bfb,\bfp,x)=\frac{1}{2}[\rho^{\nu[\gamma^+]}(\bfb,\bfp,x;\Lambda \hat{S}_z) +\lambda \rho^{\nu[\gamma^+\gamma^5]}(\bfb,\bfp,x;\Lambda \hat{S}_z)],
\label{rho_Lamlam0}
\ee
which can be decomposed as
\be 
\rho^\nu_{\Lambda \lambda}(\bfb,\bfp,x)&=&\frac{1}{2}[\rho^{\nu}_{UU}(\bfb,\bfp,x) +\Lambda \rho^{\nu}_{LU}(\bfb,\bfp,x)\nonumber\\
&&+ \lambda\rho^{\nu}_{UL}(\bfb,\bfp,x) +\Lambda\lambda \rho^{\nu}_{LL}(\bfb,\bfp,x)],
\label{rho_Lamlam}
\ee
corresponding to $\Lambda=\uparrow, \downarrow $ and  $\lambda=\uparrow,\downarrow$ (where $\uparrow$ and $\downarrow$ are corresponding to  $+1$ and $-1$ for longitudinal polarizations respectively).
Similarly Wigner distribution for a quark with transverse polarizations $\lambda_T=\Uparrow,\Downarrow$ in a proton with transverse polarizations $\Lambda_T=\Uparrow,\Downarrow$ can be written as 
\be 
\rho^{i\nu}_{\Lambda_T \lambda_T}(\bfb,\bfp,x)=\frac{1}{2}[\rho^{\nu[\gamma^+]}(\bfb,\bfp,x;\Lambda_T \hat{S}_i) +\Lambda_T \rho^{\nu[i\sigma^{i+}\gamma^5]}(\bfb,\bfp,x;\Lambda_T \hat{S}_i)],
\label{rho_Lamlam0_T}
\ee
which can be decomposed as
\be 
\rho^{i\nu}_{\Lambda_T \lambda_T}(\bfb,\bfp,x)&=&\frac{1}{2}[\rho^{\nu}_{UU}(\bfb,\bfp,x) +\Lambda_T \rho^{i\nu}_{TU}(\bfb,\bfp,x)\nonumber\\
&&+ \lambda_T\rho^{\nu i}_{UT}(\bfb,\bfp,x) +\Lambda_T\lambda_T \rho^{q}_{TT}(\bfb,\bfp,x)].
\label{rho_Lamlam_T}
\ee  
The distribution $\rho^{\nu}_{\Lambda \lambda}(\bfb,\bfp,x)$ in the transverse planes are shown in Fig.\ref{fig_Lamlam_u} and Fig.\ref{fig_Lamlam_d} for $u$ and $d$  quarks respectively. The transverse Wigner distribution  $\rho^{i\nu}_{\Lambda_T \lambda_T}(\bfb,\bfp,x)$ are shown in Fig.(\ref{fig_LamlamT_u}) and in Fig.(\ref{fig_LamlamT_d}) with $i=1$ (e,i. polarization along the x-axis). $\rho^{\nu}_{\Lambda \lambda}(\bfb,\bfp,x)$ and $\rho^{i\nu}_{\Lambda_T \lambda_T}(\bfb,\bfp,x)$ provide information about the correlations between proton spin and quark spin in the longitudinal direction and in the transverse direction respectively.
We also can define the Wigner distributions for longitudinally polarized quarks in an transversely polarized proton,  $\rho^{i\nu}_{\Lambda_T \lambda}(\bfb,\bfp,x)$ and transversely polarized quarks in an longitudinally polarized proton $\rho^{\nu j}_{\Lambda \lambda_T}(\bfb,\bfp,x)$ as  
\be 
\rho^{i\nu}_{\Lambda_T \lambda}(\bfb,\bfp,x)&=&\frac{1}{2}[\rho^{\nu}_{UU}(\bfb,\bfp,x) +\Lambda_T \rho^{i\nu}_{TU}(\bfb,\bfp,x) + \lambda\rho^{\nu}_{UL}(\bfb,\bfp,x) +\Lambda_T\lambda \rho^{i\nu}_{TL}(\bfb,\bfp,x)],\nonumber\\
\label{rho_Lamlam_TL}\\ 
\rho^{\nu j}_{\Lambda \lambda_T}(\bfb,\bfp,x)&=&\frac{1}{2}[\rho^{\nu}_{UU}(\bfb,\bfp,x) +\Lambda \rho^{\nu}_{LU}(\bfb,\bfp,x)+ \lambda_T\rho^{\nu j}_{UT}(\bfb,\bfp,x) +\Lambda\lambda_T \rho^{\nu j}_{LT}(\bfb,\bfp,x)].\nonumber
\\\label{rho_Lamlam_LT}
\ee  

The Wigner correlator, Eq.(\ref{wig_rho})can be parametrized in terms of GTMDs\cite{meissner09} as \\
(i) for unpolarized proton
\be
\rho^\nu_{UU}(\bfb,\bfp,x)&=& \mathcal{F}^\nu_{1,1}(x,0,\bfp^2,\bfp.\bfb,\bfb^2), \label{rhoUU_F}\\
\rho^\nu_{UL}(\bfb,\bfp,x)&=&\frac{1}{M^2}\epsilon^{ij}_\perp p^i_\perp \frac{\partial}{\partial b^j_\perp} \mathcal{G}^\nu_{1,1}(x,0,\bfp^2,\bfp.\bfb,\bfb^2),\label{rhoUL_G}\\
\rho^{\nu j}_{UT}(\bfb,\bfp,x)&=& -i\frac{1}{M}\epsilon^{ij}_\perp p^i_\perp \mathcal{H}^\nu_{1,1}(x,0,\bfp^2,\bfp.\bfb,\bfb^2)+ \frac{1}{M}\epsilon^{ij}_\perp \frac{\partial}{\partial b^i_\perp} \mathcal{H}^\nu_{1,2}(x,0,\bfp^2,\bfp.\bfb,\bfb^2), \label{rhoUT_H}
\ee
(ii) for longitudinally polarized proton
\be 
\rho^\nu_{LU}(\bfb,\bfp,x)&=&-\frac{1}{M^2}\epsilon^{ij}_\perp p^i_\perp \frac{\partial}{\partial b^j_\perp} \mathcal{F}^\nu_{1,4}(x,0,\bfp^2,\bfp.\bfb,\bfb^2),\label{rhoLU_F} \\
\rho^\nu_{LL}(\bfb,\bfp,x)&=&  \mathcal{G}^\nu_{1,4}(x,0,\bfp^2,\bfp.\bfb,\bfb^2),\label{rhoLL_G}\\
\rho^{\nu j}_{LT}(\bfb,\bfp,x)&=& \frac{p^j_\perp}{M}\mathcal{H}^\nu_{1,7}(x,0,\bfp^2,\bfp.\bfb,\bfb^2)+ i\frac{1}{M} \frac{\partial}{\partial b^j_\perp} \mathcal{H}^\nu_{1,8}(x,0,\bfp^2,\bfp.\bfb,\bfb^2), \label{rhoLT_H}
\ee
(iii) for transversely polarized proton
\be
\rho^{i\nu}_{TU}(\bfb,\bfp,x)&=&\frac{1}{2M}\epsilon^{ij}_\perp \frac{\partial}{\partial b^j_\perp} \bigg[\mathcal{F}^\nu_{1,1}(x,0,\bfp^2,\bfp.\bfb,\bfb^2)-2\mathcal{F}^\nu_{1,3}(x,0,\bfp^2,\bfp.\bfb,\bfb^2)\bigg] \nonumber\\
&&+ i \frac{1}{M}\epsilon^{ij}_\perp p^j_\perp \mathcal{F}^\nu_{1,2}(x,0,\bfp^2,\bfp.\bfb,\bfb^2),\label{rhoTU_F}\\
\rho^{i\nu}_{TL}(\bfb,\bfp,x)&=& \frac{1}{2M^3} \epsilon^{ij}_\perp \epsilon^{kl}_\perp p^k_\perp \frac{\partial}{\partial b^j_\perp}\frac{\partial}{\partial b^l_\perp} \mathcal{G}_{1,1}(x,0,\bfp^2,\bfp.\bfb,\bfb^2)\nonumber\\
&& +\frac{p^i_\perp}{M}\mathcal{G}^\nu_{1,2}(x,0,\bfp^2,\bfp.\bfb,\bfb^2) + i\frac{1}{M} \frac{\partial}{\partial b^i_\perp}\mathcal{G}^\nu_{1,3}(x,0,\bfp^2,\bfp.\bfb,\bfb^2), \label{rhoTL_G}\\
\rho^{\nu}_{TT}(\bfb,\bfp,x)&=& \epsilon^{ij}(-1)^i\bigg[-i\frac{p^j_\perp}{2M^2}\frac{\partial}{\partial b^j_\perp}  \mathcal{H}_{1,1}(x,0,\bfp^2,\bfp.\bfb,\bfb^2) +\frac{1
}{2M^2} \frac{\partial^2}{\partial b^{j2}_\perp} \mathcal{H}_{1,2}(x,0,\bfp^2,\bfp.\bfb,\bfb^2)\bigg]\nonumber\\
&&+ \mathcal{H}^\nu_{1,3}(x,0,\bfp^2,\bfp.\bfb,\bfb^2)+ \delta_{ij}\bigg[\frac{p^i_\perp p^j_\perp}{M^2}\mathcal{H}^\nu_{1,4}(x,0,\bfp^2,\bfp.\bfb,\bfb^2)\nonumber\\
&&+\frac{1}{M^2}p^i_\perp\frac{\partial}{\partial b^j_\perp}\mathcal{H}^\nu_{1,5}(x,0,\bfp^2,\bfp.\bfb,\bfb^2)- \frac{1}{M^2}\frac{\partial}{\partial b^i_\perp}\frac{\partial}{\partial b^j_\perp}\mathcal{H}^\nu_{1,6}(x,0,\bfp^2,\bfp.\bfb,\bfb^2)\bigg]. \label{rhoTT_H}\nonumber\\
\ee
The pretzelous distribution is parametrized as
\be 
\rho^{\perp\nu}_{TT}(\bfb,\bfp,x)&=& \frac{\epsilon^{ij}_\perp}{2M^2}\bigg[ ip^i\frac{\partial}{\partial b^j_\perp}\bigg(-\mathcal{H}^\nu_{1,1}(x,0,\bfp^2,\bfp.\bfb,\bfb^2)+2\mathcal{H}^\nu_{1,5}(x,0,\bfp^2,\bfp.\bfb,\bfb^2)\bigg)\nonumber\\
&&+\frac{\partial}{\partial b^i_\perp}\frac{\partial}{\partial b^j_\perp}\bigg(\mathcal{H}^\nu_{1,2}(x,0,\bfp^2,\bfp.\bfb,\bfb^2)-2\mathcal{H}^\nu_{1,6}(x,0,\bfp^2,\bfp.\bfb,\bfb^2)\bigg)\nonumber\\
&&+ 2p^ip^j\mathcal{H}^\nu_{1,4}(x,0,\bfp^2,\bfp.\bfb,\bfb^2)\bigg]. \label{rhoTTperp_H}
\ee
Where the $\chi^\nu = \mathcal{F}^\nu_{1,1}, \mathcal{F}^\nu_{1,4}, \mathcal{G}^\nu_{1,1}, \mathcal{G}^\nu_{1,4}$ and $\mathcal{H}_{1,n}$(n=1,2,3...8) can be expressed as Fourier transform of GTMDs $X^\nu= F^\nu_{1,1}, F^\nu_{1,4}, G^\nu_{1,1}, G^\nu_{1,4}$ and $H_{1,n}$ respectively.
\be 
\chi^\nu(x,0,\bfp^2,\bfp.\bfb,\bfb^2) = \int\frac{d^2\Dp}{(2\pi)^2} e^{-i\Dp.\bfb} X^\nu(x,0,\bfp^2,\bfp.\Dp,\Dp^2),\label{chi_GTMDs}
\ee
There are altogether 16 GTMDs at the leading twist. At  $\Delta=0$ the GTMDs reduces to transverse momentum dependent distributions(TMDs) which are functions of longitudinal momentum fraction $x$ and transverse momentum $\bfp$ carried by quark. There are altogether 8 TMDs at the leading twist.  

%==========================
\section{Orbital angular momentum}\label{oam}
%============================
The canonical orbital angular momentum(OAM) operator for quark is defined as
\be
\hat{\ell}^\nu_z(b^-,\bfb,p^+,\bfp)=\frac{1}{4}\int \frac{dz^-d^2\textbf{z}_\perp}{(2\pi)^3}e^{-ip.z}\bar{\psi}^\nu(b^-,\textbf{b}_\perp)\gamma^+(\textbf{b}_\perp \times(-i\partial_\perp))\psi^\nu(b^--z^-,\textbf{b}_\perp). 
\ee  
The OAM density operator can be expressed in terms of Wigner operator as
\be
 \hat{\ell}^\nu_z=(\textbf{b}_\perp \times \bfp)\hat{W}^{\nu[\gamma^+]}.
\ee
Thus in Light-front gauge the average canonical OAM for quark is written in terms of Wigner distribution as.
\be 
\ell^\nu_z &=& \int\frac{d\Delta^+ d^2\Dp}{2P^+(2\pi)^3}\langle P^{\prime\prime};S|\hat{\ell}^\nu_z|P^\prime;S\rangle \nonumber\\
&=&\int dx d^2\bfp d^2\bfb (\bfb\times\bfp)_z\rho^{\nu[\gamma^+]}(\bfb,\bfp,x,\hat{S}_z).\label{OAM_ell_def}
\ee
Where, the distribution $\rho^{\nu[\gamma^+]}(\bfb,\bfp,x,\hat{S}_z)$ can be written from Eqs.(\ref{rho_UU_def},\ref{rho_LU_def}) as:
\be
\rho^{\nu[\gamma^+]}(\bfb,\bfp,x,+\hat{S}_z)=\rho^\nu_{UU}(\bfb,\bfp,x)+\rho^\nu_{LU}(\bfb,\bfp,x).
\ee
From Eq.(\ref{rhoUU_F}) we see that
\be 
\int dx d^2\bfp d^2\bfb (\bfb\times\bfp)_z \rho^\nu_{UU}(\bfb,\bfp,x)=0,
\label{OAM_rhoUU}
\ee
 which satisfies the angular momentum sum rule for unpolarized proton-- the total angular momentum of constituents sum up to zero. From Eq.(\ref{rhoLU_F}) and Eq.(\ref{chi_GTMDs}), the twist-2 canonical quark OAM in the light-front gauge is  written in terms of GTMDs as
\be 
\ell^\nu_z &=&-\int dx d^2\bfp \frac{\bfp^2}{M^2}F^\nu_{1,4}(x,0,\bfp^2,0,0). \label{OAM_ell}
\ee
The correlation between proton spin and quark OAM is understood from $\ell^\nu_z $. If $\ell^\nu_z >0$ quark OAM is parallel to proton spin and $\ell^\nu_z <0$ indicates the quark OAM is anti-parallel to proton spin.  

The spin-orbit correlation of a quark is given by the operator
\be
C^\nu_z(b^-,\bfb,p^+,\bfp)=\frac{1}{4}\int \frac{dz^-d^2\textbf{z}_\perp}{(2\pi)^3}e^{-ip.z}\bar{\psi}^\nu(b^-,\textbf{b}_\perp)\gamma^+\gamma^5(\textbf{b}_\perp \times(-i\partial_\perp))\psi^\nu(b^--z^-,\textbf{b}_\perp).
\ee 
The correlation between quark spin and quark OAM can be expressed  with Wigner distributions $\rho^\nu_{UL}$ and equivalently in terms of GTMD as:
\be 
C^\nu_z&=&\int dx d^2\bfp d^2\bfb (\bfb\times\bfp)_z \rho^\nu_{UL}(\bfb,\bfp,x) \nonumber \\
&=& \int dx d^2\bfp \frac{\bfp^2}{M^2}G^\nu_{1,1}(x,0,\bfp^2,0,0). \label{Cqz}
\ee 
Where $C^\nu_z>0$ implies the quark spin and OAM tend to be aligned  and $C^\nu_z<0$ implies they are anti-aligned. 

The spin contribution of the quark to the proton spin is defined\cite{Lorce11} as
\be s^\nu_z=\frac{1}{2}g^\nu_A=\frac{1}{2}\int dx \tilde{H}^\nu(x,0,0)=\frac{1}{2}\int dx d^2p_\perp G_{1,4}^\nu(x,0,\bfp^2,0,0)
\ee
where $g^\nu_A$ is the axial charge.
%In the diquark model the momentum sum rule can be written separately for scalar diquark and for vector diquark as\cite{Liu15}
%\be 
%\ell^{(S)}_q + s^{(S)}_q + \ell^{(S)}_D + s^{(S)}_D &=& \frac{1}{2} \label{Eq_Sum_S}\\
%\ell^{(A)}_q + s^{(A)}_q + \ell^{(A)}_D + s^{(A)}_D &=& \frac{1}{2}\label{Eq_Sum_V},
%\ee
%with $s^{(S)}_D =0$ and $s^{(A)}_D =1$.

%%%%%%%%%%%%%%%%%%%%%%%%%%%%%%%%%%%%%%%%%%%%%%%
\section{Results}\label{results}
%%%%%%%%%%%%%%%%%%%%%%%%%%%%%%%%%%%%%%%%%%%%%%%
The quark Wigner distributions are evaluated in the light-front quark-diquark model constructed from the AdS/QCD correspondence. 
Using the two particle Fock states expression of proton for both the scalar and vector diquark respectively in Eq.(\ref{wigner_W}), we can express the quark-quark correlator, $W^{\nu [\Gamma]}(\Dp,\bfp,x;S)$ in terms of LFWFs. For the scalar diquark the expansion of $W^{\nu [\Gamma]}(\Dp,\bfp,x;S)$ is given by
\be
W^{\nu [\gamma^+]}_S(\Dp,\bfp,x;\pm\hat{S}_z)&=&\frac{1}{16\pi^3}\bigg[\psi^{\pm\dagger}_{\nu +}(x,\bfp^{\prime\prime})\psi^{\pm}_{\nu +}(x,\bfp^{\prime}) +\psi^{\pm\dagger}_{\nu -}(x,\bfp^{\prime\prime})\psi^{\pm}_{\nu -}(x,\bfp^{\prime}) \bigg],\label{s1}\\
W^{\nu [\gamma^+\gamma^5]}_S(\Dp,\bfp,x;\pm\hat{S}_z)&=&\frac{1}{16\pi^3}\bigg[\psi^{\pm\dagger}_{\nu +}(x,\bfp^{\prime\prime})\psi^{\pm}_{\nu +}(x,\bfp^{\prime}) -\psi^{\pm\dagger}_{\nu -}(x,\bfp^{\prime\prime})\psi^{\pm}_{\nu -}(x,\bfp^{\prime}) \bigg],\label{s2} \\
W^{\nu [i\sigma^{j+}\gamma^5]}_S(\Dp,\bfp,x;\pm\hat{S}_z)&=&\frac{1}{16\pi^3}\epsilon^{ij}_\perp\bigg[(-i)^i\psi^{\pm\dagger}_{\nu +}(x,\bfp^{\prime\prime})\psi^{\pm}_{\nu -}(x,\bfp^{\prime}) + (i)^i\psi^{\pm\dagger}_{\nu -}(x,\bfp^{\prime\prime})\psi^{\pm}_{\nu +}(x,\bfp^{\prime}) \bigg].\label{s3}\nonumber\\
\ee
For the vector diquark, the expressions read as
\be
W^{\nu [\gamma^+]}_A(\Dp,\bfp,x;\pm\hat{S}_z)&=&\frac{1}{16\pi^3}\bigg[\psi^{\pm\dagger}_{\nu ++}(x,\bfp^{\prime\prime})\psi^{\pm}_{\nu ++}(x,\bfp^{\prime}) +\psi^{\pm\dagger}_{\nu -+}(x,\bfp^{\prime\prime})\psi^{\pm}_{\nu -+}(x,\bfp^{\prime})\nonumber\\
&+&\psi^{\pm\dagger}_{\nu +0}(x,\bfp^{\prime\prime})\psi^{\pm}_{\nu +0}(x,\bfp^{\prime}) +\psi^{\pm\dagger}_{\nu -0}(x,\bfp^{\prime\prime})\psi^{\pm}_{\nu -0}(x,\bfp^{\prime}) \nonumber\\
&+&\psi^{\pm\dagger}_{\nu +-}(x,\bfp^{\prime\prime})\psi^{\pm}_{\nu +-}(x,\bfp^{\prime}) +\psi^{\pm\dagger}_{\nu --}(x,\bfp^{\prime\prime})\psi^{\pm}_{\nu --}(x,\bfp^{\prime})\bigg],\label{v1}\\
W^{\nu [\gamma^+\gamma^5]}_A(\Dp,\bfp,x;\pm\hat{S}_z)&=&\frac{1}{16\pi^3}\bigg[\psi^{\pm\dagger}_{\nu ++}(x,\bfp^{\prime\prime})\psi^{\pm}_{\nu ++}(x,\bfp^{\prime}) - \psi^{\pm\dagger}_{\nu -+}(x,\bfp^{\prime\prime})\psi^{\pm}_{\nu -+}(x,\bfp^{\prime})\nonumber\\
&+&\psi^{\pm\dagger}_{\nu +0}(x,\bfp^{\prime\prime})\psi^{\pm}_{\nu +0}(x,\bfp^{\prime}) - \psi^{\pm\dagger}_{\nu -0}(x,\bfp^{\prime\prime})\psi^{\pm}_{\nu -0}(x,\bfp^{\prime}) \nonumber\\
&+&\psi^{\pm\dagger}_{\nu +-}(x,\bfp^{\prime\prime})\psi^{\pm}_{\nu +-}(x,\bfp^{\prime}) - \psi^{\pm\dagger}_{\nu --}(x,\bfp^{\prime\prime})\psi^{\pm}_{\nu --}(x,\bfp^{\prime})\bigg],\label{v2} \\
W^{\nu [i\sigma^{j+}\gamma^5]}_A(\Dp,\bfp,x;\pm\hat{S}_z)&=&\frac{1}{16\pi^3}\epsilon^{ij}_\perp\bigg[(-i)^i\psi^{\pm\dagger}_{\nu ++}(x,\bfp^{\prime\prime})\psi^{\pm}_{\nu -+}(x,\bfp^{\prime}) + (i)^i\psi^{\pm\dagger}_{\nu -+}(x,\bfp^{\prime\prime})\psi^{\pm}_{\nu ++}(x,\bfp^{\prime})\nonumber\\
&+&(-i)^i\psi^{\pm\dagger}_{\nu +0}(x,\bfp^{\prime\prime})\psi^{\pm}_{\nu -0}(x,\bfp^{\prime}) + (i)^i\psi^{\pm\dagger}_{\nu -0}(x,\bfp^{\prime\prime})\psi^{\pm}_{\nu +0}(x,\bfp^{\prime}) \nonumber\\
&+&(-i)^i\psi^{\pm\dagger}_{\nu +-}(x,\bfp^{\prime\prime})\psi^{\pm}_{\nu --}(x,\bfp^{\prime}) + (i)^i\psi^{\pm\dagger}_{\nu --}(x,\bfp^{\prime\prime})\psi^{\pm}_{\nu +-}(x,\bfp^{\prime})\bigg],\label{v3}
\ee
with the Dirac structures $\Gamma=\gamma^+,\gamma^+\gamma^5$ and $i\sigma^{j+}\gamma^5$. Where the initial and final momentums of the struck quark are 
\be 
\bfp^{\prime}=\bfp-(1-x)\frac{\Dp}{2}, \quad \quad
\bfp^{\prime\prime}=\bfp+(1-x)\frac{\Dp}{2},
\ee
respectively. Using the light-front wavefunctions from Eqs.(\ref{LFWF_S},\ref{LFWF_Vp}, and \ref{LFWF_Vm}) in Eqs.(\ref{s1}-\ref{v3})  at the initial scale $\mu_0$, we explicitly calculate all the quark-quark correlators which give the expressions of Wigner distributions in the following forms 
\be 
\rho^{\nu(S)}_{UU}&=&N_S^2 \rho^\nu _1(\bfb,\bfp,x), \quad\quad \rho^{\nu(A)}_{UU}=\Big(\frac{1}{3}N^{(\nu)2}_0+\frac{2}{3}N^{(\nu)2}_1\Big) \rho^\nu _1(\bfb,\bfp,x),
\label{rhoUU_nu} \\
\rho^{\nu(S)}_{UL}&=&-N_S^2 \rho^\nu _2(\bfb,\bfp,x), \quad\quad  \rho^{\nu(A)}_{UL}=-\Big(\frac{1}{3}N^{(\nu)2}_0+\frac{2}{3}N^{(\nu)2}_1\Big) \rho^\nu _2(\bfb,\bfp,x),
\label{rhoUL_nu} \\
\rho^{\nu j(S)}_{UT}&=&N_S^2 \rho^{\nu j} _3(\bfb,\bfp,x), \quad\quad \rho^{\nu j(A)}_{UT}=\Big(\frac{1}{3}N^{(\nu)2}_0+\frac{2}{3}N^{(\nu)2}_1\Big) \rho^{\nu j} _3(\bfb,\bfp,x),
\label{rhoUT_nu}\\
\rho^{\nu(S)}_{LU}&=&N_S^2 \rho^\nu _2(\bfb,\bfp,x), \quad\quad \rho^{\nu(A)}_{LU}=\Big(\frac{1}{3}N^{(\nu)2}_0-\frac{2}{3}N^{(\nu)2}_1\Big) \rho^\nu _2(\bfb,\bfp,x),\label{rhoLU_nu}\\
\rho^{\nu(S)}_{LL}&=&N_S^2 \rho^\nu _4(\bfb,\bfp,x), \quad\quad \rho^{\nu(A)}_{LL}=\Big(\frac{1}{3}N^{(\nu)2}_0-\frac{2}{3}N^{(\nu)2}_1\Big) \rho^\nu _4(\bfb,\bfp,x),\label{rhoLL_nu}\\
\rho^{\nu j(S)}_{LT}&=&-N_S^2 \rho^{\nu j} _5(\bfb,\bfp,x), \quad\quad \rho^{\nu j(A)}_{LT}=-\Big(\frac{1}{3}N^{(\nu)2}_0-\frac{2}{3}N^{(\nu)2}_1\Big) \rho^{\nu j} _5(\bfb,\bfp,x),\label{rhoLT_nu}\\
\rho^{i\nu(S)}_{TU}&=&N_S^2 \rho^{i\nu} _3(\bfb,\bfp,x), \quad\quad \rho^{i\nu(A)}_{TU}=\frac{1}{3}N^{(\nu)2}_0 \rho^{i\nu}_3(\bfb,\bfp,x),\label{rhoTU_nu}\\
\rho^{i\nu(S)}_{TL}&=&N_S^2 \rho^{i\nu} _5(\bfb,\bfp,x), \quad\quad \rho^{i\nu(A)}_{TU}=-\frac{1}{3}N^{(\nu)2}_0 \rho^{i\nu} _5(\bfb,\bfp,x),\label{rhoTL_nu}\\
\rho^{\nu(S)}_{TT}&=&N_S^2 \rho^\nu _6(\bfb,\bfp,x), \quad\quad \rho^{\nu(A)}_{TT}=-\frac{1}{3}N^{(\nu)2}_0 \rho^\nu _6(\bfb,\bfp,x),\label{rhoTT_nu}\\
\rho^{\perp qS}_{TT}&=&-N_S^2 \rho^\nu _7(\bfb,\bfp,x), \quad\quad \rho^{\perp qA}_{TT}=\frac{1}{3}N^{(\nu)2}_0 \rho^\nu _7(\bfb,\bfp,x),\label{rhoTTp_nu}
\ee
where the label $S$ represents the scalar and $A$ denotes the isoscalar-vector(V) diquark corresponding to $u$ quark and isovector-vector(VV) diquark corresponding to $d$  quark. Combining the contributions from scalar and vector parts, one can write the distributions for $u$ and $d$ as 
\be 
\rho^u_{N N'}(\bfb,\bfp,x)&=&C_S^2 ~\rho^{u(S)}_{N N'}+C_V^2 ~\rho^{u(A)}_{N N'},\\
\rho^d_{N N'}(\bfb,\bfp,x)&=&C_{VV}^2 ~\rho^{d(A)}_{N N'},
\ee
where $N(N')$ implies the proton(quark) polarization. Now, integrating over the light-front momentum fraction $x$, we display the behavior of the Wigner distributions in the remaining four dimensions i.e. in the transverse coordinate space with a definite transverse momentum and in the transverse momentum space with a definite coordinate.

The distribution functions $\rho^\nu _i(\bfb,\bfp,x)$ are given by
\be 
\rho^\nu _1(\bfb,\bfp,x)&=&\frac{1}{16\pi^3}\int\frac{d\Delta_\perp}{2\pi} \Delta_\perp \rm{J}_0(|\Delta_\perp||b_\perp|)\exp\big(-2\tilde{a}(x)\tilde{\textbf{p}}^2_\perp \big) \nonumber \\
&&~~~\times\bigg[|A^\nu_1(x)|^2 +
\bigg(\bfp^2 - \frac{\Dp^2}{4}(1-x)^2\bigg)\frac{1}{M^2x^2}|A^{\nu}_2(x)|^2 \bigg],\label{rho_1}
\\
\rho^\nu _2(\bfb,\bfp,x)&=&\frac{1}{M^2}\epsilon^{ij}_\perp p^i_\perp \frac{\partial}{\partial b^j_\perp}\bigg[\frac{1}{16\pi^3}\int\frac{d\Delta_\perp}{2\pi} \Delta_\perp \rm{J}_0(|\Delta_\perp||b_\perp|)\exp\big(-2\tilde{a}(x)\tilde{\textbf{p}}^2_\perp \big) \nonumber \\
&&~~~\times\frac{(1-x)}{x^2}|A^{\nu}_2(x)|^2\bigg],\label{rho_2}
\\
\rho^{\nu j}_{3}(\bfb,\bfp,x)&=&\epsilon^{ij}_\perp \frac{\partial}{\partial b^i_\perp}\frac{1}{16\pi^3}\int\frac{d\Delta_\perp}{2\pi} \Delta_\perp \rm{J}_0(|\Delta_\perp||b_\perp|)\exp\big(-2\tilde{a}(x)\tilde{\textbf{p}}^2_\perp \big)\nonumber\\
&&~~~\times \frac{(1-x)}{xM}A_1^{\nu}(x) A_2^{\nu} (x),\label{rho_3}
\\
\rho^\nu _{4}(\bfb,\bfp,x)&=&\frac{1}{16\pi^3}\int\frac{d\Delta_\perp}{2\pi} \Delta_\perp \rm{J}_0(|\Delta_\perp||b_\perp|)\exp\big(-2\tilde{a}(x)\tilde{\textbf{p}}^2_\perp \big) \nonumber \\
&&~~~\times\bigg[|A^\nu_1(x)|^2 -
\bigg(\bfp^2 - \frac{\Dp^2}{4}(1-x)^2\bigg)\frac{1}{M^2x^2}|A^{\nu}_2(x)|^2 \bigg],\label{rho_4}
\ee \be
\rho^{\nu j}_{5}(\bfb,\bfp,x)&=& \frac{1}{16\pi^3}\int\frac{d\Delta_\perp}{2\pi} \Delta_\perp \rm{J}_0(|\Delta_\perp||b_\perp|)\frac{2 p^j}{xM}A_1^{\nu}(x) A_2^{\nu} (x)\exp\big(-2\tilde{a}(x)\tilde{\textbf{p}}^2_\perp \big),\label{rho_5}
\\
\rho^\nu_{6}(\bfb,\bfp,x)&=& \frac{1}{16\pi^3}\int\frac{d^2\Delta_\perp}{(2\pi)^2} e^{-i \Dp.\bfb}\bigg[|A_1^{\nu}(x)|^2 + \bigg(\bfp^2 - \frac{\Dp^2}{4}(1-x)^2\bigg)\frac{1}{x^2M^2}|A^\nu_2(x)|^2 \nonumber \\
&&-\delta_{ij}\bigg(p^ip^j - \frac{\Delta^i\Delta^j}{4}(1-x)^2\bigg)\frac{2}{x^2M^2}|A^\nu_2(x)|^2 \bigg]\exp\big(-2\tilde{a}(x)\tilde{\textbf{p}}^2_\perp \big),\label{rho_6}\\
\rho^\nu_{7}(\bfb,\bfp,x)&=& \frac{1}{16\pi^3}\int\frac{d^2\Delta_\perp}{(2\pi)^2} e^{-i \Dp.\bfb} \epsilon^{ij}_\perp \bigg(p^i p^j-\frac{\Delta^i\Delta^j}{4}(1-x)^2 \bigg) \nonumber \\
&& \hspace{5cm} \times \frac{2}{x^2M^2}|A^\nu_2(x)|^2\exp\big(-2\tilde{a}(x)\tilde{\textbf{p}}^2_\perp \big)\label{rho_7},
\ee
with 
\be 
A^{\nu}_i(x)&=&\frac{4\pi}{\kappa}\sqrt{\frac{\log(1/x)}{(1-x)}}x^{a^{\nu}_i}(1-x)^{b^{\nu}_i},\\
\tilde{a}(x)&=& \frac{\log(1/x)}{2\kappa^2(1-x)^2},\\
\tilde{\textbf{p}}^2_\perp &=&\bfp^2+\frac{\Dp^2}{4}(1-x)^2.
\ee
Note that there are no implicit sum over $i$ and $j$, in the expression of $\rho^\nu_6$ and $\rho^\nu_7$.
%%%%%%%%%%%%%%%%%%%%%%%%%%%%%%%%%%
\subsection{Unpolarized proton}\label{unpol}
%%%%%%%%%%%%%%%%%%%%%%%%%%%%%%%
\begin{figure}[ht]
\centering
\subfigure[]{\includegraphics[width=5.cm,height=4.cm]{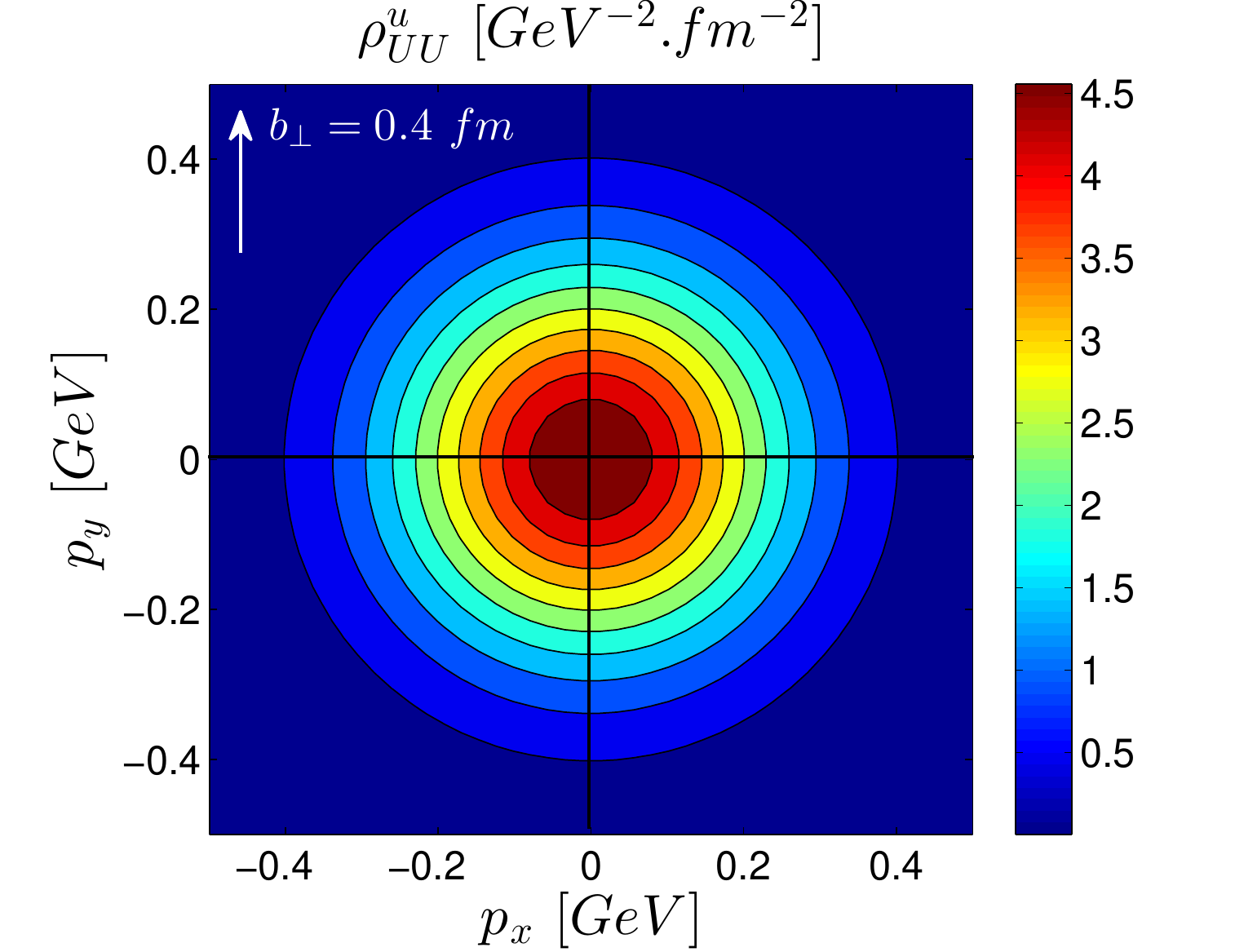}}
\subfigure[]{\includegraphics[width=5.cm,height=4.cm]{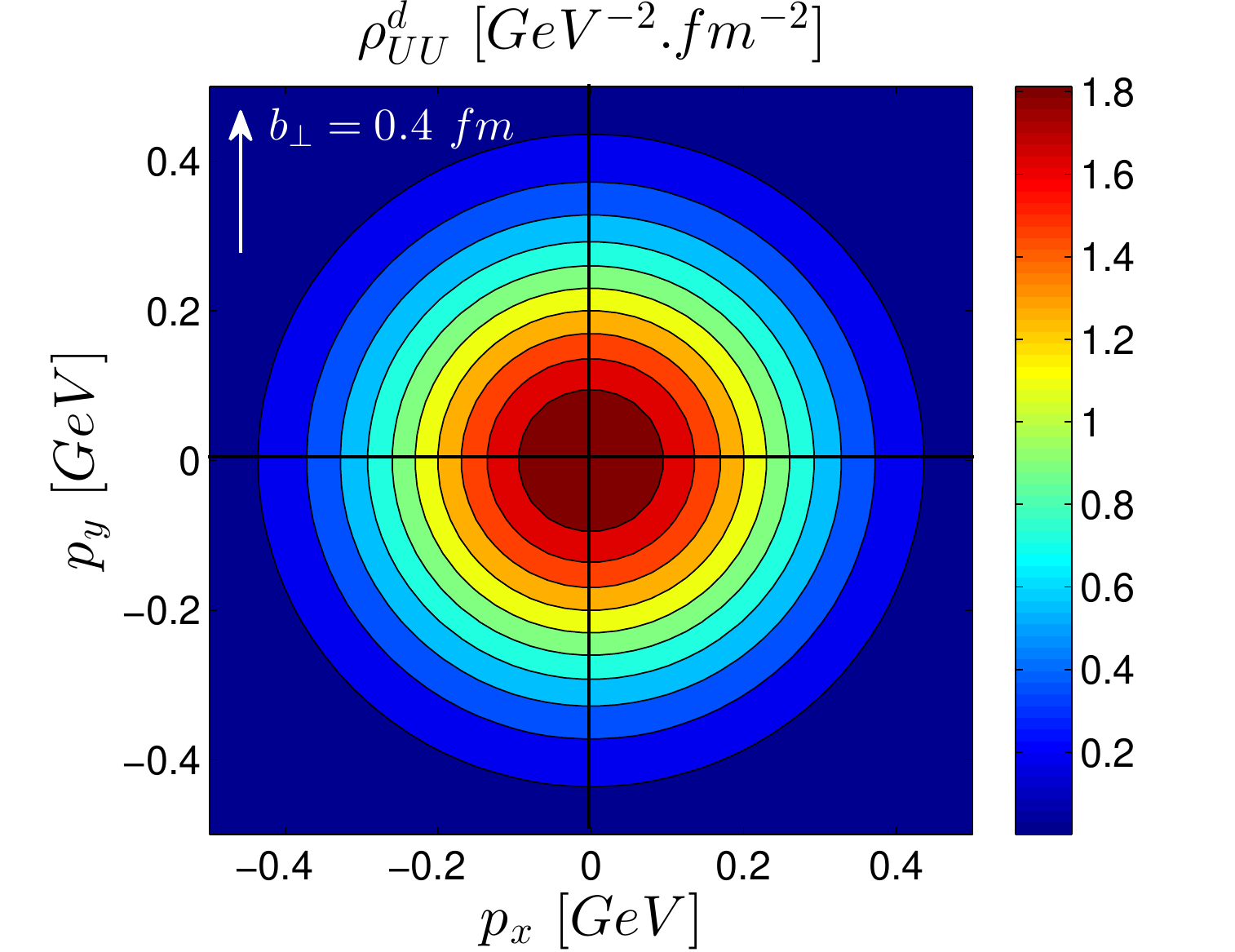}}\\
\subfigure[]{\includegraphics[width=5.cm,height=4.cm]{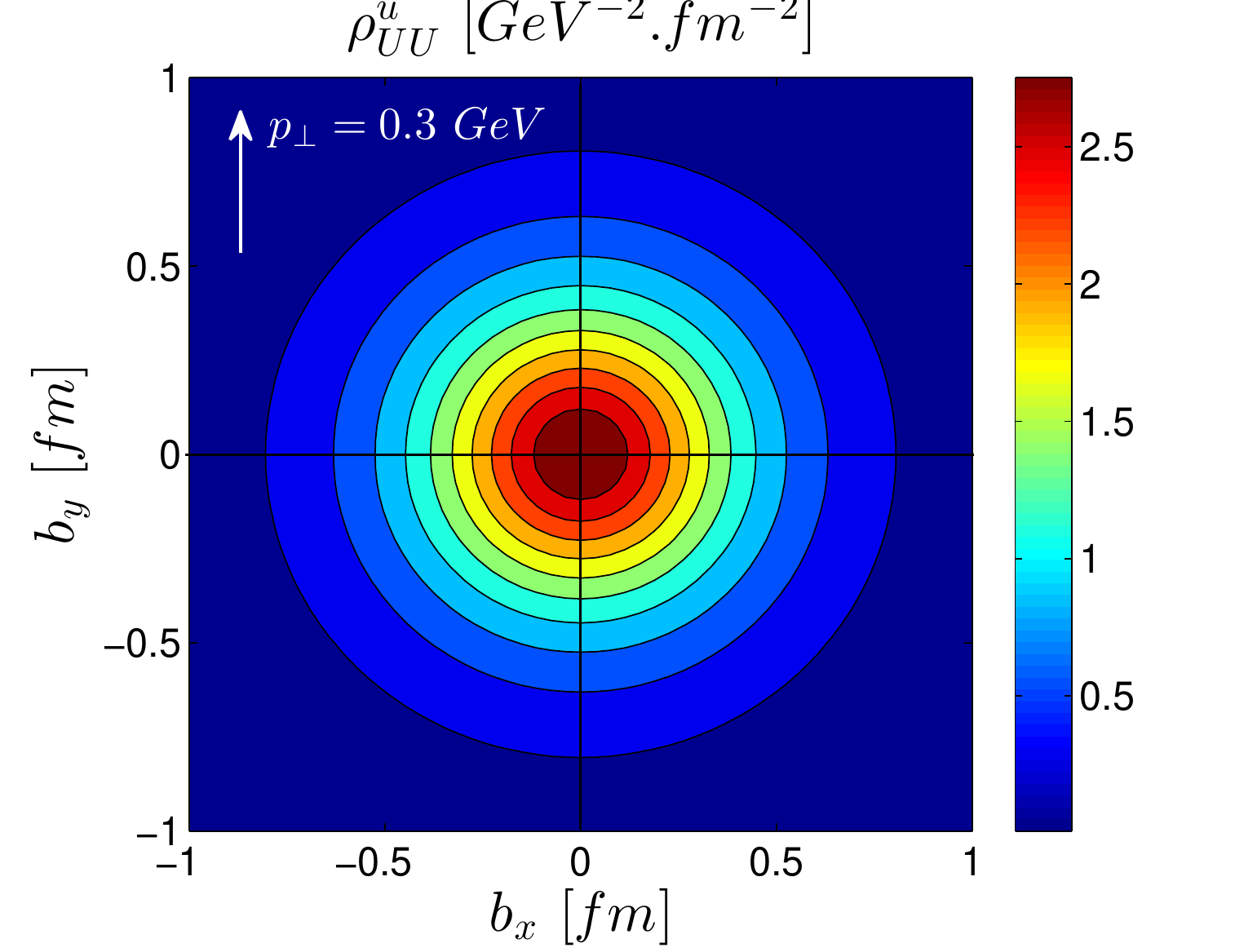}}
\subfigure[]{\includegraphics[width=5.cm,height=4.cm]{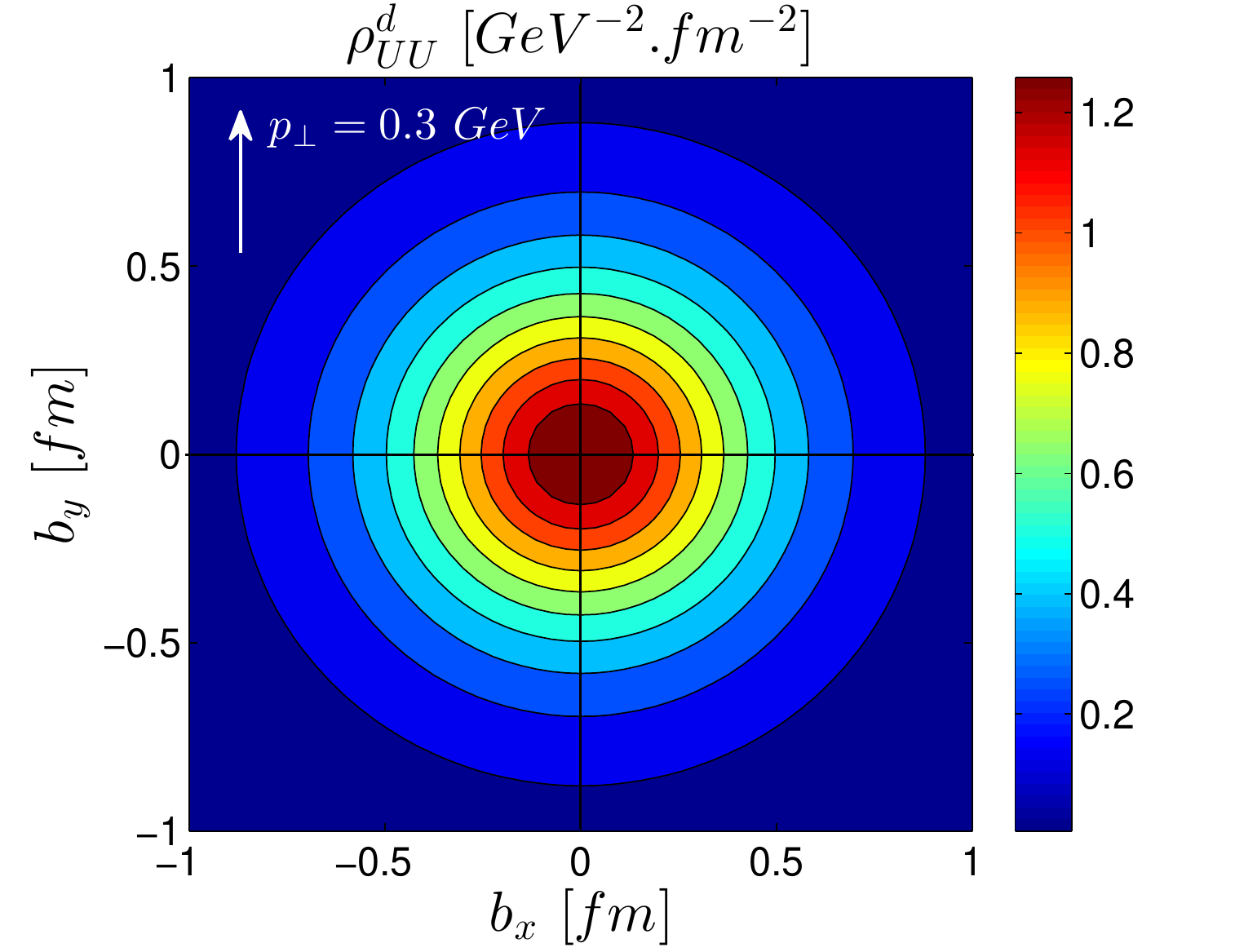}}\\
\subfigure[]{\includegraphics[width=5.cm,height=4.cm]{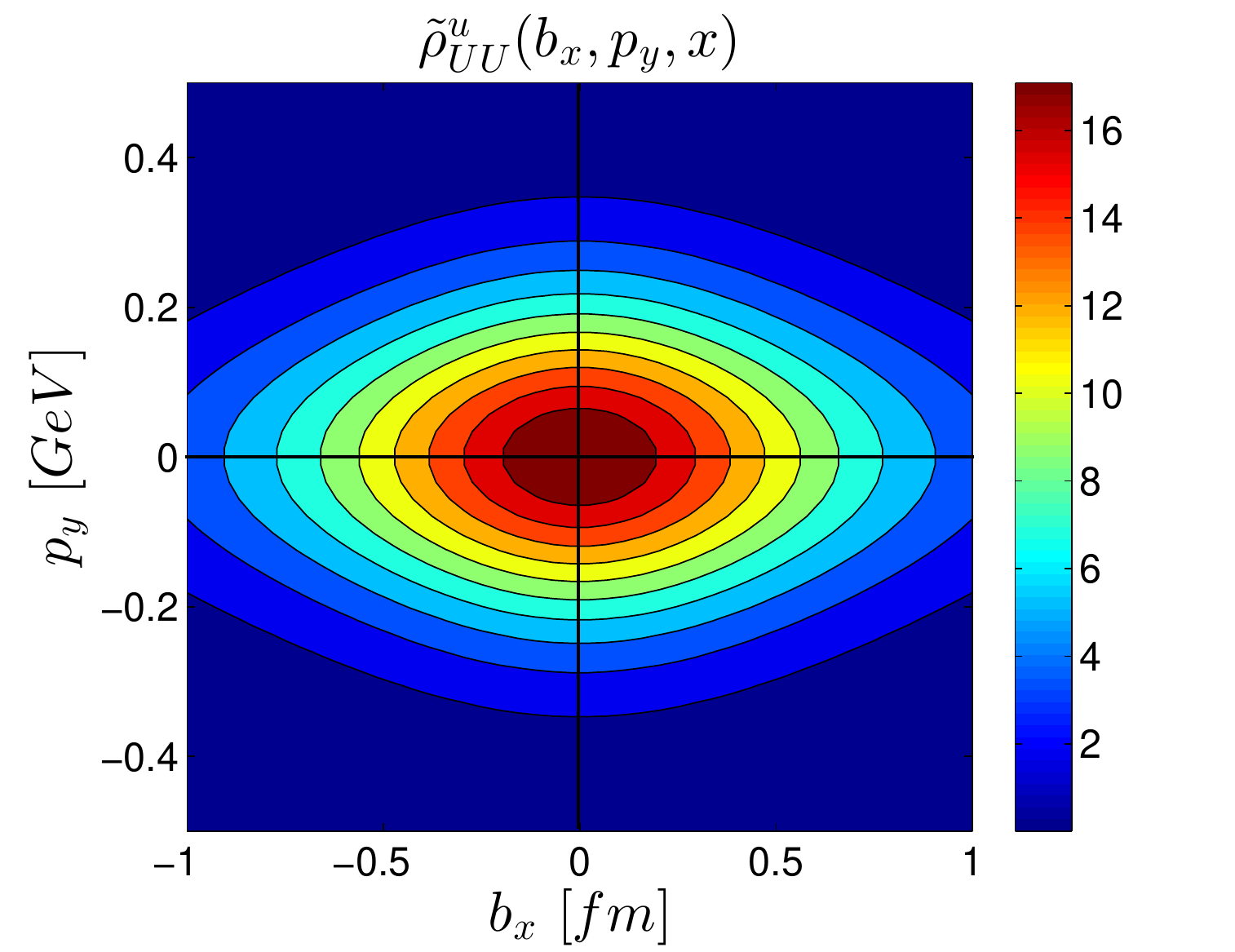}}
\subfigure[]{\includegraphics[width=5.cm,height=4.cm]{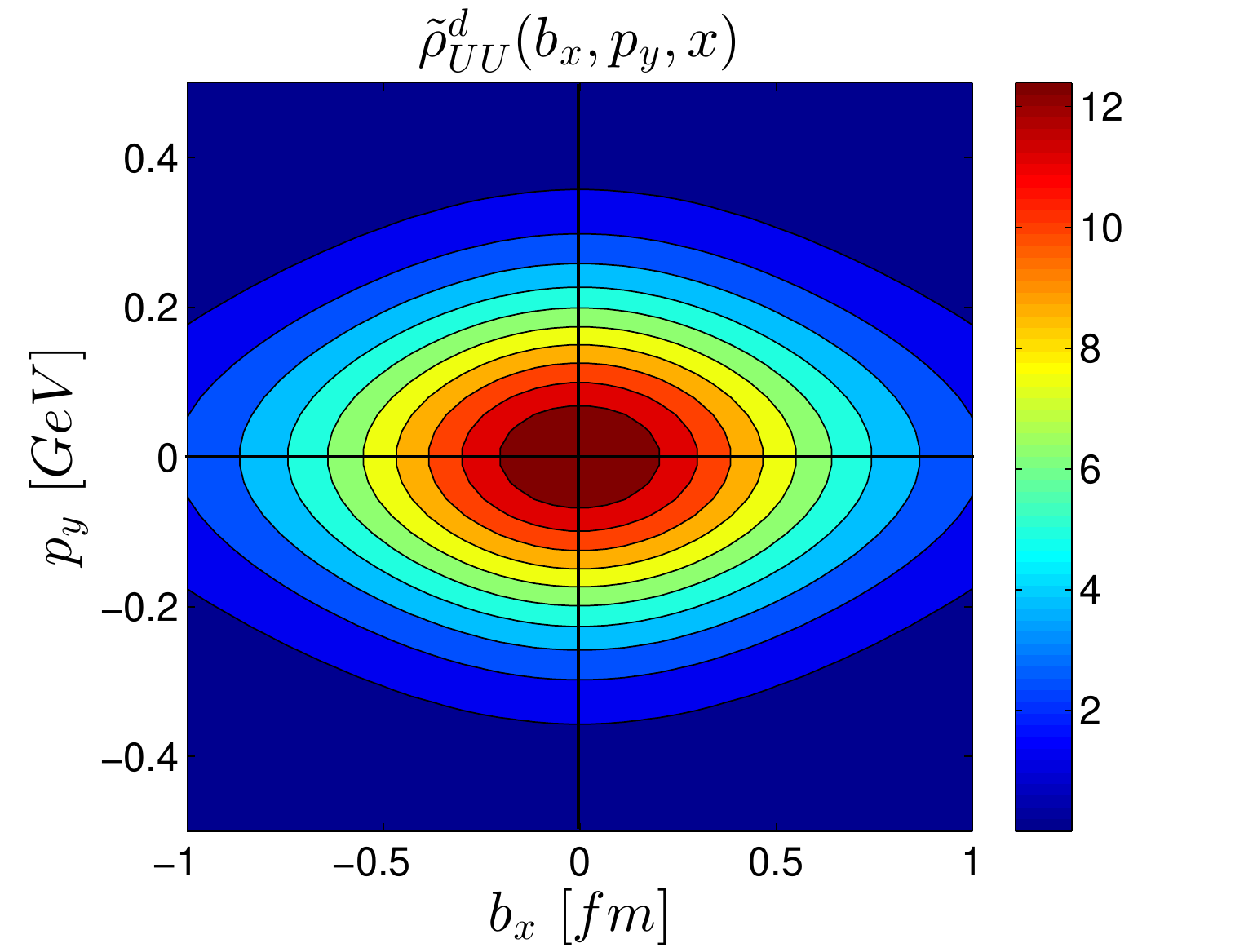}}
\caption{\label{fig_UU}The distributions $\rho_{UU}$ are shown in the transverse momentum plane, transverse coordinate plane and the mixed plane for $u$ and $d$  quarks. The distributions in the mixed plane are given in $GeV^0 fm^0$.}
\end{figure}
\begin{figure}[ht]
\centering
\subfigure[]{\includegraphics[width=5.cm,height=4.cm]{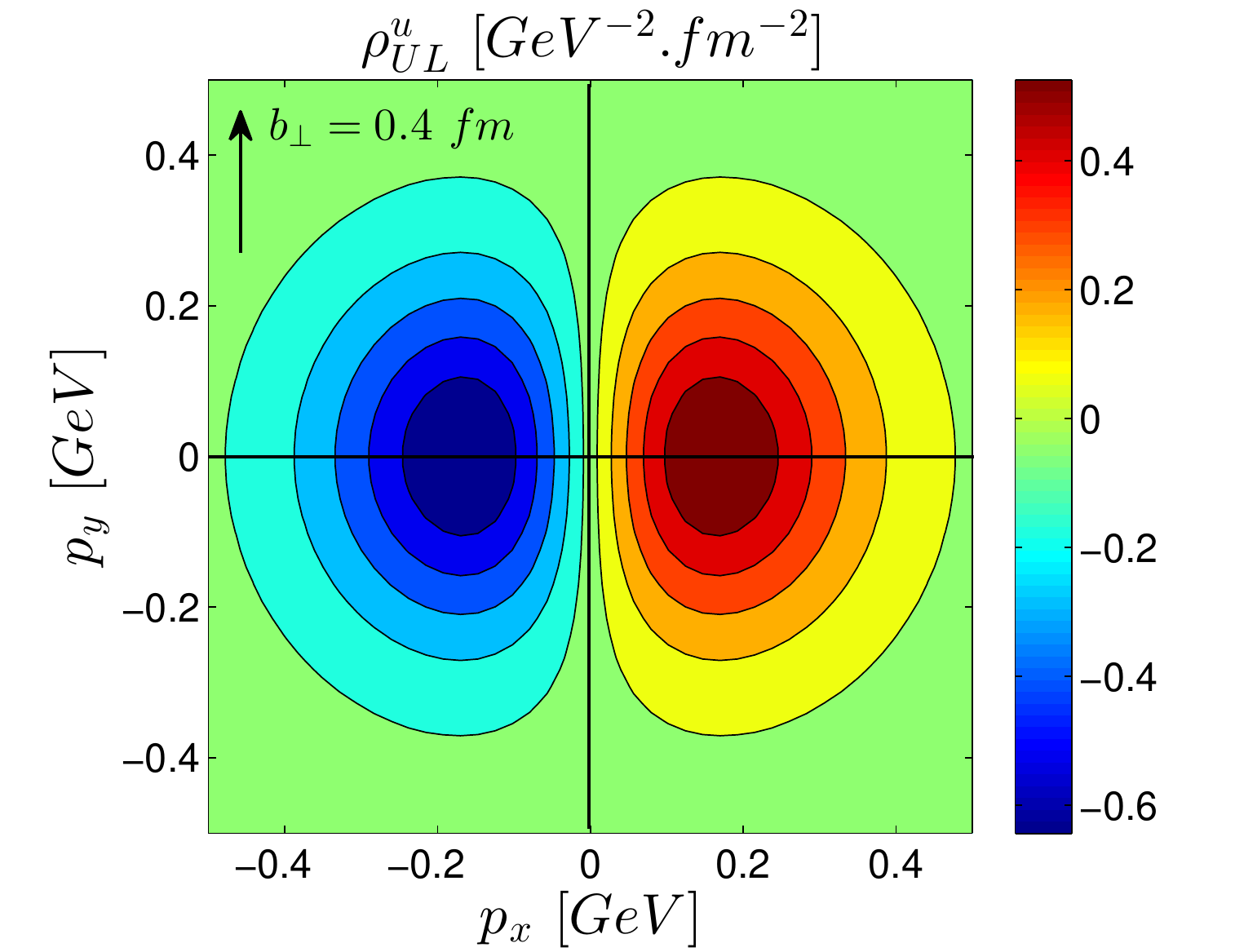}}
\subfigure[]{\includegraphics[width=5.cm,height=4.cm]{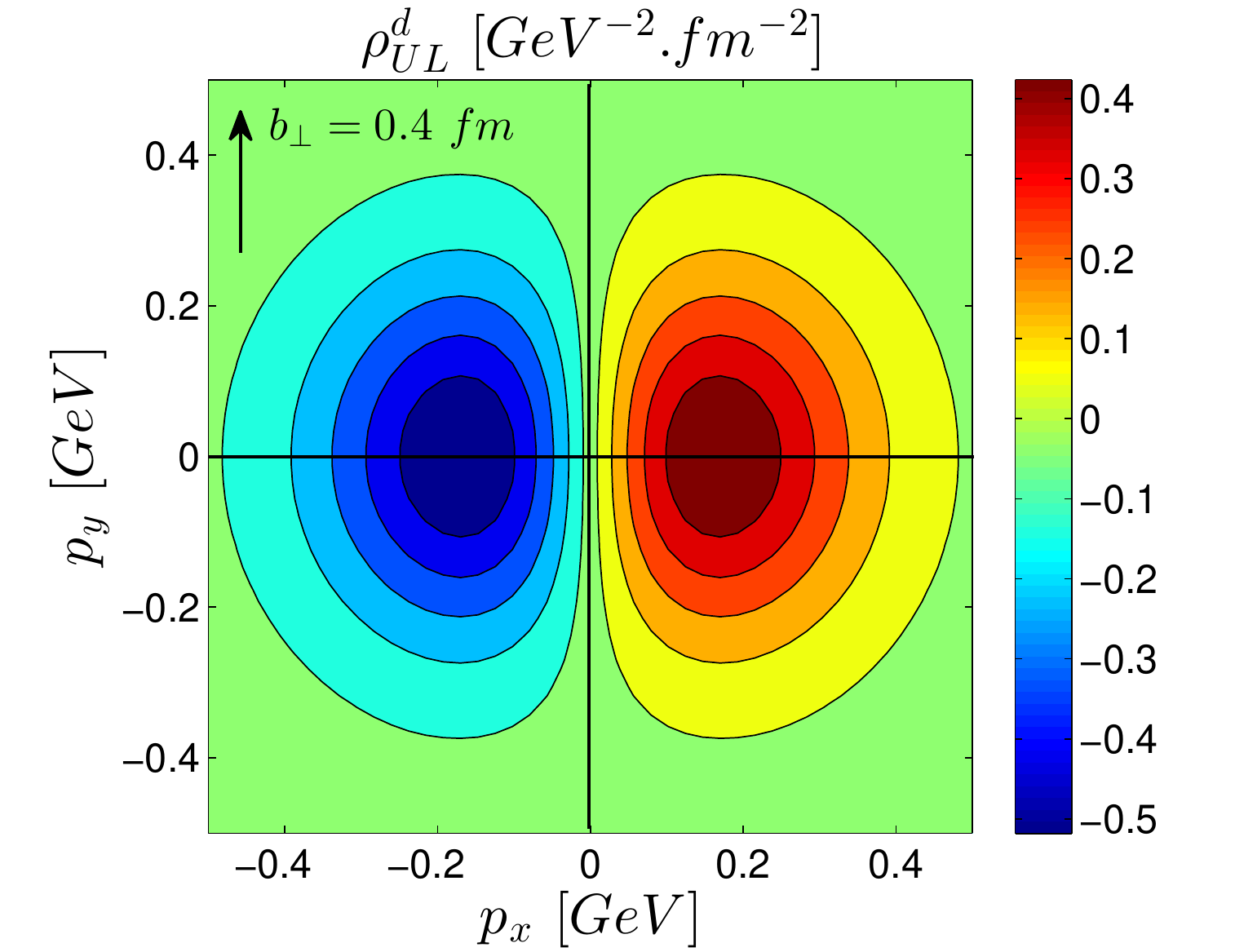}}\\
\subfigure[]{\includegraphics[width=5.cm,height=4.cm]{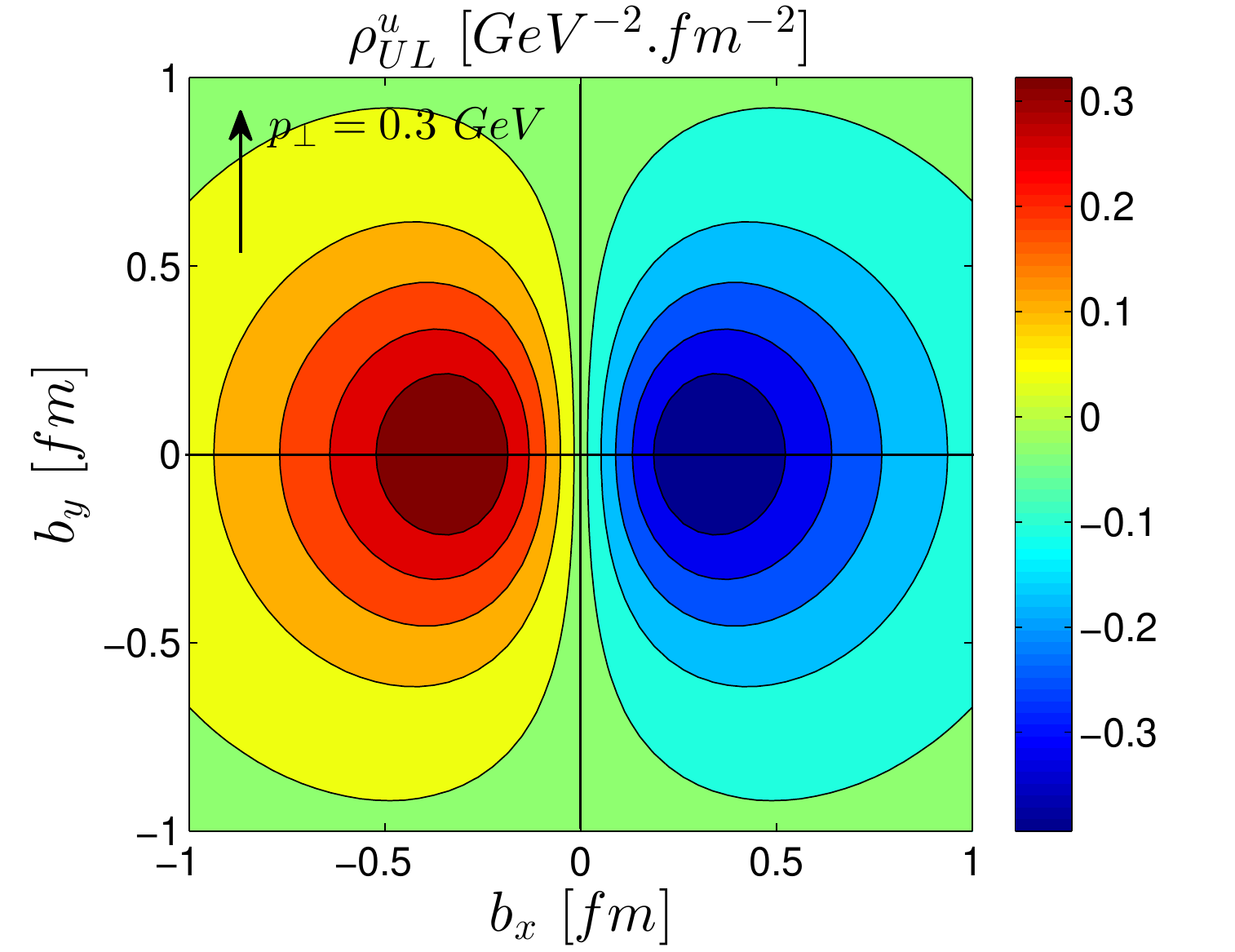}}
\subfigure[]{\includegraphics[width=5.cm,height=4.cm]{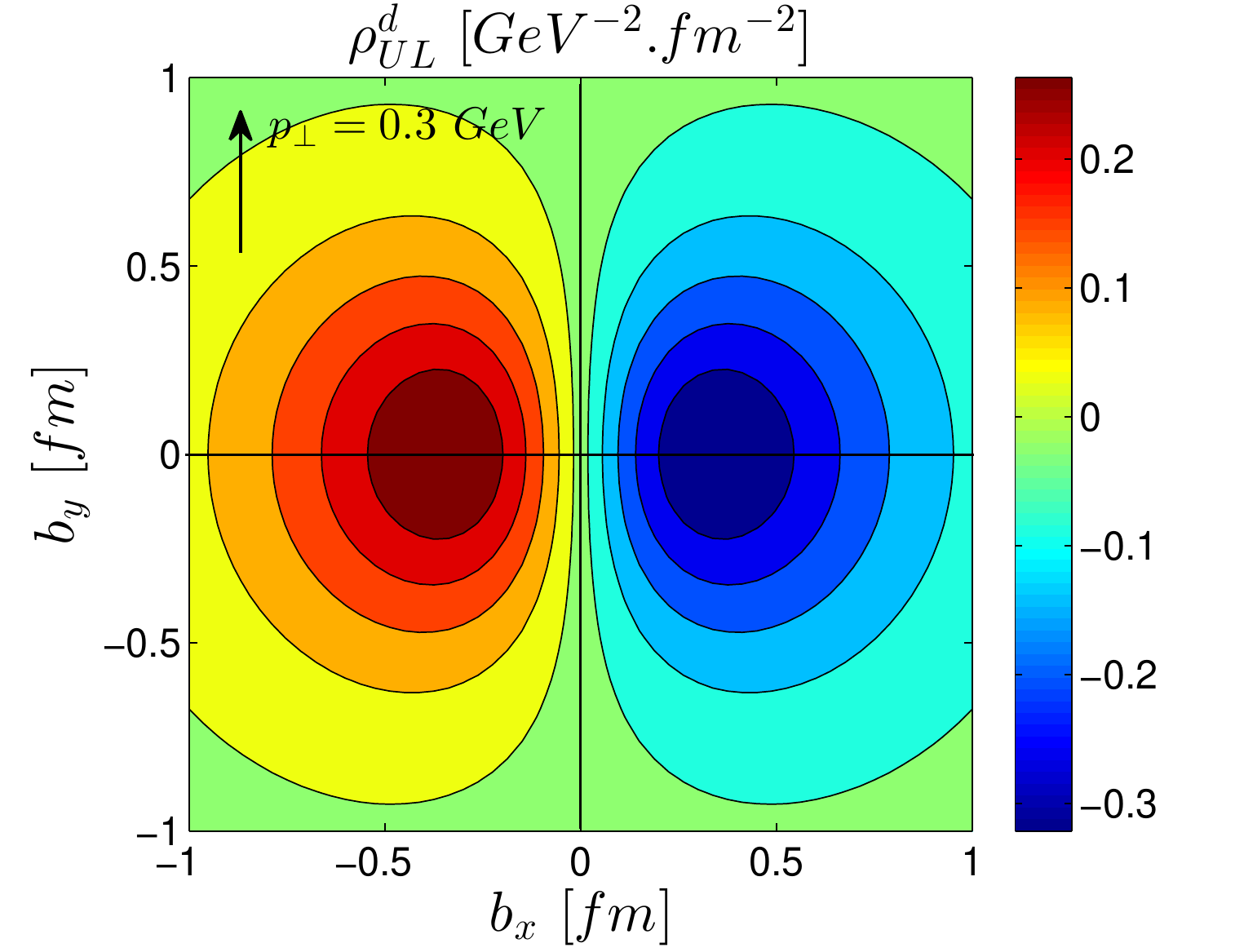}}\\
\subfigure[]{\includegraphics[width=5.cm,height=4.cm]{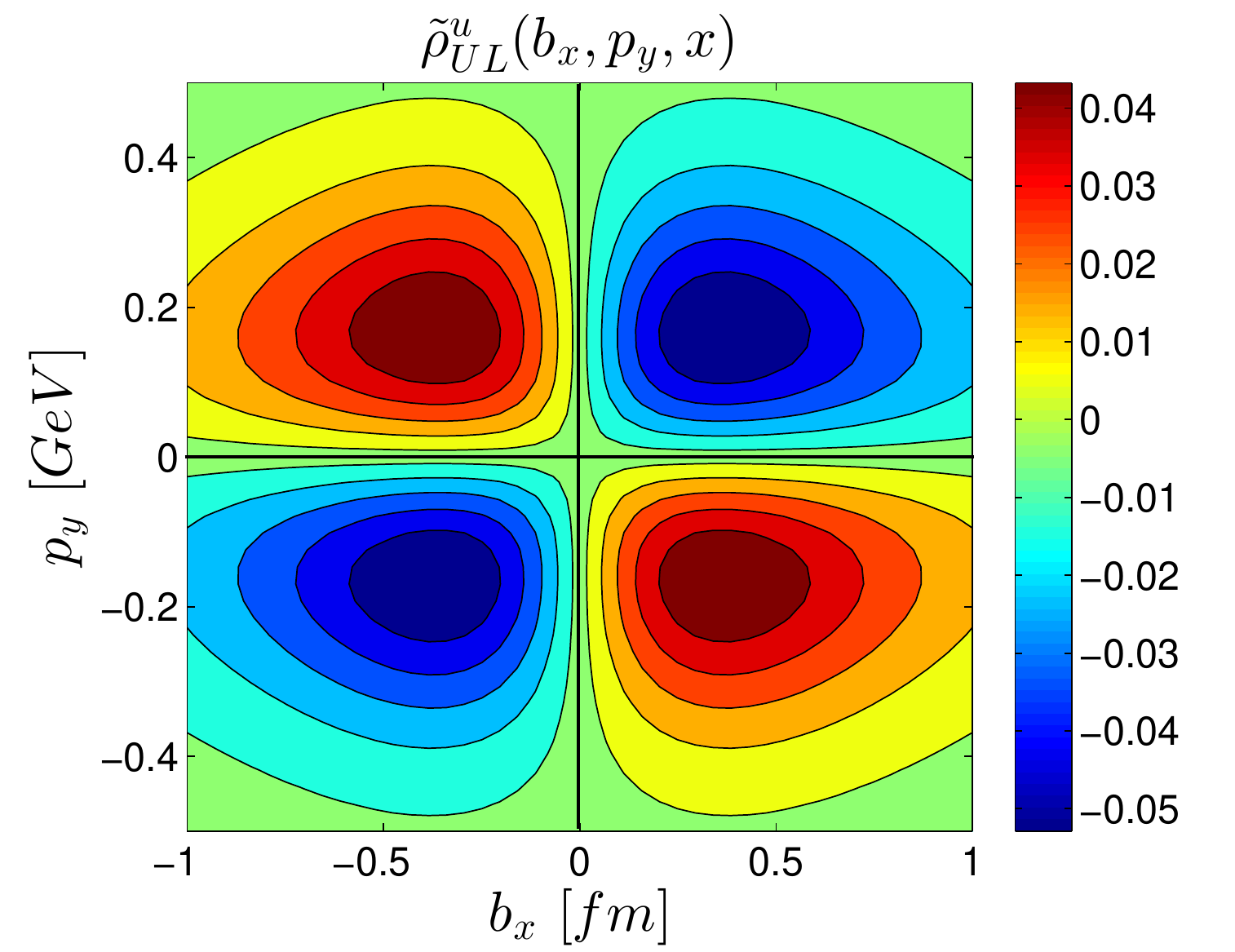}}
\subfigure[]{\includegraphics[width=5.cm,height=4.cm]{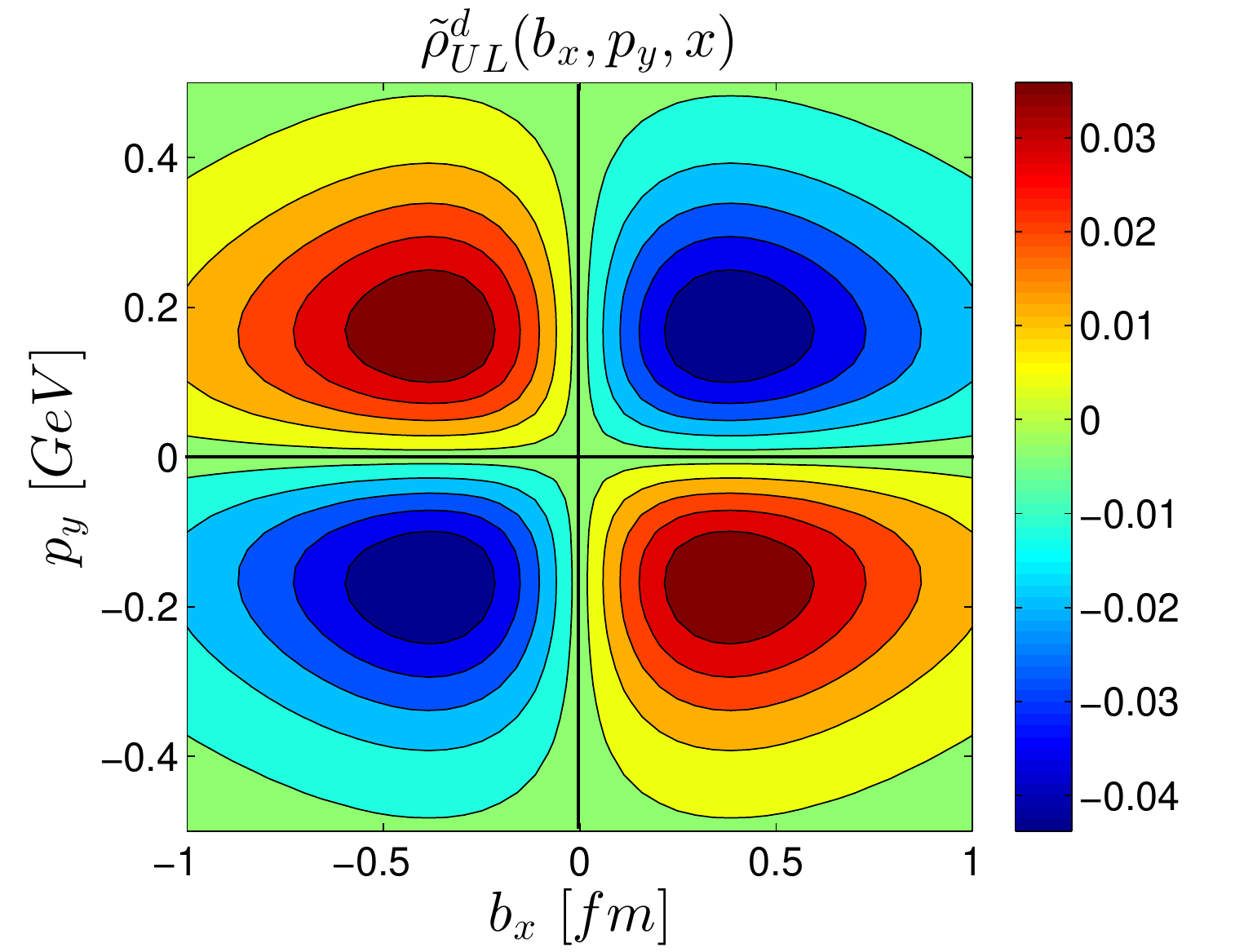}}
\caption{\label{fig_UL} The distributions $\rho_{UL}$ are presented in the  transverse momentum plane, transverse coordinate plane and the mixed plane for $u$ and $d$  quarks. The distributions in the mixed plane are given in $GeV^0 fm^0$.}
\end{figure}
\begin{figure}[ht]
\centering
\subfigure[]{\includegraphics[width=5.cm,height=4.cm]{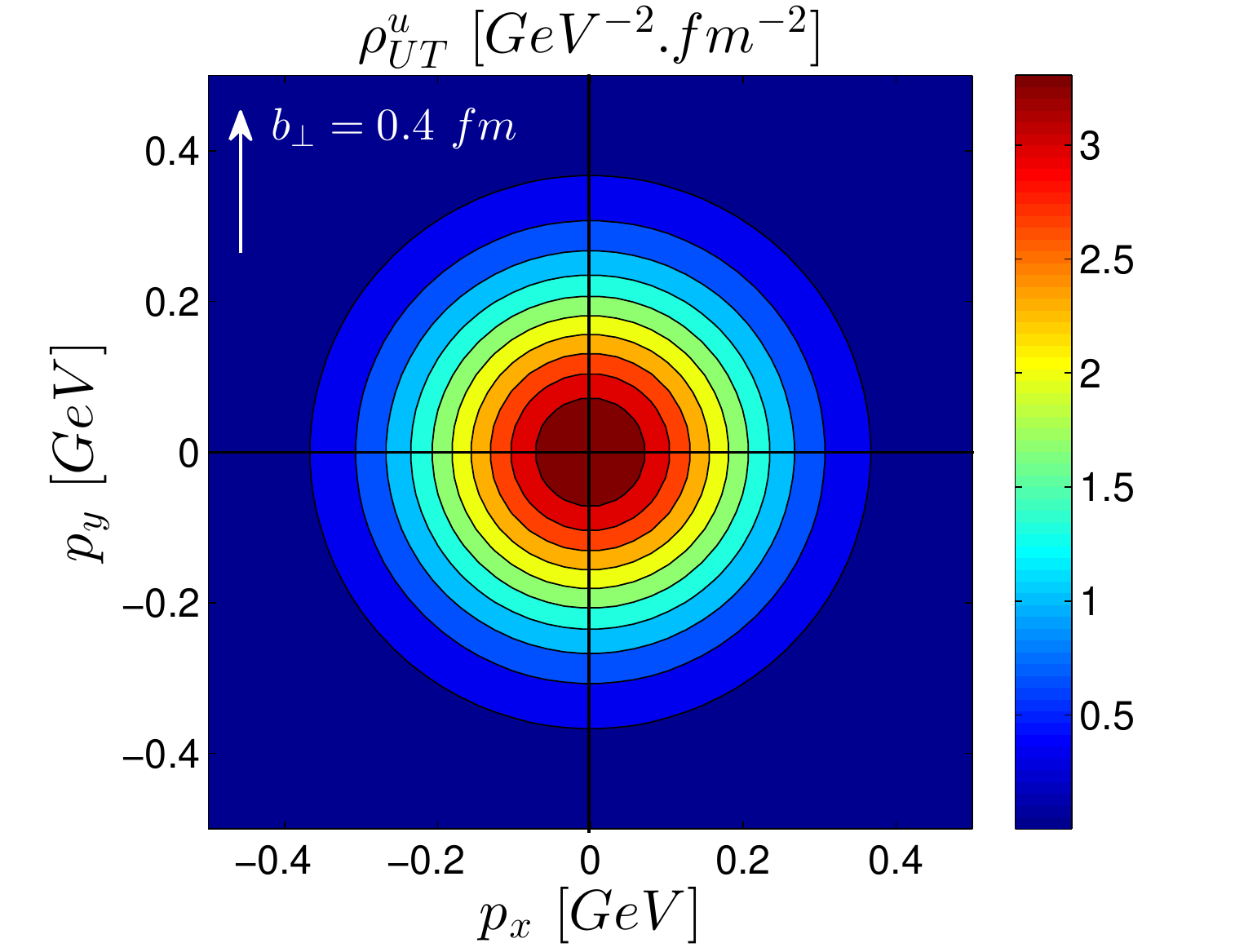}}
\subfigure[]{\includegraphics[width=5.cm,height=4.cm]{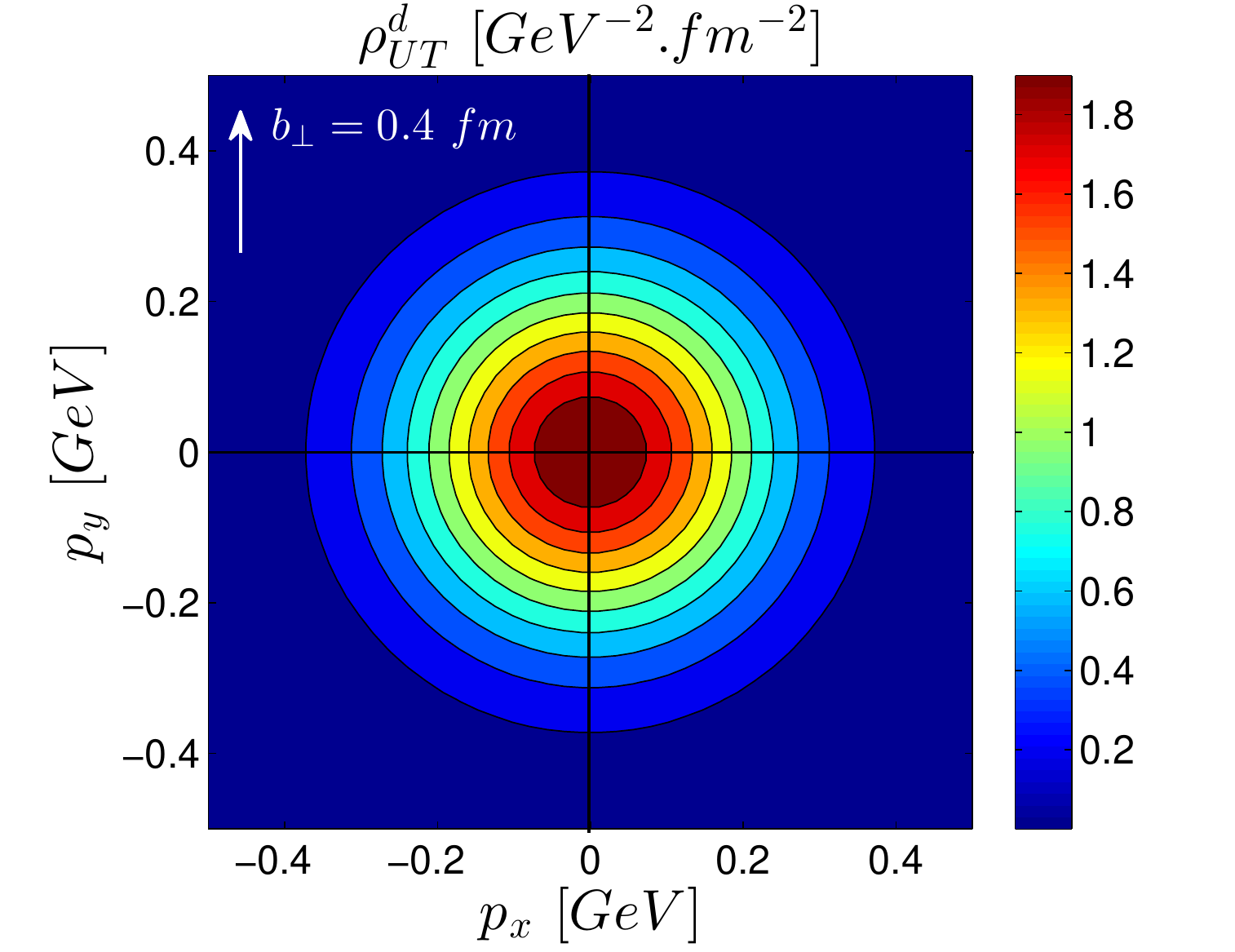}}\\
\subfigure[]{\includegraphics[width=5.cm,height=4.cm]{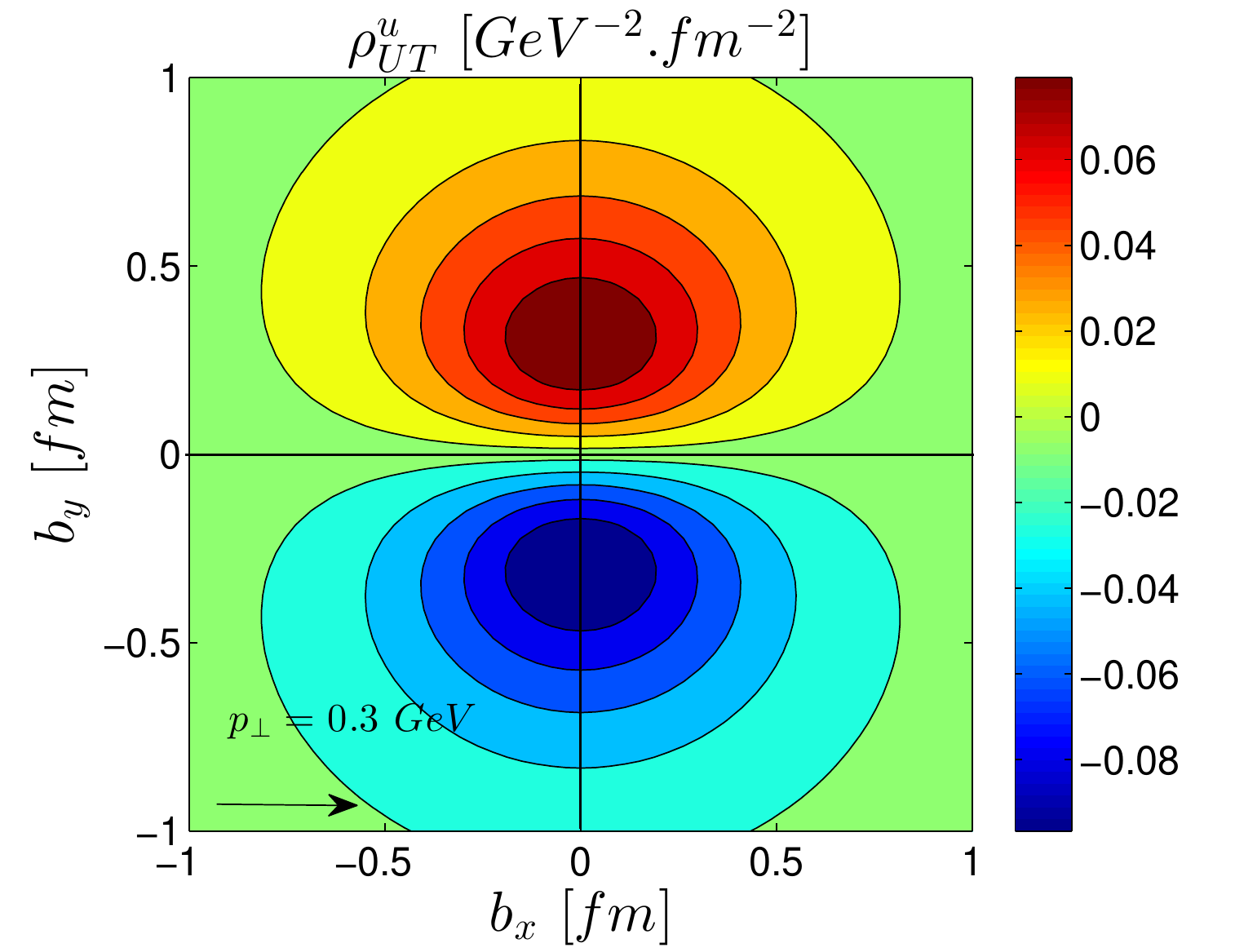}}
\subfigure[]{\includegraphics[width=5.cm,height=4.cm]{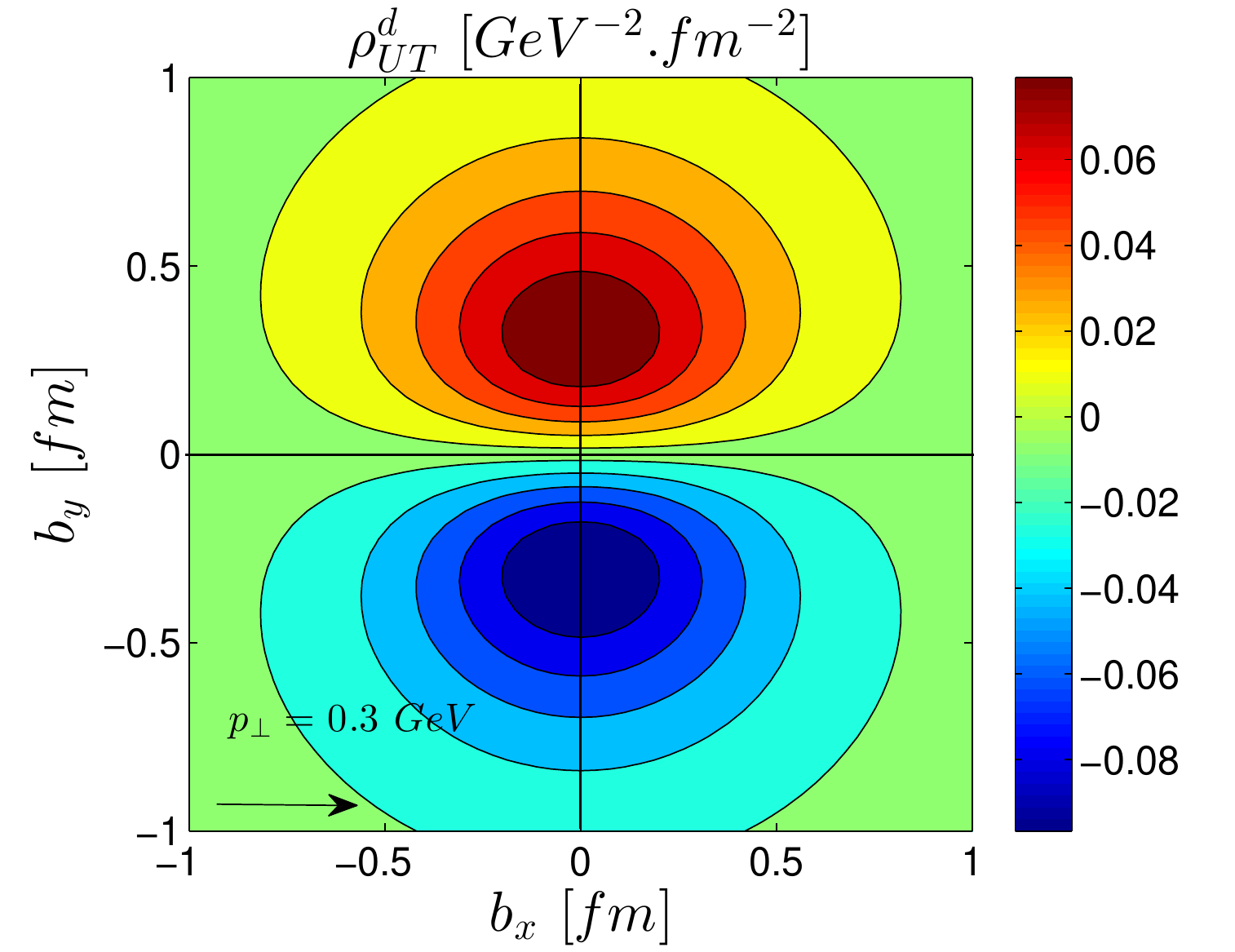}}\\
\subfigure[]{\includegraphics[width=5.cm,height=4.cm]{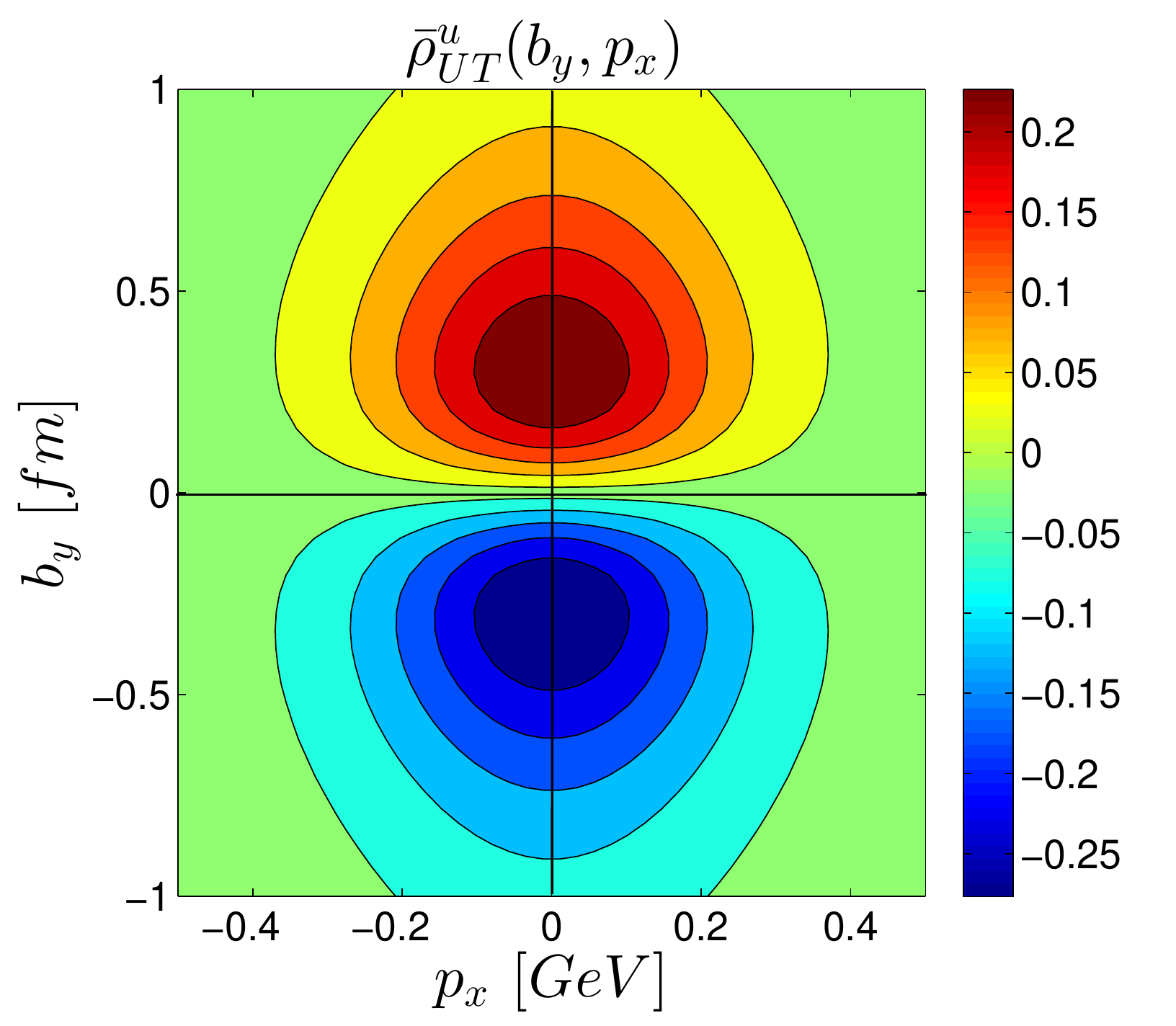}}
\subfigure[]{\includegraphics[width=5.cm,height=4.cm]{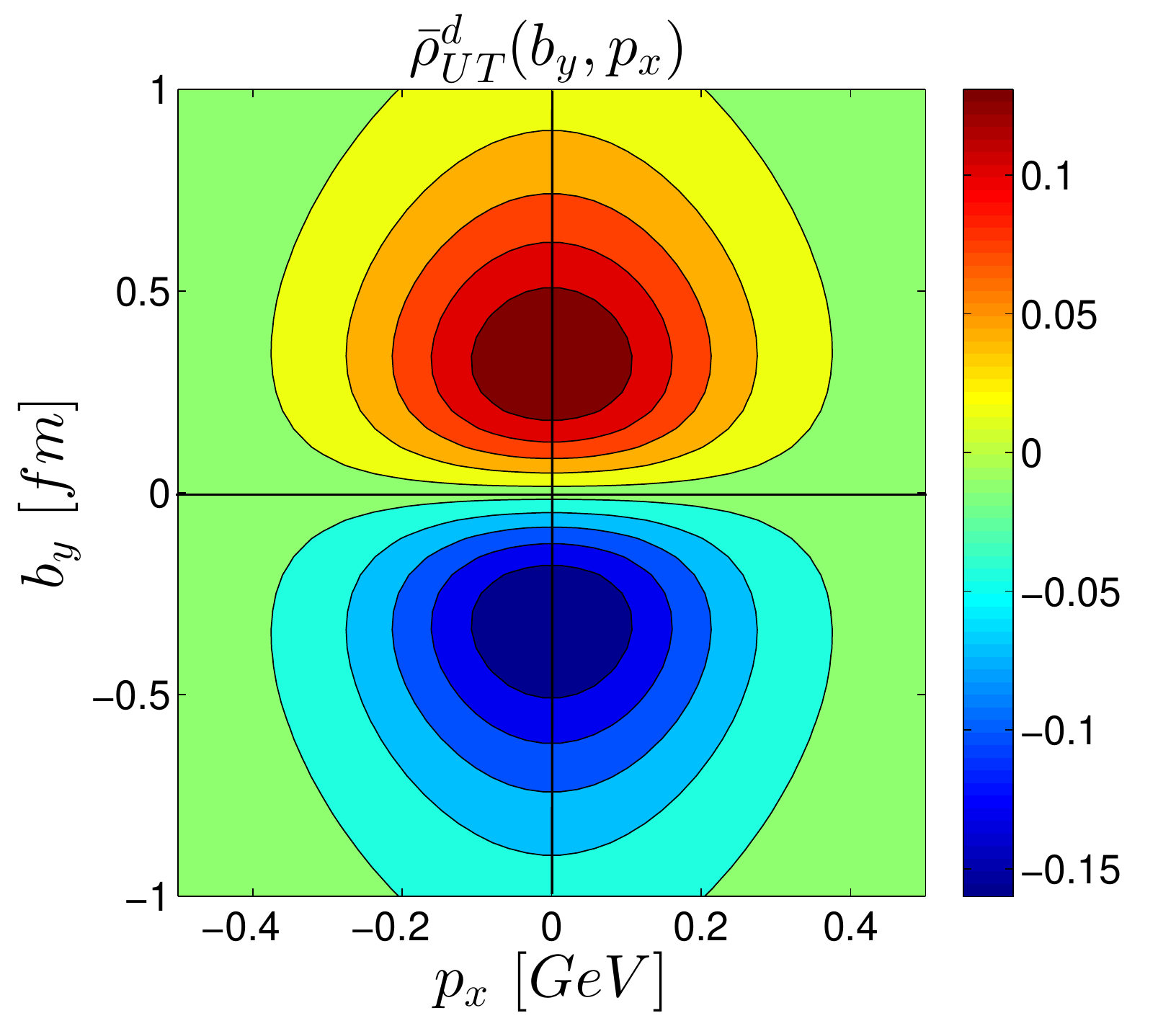}}
\caption{\label{fig_UT}The distribution $\rho_{UT}$ are shown in the transverse momentum plane(a,d) with $\bfb=0.4~\hat{y}~fm$,  in the transverse coordinate plane(b,e) with $\bfp=0.3~\hat{x}~GeV$ and  in mixed plane(c,f) for $u$ and $d$  quarks. The distributions in the mixed planes are given in $GeV^0 fm^0$. The quark is transversely polarised along x-axis.}
\end{figure}
We plot the first Mellin moment of unpolarized Wigner distribution, $\rho^\nu_{UU}(\bfb,\bfp,x)$ and mixing distributions, $\tilde{\rho}^\nu_{UU}(b_x,p_y,x)$ for $u$ and $d$ quark in Fig.\ref{fig_UU}. The first Mellin moment of unpolarized Wigner distributions represent the transverse phase-space distribution of the unpolarized quark in an unpolarized proton. Fig.\ref{fig_UU}(a) and Fig.\ref{fig_UU}(d) show the distributions in transverse momentum plane for $u$ quark and $d$ quark respectively with fixed impact parameter $\bfb$ along $\hat{y}$ and $b_y=0.4~fm$ whereas the variation of the distributions in the transverse impact parameter plane are shown in Fig.\ref{fig_UU}(b) and Fig.\ref{fig_UU}(e) with fixed transverse momentum $\bfp$ along $\hat{y}$ for $p_y=0.3~GeV$. Fig.\ref{fig_UU}(c) and Fig.\ref{fig_UU}(f) represent the mixing distributions for $u$ quark and $d$ quark respectively. 

The average quadrupole distortions $Q^{ij}_b(\bfp)$ and $Q^{ij}_p(\bfb)$ are defined as\cite{Lorce11}
\be 
Q^{ij}_b(\bfp)= \frac{\int d^2\bfb (2b^i_\perp b^j_\perp-\delta^{ij}\bfb^2)\rho_{UU}(\bfb,\bfp)}{\int d^2\bfb \bfb^2 \rho_{UU}(\bfb,\bfp)}\\
Q^{ij}_p(\bfb)= \frac{\int d^2\bfp (2p^i_\perp p^j_\perp-\delta^{ij}\bfp^2)\rho_{UU}(\bfb,\bfp)}{\int d^2\bfp \bfp^2 \rho_{UU}(\bfb,\bfp)}.\label{Eq_Qdis}
\ee
Since the wave functions in soft-wall AdS/QCD model are of Gaussian type, the average quadrupole distortion is found to be zero for $\rho^\nu_{UU}$. Similarly $\rho^\nu_{LL}$ have zero quadrupole distortion.
Therefore, the distributions $\rho^\nu_{UU}$ in transverse momentum plane as well as transverse impact parameter plane are circularly symmetric but the distributions in mixed space are axially symmetric. Thus, there is no favored configuration between $\bfb\perp\bfp$ and $\bfb\parallel\bfp$ in mixed space unlike the light-cone constituent quark model(LCCQM) \cite{Lorce12} or chiral quark soliton model($\chi$QSM) \cite{Lorce11}. Comparing the behaviors of the $u$ quark and the $d$ quark, one finds in this model that for both $u$ and $d$ quarks, the distributions have positive maxima at the center $(p_x=p_y=0)$, $(b_x=b_y=0)$ in both planes and gradually decrease towards periphery. The peak of the distributions for $u$ quark are large compared to $d$ quark but the $u$ quark distributions are little concentrating in the center relative to the $d$ quark in both planes. The distributions have a similar spread behaviors in the mixed plane for $u$ and $d$ quarks. 

In Fig.\ref{fig_UL}, we plot unpolarized-longitudinal Wigner distribution which represents the transverse phase-space distribution of the longitudinally polarized quark in an unpolarized proton. The transverse Wigner distributions $\rho^\nu_{UL}(\bfb,\bfp)$, in the transverse momentum plane with fixed impact parameter $\bfb$ along $\hat{y}$, are presented in Fig.\ref{fig_UL}(a) and Fig.\ref{fig_UL}(d) for $u$ and $d$ quarks respectively. The Fig.\ref{fig_UL}(b) and \ref{fig_UL}(e) show the same distributions in transverse impact parameter plane, for $u$ and $d$ quark with fixed $\bfp=p_y\hat{y}=0.3 ~GeV$. We find in this model that in both planes $\rho^\nu_{UL}$ exhibit dipolar structures having same polarity for $u$ and $d$ quarks but the polarity in momentum space is opposite from coordinate space for each quark. $\rho^\nu_{UL}$ in transverse momentum plane is more concentrating in the center relative to that in transverse coordinate plane.
%The quadrapole structures appear when we plot the distribution in the transverse mixed plane as shown in Fig.\ref{fig_UL}(e) and Fig.\ref{fig_UL}(f) for $u$ and $d$ quark respectively.
The mixing distribution $\tilde{\rho}^\nu_{UL}(b_x,p_y)$ in the transverse mixed plane are shown in Fig.\ref{fig_UL}(c) and Fig.\ref{fig_UL}(f) for $u$ and $d$ quark respectively which display the quadrupole structures with same polarity for both quarks. These distributions essentially reflect quark spin-orbit correlations. 
From Eq.(\ref{Cqz}), we calculate  $C^\nu_z$ at $\mu^2=1GeV^2$ and the values are $C^u_z=-0.55$ and $C^d_z=-0.75$ for $u$ and $d$  quarks. $C^\nu_z<0$ implies the quark OAM is anti parallel to the quark spin as observed in scalar diquark model \cite{WD_SD}, whereas in light-cone constituent quark model \cite{Lorce11} $C^\nu_z$ are found to be positive for both $u$ and $d$ quarks.
 % for $u$ and $d$  quarks.

%\subsection{Unpolarised-transverse Wigner distributions}
The Wigner distribution $\rho_{UT}(\bfb,\bfp)$ and the mixing distribution $\bar{\rho}_{UT}(b_y,p_x)$ are shown in fig.\ref{fig_UT}. From Eq.(\ref{rhoUT_nu}) it is clear that this distribution vanishes if the quark transverse coordinate is parallel to the polarization.  Here the plots are shown for $j=1$, the quark is polarized along x-direction. The figs.\ref{fig_UT}(a) and (d) represent the distribution in the transverse momentum plane, with $\bfb=0.4~\hat{y}~fm$, for $u$ and $d$  quarks respectively. This is circularly symmetric in transverse momentum space. The fig.\ref{fig_UT}(b) and (e) show the distribution in transverse impact parameter plane with $\bfp=0.3~\hat{x}~GeV$ for $u$ and $d$  quarks respectively. We see a dipolar distribution in the impact parameter plane. 
The mixing distribution $\bar{\rho}_{UT}(b_y,p_x)$ is shown in fig.\ref{fig_UT}(c) and (f) for $u$ and $d$  quarks respectively. Since this distribution is symmetric in transverse momentum plane, it shows a dipolar behavior  in the mixed plane(unlike $\bar{\rho}_{UL}(b_x,p_y)$ shows a quadrupole distribution). Because of the dipolar symmetry in impact parameter plane, the other class of mixing distributions $\tilde{\rho}_{UT}(b_x,p_y)$ vanishes.

In certain kinematical limit(see Eq.(\ref{rhoUT_H})), $\rho_{UT}(\bfb,\bfp,x)$ reduces to the  Boer-Mulders function $h^\perp_1(x,\bfp)$ which is one of the T-odd TMDs at leading twist. Here we consider the T-even leading twist TMDs only. The one gluon final sate interaction(FSI) is needed to calculate the T-odd TMDs. So, here we have no contribution from this distribution at the TMD limit.
%Since we consider the Willson line as unity, this distribution does not have any contribution to the TMDs in this model.
At the impact parameter distribution limit the distribution is related to $\tilde{H}_T$ GPD together with some other distributions\cite{meissner09}.

%------------------------------------
\begin{figure}[ht]
\centering
\subfigure[]{\includegraphics[width=5.cm,height=4.cm]{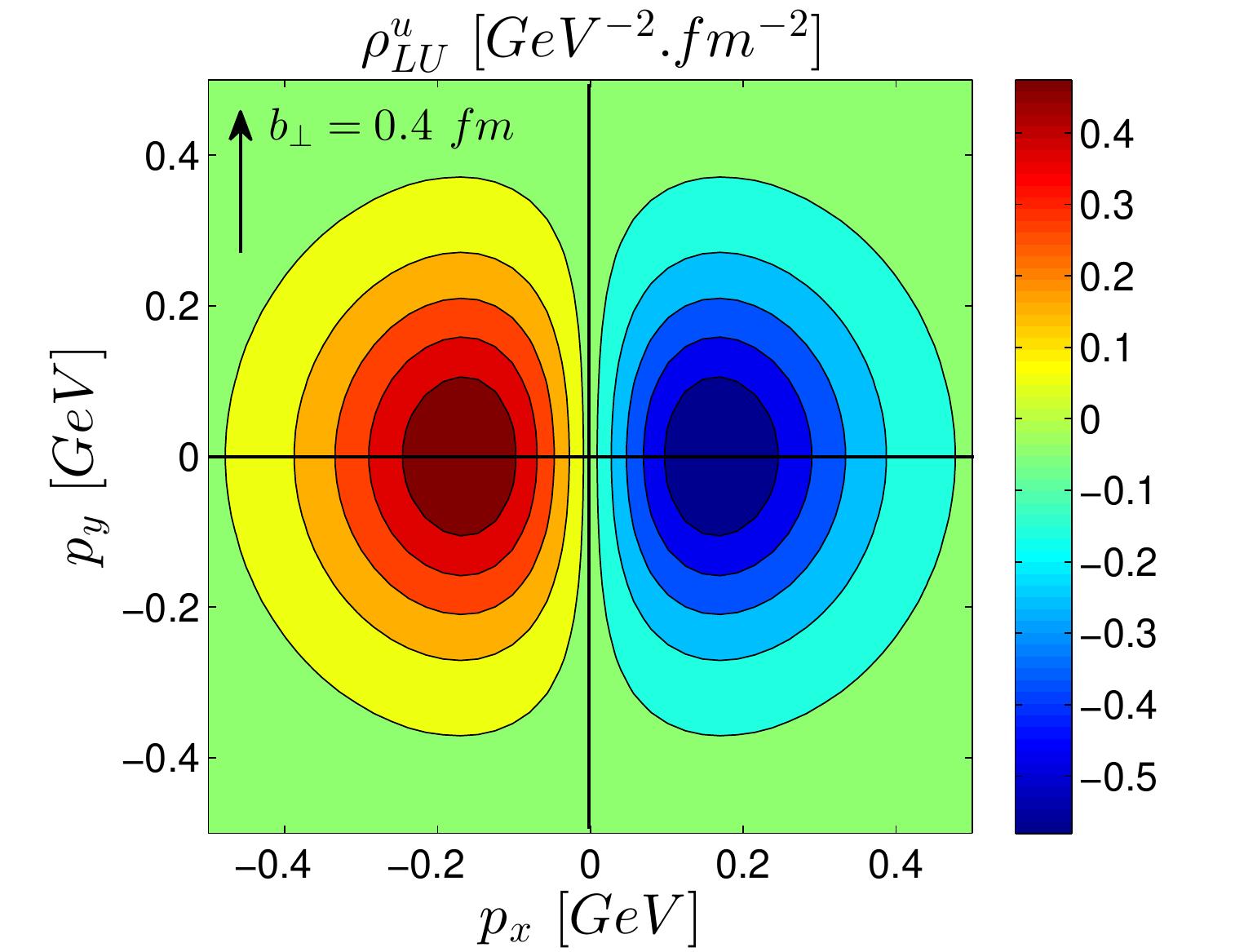}}
\subfigure[]{\includegraphics[width=5.cm,height=4.cm]{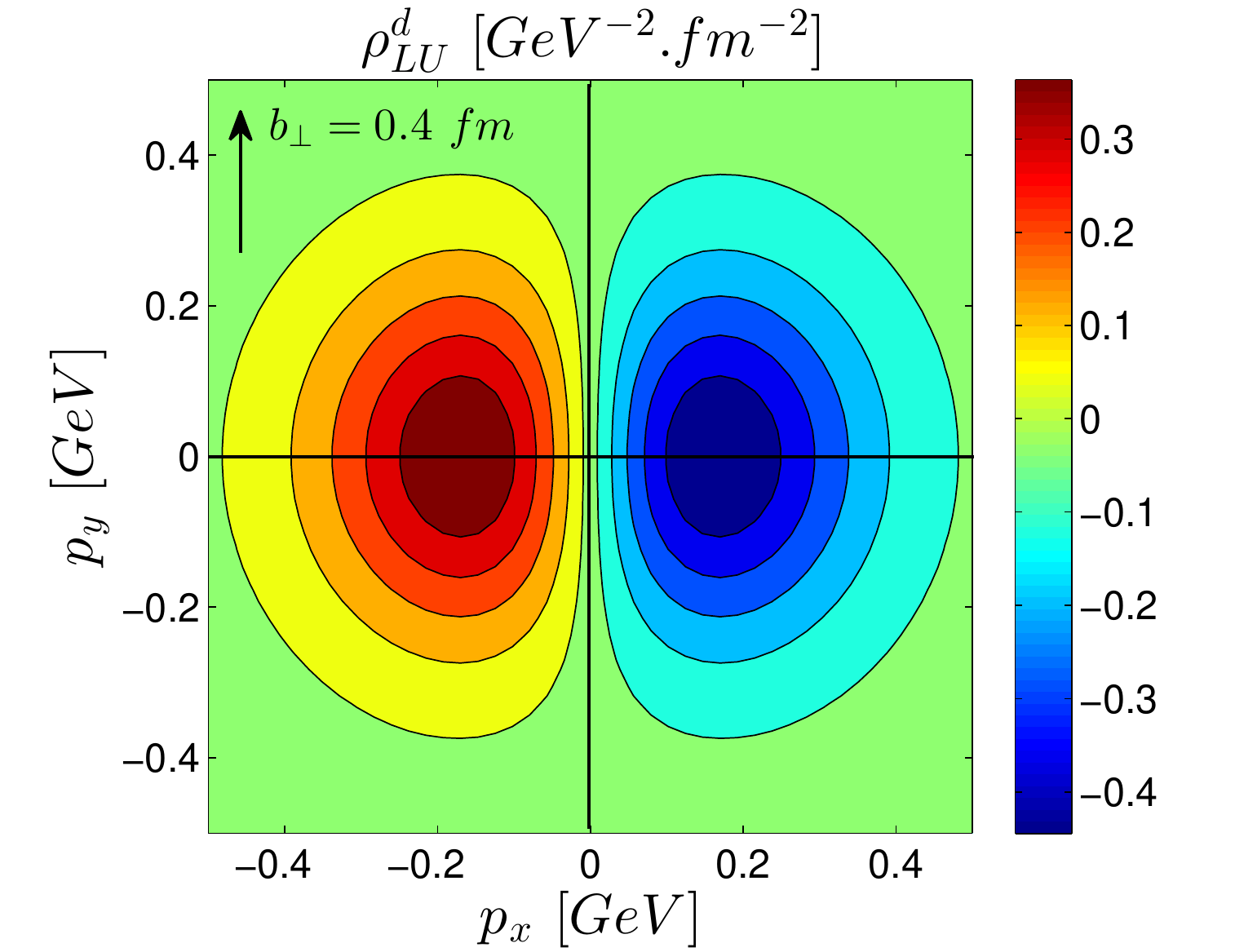}}\\
\subfigure[]{\includegraphics[width=5.cm,height=4.cm]{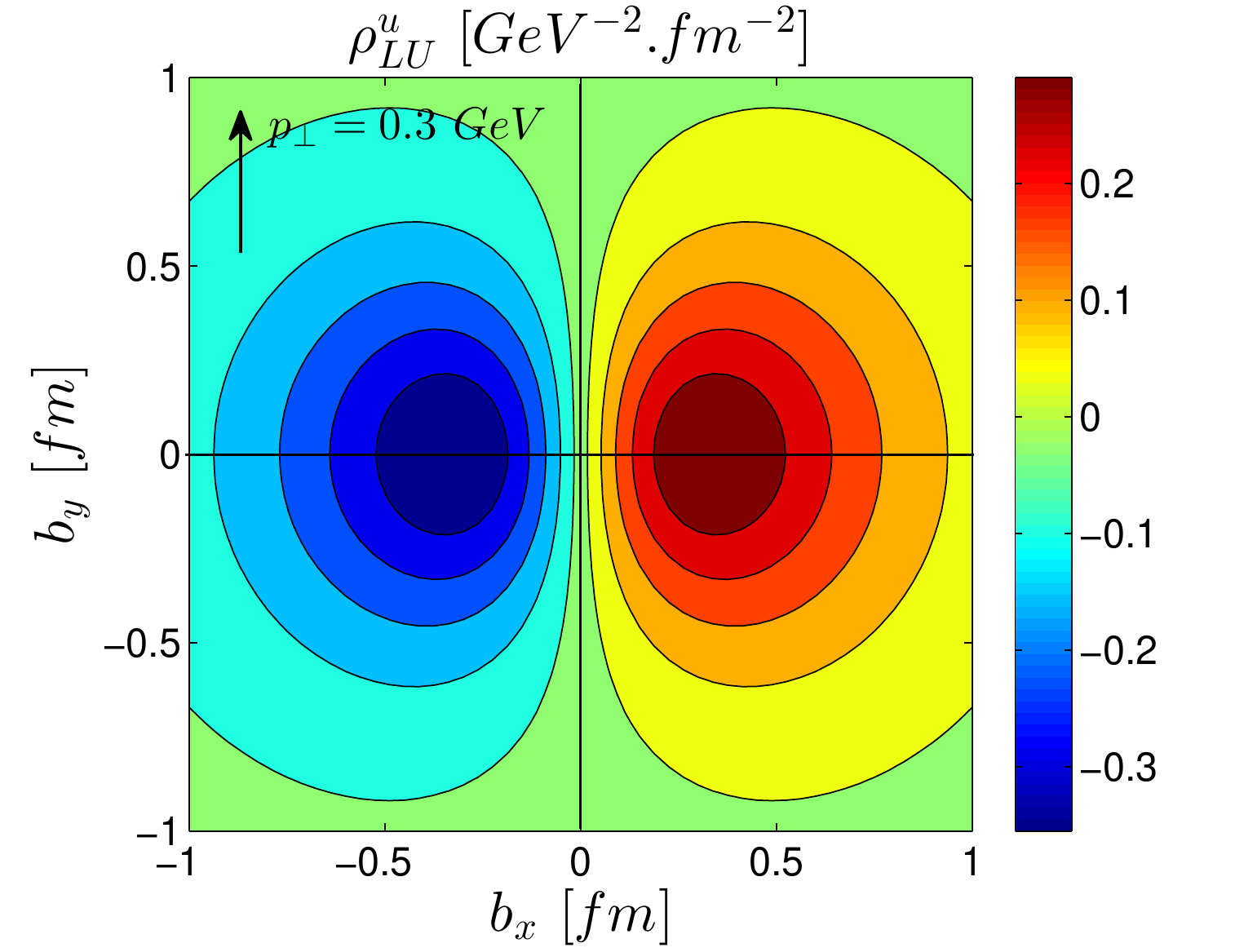}}
\subfigure[]{\includegraphics[width=5.cm,height=4.cm]{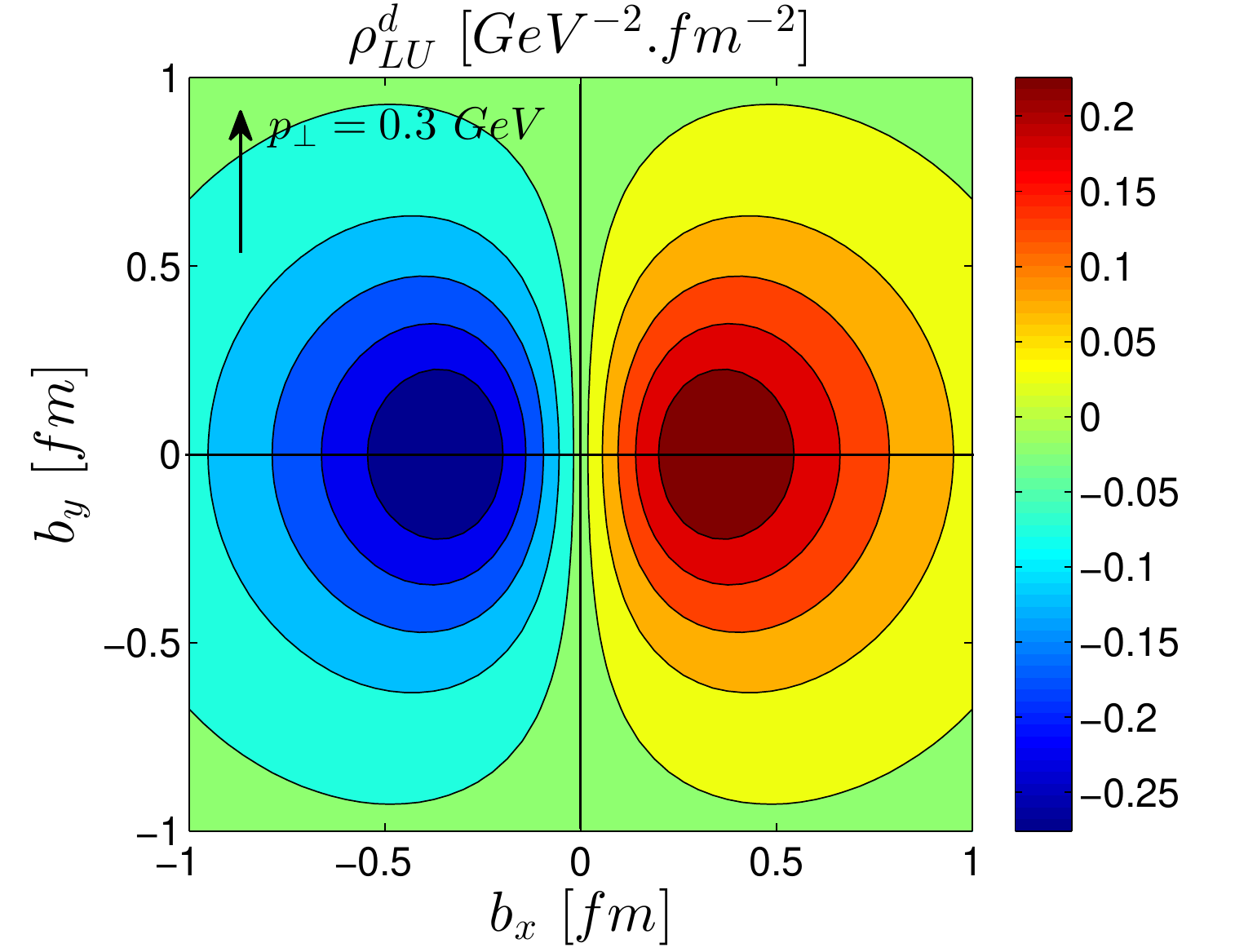}}\\
\subfigure[]{\includegraphics[width=5.cm,height=4.cm]{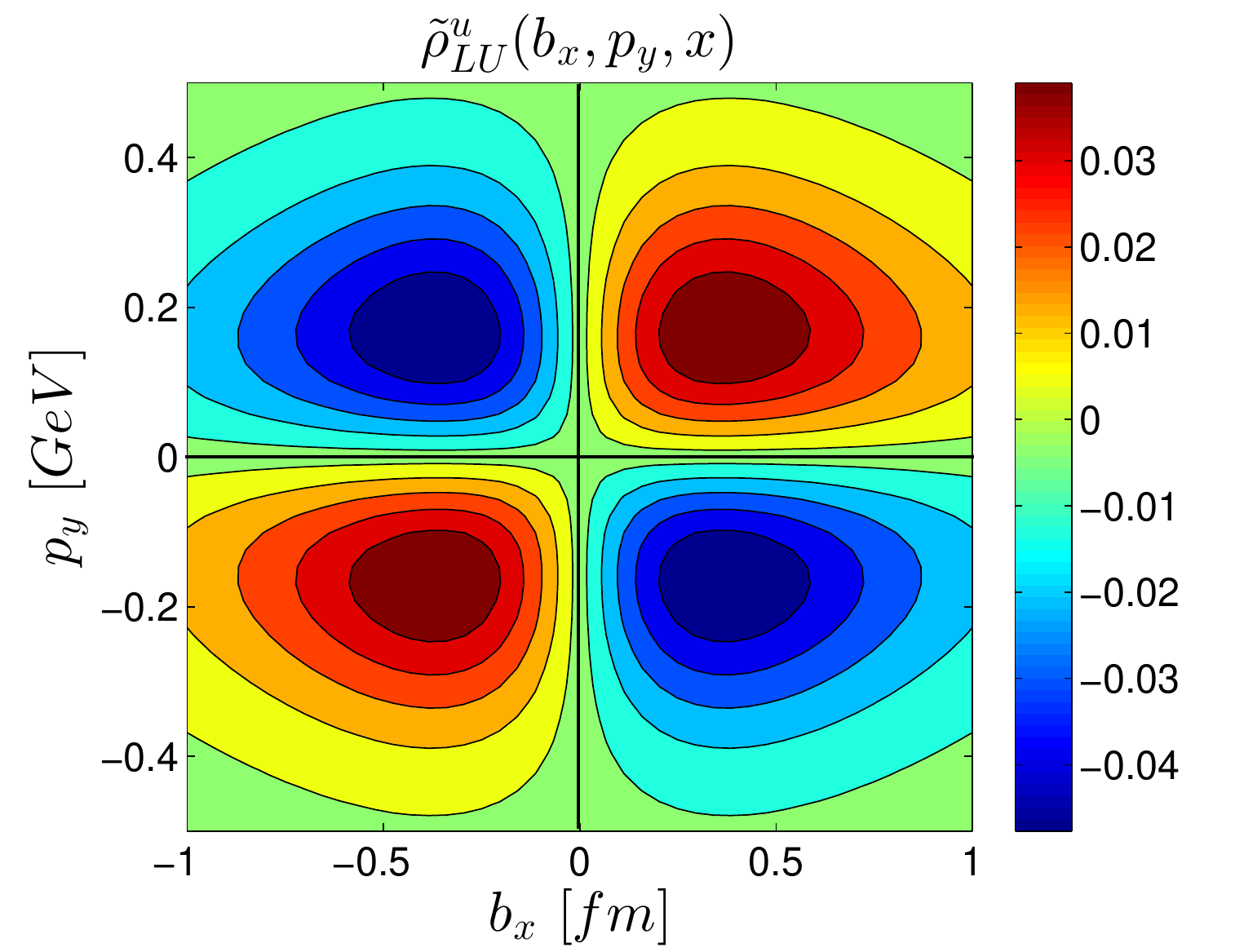}}
\subfigure[]{\includegraphics[width=5.cm,height=4.cm]{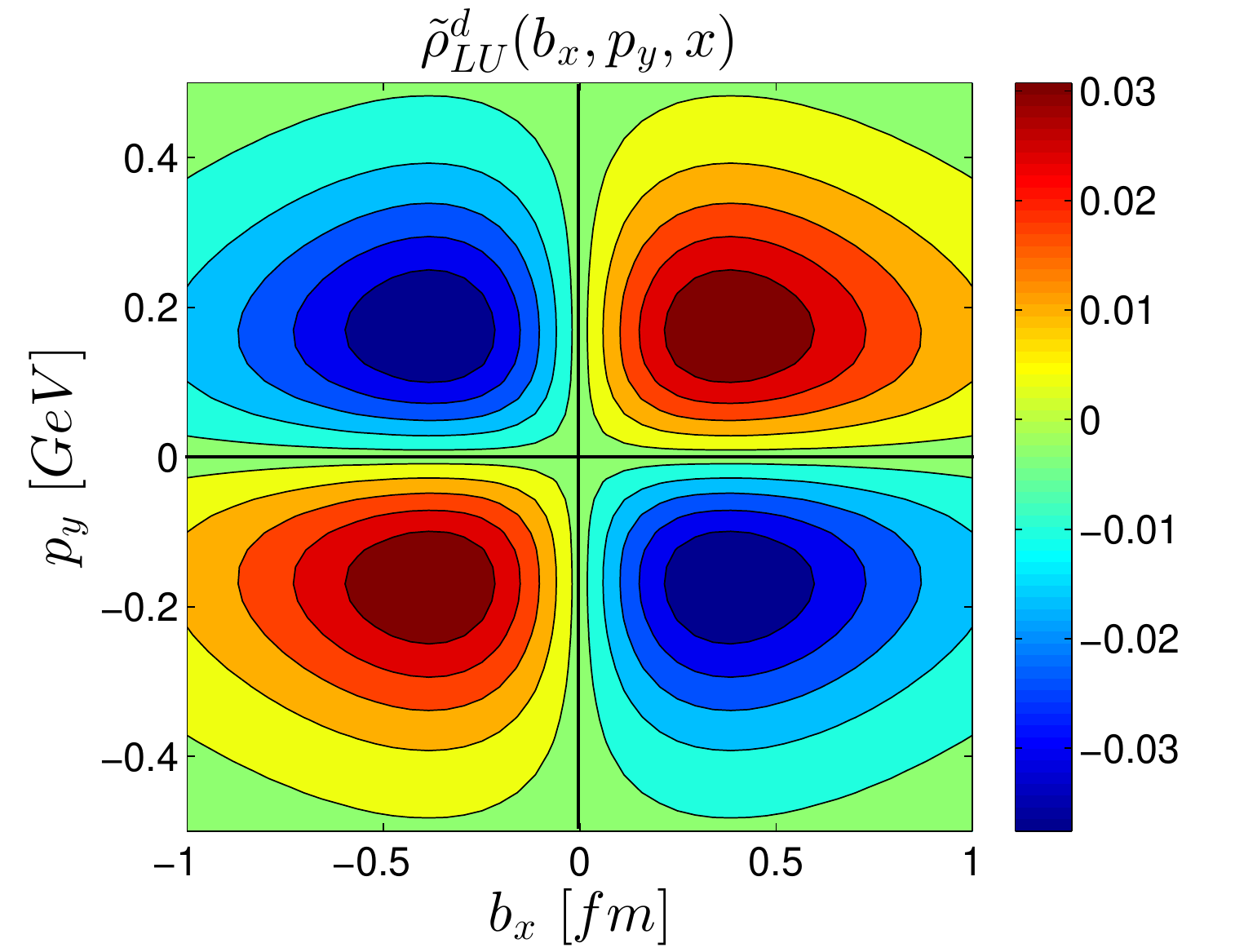}}
\caption{\label{fig_LU}The distributions $\rho_{LU}$ are shown in the transverse momentum plane, transverse coordinate plane and in the mixed plane for $u$ and $d$  quarks. The distributions in the mixed planes are given in $GeV^0 fm^0$.}
\end{figure}
\begin{figure}[ht]
\centering
\subfigure[]{\includegraphics[width=5.cm,height=4.cm]{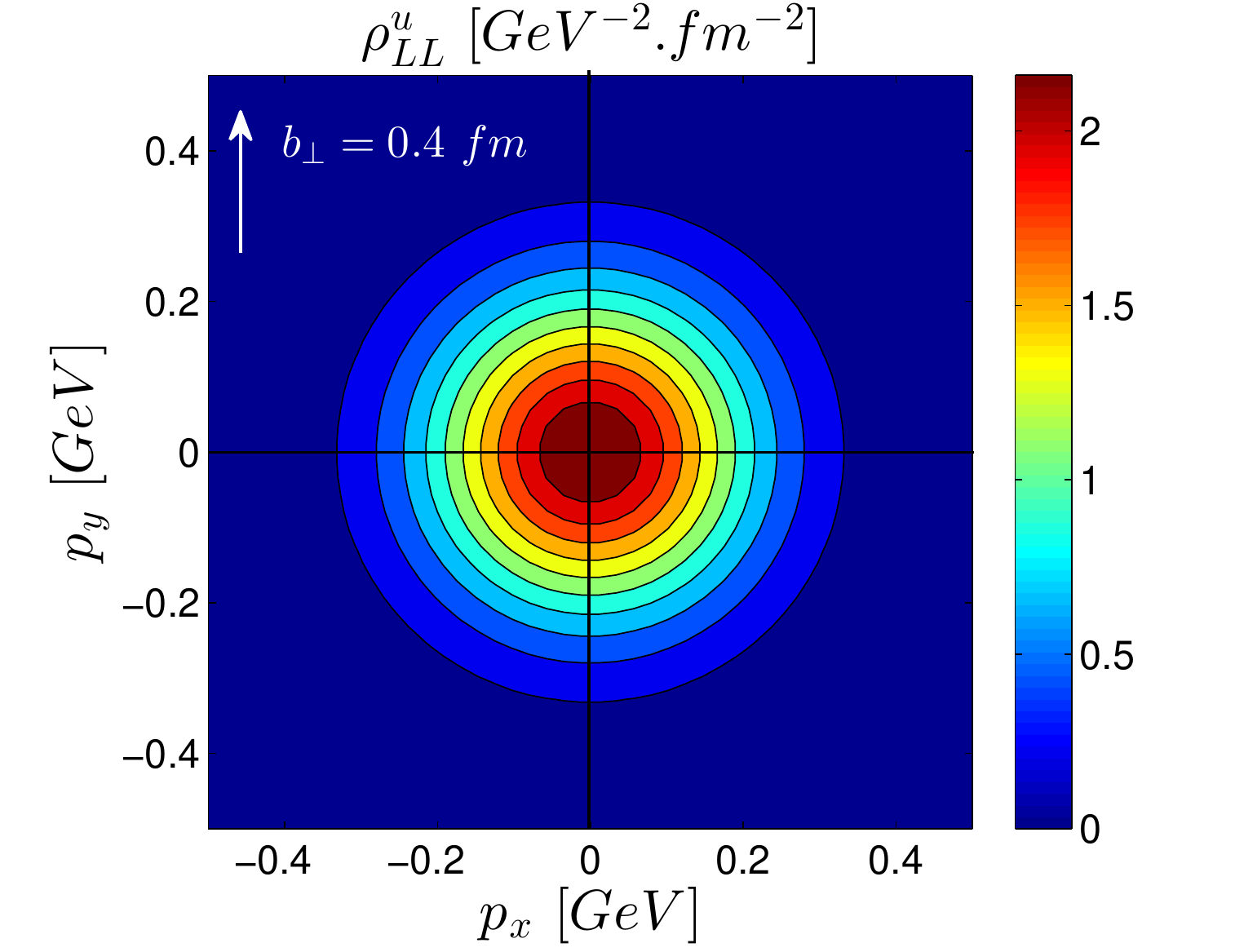}}
\subfigure[]{\includegraphics[width=5.cm,height=4.cm]{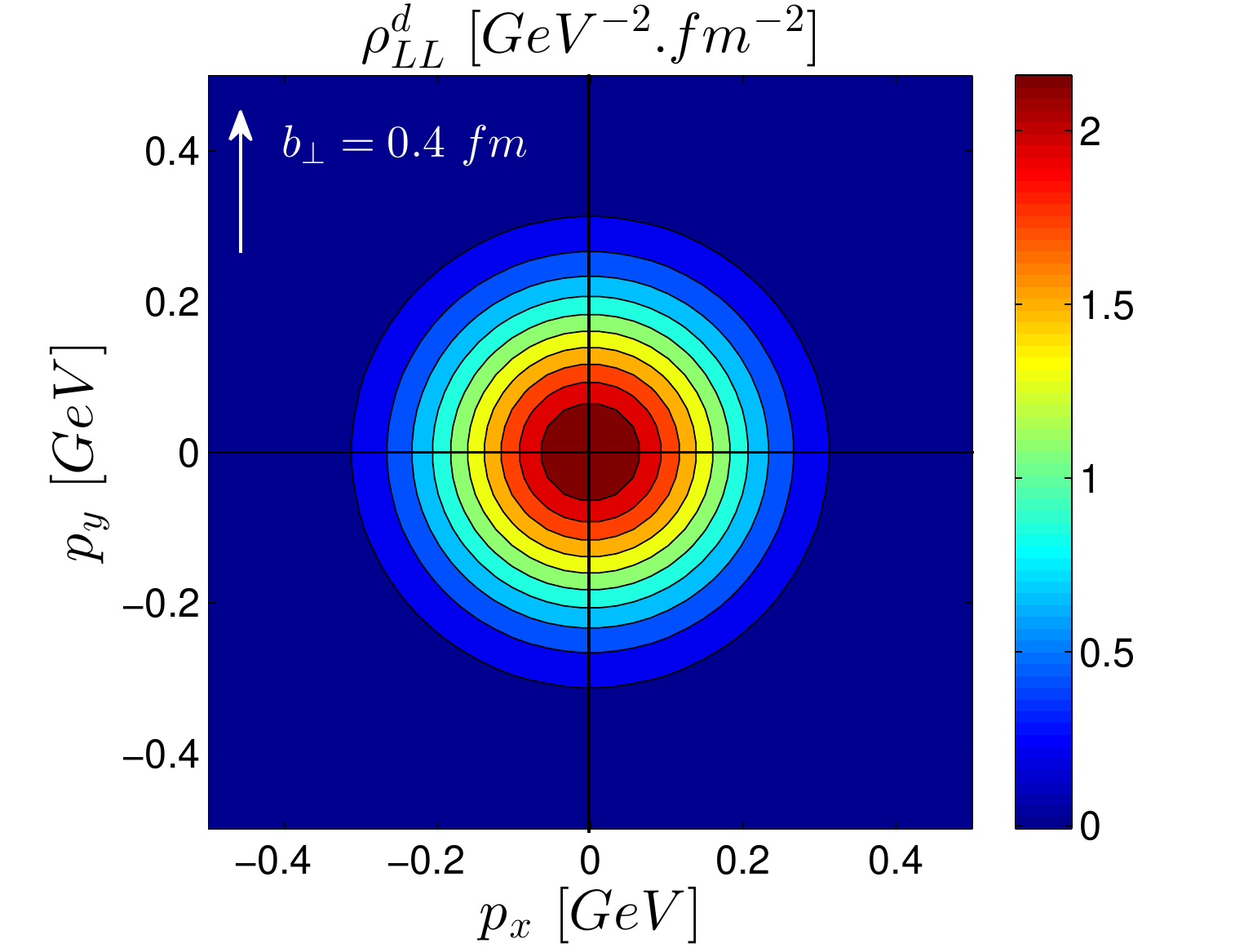}}\\
\subfigure[]{\includegraphics[width=5.cm,height=4.cm]{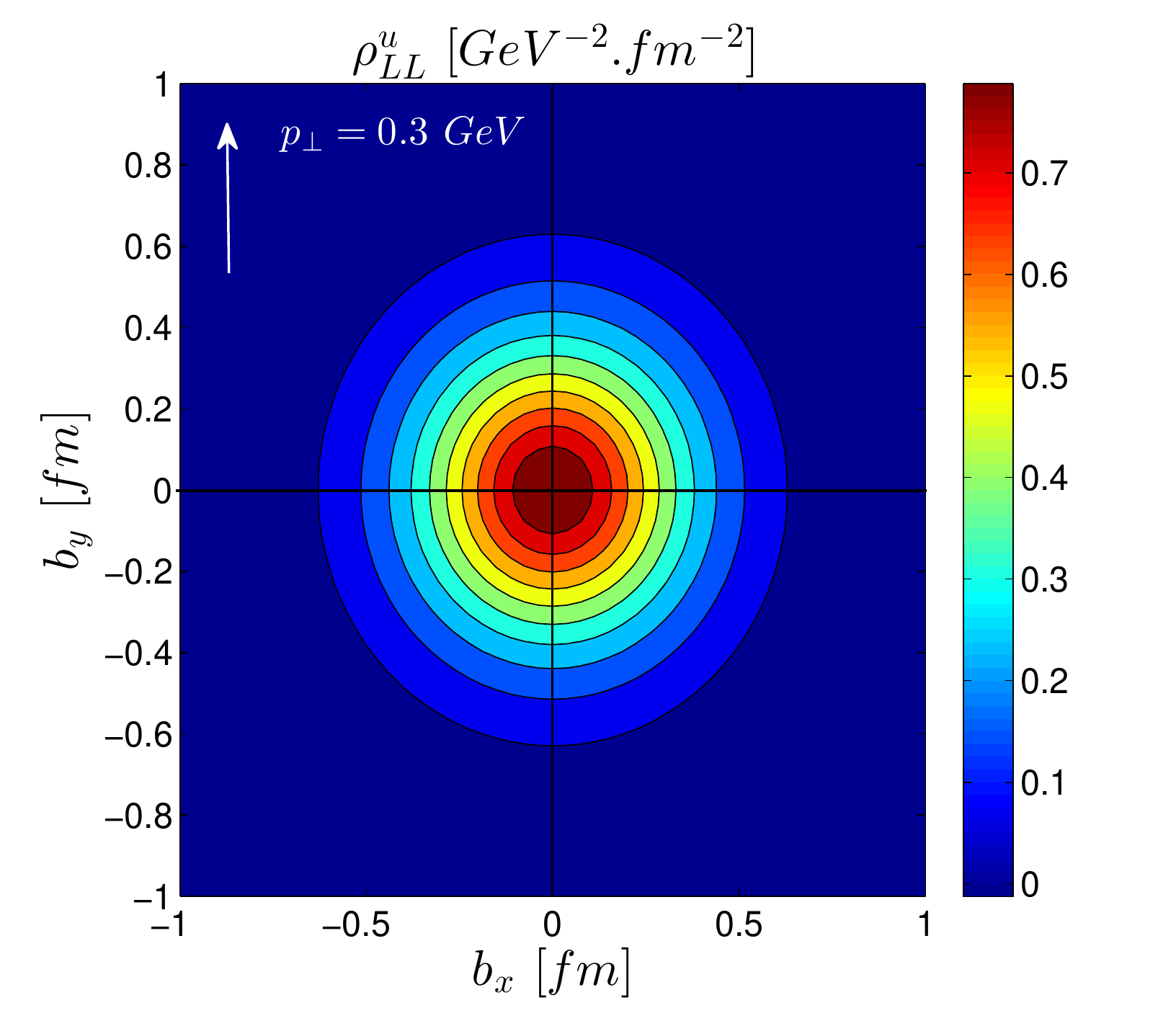}}
\subfigure[]{\includegraphics[width=5.cm,height=4.cm]{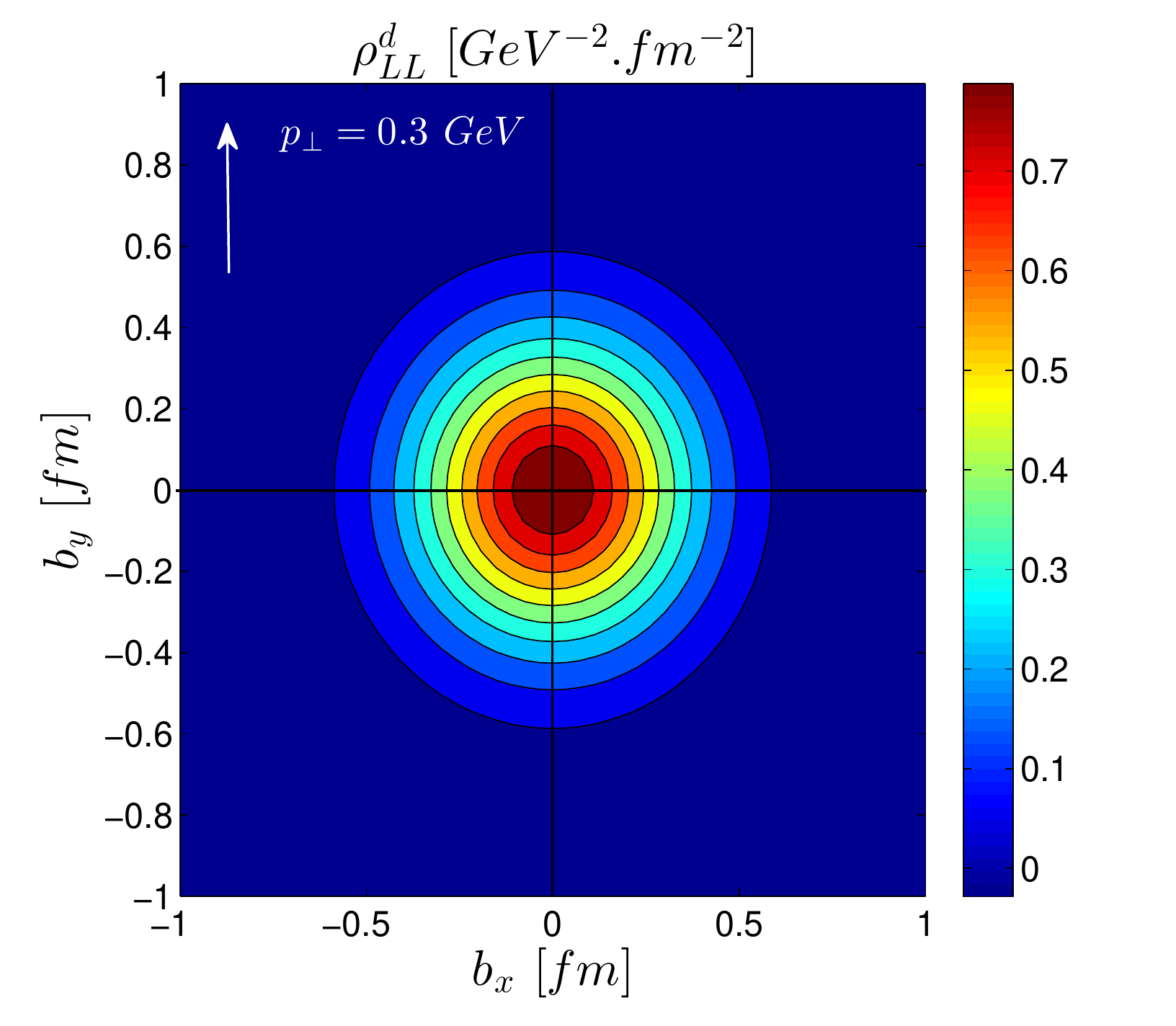}}\\
\subfigure[]{\includegraphics[width=5.cm,height=4.cm]{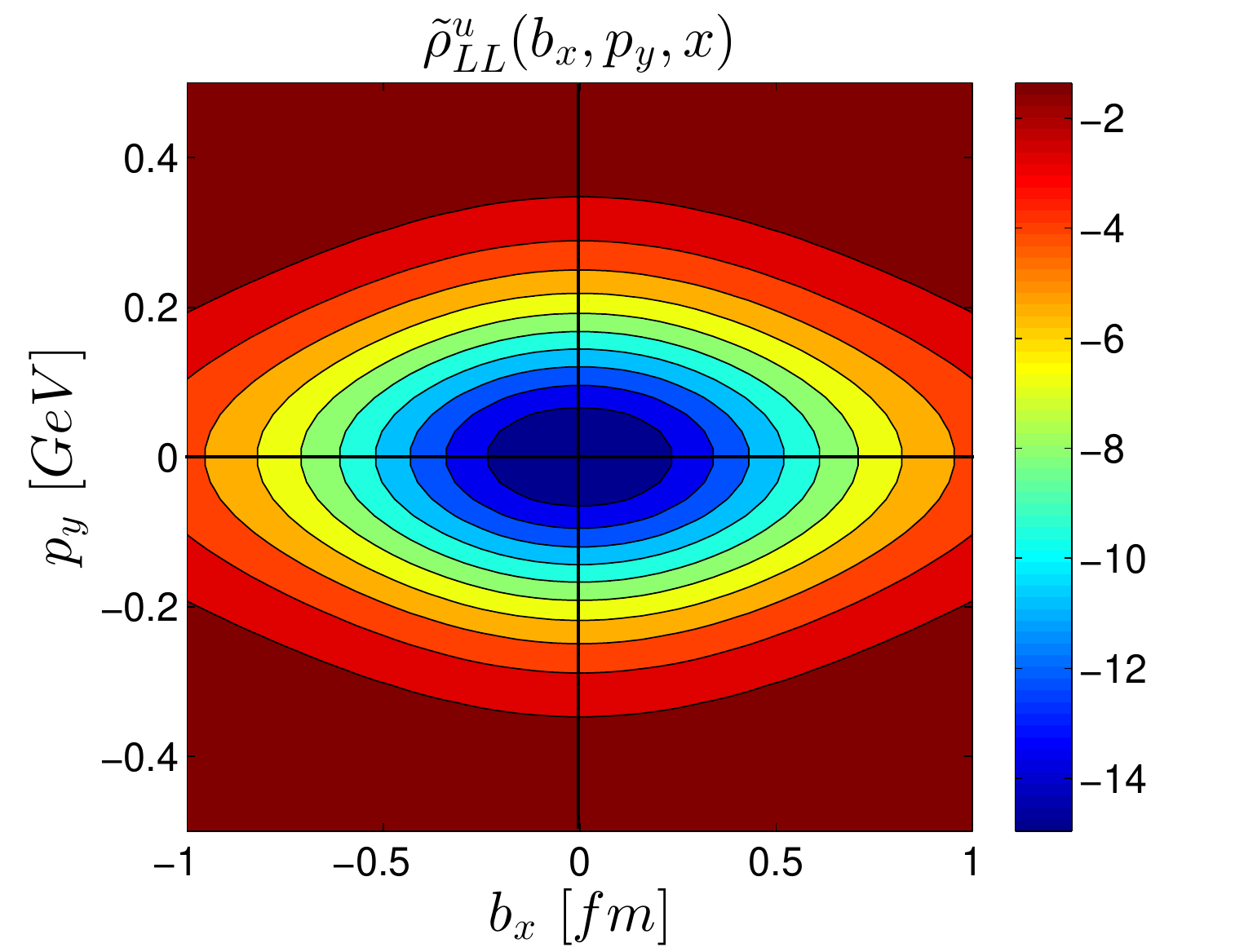}}
\subfigure[]{\includegraphics[width=5.cm,height=4.cm]{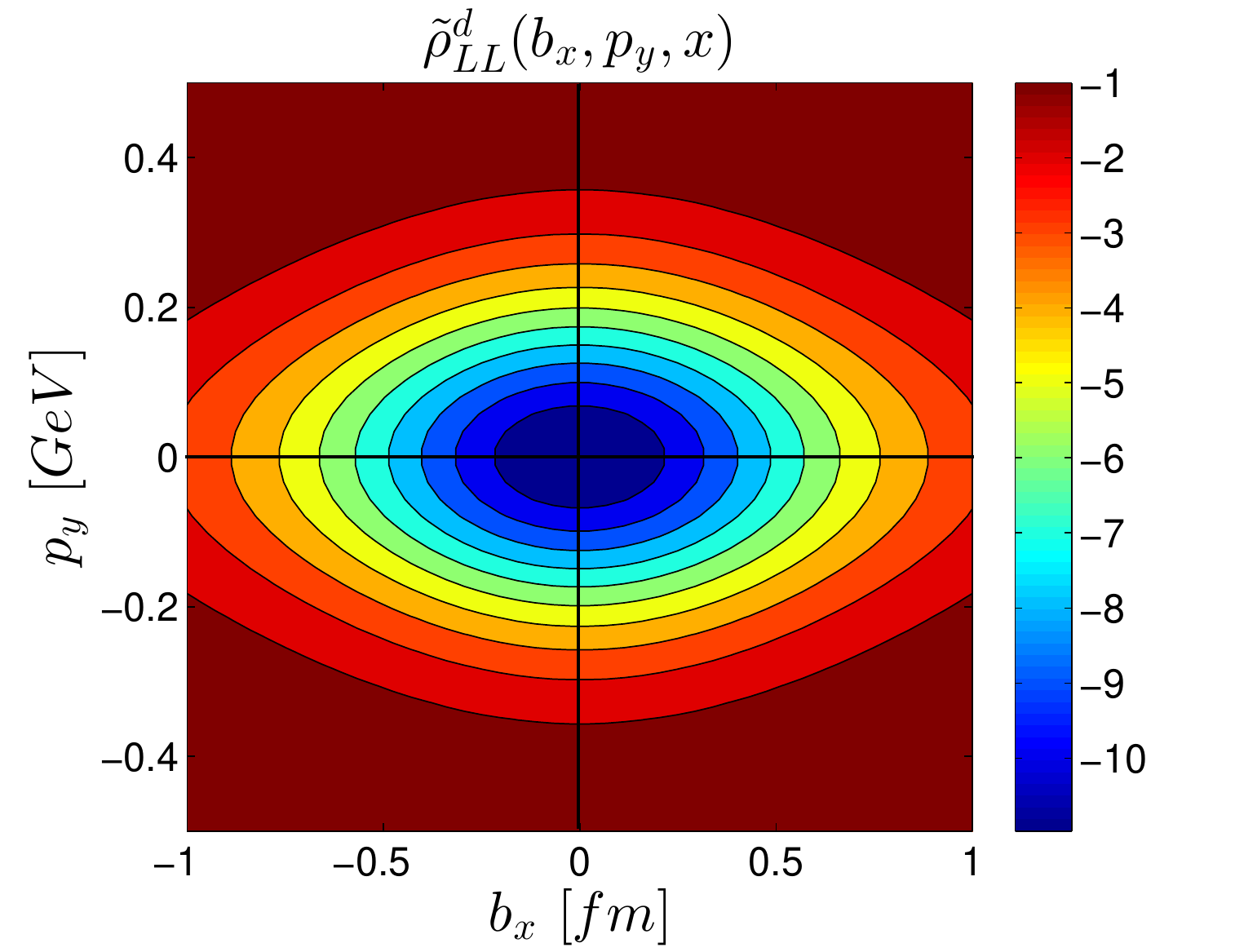}}
\caption{\label{fig_LL}The distributions $\rho_{LL}$ are shown in the transverse momentum plane, transverse coordinate plane and in the mixed plane for $u$ and $d$  quarks. The distributions in the mixed planes are given in $GeV^0 fm^0$.}
\end{figure}
\begin{figure}[ht]
\centering
\subfigure[]{\includegraphics[width=5.cm,height=4.cm]{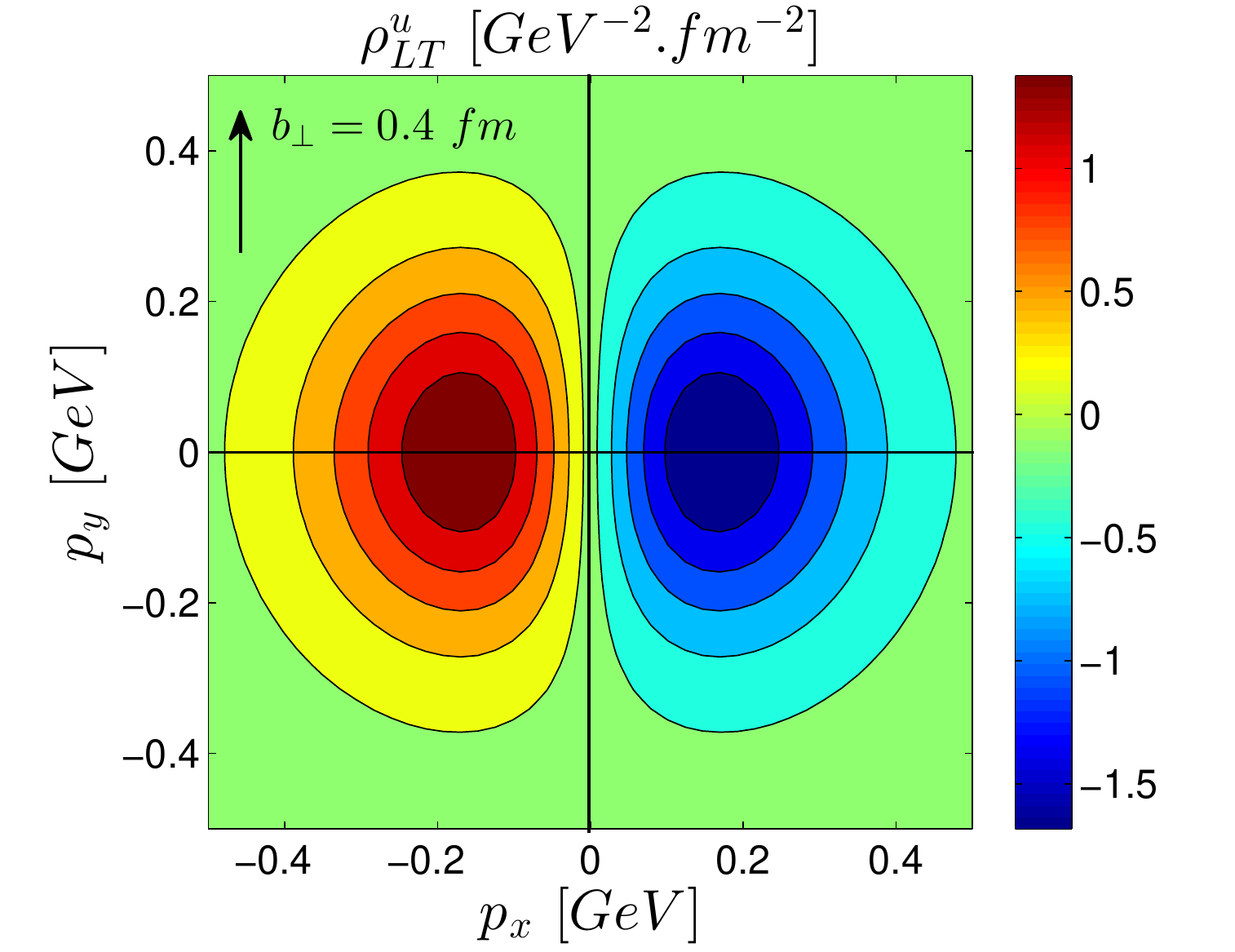}}
\subfigure[]{\includegraphics[width=5.cm,height=4.cm]{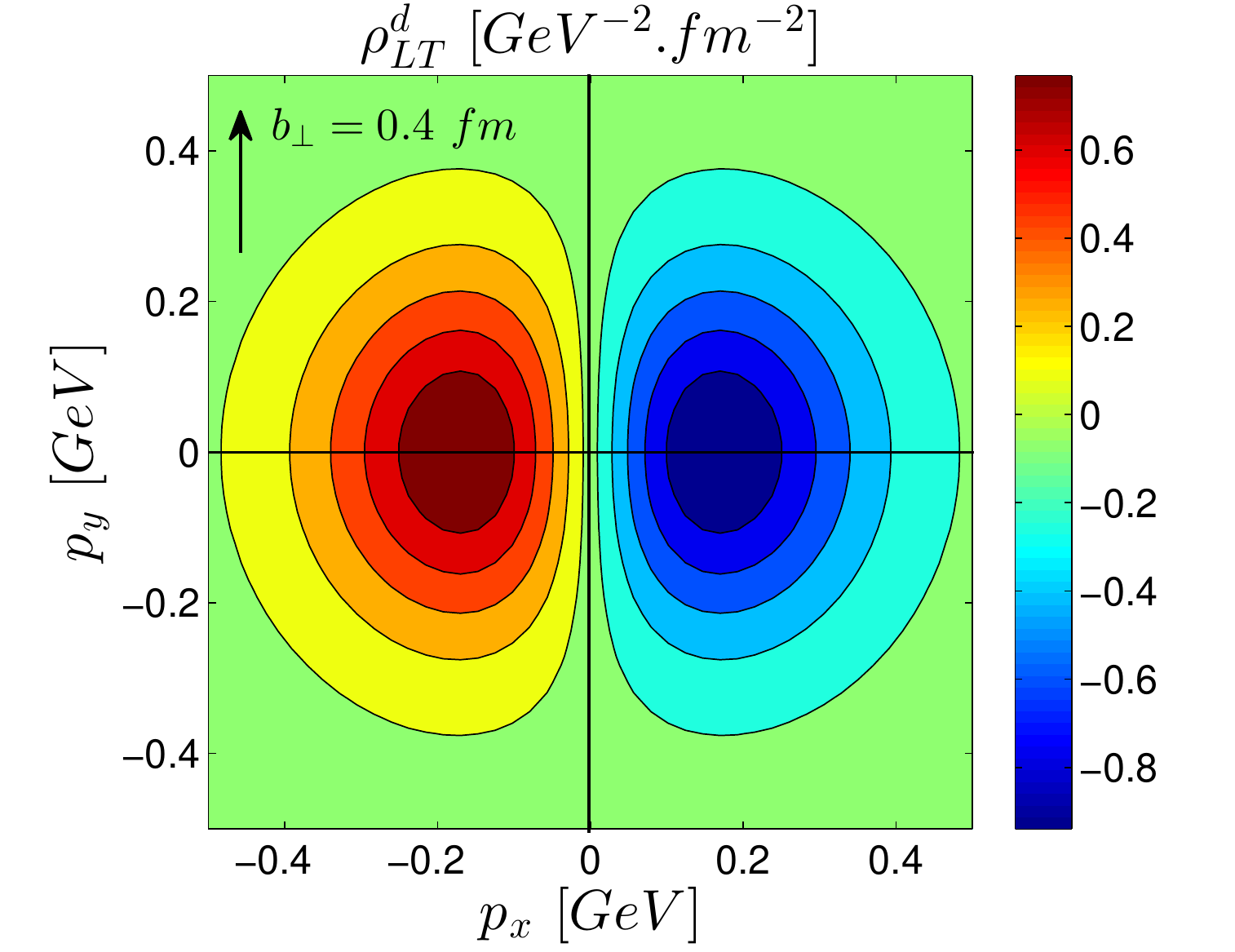}}\\
\subfigure[]{\includegraphics[width=5.cm,height=4.cm]{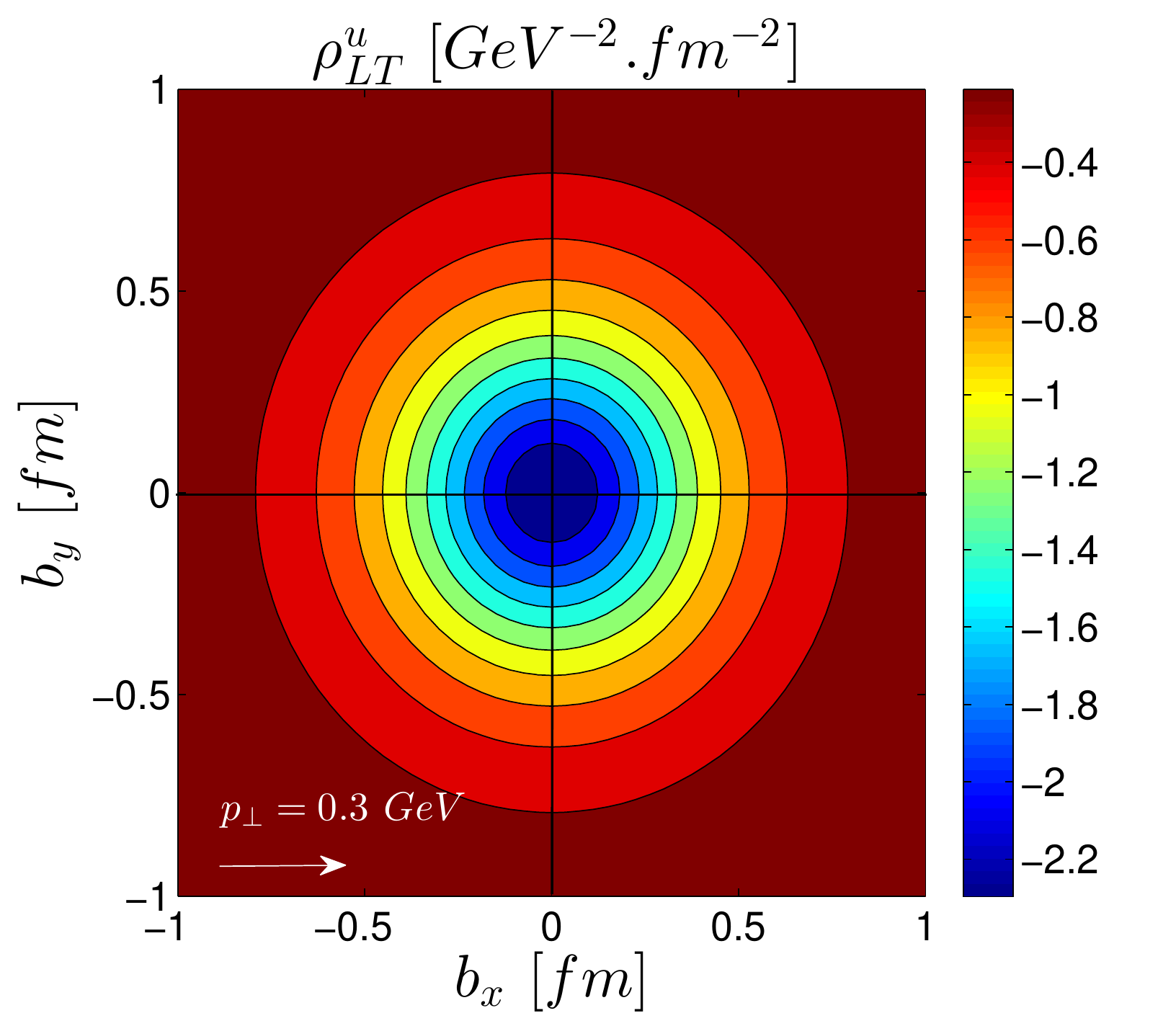}}
\subfigure[]{\includegraphics[width=5.cm,height=4.cm]{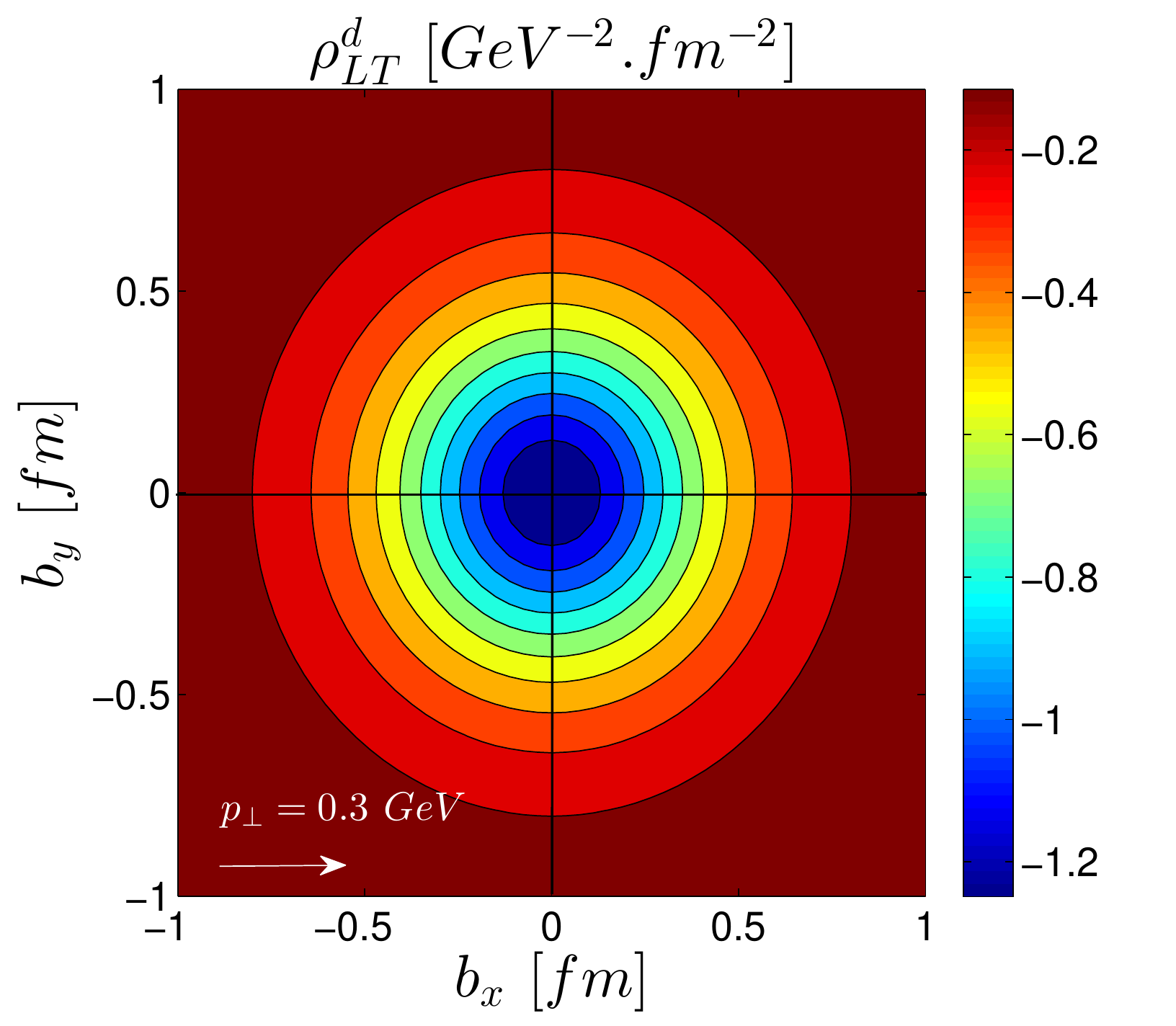}}\\
\subfigure[]{\includegraphics[width=5.cm,height=4.cm]{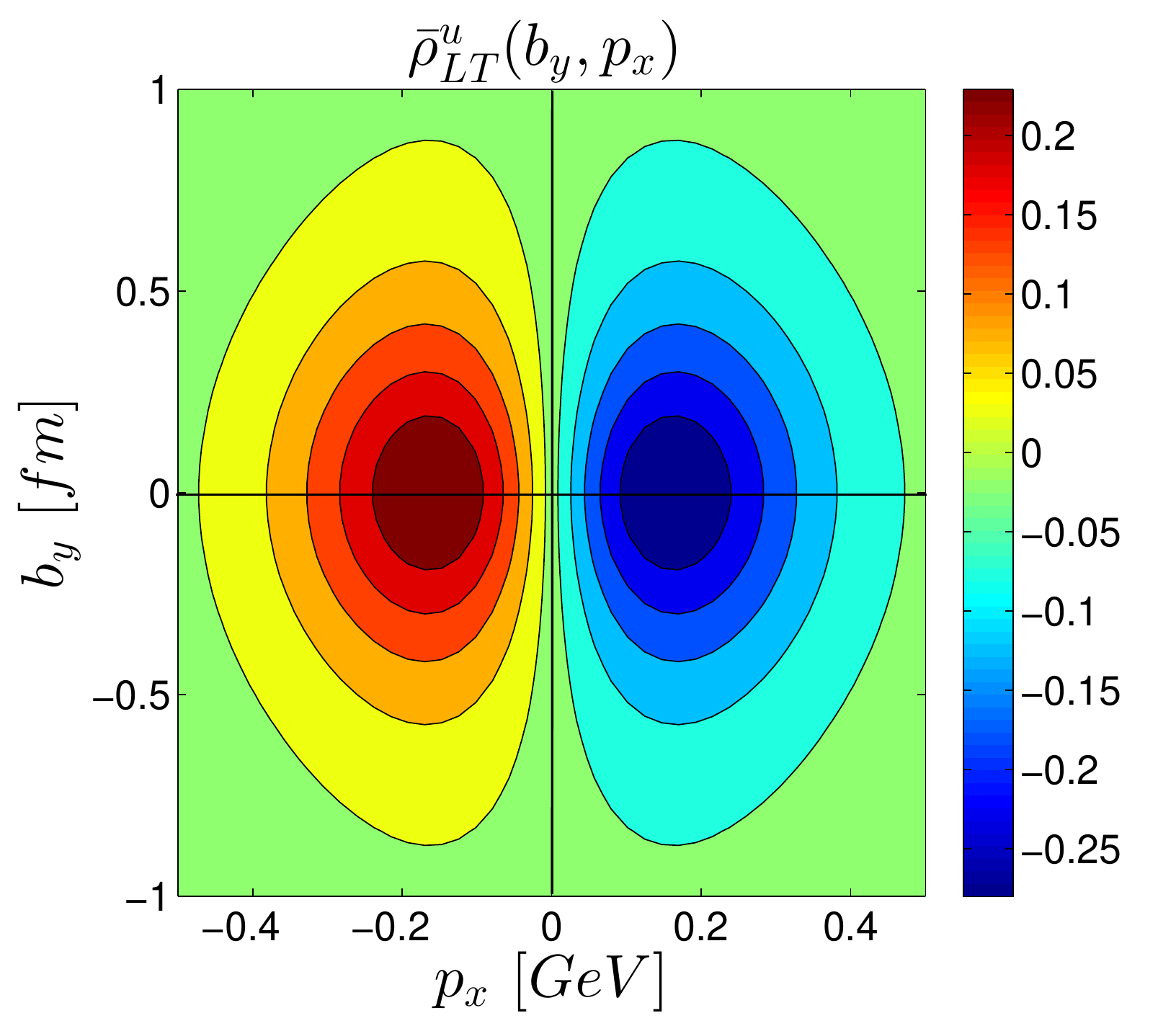}}
\subfigure[]{\includegraphics[width=5.cm,height=4.cm]{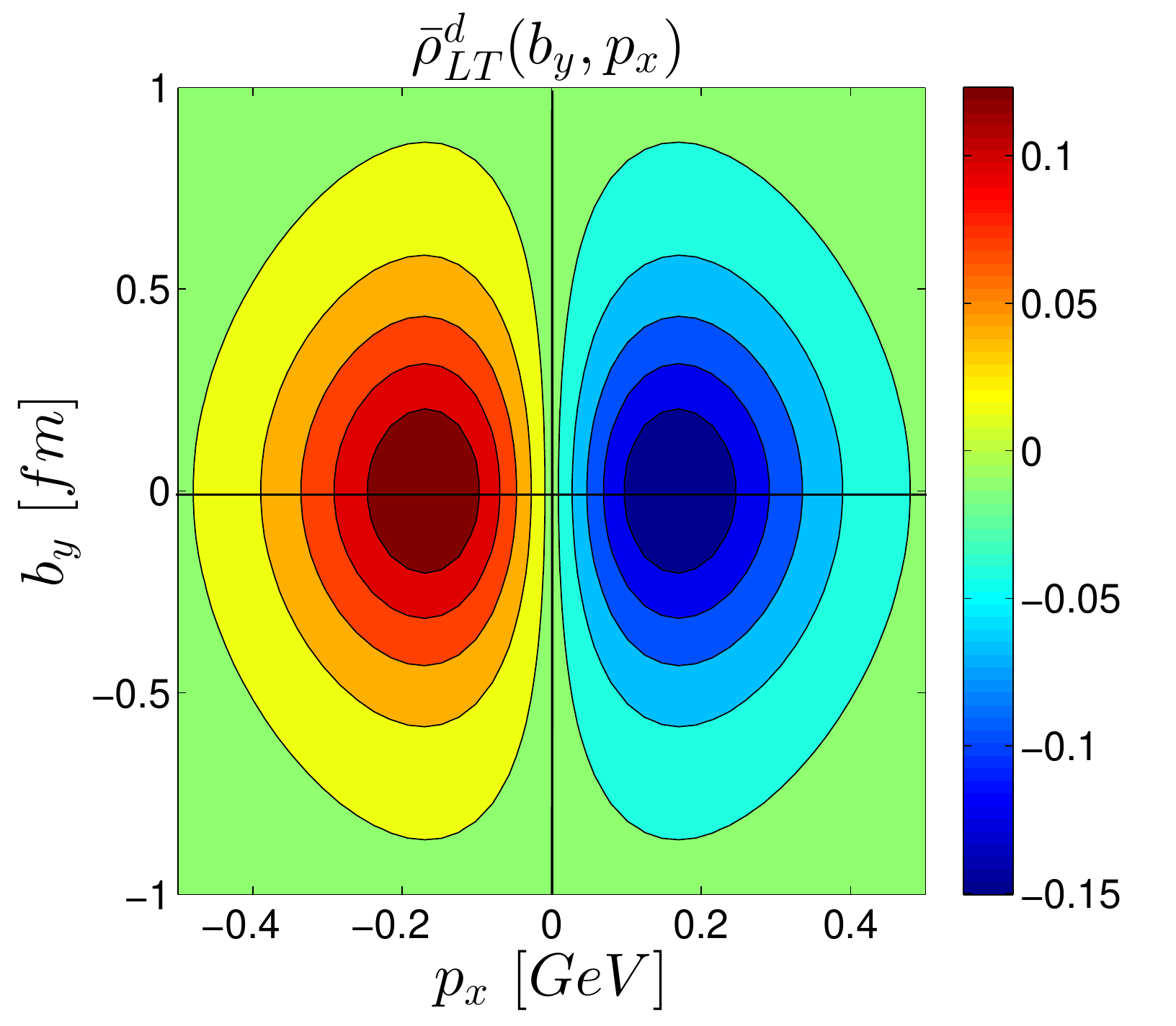}}
\caption{\label{fig_LT}The distributions $\rho_{LT}$ are shown in the transverse momentum plane(a,d) with $\bfb=0.4~\hat{y}~fm$,  in the transverse coordinate plane(b,e) with $\bfp=0.3~\hat{x}~GeV$ and  in mixed plane(c,f) for $u$ and $d$  quarks. The quark is polarised along x-axis. The distributions in the mixed planes are given in $GeV^0 fm^0$. }
\end{figure}

%%%%%%%%%%%%%%%%%%%%%%%%%%%%%%%%%%%%%%%%%%%%
\subsection{Longitudinally polarized proton}\label{longpol}
%%%%%%%%%%%%%%%%%%%%%%%%%%%%%%%%%%%%%%%%%%%%
In Fig. \ref{fig_LU}, we show the longitudinal-unpolarized Wigner distributions $\rho^\nu_{LU}(\bfb,\bfp)$ and mixing distributions $\tilde{\rho}^\nu_{LU}(b_x,p_y)$ which describe the unpolarized quark phase-space distributions in a longitudinal polarized proton. 
Fig.\ref{fig_LU}(a) and Fig.\ref{fig_LU}(d) display the variation of  $\rho^\nu_{LU}(\bfb,\bfp)$ in transverse momentum plane for $u$ and $d$ quarks respectively with fixed $\bfb$ along $\hat{y}$ and $b_y=0.4~fm$ and the variation of $\rho^\nu_{LU}(\bfb,\bfp)$ in transverse impact parameter plane are shown in Fig.\ref{fig_LU}(b) and Fig.\ref{fig_LU}(e) with fixed $\bfp$ along $\hat{y}$, $p_y=0.3~ GeV$. In this model,
The distributions $\rho^\nu_{LU}(\bfb,\bfp)$ are quite similar with $\rho^\nu_{UL}(\bfb,\bfp)$ in both transverse momentum as well as transverse impact parameter planes but the polarity of the dipolar structures of $\rho^\nu_{LU}$ is opposite to the polarity of $\rho^\nu_{UL}$. Again, the quadrupole structures appear when we plot the distribution in the transverse mixed plane as shown in Fig.\ref{fig_LU}(c) and Fig.\ref{fig_LU}(f) for $u$ and $d$ quark respectively which are very similar to $\tilde{\rho}^\nu_{UL}(b_x,p_y)$ with opposite sign. These distributions essentially reflect the correlations between quark OAM and proton spin. 
 In this model, the quark OAM $\ell^u_z=0.49$ for $u$ quark and $\ell^d_z=0.58$ for $d$  quark at $\mu^2=1~ GeV^2$.
Therefore quark OAM is parallel to proton spin for both $u$ and $d$ quarks. Note that also in scalar diquark model with AdS/QCD wave functions the OAMs are found to be positive for both quarks. This result is model dependent and may be due to the particular form of the AdS/QCD wave functions. 
%\begin{table}[ht]
%\centering % used for centering table 
%\begin{tabular}{|c | c c c|}
%    \hline Quark  &~~ $\ell_z^\nu$ ~~&~~ $s_z^\nu$ ~~&~~ $C_z^\nu$ \\ \hline
%     $u$ &  0.49 & 0.35 & -0.55\\
%     $d$ &  0.58 & -0.27 & -0.75\\ \hline
%\end{tabular} 
%\caption{ OAM $\ell^\nu_z$, quark spin $s^\nu_z$ and $C^\nu_z$  for $u$ and $d$ quark at scale $\mu^2=1~GeV^2$.}
%\label{tab_OAM} 
%\end{table}

The longitudinal-longitudinal Wigner distributions $\rho^\nu_{LL}(\bfb,\bfp)$ and mixing distributions $\tilde{\rho}^\nu_{LL}(b_x,p_y)$ are presented in Fig.\ref{fig_LL}. These Wigner distributions describe the phase-space distributions of longitudinal polarized quark in a longitudinal polarized proton, and after integrating over transverse variables they correspond to the axial charge ($\Delta q$) which is positive for $u$ quark but negative for $d$ quark at large scales.
The distributions $\rho^\nu_{LL}(\bfb,\bfp)$ in transverse momentum plane for $u$ and $d$ quark are plotted in Fig.\ref{fig_LL}(a) and Fig.\ref{fig_LL}(d) respectively whereas $\rho^\nu_{LL}(\bfb,\bfp)$ in transverse coordinate plane are shown in Fig.\ref{fig_LL}(b) and Fig.\ref{fig_LL}(e). In this model we find that in two planes the distributions are positive for $u$ quark in consistence with the sign of $\Delta u$ but for $d$ quark,  the distributions are also positive whereas  the axial charge $\Delta d$ is known to be negative. One should note that the axial charges are  highly scale dependent and are measured only at high energies whereas  the model here have a very low initial scale $\mu_0 = 0.313 ~ GeV$. So, we need to consider the scale evolution of the distributions before comparing with the measured data.  For $d$ quark, the axial charge is known to be negative at larger scales. 
%In this model, as we evolve the scale\cite{TM_VD} the axial charge becomes negative for $d$  quark. 
The scale evolutions of axial charges in this model are shown in \cite{MC}. Where it is shown  that the axial charge for $d$ quark becomes negative for $\mu^2 \geq 0.15~ GeV^2 $. At $\mu^2=1~GeV^2$ the axial charges for the quarks are found to be $g^u_A=0.73$ and $g^d_A=-0.54$ which are consistent with the measured data.
 The distributions are circularly symmetric for $u$ and $d$ quarks in both the planes and they are more concentrated in the center in $\bfb$ plane relative to $\bfp$ plane. The peaks of the distributions in $\bfp$ plane are larger than that in $\bfb$ plane. The mixing distributions $\tilde{\rho}^\nu_{LL}(b_x,p_y)$ for $u$ and $d$ quark are shown in Fig.\ref{fig_LL}(c) and Fig.\ref{fig_LL}(f) respectively. They are axially symmetric in mixed plane. $\tilde{\rho}^\nu_{LL}(b_x,p_y)$ show quite similar behavior of $\tilde{\rho}^\nu_{UU}(b_x,p_y)$ but with opposite sign and with much lower peak at the center.
%\TM{The quark OAM and quark spin are given in Table.\ref{tab_OAM_SV} for isoscalar-calar(S), isoscalar-vector(V) and  and isovector-vector(VV) diquarks. Therefore from Eq.(\ref{Eq_Sum_S},\ref{Eq_Sum_V}), the diquark contribution to the OAM are $\ell^{(S)}_D=-0.31$ and  $\ell^{(A)}_D=-0.93$ for spin-0 and spin-1 diquark respectively. }
%\begin{table}[ht]
%\centering % used for centering table 
%\begin{tabular}{| c c c c c c |}
%    \hline $\ell^{(S)}_q$  &~~ $\ell^{(V)}_q$ ~~&~~ $\ell^{(VV)}_q$ ~~&~~ $s^{(S)}_q$  &~~ $s^{(V)}_q$ ~~&~~ $s^{(VV)}_q$  \\ \hline
%     0.33 &  0.16 & 0.58 & 0.48 & 0.23 & -0.54 \\ \hline 
%\end{tabular} 
%\caption{ OAM $\ell_q$ and quark spin $s_q$ corresponding to the scalar-isoscalar diquark(S), vector-icoscalar diquark(V) and vector-isovector diquark(VV) at scale $\mu^2=1~GeV^2$.}
%\label{tab_OAM_SV} 
%\end{table}

%+++++++++++++++++++++++++
%\subsection{Longiwudinal-transverse WD}
%+++++++++++++++++++++++

The wigner distribution with a transversely polarized quark in a longitudinally polarized proton, $\rho_{LT}(\bfb,\bfp)$, is shown in fig.\ref{fig_LT}. The fig.\ref{fig_LT}(a) and (d) represent the distribution in transverse momentum plane, with $\bfb=0.4~\hat{y}~fm$, for $u$ and $d$  quark respectively. We see a dipolar distribution as expected from the Eq.(\ref{rhoLT_nu}). The fig.\ref{fig_LT}(b) and (e) show the distribution in transverse coordinate space with $\bfp=0.3~\hat{x}~GeV$ for $u$ and $d$  quarks respectively. The distribution is circularly symmetric with negative peak at the center of the coordinate space.
The distribution vanishes if the quark transverse momentum is perpendicular to the polarization. This reflects there is a strong correlation between the quark transverse momentum and quark transverse polarization.
The mixing distribution $\bar{\rho}_{LT}(b_y,p_x)$ is shown in fig.\ref{fig_LT}(c) and (f) for $u$ and $d$  quarks respectively. Because of the dipolar structure in transverse momentum plane, the other class of mixing distributions $\tilde{\rho}_{LT}(b_x,p_y)$ vanishes for the quark with a polarization along x-axis.

At the TMD limit, $\rho_{LT}(\bfb,\bfp,x)$ reduces to $h^\perp_{1L}(x,\bfp)$\cite{meissner09}, one of the eight T-even TMDs at leading twist. At the impact parameter distribution limit the distribution is related to the $H_T$ and $\tilde{H}_T$ GPD together with some other distributions.

%====================
%Transverse Part
%=====================
\begin{figure}[ht]
\centering
\subfigure[]{\includegraphics[width=5.cm,height=4.cm]{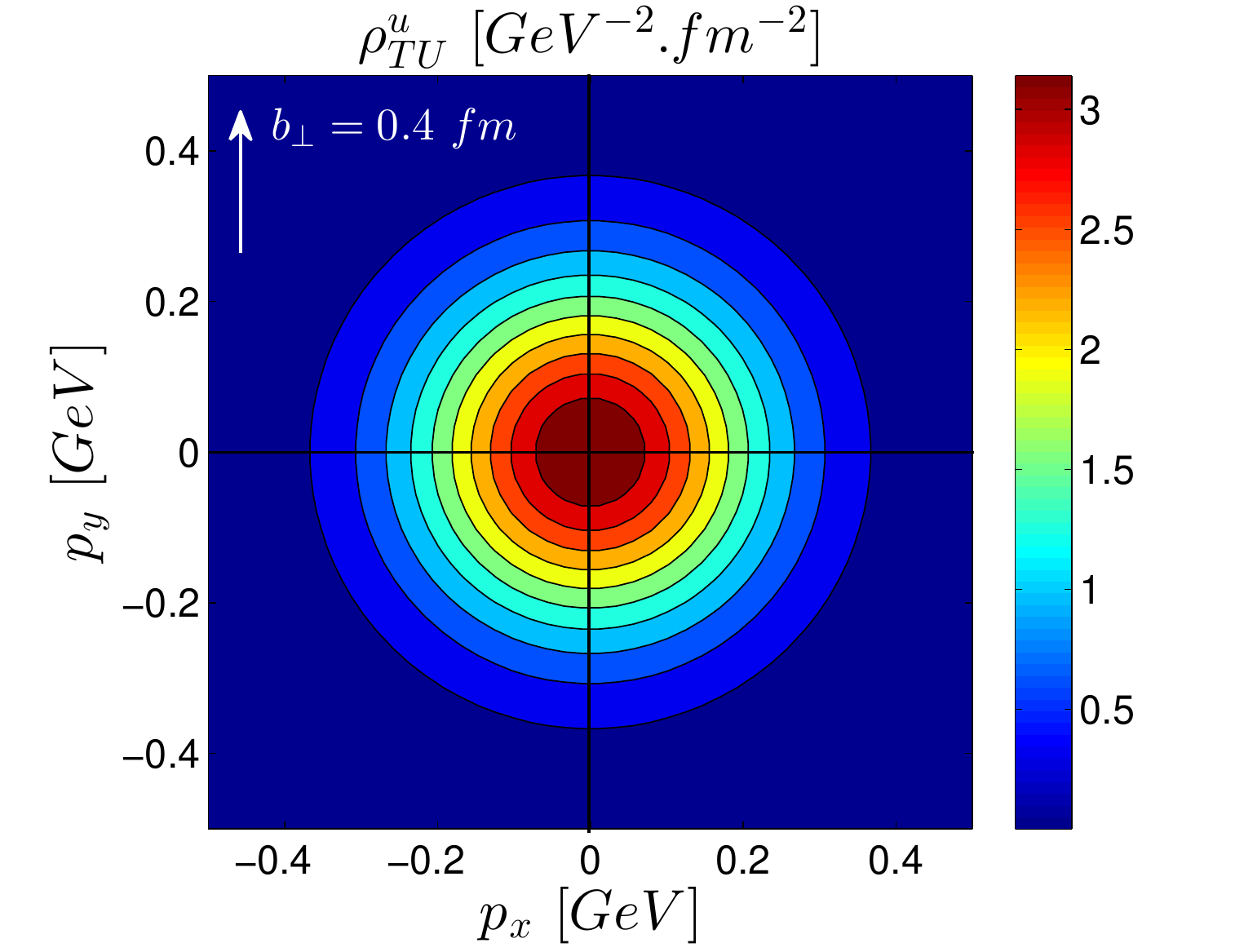}}
\subfigure[]{\includegraphics[width=5.cm,height=4.cm]{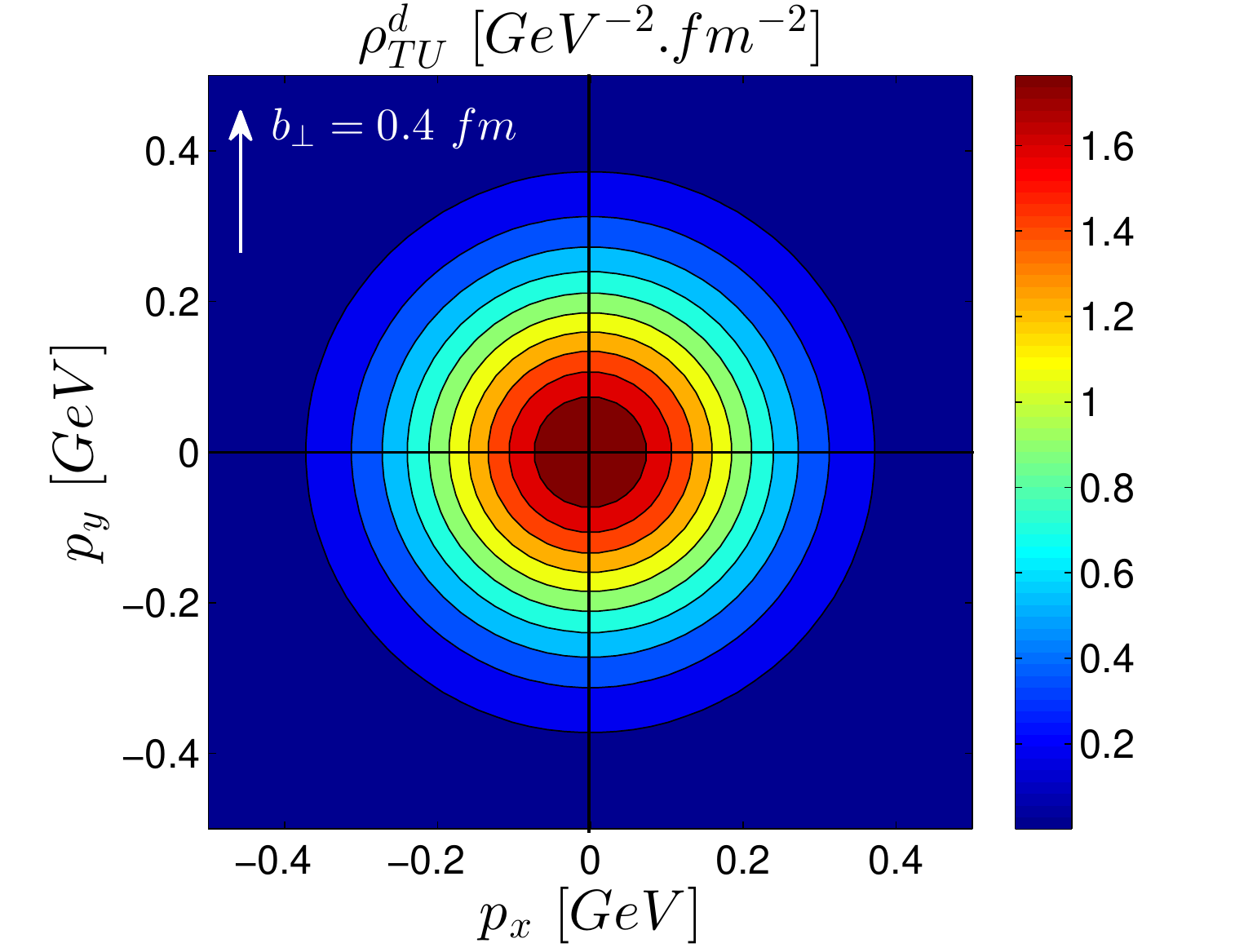}}\\
\subfigure[]{\includegraphics[width=5.cm,height=4.cm]{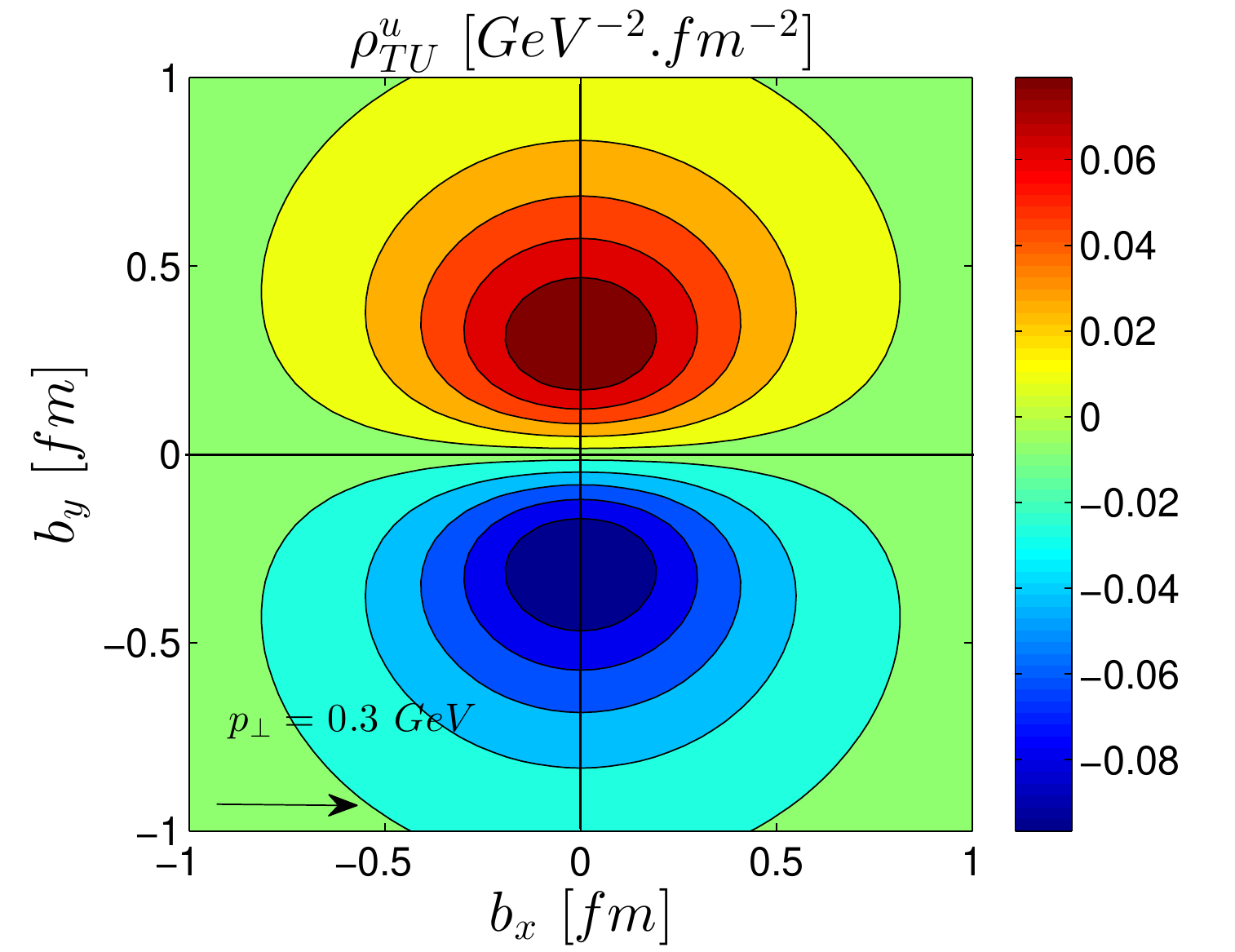}}
\subfigure[]{\includegraphics[width=5.cm,height=4.cm]{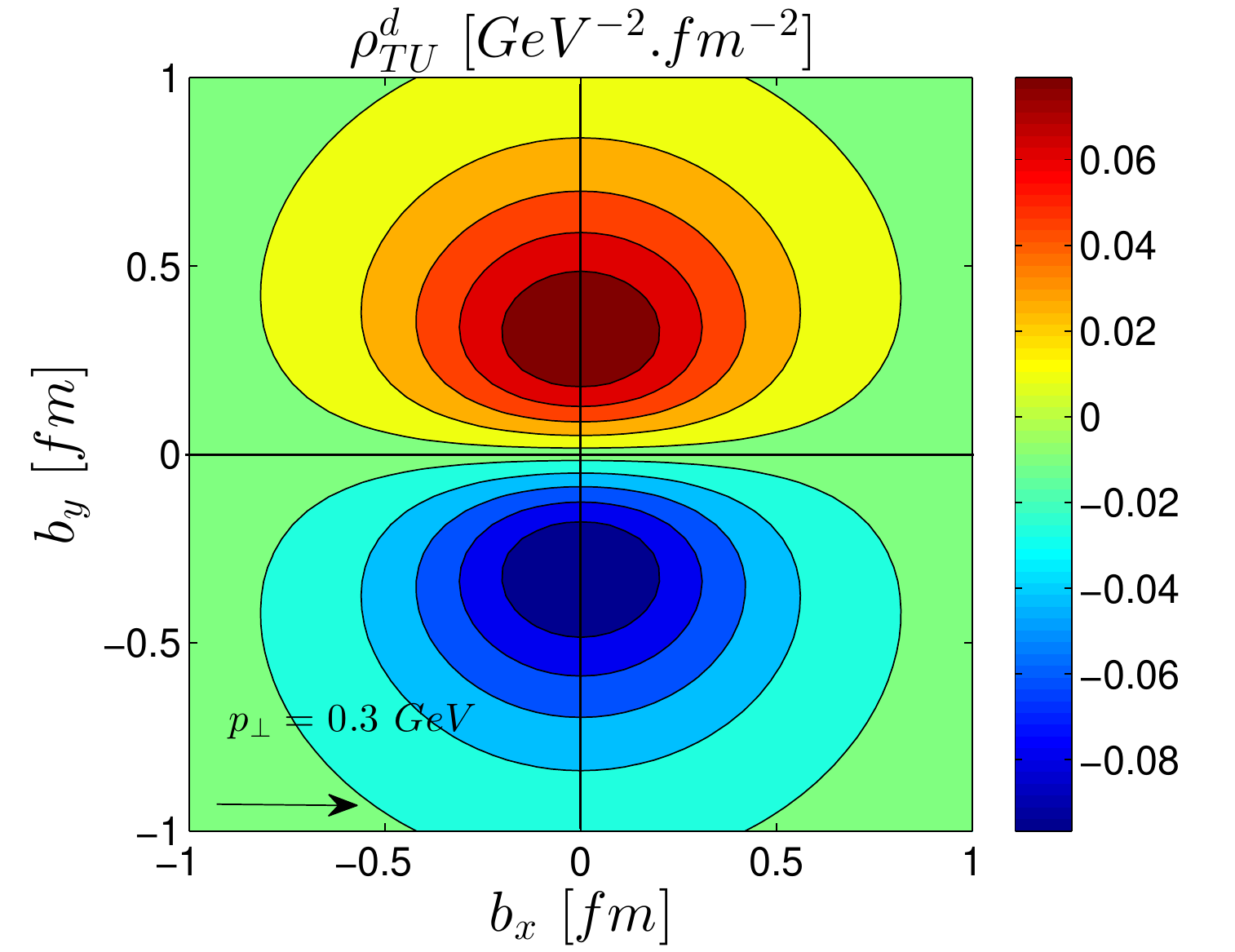}}\\
\subfigure[]{\includegraphics[width=5.cm,height=4.cm]{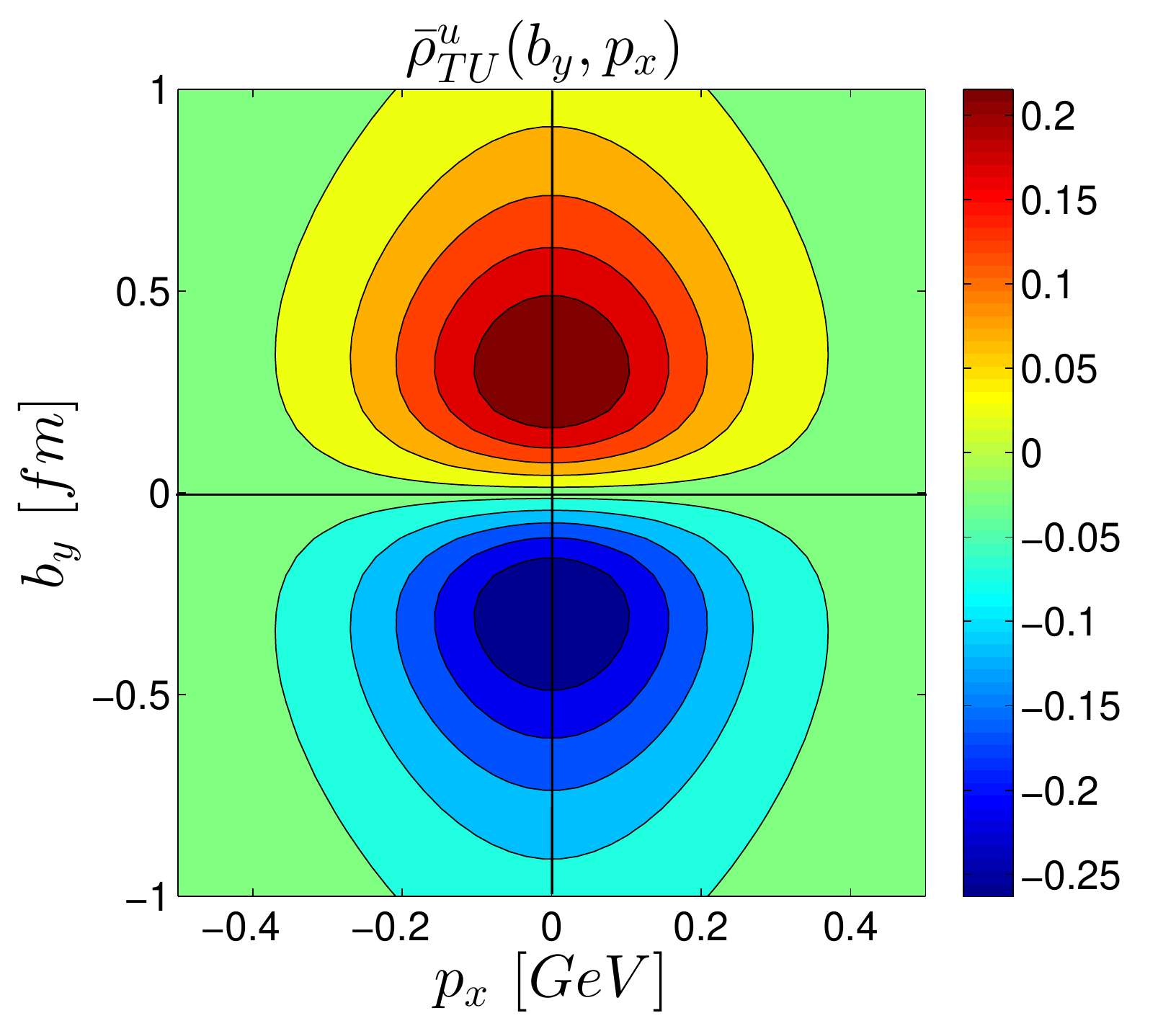}}
\subfigure[]{\includegraphics[width=5.cm,height=4.cm]{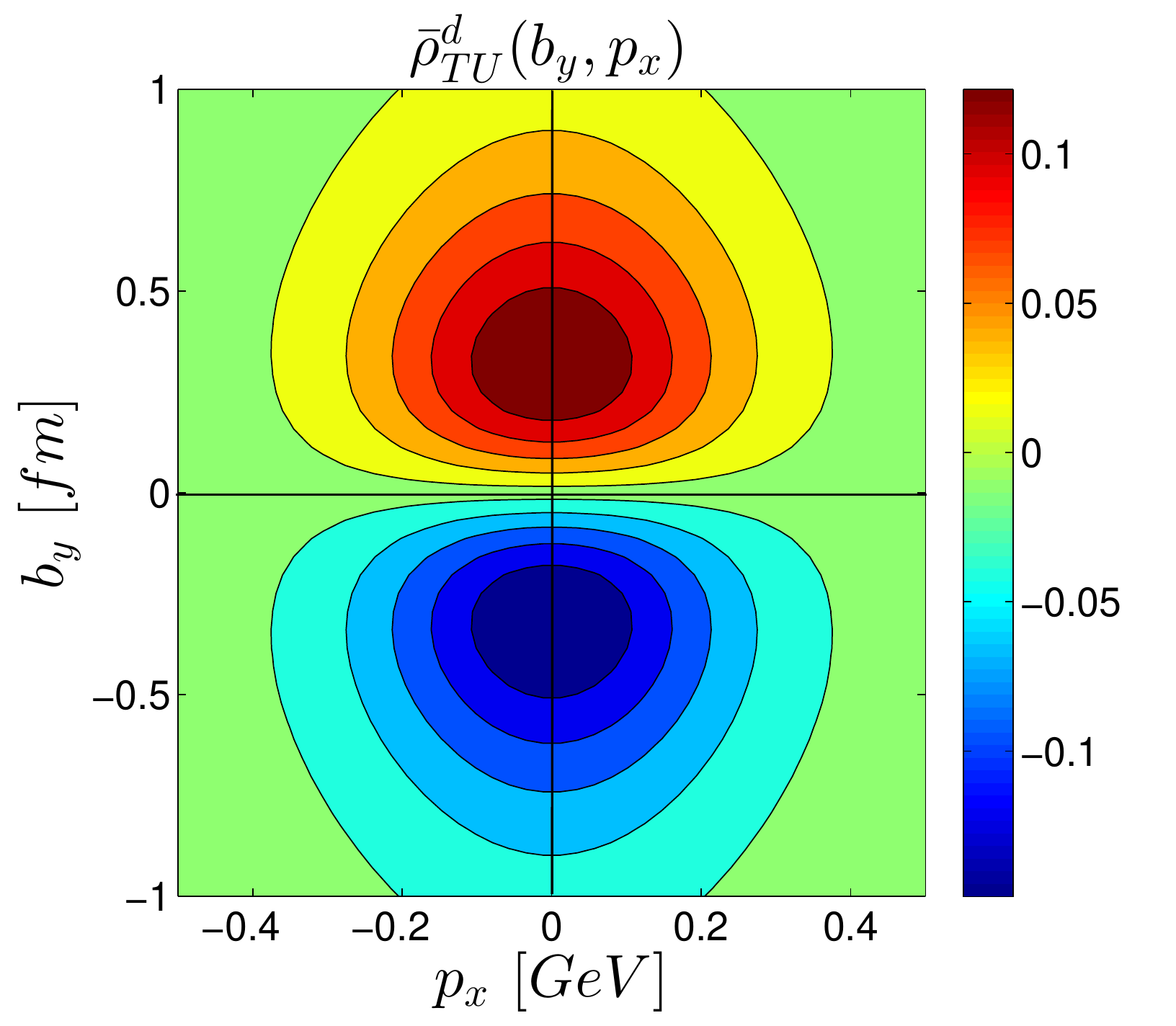}}
\caption{\label{fig_TU} The distributions $\rho_{TU}(\bfb,\bfp)$ are presented in the transverse momentum plane(a,d) with $\bfb=0.4~\hat{y}~fm$,  in the transverse coordinate plane(b,e) with $\bfp=0.3~\hat{x}~GeV$ and  in mixed plane(c,f) for $u$ and $d$  quarks. The distributions in the mixed planes are given in $GeV^0 fm^0$.}
\end{figure}
\begin{figure}[ht]
\centering
\subfigure[]{\includegraphics[width=5.cm,height=4.cm]{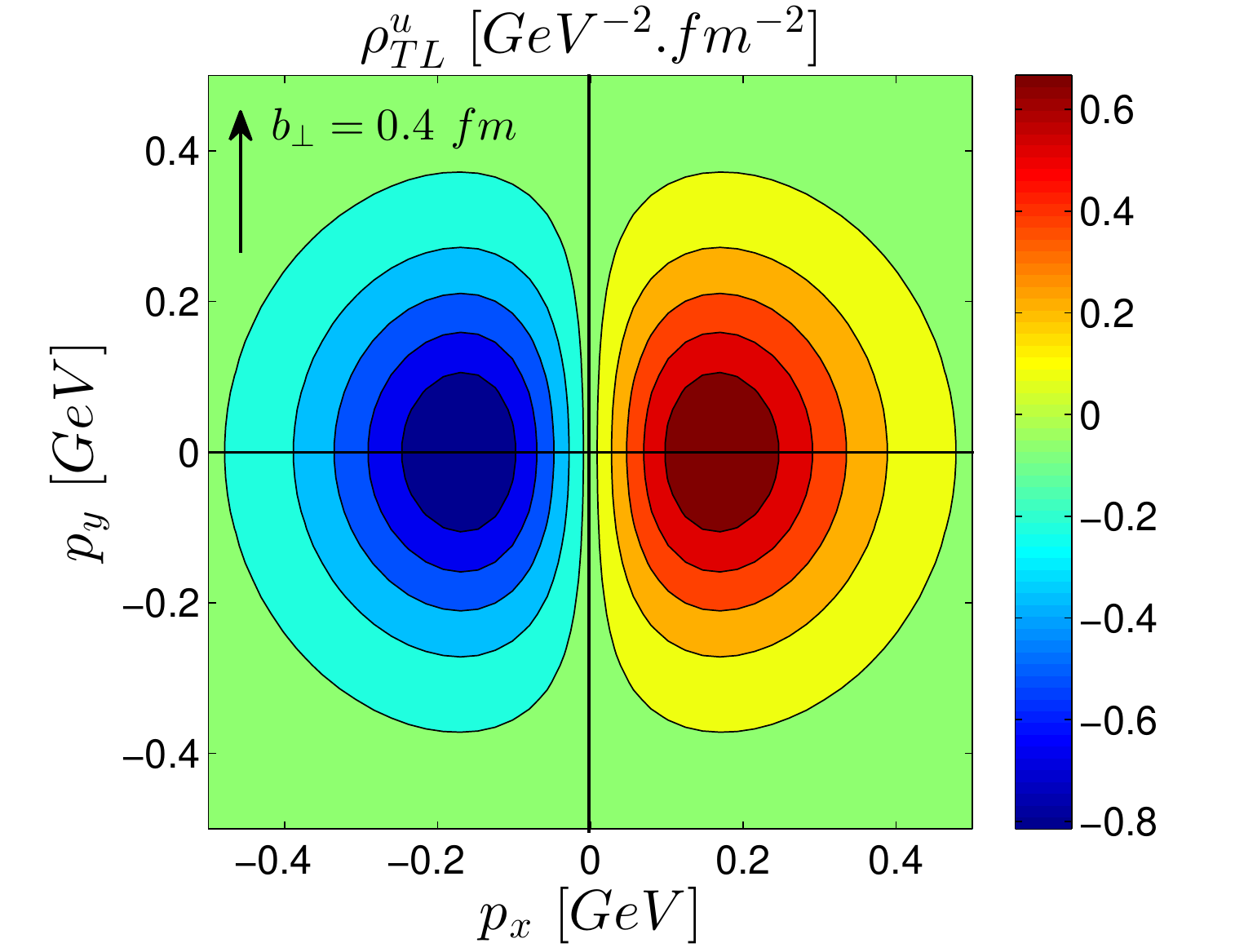}}
\subfigure[]{\includegraphics[width=5.cm,height=4.cm]{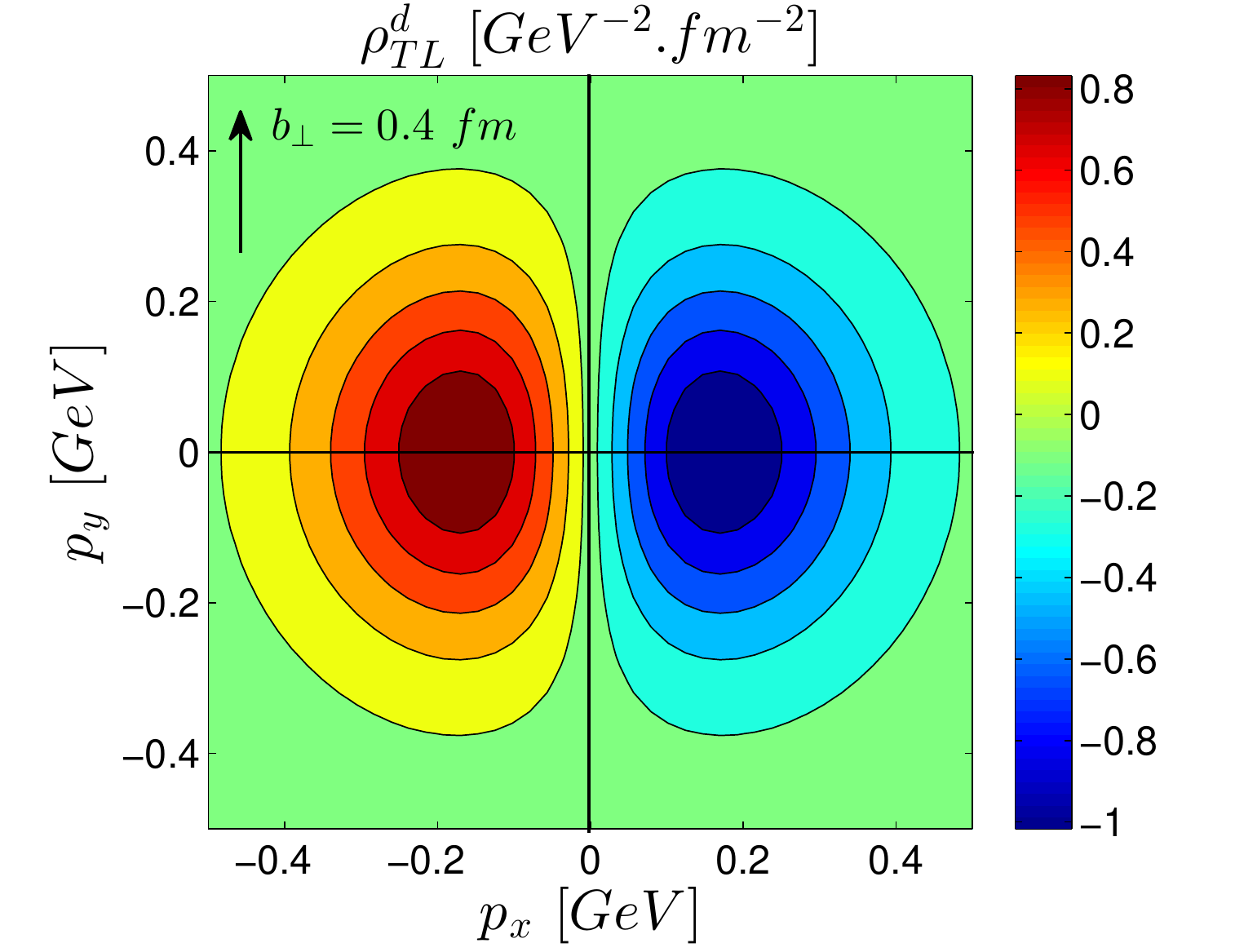}}\\
\subfigure[]{\includegraphics[width=5.cm,height=4.cm]{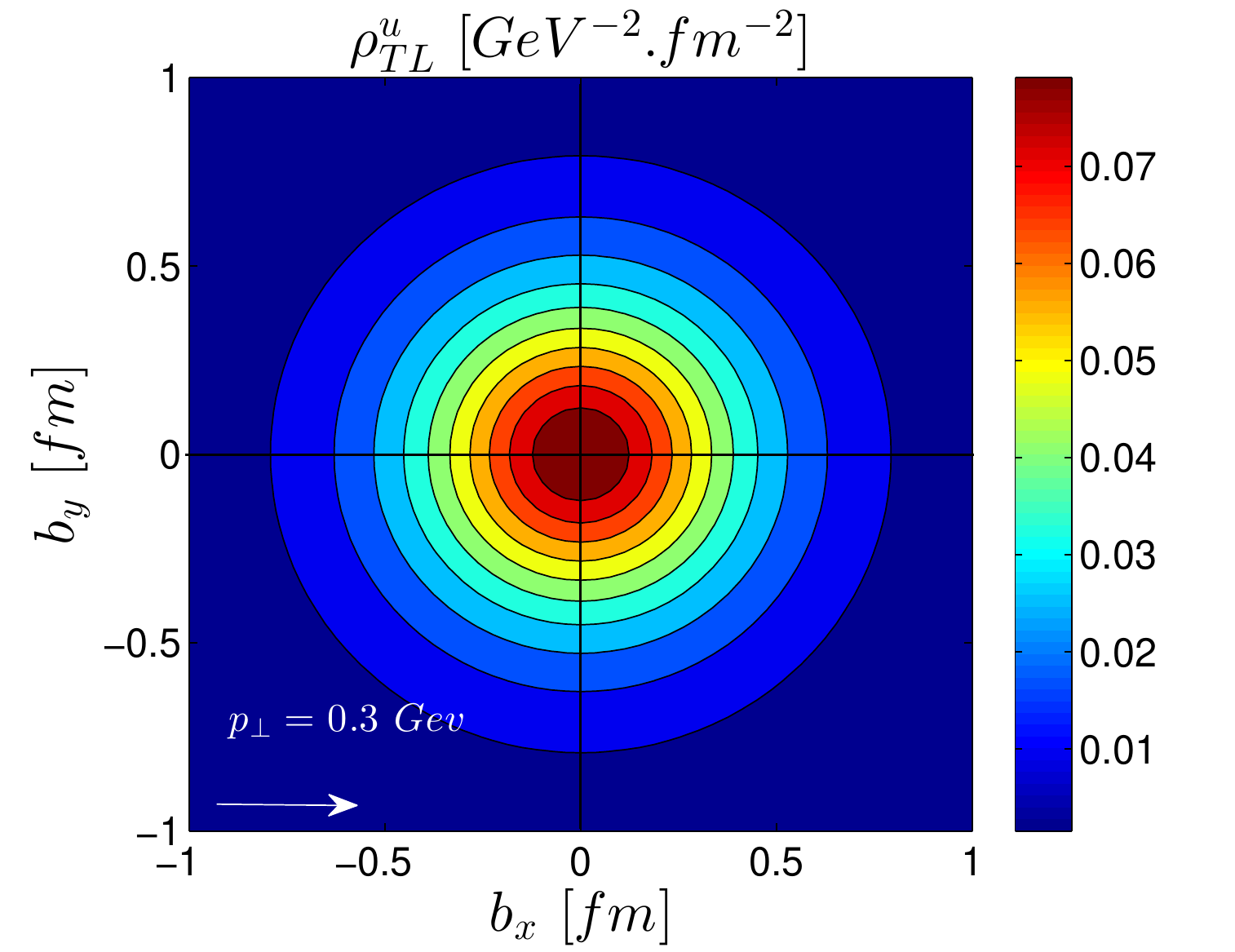}}
\subfigure[]{\includegraphics[width=5.cm,height=4.cm]{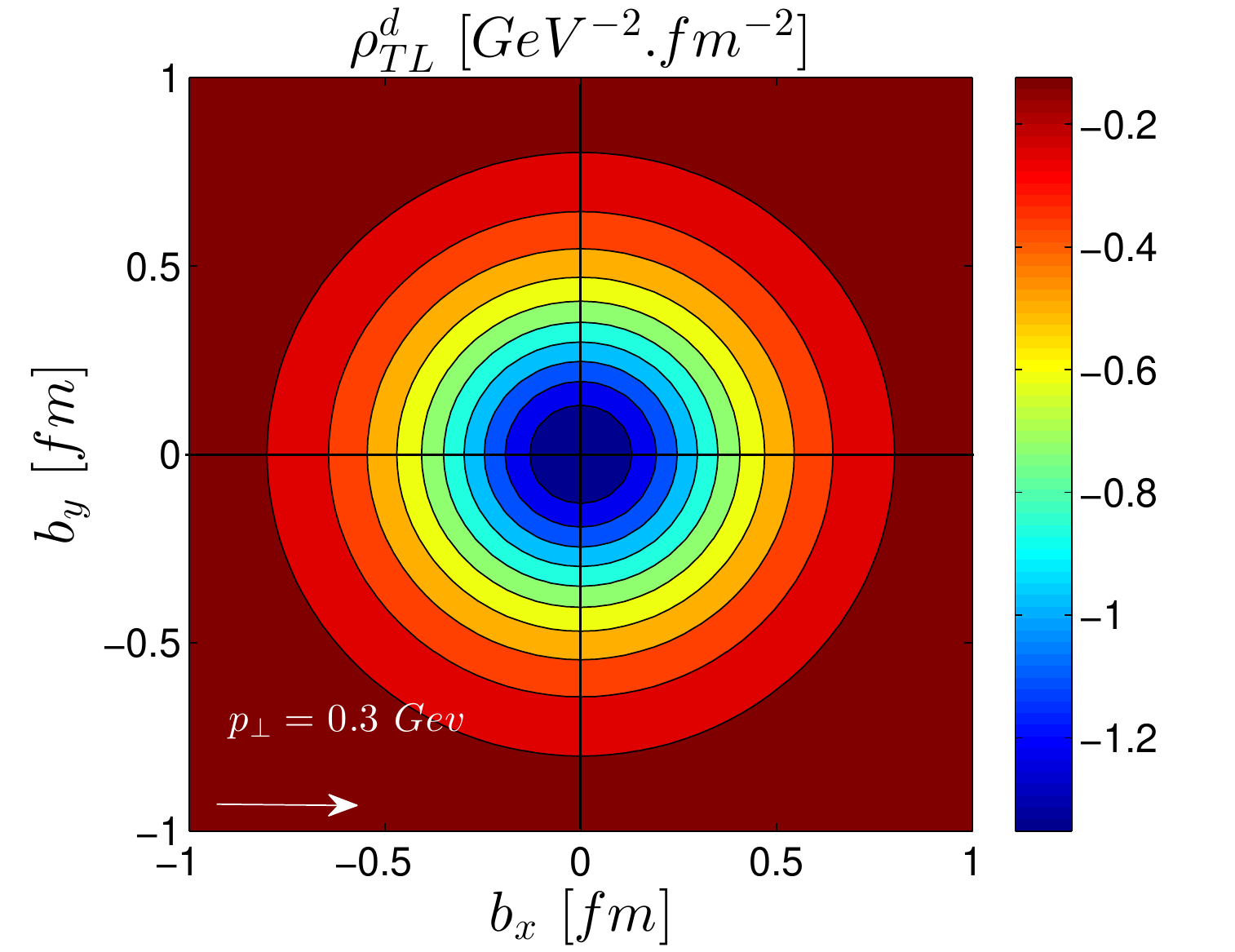}}\\
\subfigure[]{\includegraphics[width=5.cm,height=4.cm]{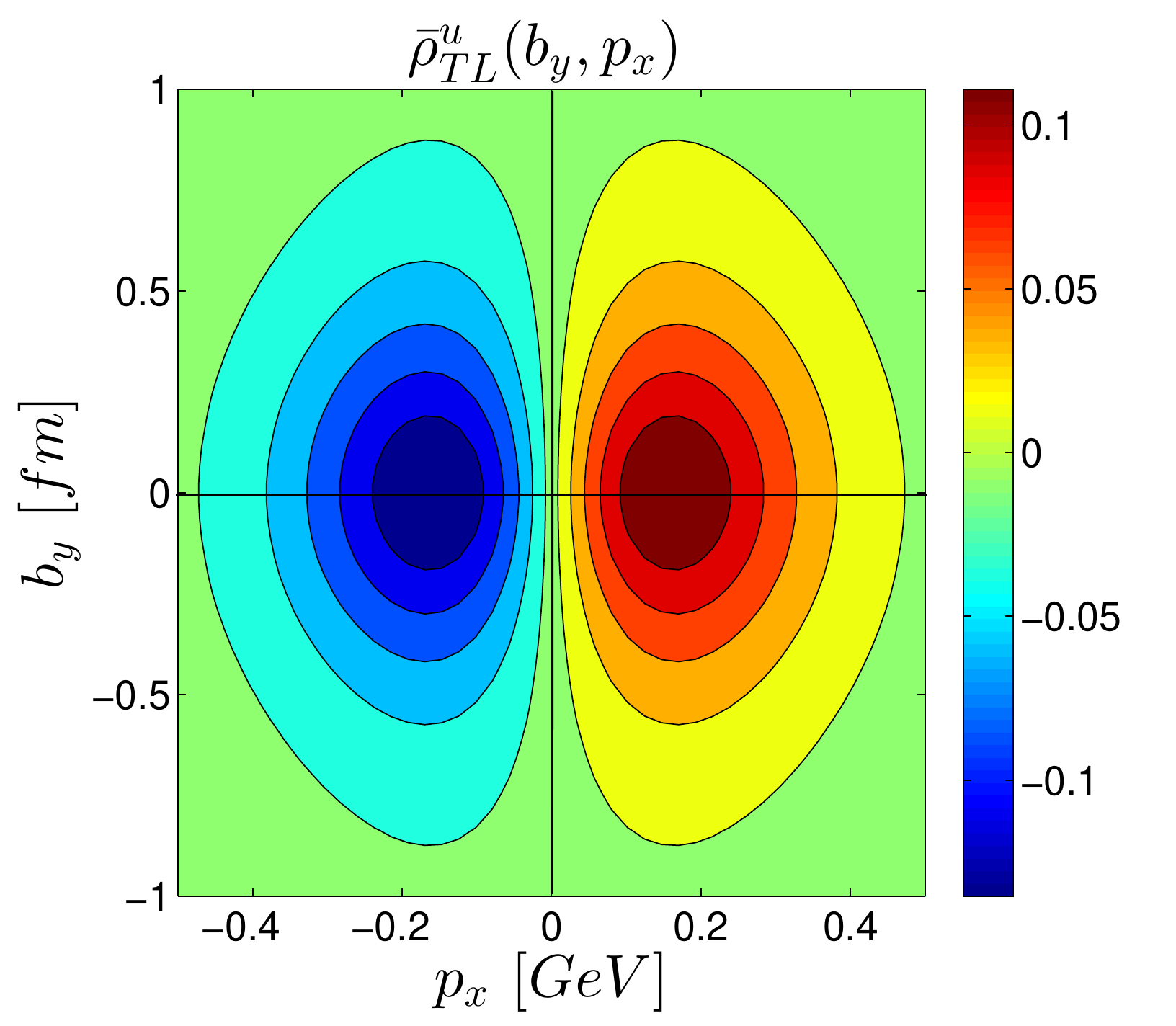}}
\subfigure[]{\includegraphics[width=5.cm,height=4.cm]{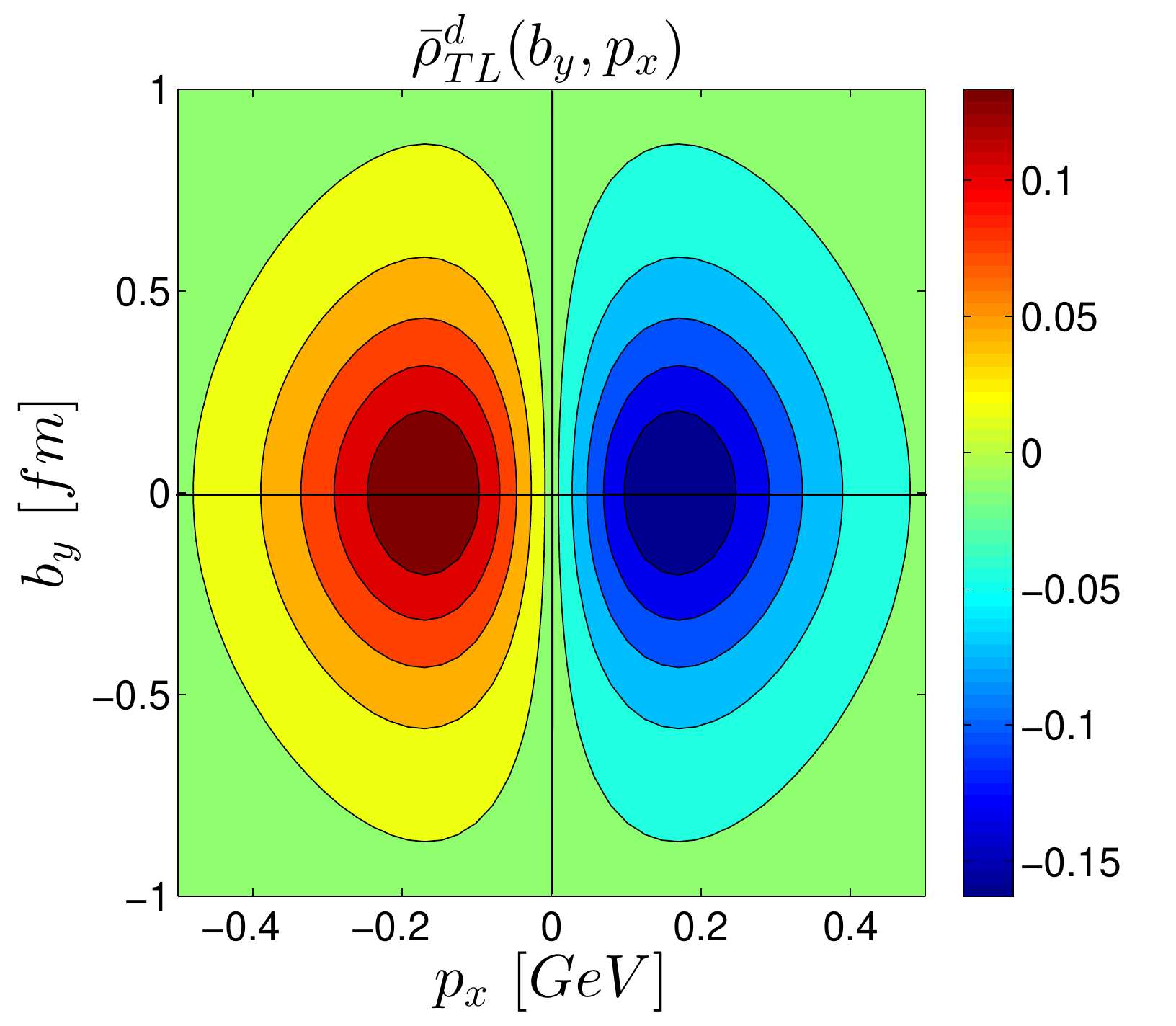}}
\caption{\label{fig_TL}The distributions $\rho_{TL}(\bfb,\bfp)$ are shown in the transverse momentum plane(a,d) with $\bfb=0.4~\hat{y}~fm$,  in the transverse coordinate plane(b,e) with $\bfp=0.3~\hat{x}~GeV$ and  in mixed plane(c,f) for $u$ and $d$  quarks. The polarization of proton is taken along x-axis. The distributions in the mixed planes are given in $GeV^0 fm^0$.}
\end{figure}
\begin{figure}[ht]
\centering
\subfigure[]{\includegraphics[width=5.cm,height=4.cm]{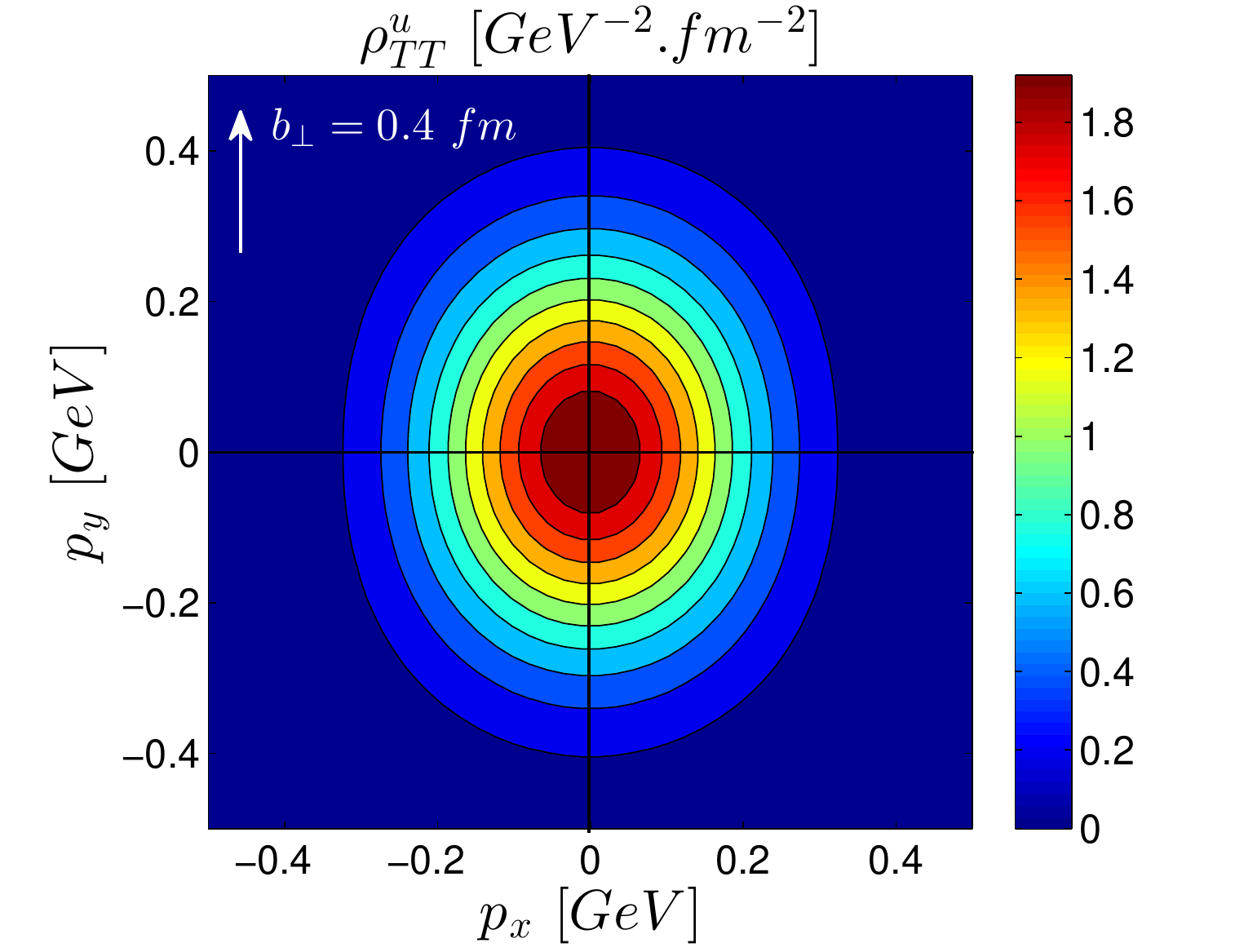}}
\subfigure[]{\includegraphics[width=5.cm,height=4.cm]{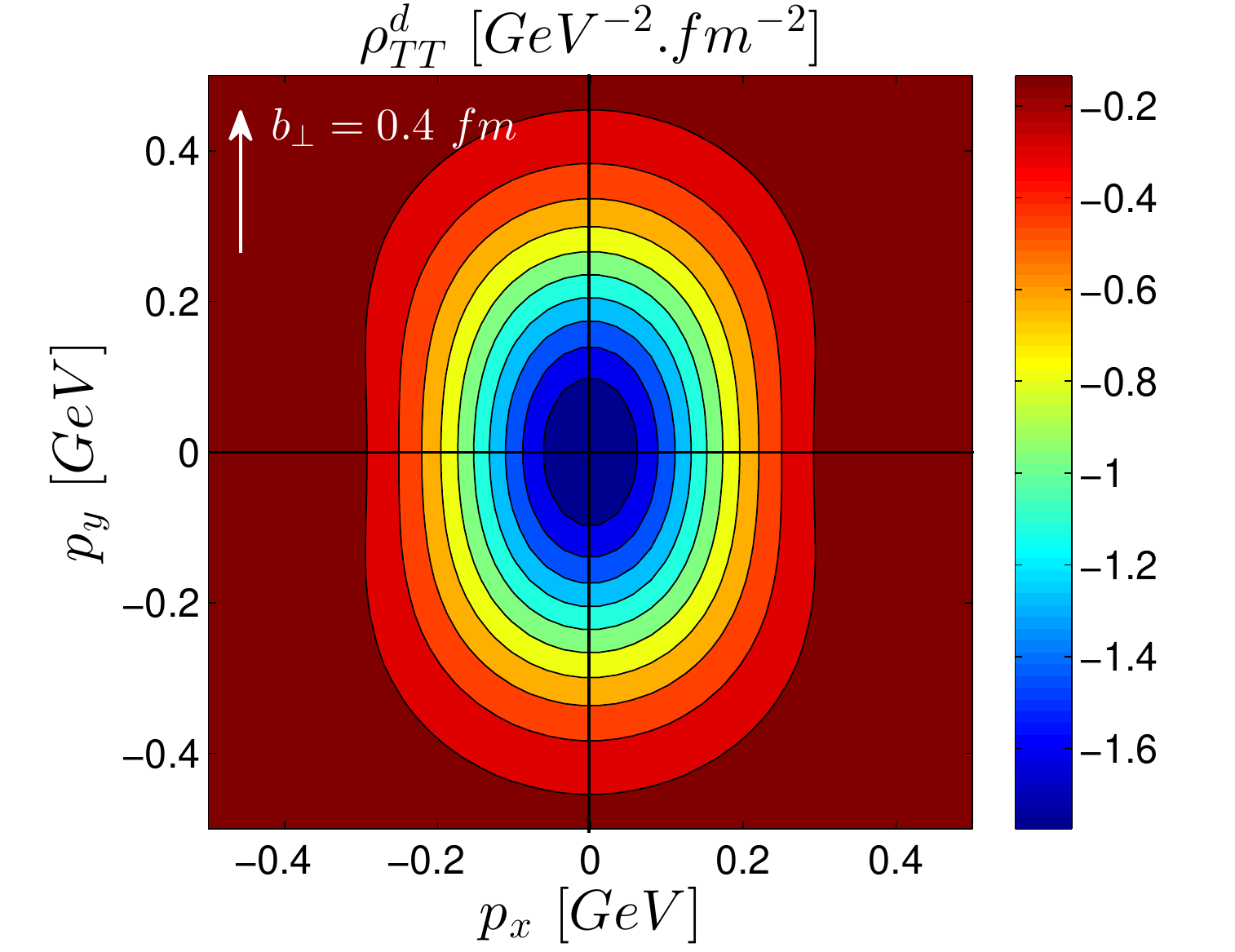}}\\
\subfigure[]{\includegraphics[width=5.cm,height=4.cm]{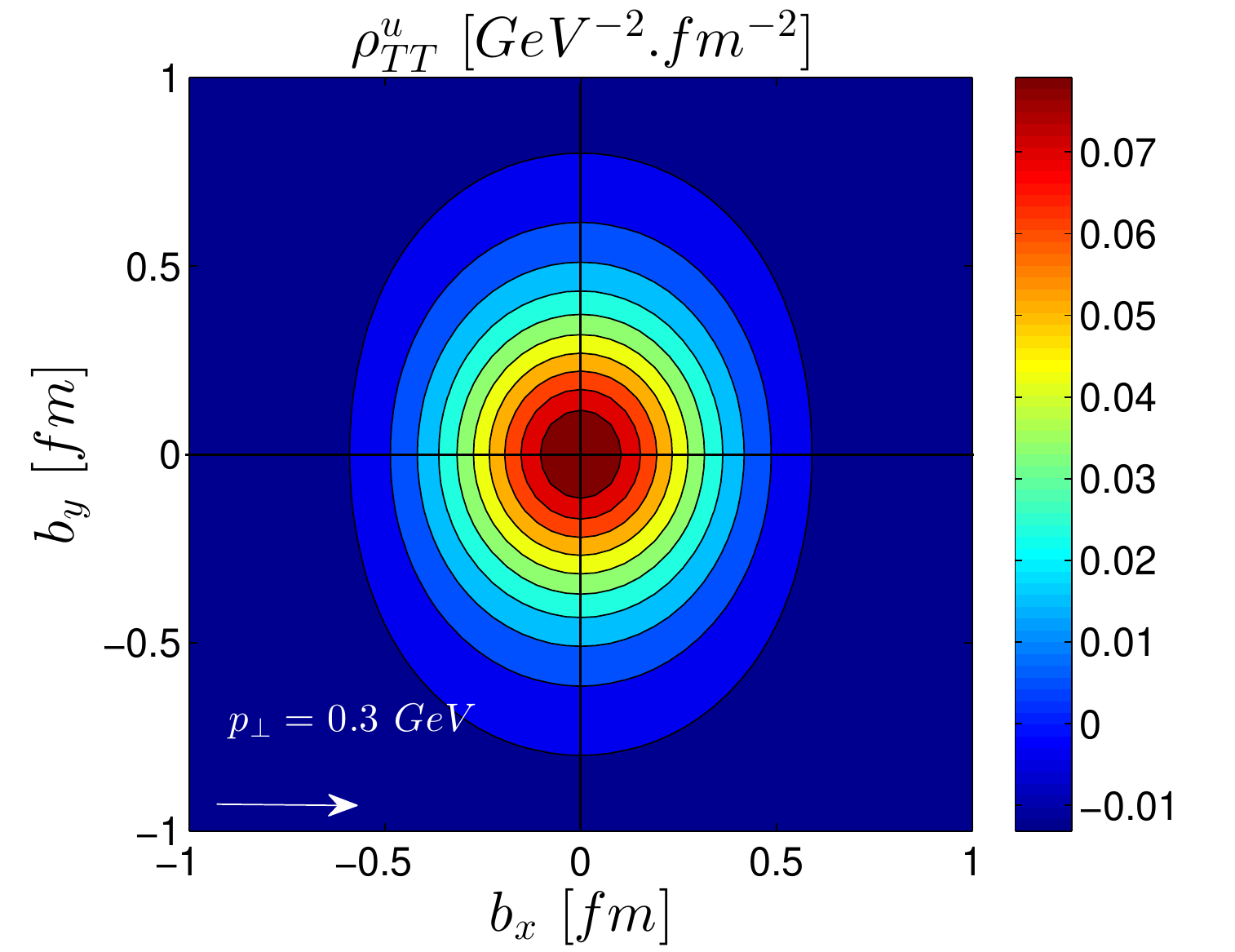}}
\subfigure[]{\includegraphics[width=5.cm,height=4.cm]{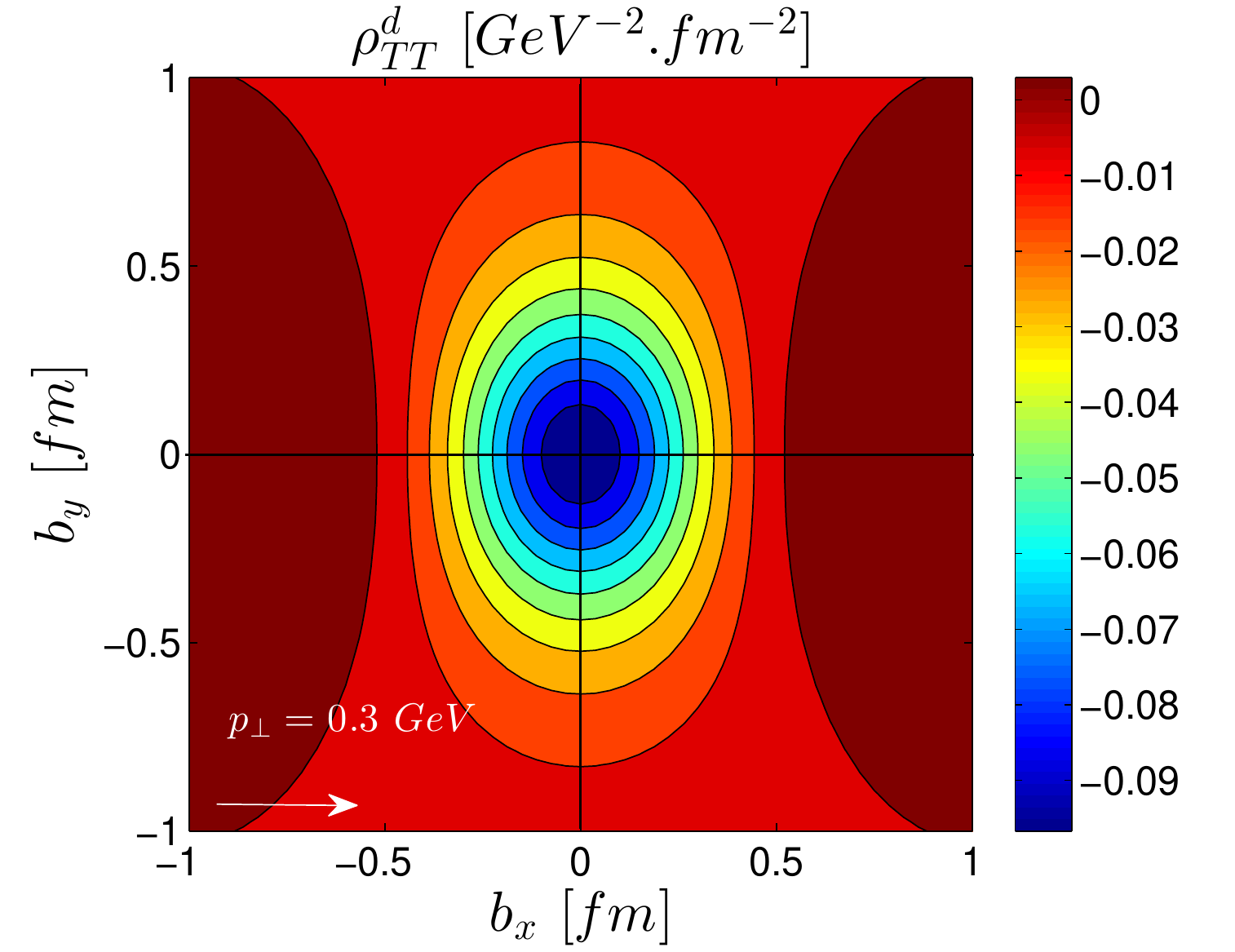}}
\caption{\label{fig_TT}The distributions $\rho_{TT}(\bfb,\bfp)$ are shown in the transverse momentum plane(a,c) with $\bfb=0.4~\hat{y}~fm$ and  in the transverse coordinate plane(b,d) with $\bfp=0.3~\hat{x}~GeV$ for $u$ and $d$  quarks. The proton polarization and the quark polarization are taken along x-axis.}
\end{figure}
\begin{figure}[ht]
\centering
\subfigure[]{\includegraphics[width=5.cm,height=4.cm]{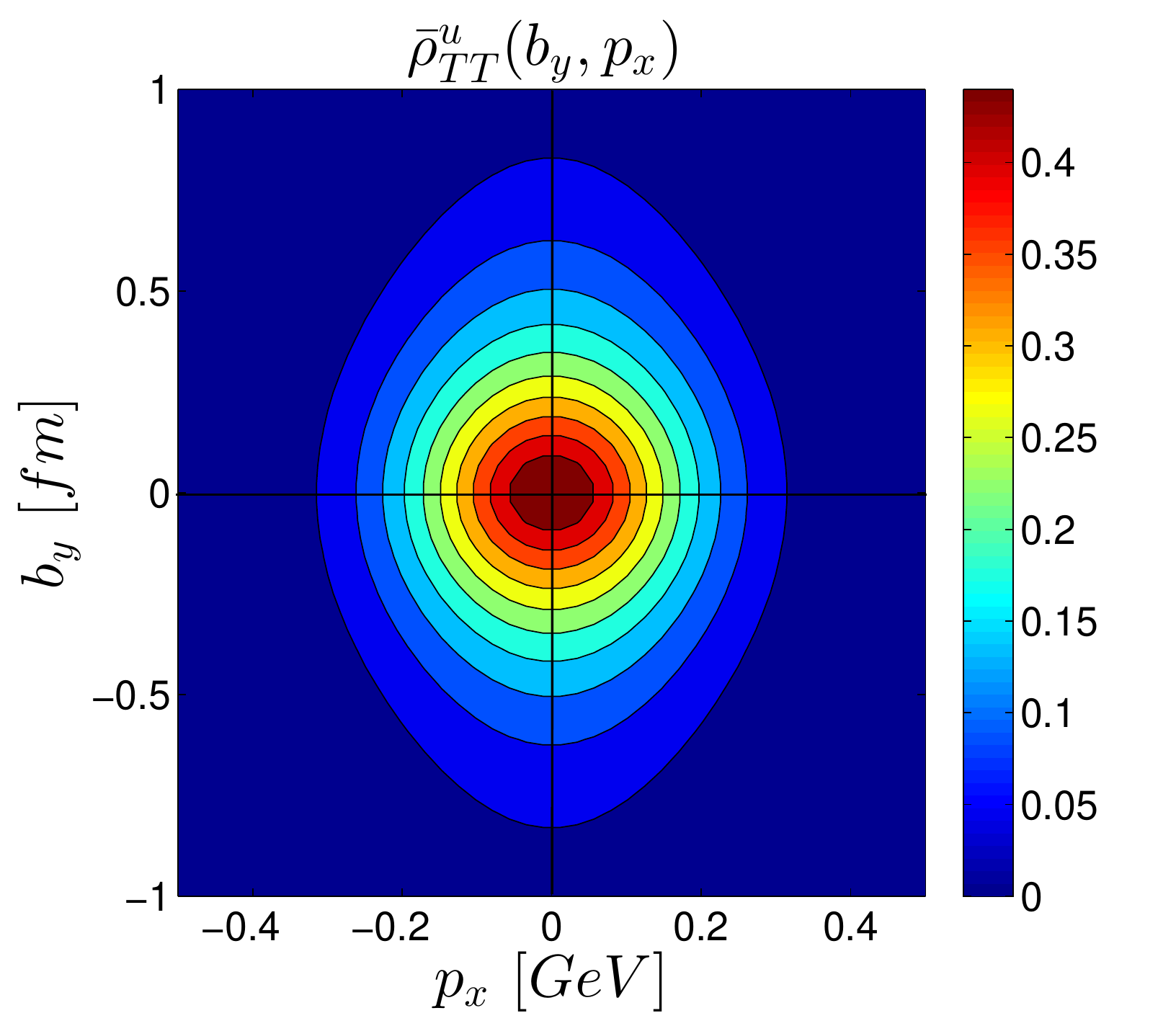}}
\subfigure[]{\includegraphics[width=5.cm,height=4.cm]{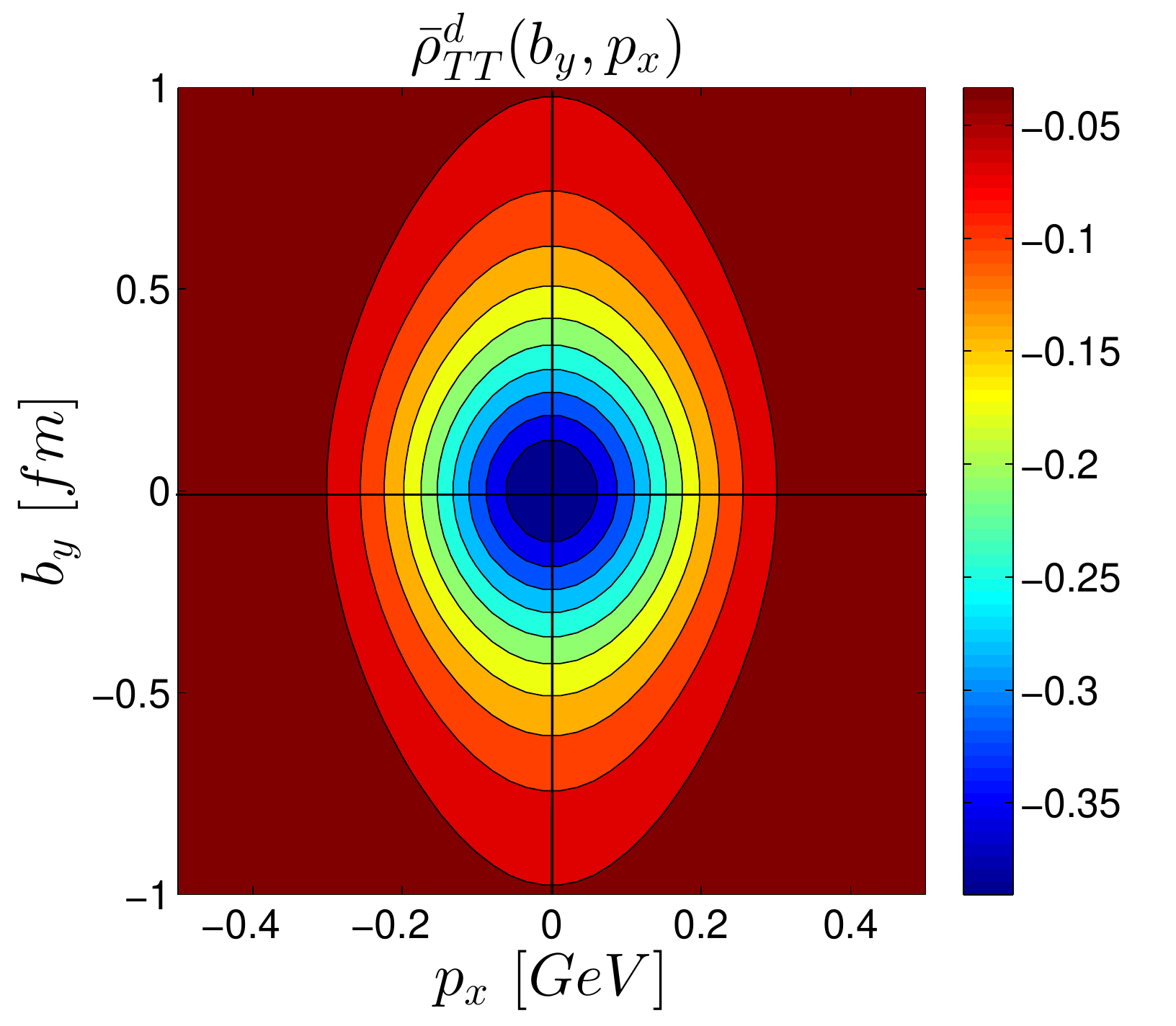}}\\
\subfigure[]{\includegraphics[width=5.cm,height=4.cm]{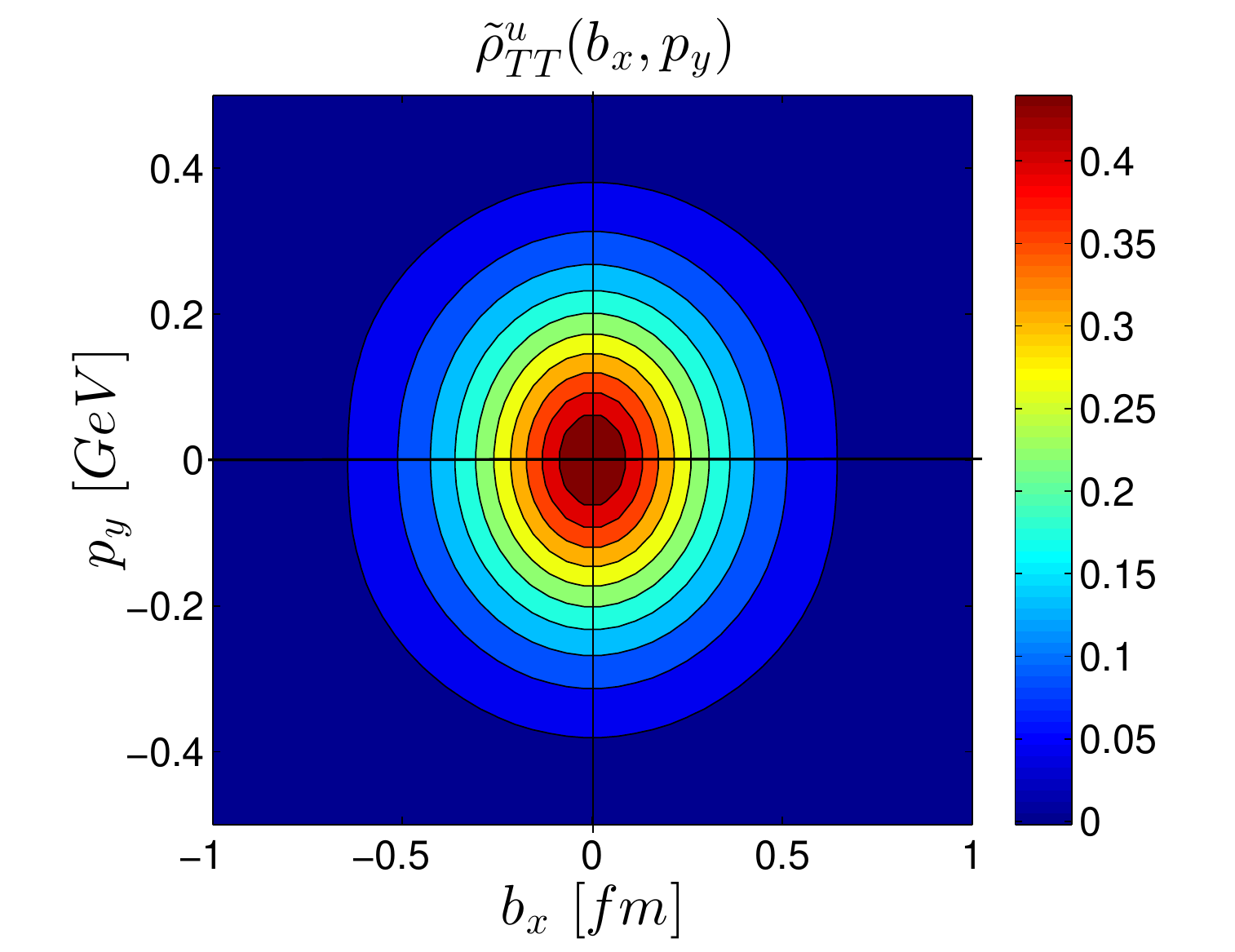}}
\subfigure[]{\includegraphics[width=5.cm,height=4.cm]{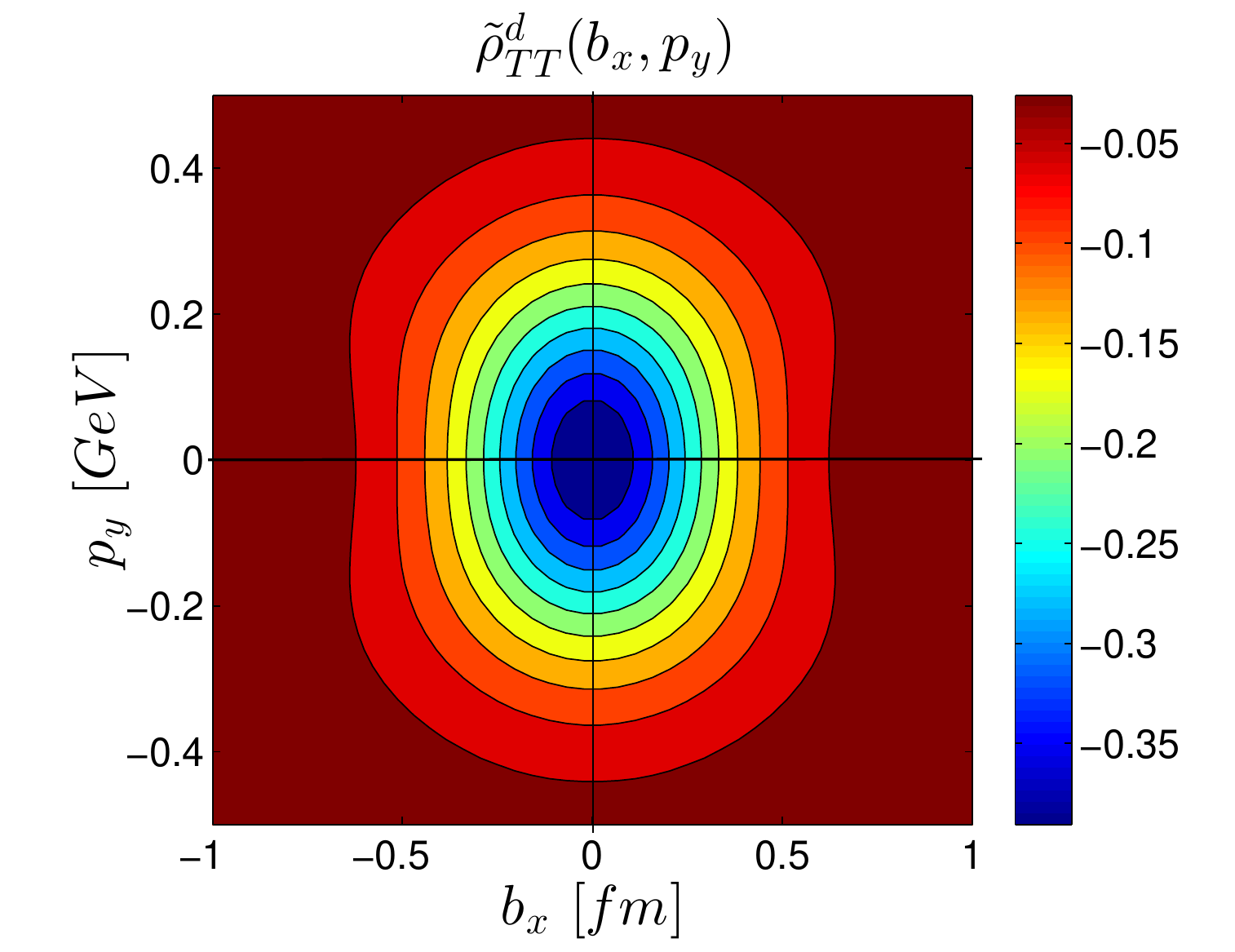}}
\caption{\label{fig_TT_bypx} The distributions $\bar{\rho}_{TT}(by,px)$ and $\tilde{\rho}_{TT}(bx,py)$ are shown in the transverse mixed planes for $u$ and $d$ quarks. The quark and proton both are polarised along x-axis. The distributions are given in $GeV^0 fm^0$.}
\end{figure}
\begin{figure}[ht]
\centering
\subfigure[]{\includegraphics[width=5.cm,height=4.cm]{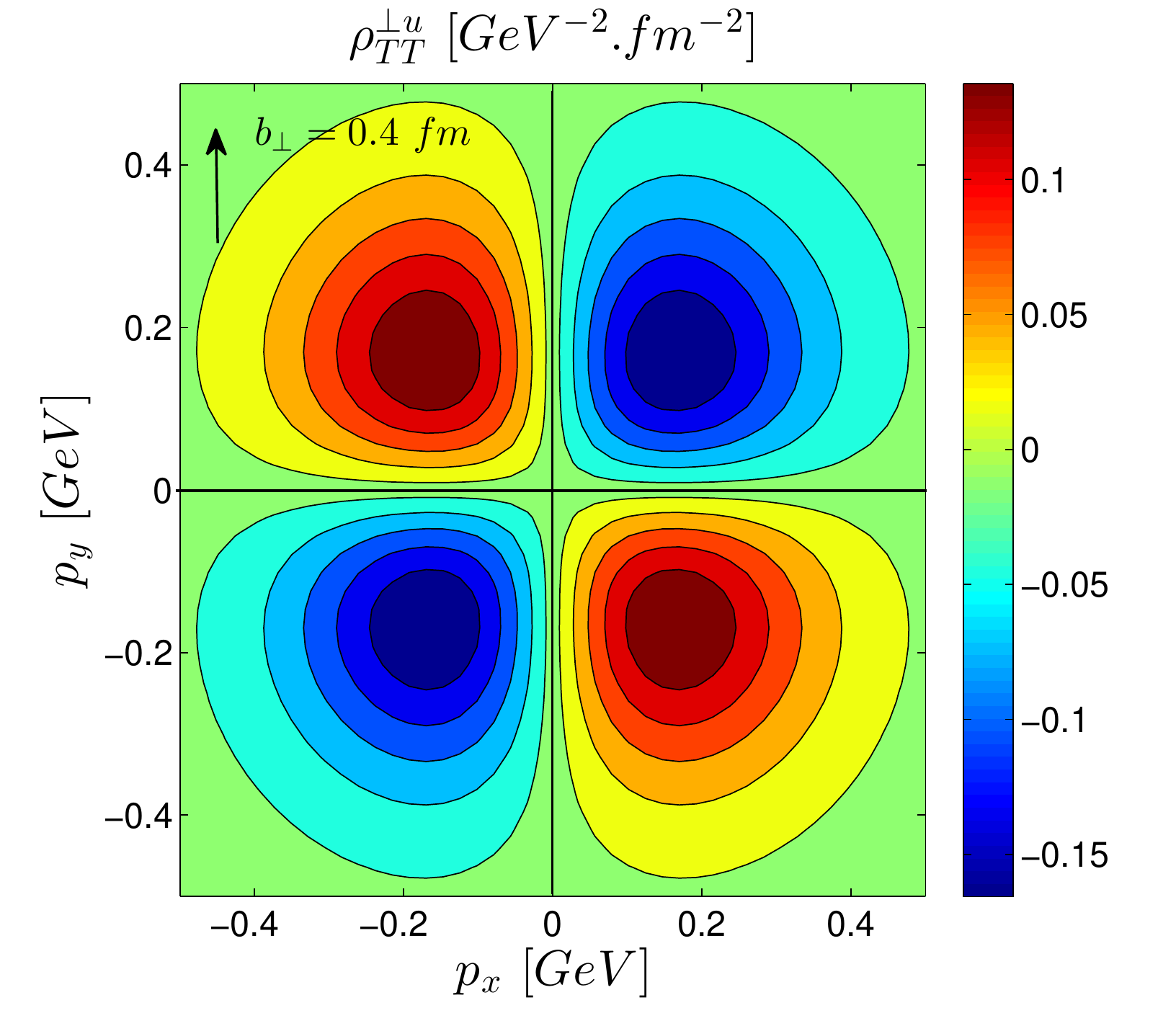}}
\subfigure[]{\includegraphics[width=5.cm,height=4.cm]{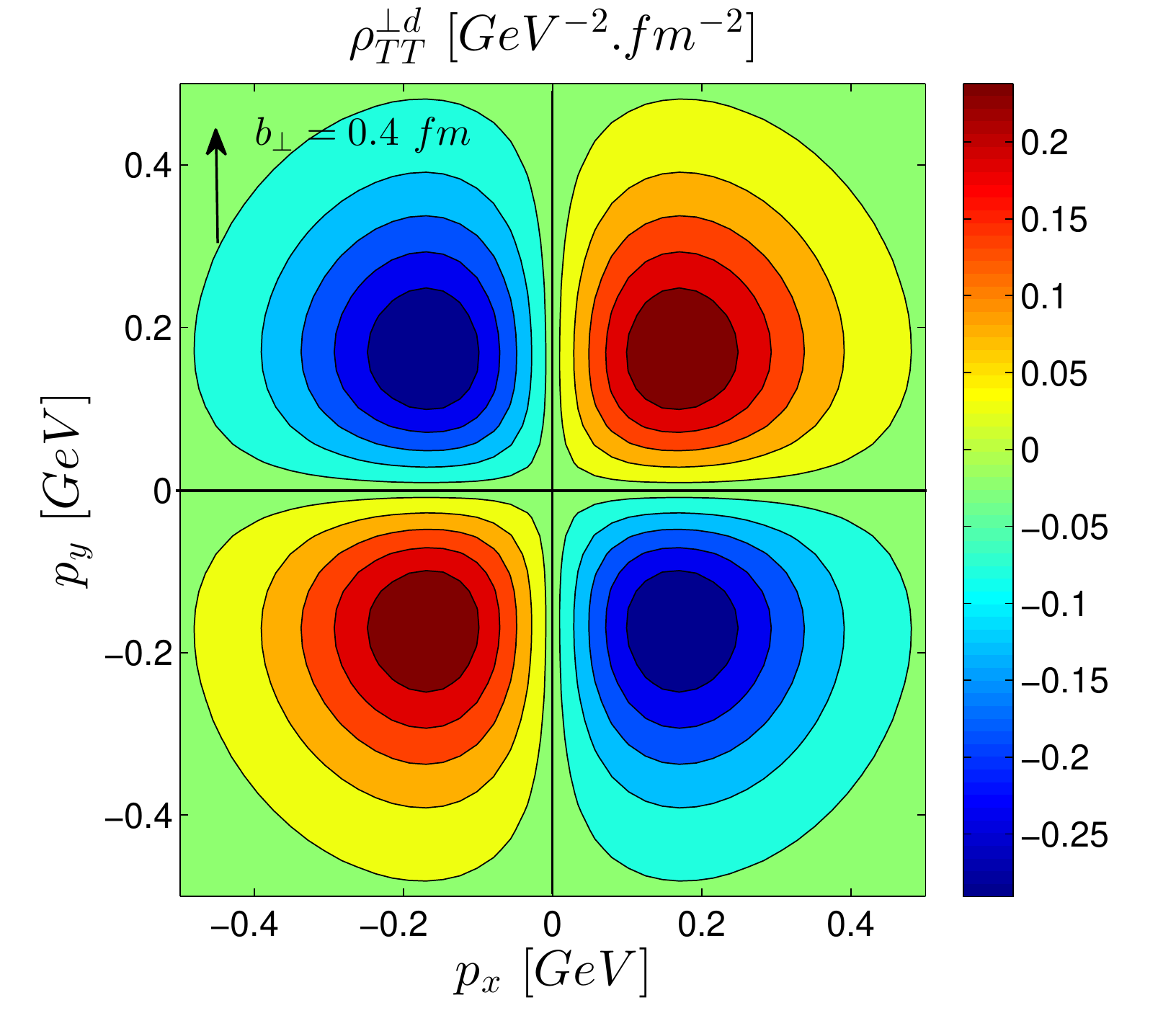}}\\
\subfigure[]{\includegraphics[width=5.cm,height=4.cm]{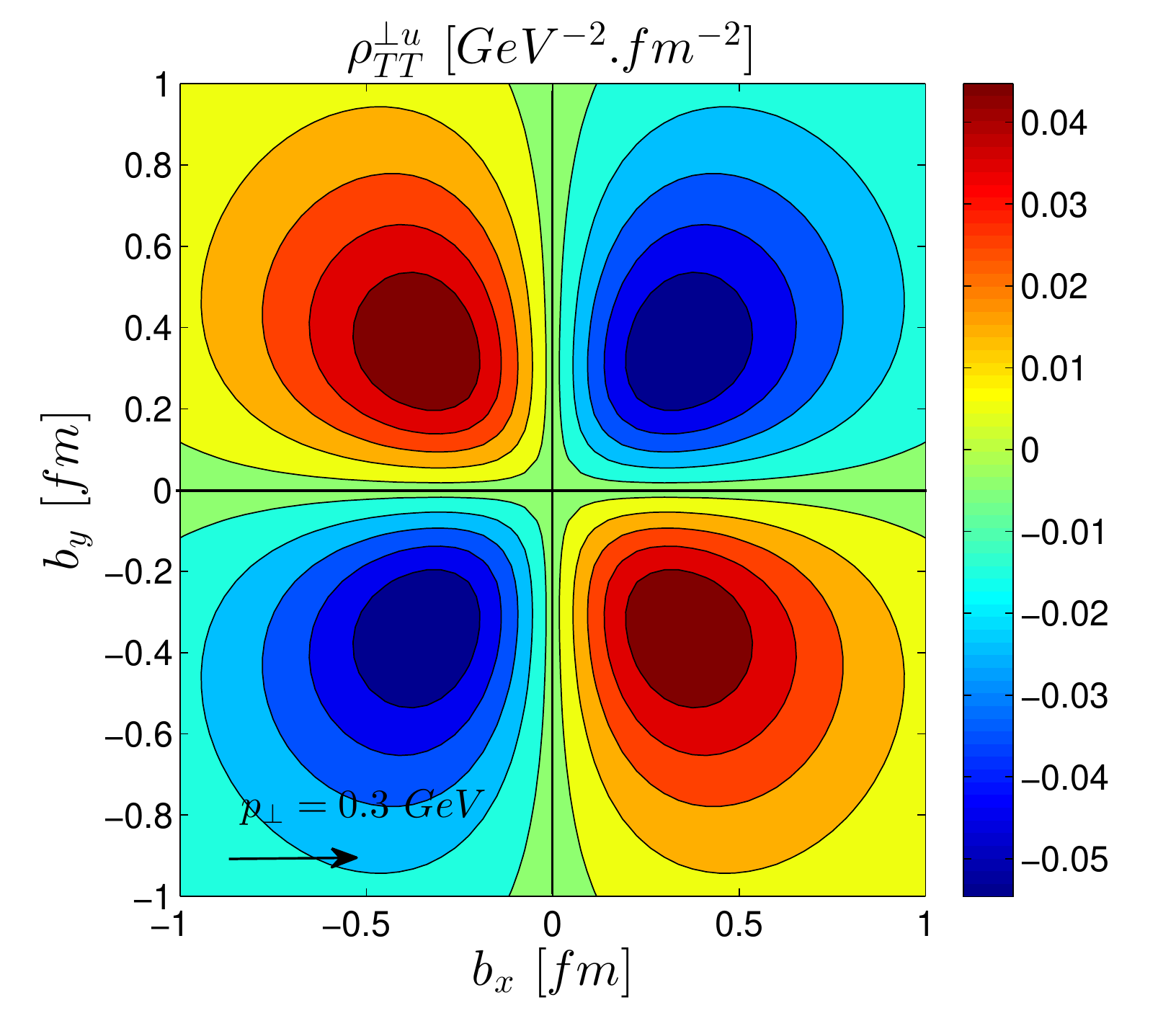}}
\subfigure[]{\includegraphics[width=5.cm,height=4.cm]{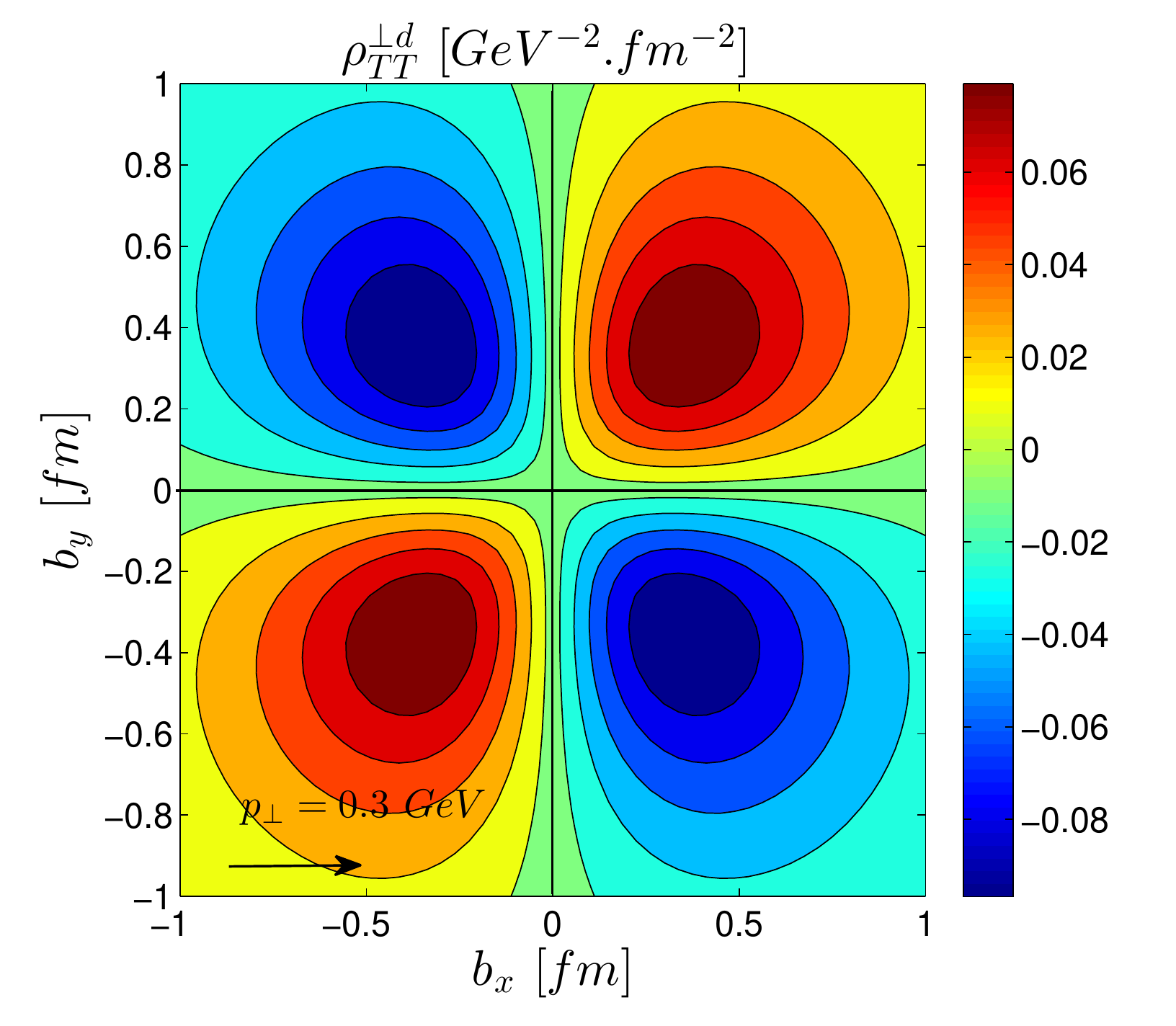}}
\caption{\label{fig_TTP}The distributions $\rho^\perp_{TT}(\bfb,\bfp)$ are shown in the transverse momentum plane and in the transverse coordinate plane for $u$ and $d$  quarks. The quark is polarised along y-axis and proton is polarised along x-axis.}
\end{figure}

%\section{Transverse polarised Wigner distributions}
%+++++++++++++++++++++++++
\subsection{Transversely polarised proton}\label{transpol}
%++++++++++++++++++++++++++++
The Wigner distribution $\rho_{TU}(\bfb,\bfp)$ in transverse plans and the mixing distribution $\bar{\rho}_{TU}(b_y,p_x)$ are shown in fig.\ref{fig_TU}. From Eq.(\ref{rhoTU_nu}) it is clear that this distribution vanishes if the quark transverse coordinate is parallel to the polarization.  Here the plots are shown for $j=1$ i.e. the quark is polarized along x-direction. The figs.\ref{fig_UT}(a) and (d) represent the distribution in the transverse momentum plane, with $\bfb=0.4~\hat{y}~fm$, for $u$ and $d$  quarks respectively. This is circularly symmetric in transverse momentum space. The fig.\ref{fig_TU}(b) and (e) show the distribution in transverse coordinate space with $\bfp=0.3~\hat{y}~GeV$ for $u$ and $d$  quarks respectively. We see a dipolar distribution in the impact parameter plane. 
The mixing distribution $\bar{\rho}_{TU}(b_y,p_x)$ is shown in fig.\ref{fig_TU}(c) and (f) for $u$ and $d$  quarks respectively. Since this distribution is symmetric in transverse momentum plane, it shows a dipolar behavior  in the mixed plane(unlike $\bar{\rho}_{UL}(b_x,p_y)$ shows a quadruple distribution). Because of the dipolar structure in coordinate space, the other class of mixing distributions $\tilde{\rho}_{TU}(b_x,p_y)$ vanishes. 
%Here the quark is polarized along x-axis.

At the TMD limit, $\rho_{TU}(\bfb,\bfp,x)$ reduces to the  Sivers function $f^\perp_{1T}(x,\bfp)$, one of the T-odd TMDs at leading twist. Since we consider T-even contributions only, the TMD limit of  $\rho_{TU}(\bfb,\bfp)$ vanishes here.
At the impact parameter distribution limit the distribution is related to the $H$ and $E$ GPDs together with some other distributions.

%++++++++++++++++++
%\subsection{Transverse-Longitudinal WD}
%++++++++++++++++++
The wigner distribution with a longitudinally polarized quark in a transversely polarized proton, $\rho_{TL}(\bfb,\bfp)$, is shown in fig.\ref{fig_TL}. The fig.\ref{fig_TL}(a) and (d) represent the distribution in transverse momentum plane, with $\bfb=0.4~\hat{y}~fm$, for $u$ and $d$  quark respectively. We see a dipolar distribution as expected from the Eq.(\ref{rhoTL_nu}). The fig.\ref{fig_LT}(b) and (e) show the distribution in transverse coordinate space with $\bfp=0.3~\hat{x}~GeV$ for $u$ and $d$  quarks respectively. The distribution is circularly symmetric  at the center of the coordinate space with  positive peak for $u$ quark and negative peak for $d$  quark.

The distribution vanishes if the quark transverse momentum is perpendicular to the polarization. This reflects that there is a strong correlation between the quark transverse momentum and quark transverse polarization.
The mixing distribution $\bar{\rho}_{TL}(b_y,p_x)$ is shown in fig.\ref{fig_TL}(c) and (f) for $u$ and $d$  quarks respectively. Because of the dipolar structure in transverse momentum plane, the other class of mixing distributions $\tilde{\rho}_{TL}(b_x,p_y)$ vanishes for the quark with a polarization along x-axis.

At the TMD limit, $\rho_{TL}(\bfb,\bfp)$ reduces to $g_{1T}(x,\bfp)$\cite{meissner09}, one of the T-even eight TMDs at leading twist. At the impact parameter distribution limit the distribution is related to $\tilde{H}$ and $\tilde{E}$ GPDs together with some other distributions.

%++++++++++++++++++
%\subsection{Transverse WD}
%++++++++++++++++++
The wigner distribution with a transversely polarized quark in a transversely polarized proton, $\rho_{TT}(\bfb,\bfp)$, is shown in fig.\ref{fig_TT}. The fig.\ref{fig_TT}(a,c) represent the distribution in transverse momentum plane with $\bfb=0.4~\hat{y}~fm$ for $u$ and $d$  quarks respectively. The distribution in transverse impact parameter plane is shown in fig.\ref{fig_TT}(b,d) with $\bfp=0.3~\hat{x}~GeV$. We observe that the peak of the distribution in both the plane are positive for $u$ quark and negative for $d$  quark. 
In fig.\ref{fig_TT_bypx}, we plot the  $\rho_{TT}(\bfb,\bfp)$ in mixed transverse plane for $u$ and $d$  quarks. The distributions are not symmetric in the mixed plane. This is due to the quadrupole contributions $p^i p^j$ and $\Delta^i \Delta^j$ appearing in the expressions for $\rho^\nu_6$(Eq.\ref{rho_6}).

%++++++++++
%Pretzelosity
%++++++++++++
The pretzelous distribution, $\rho^\perp_{TT}(\bfb,\bfp)$  is shown in Fig.\ref{fig_TTP}. This distribution is found when the quark is transversely polarized along the perpendicular direction to the transversely polarized proton. We find quadruple distributions in transverse momentum plane as well as in transverse impact parameter plane. It is also observed that the polarity of quadruple distribution changes sign for $d$  quark in both the planes.
Due to pure quadrupole contribution in $\rho^\nu_7$(Eq.\ref{rho_7}) the pretzelous distribution is identically zero in the mixed plane.

\begin{figure}[htb]
\centering
\subfigure[]{\includegraphics[width=5.cm,height=4.cm]{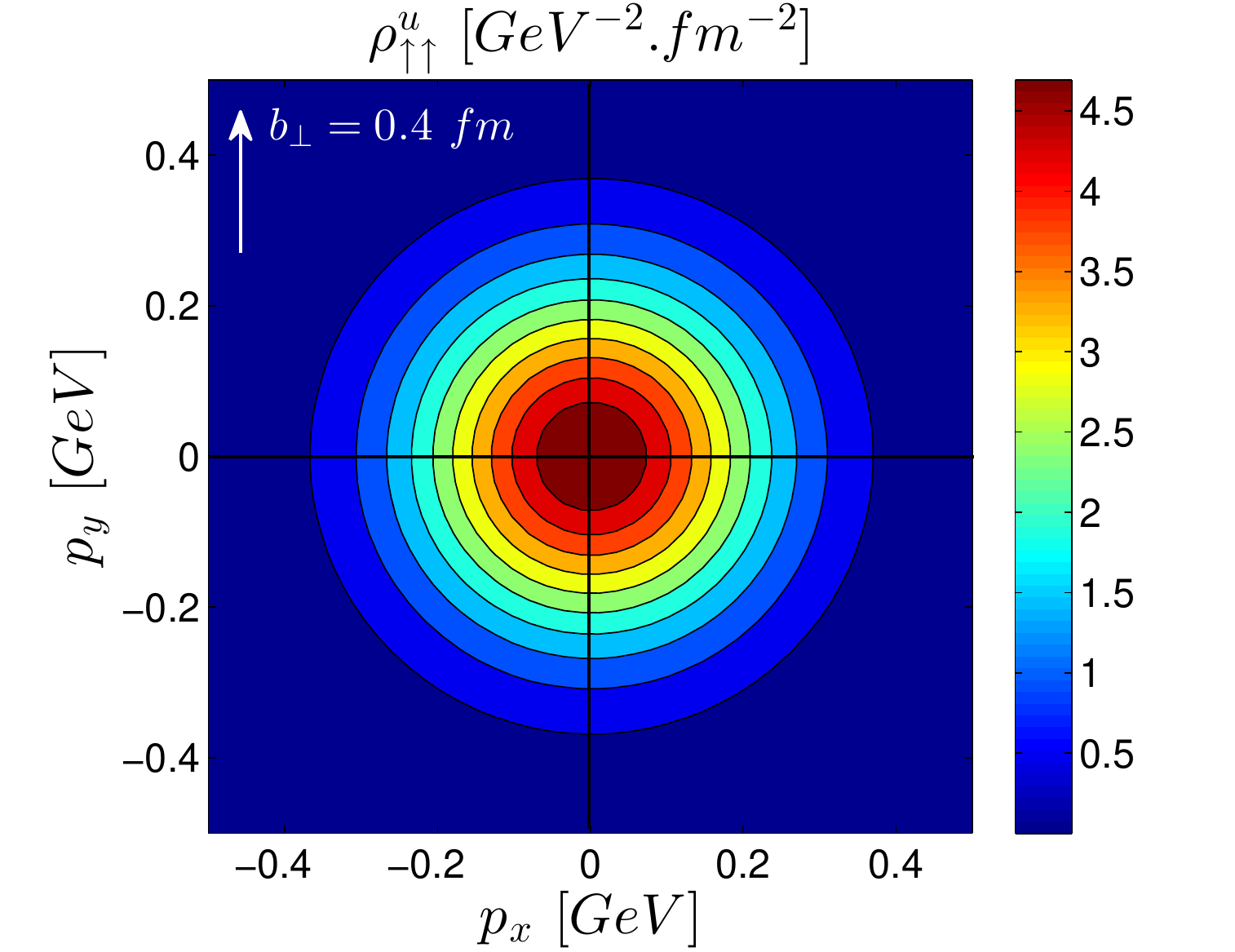}}
\subfigure[]{\includegraphics[width=5.cm,height=4.cm]{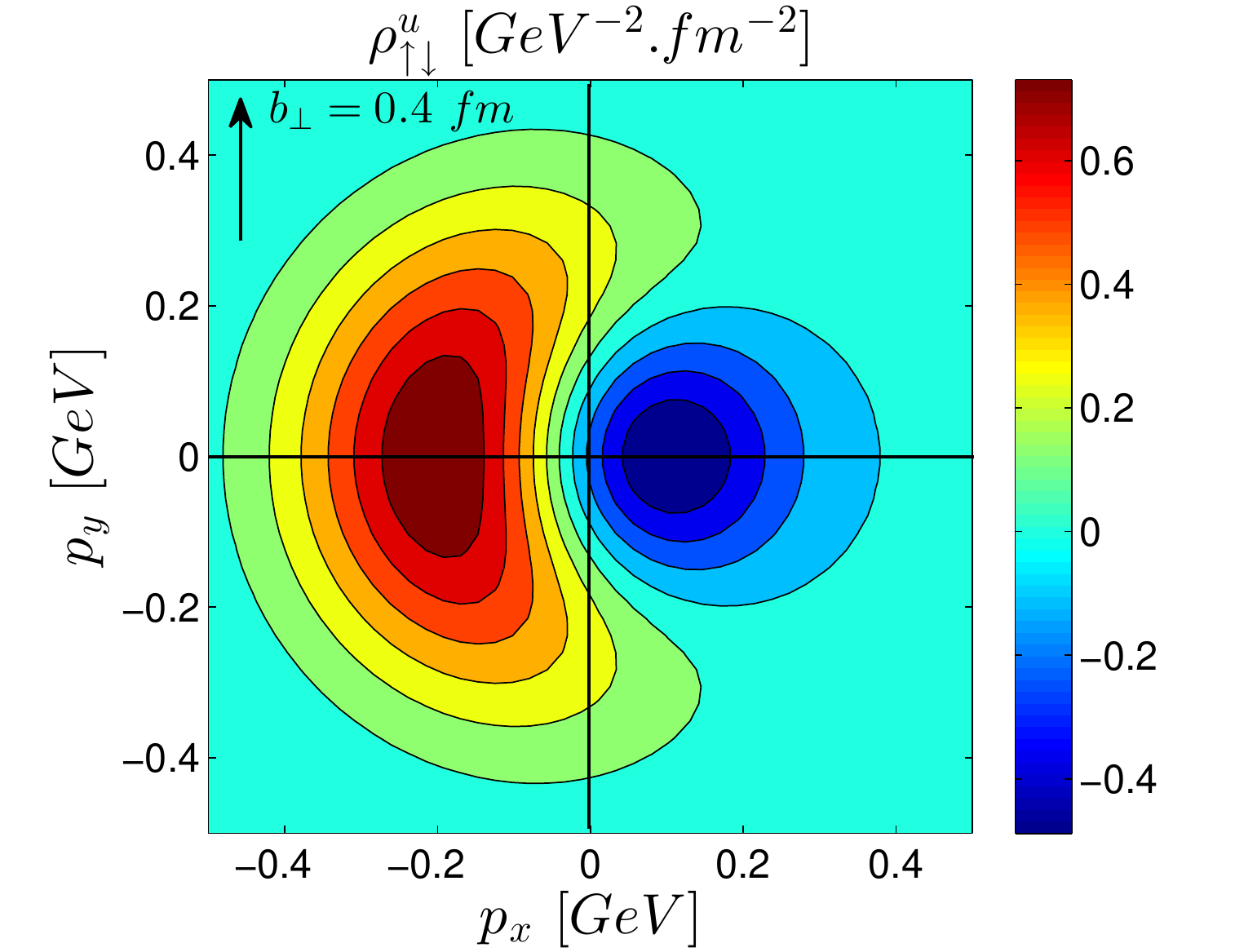}}\\
\subfigure[]{\includegraphics[width=5.cm,height=4.cm]{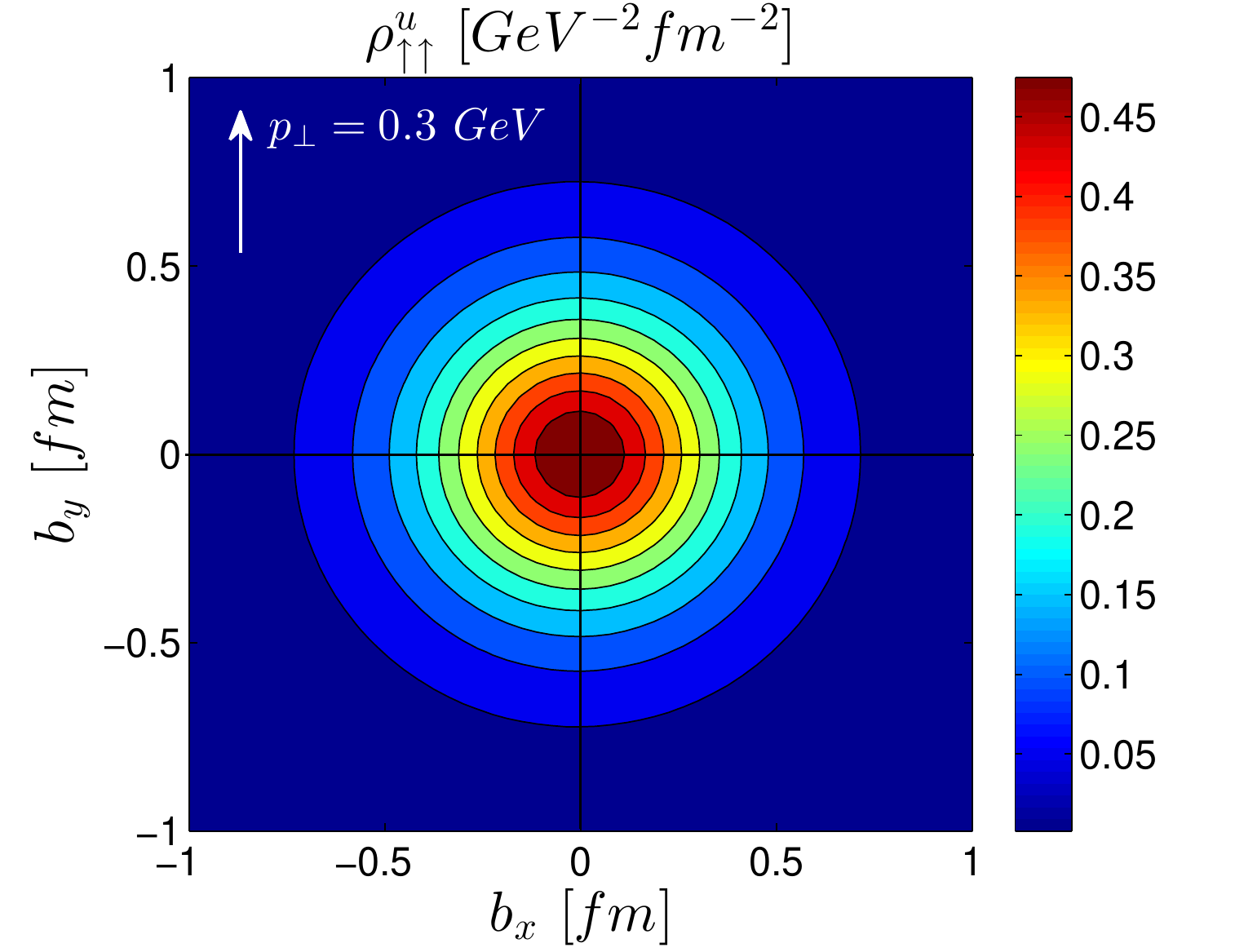}}
\subfigure[]{\includegraphics[width=5.cm,height=4.cm]{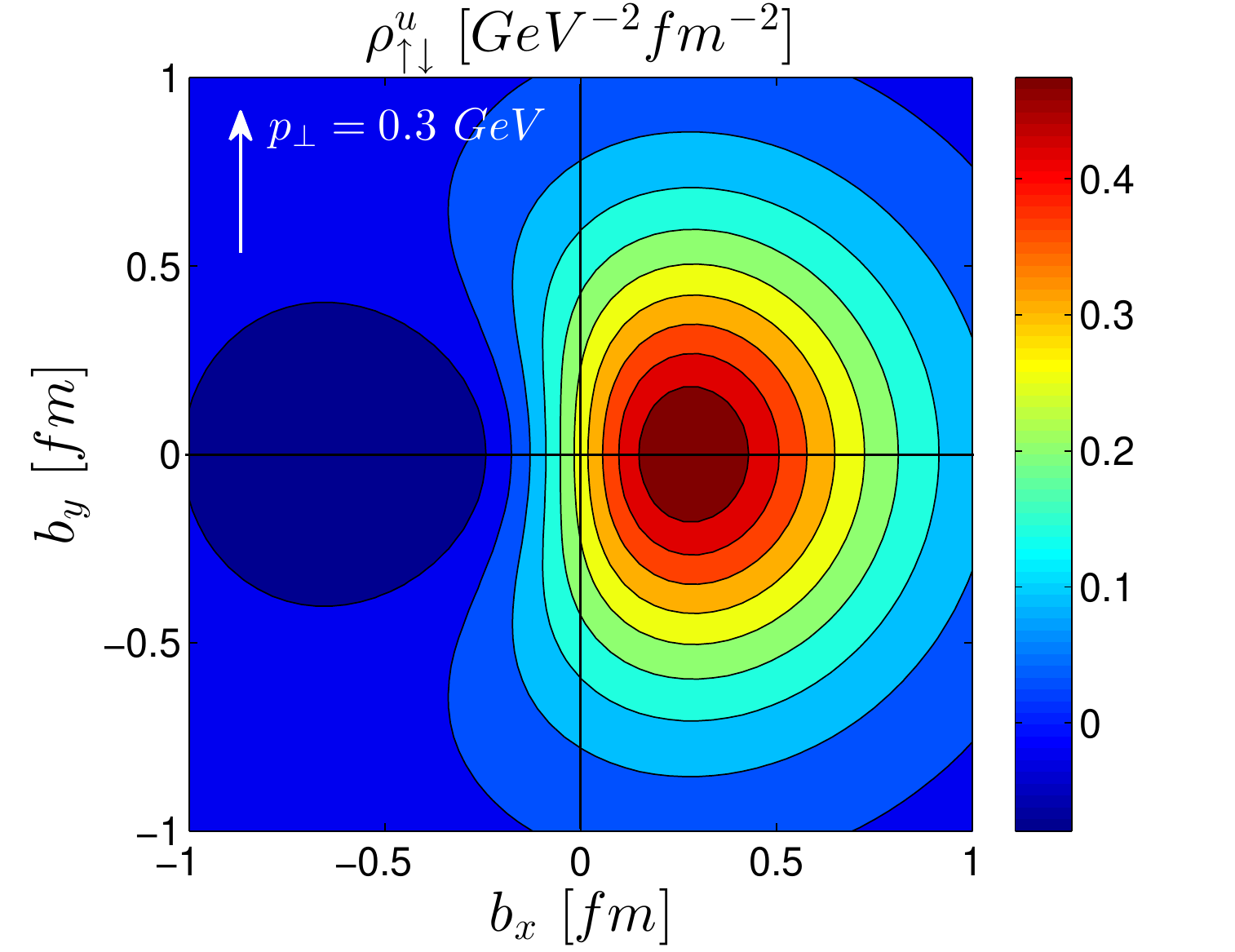}}\\
\subfigure[]{\includegraphics[width=5.cm,height=4.cm]{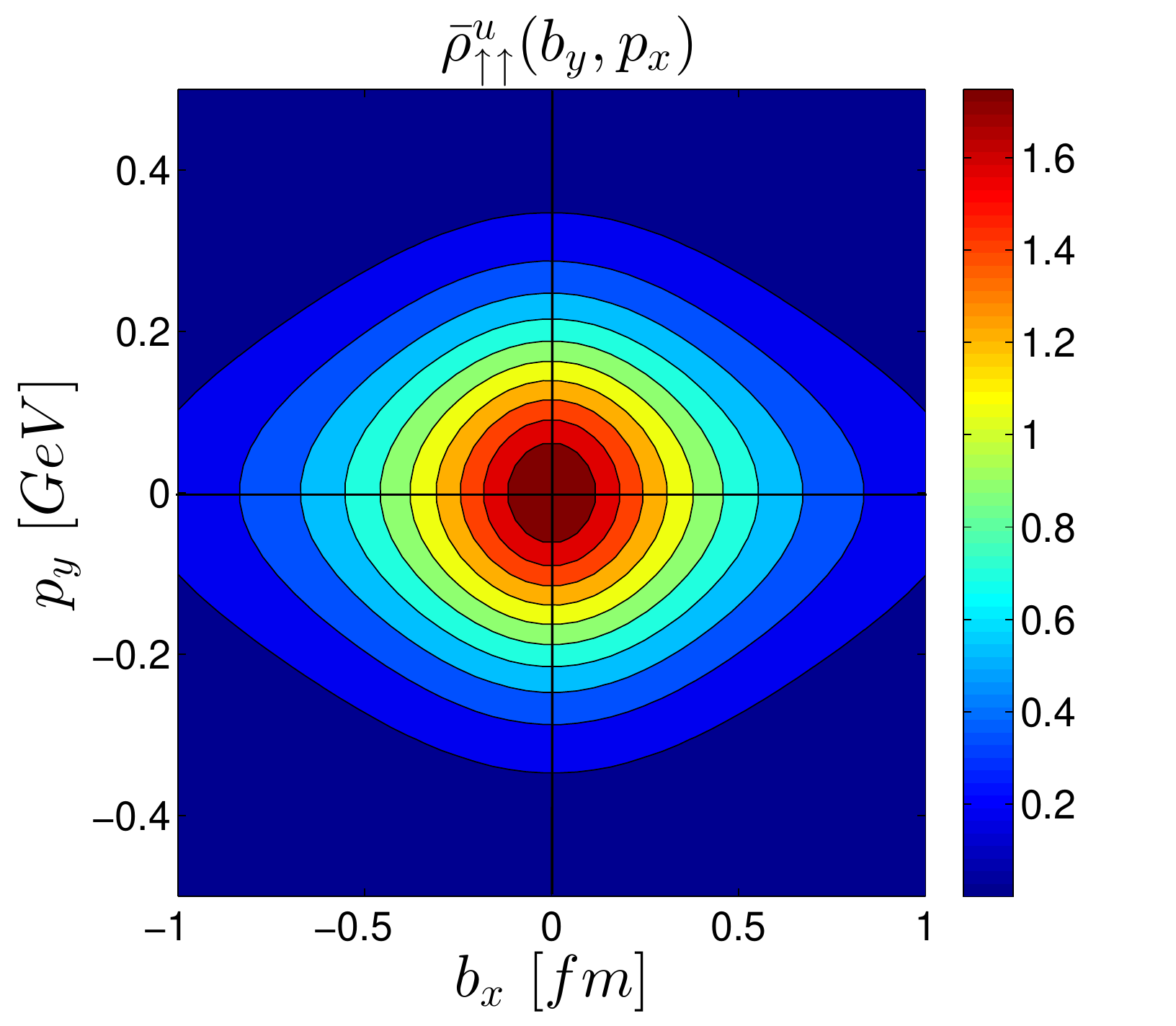}}
\subfigure[]{\includegraphics[width=5.cm,height=4.cm]{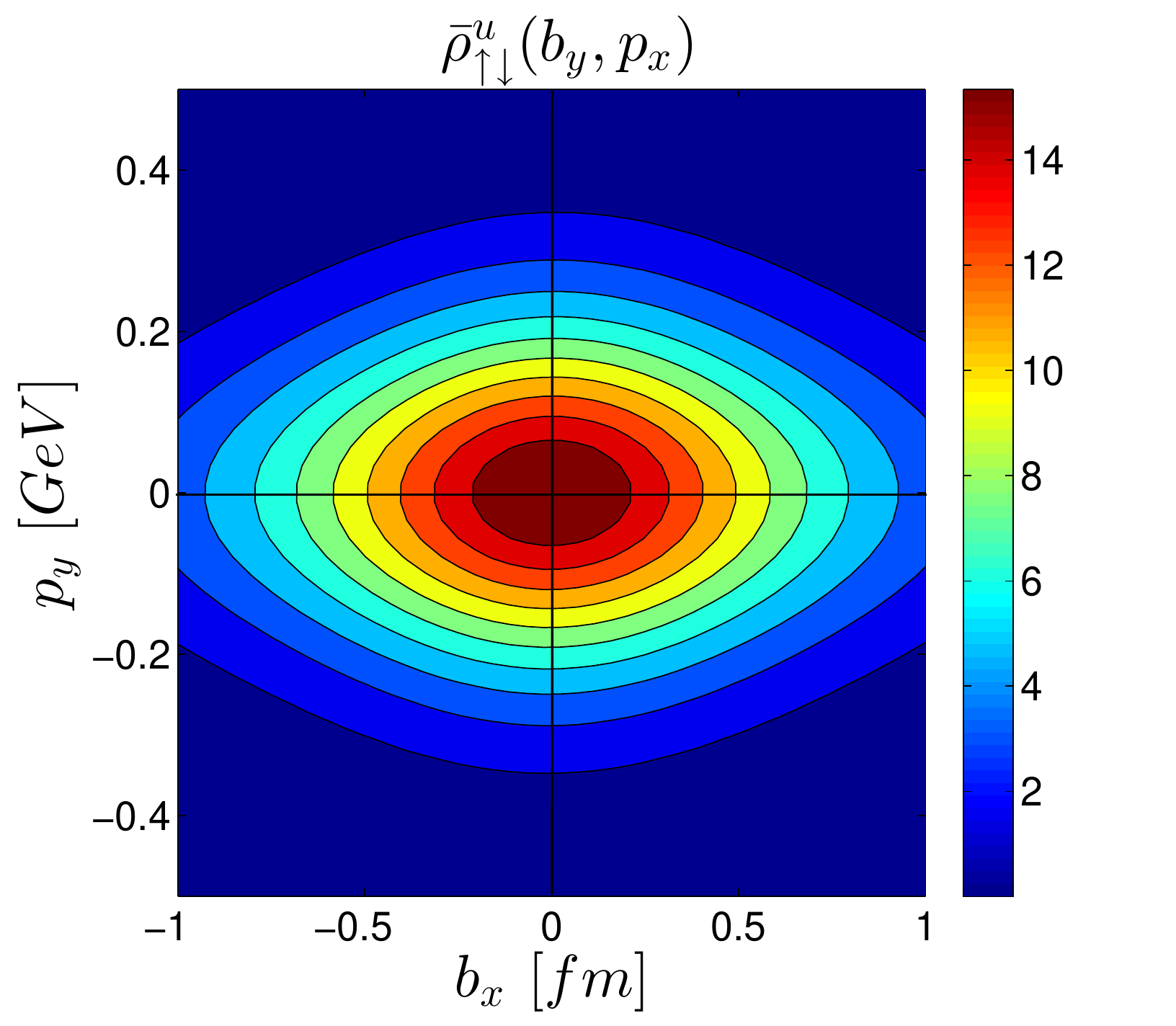}}
\caption{\label{fig_Lamlam_u} The distributions $\rho_{\Lambda\lambda}$ are shown in transverse momentum plane, transverse impact parameter plane and mixed plane for $u$ quark(Eq.(\ref{rho_Lamlam})). The distributions in the mixed plane are give in $GeV^0 fm^0$.}
\end{figure}
\begin{figure}[ht]
\centering
\subfigure[]{\includegraphics[width=5.cm,height=4.cm]{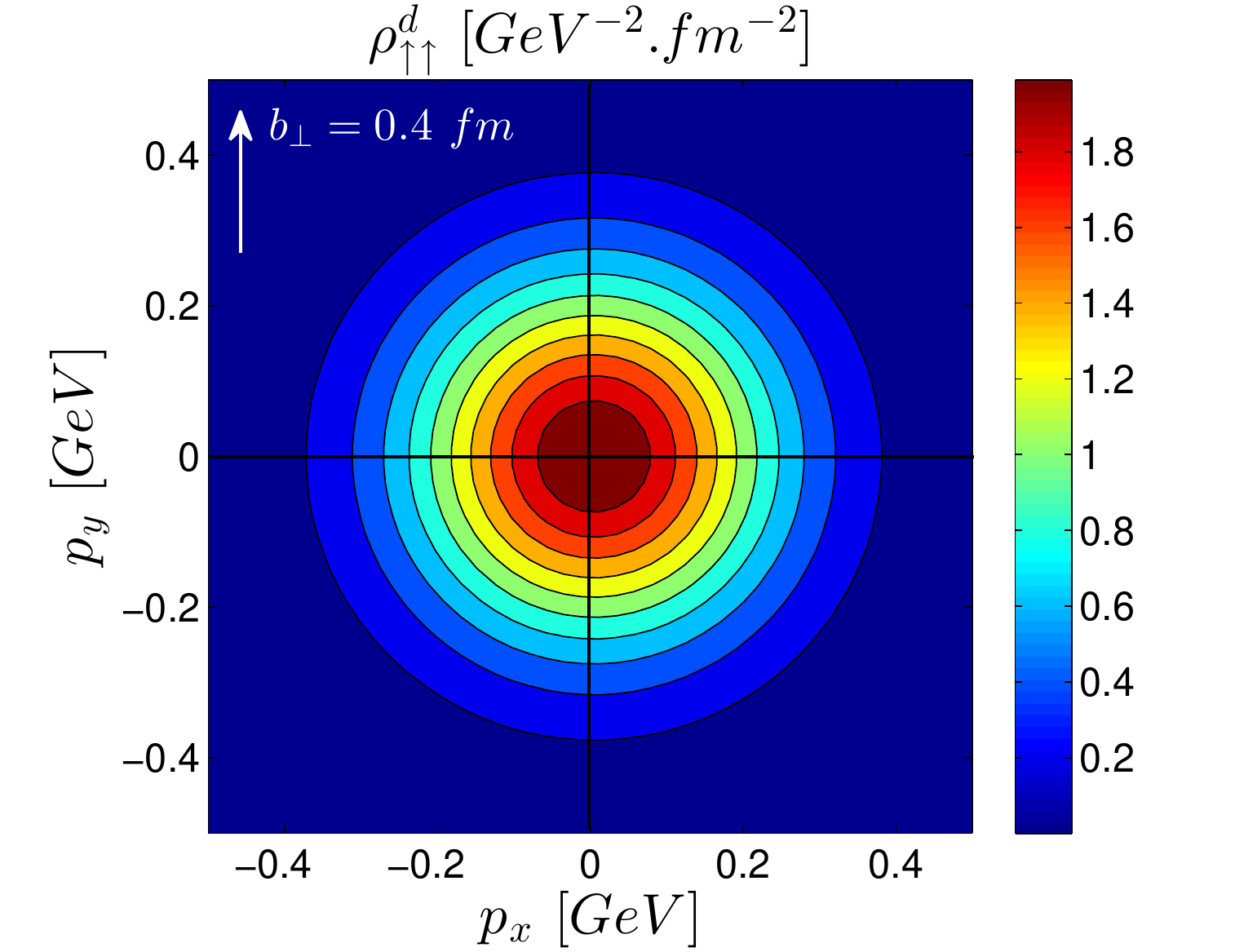}}
\subfigure[]{\includegraphics[width=5.cm,height=4.cm]{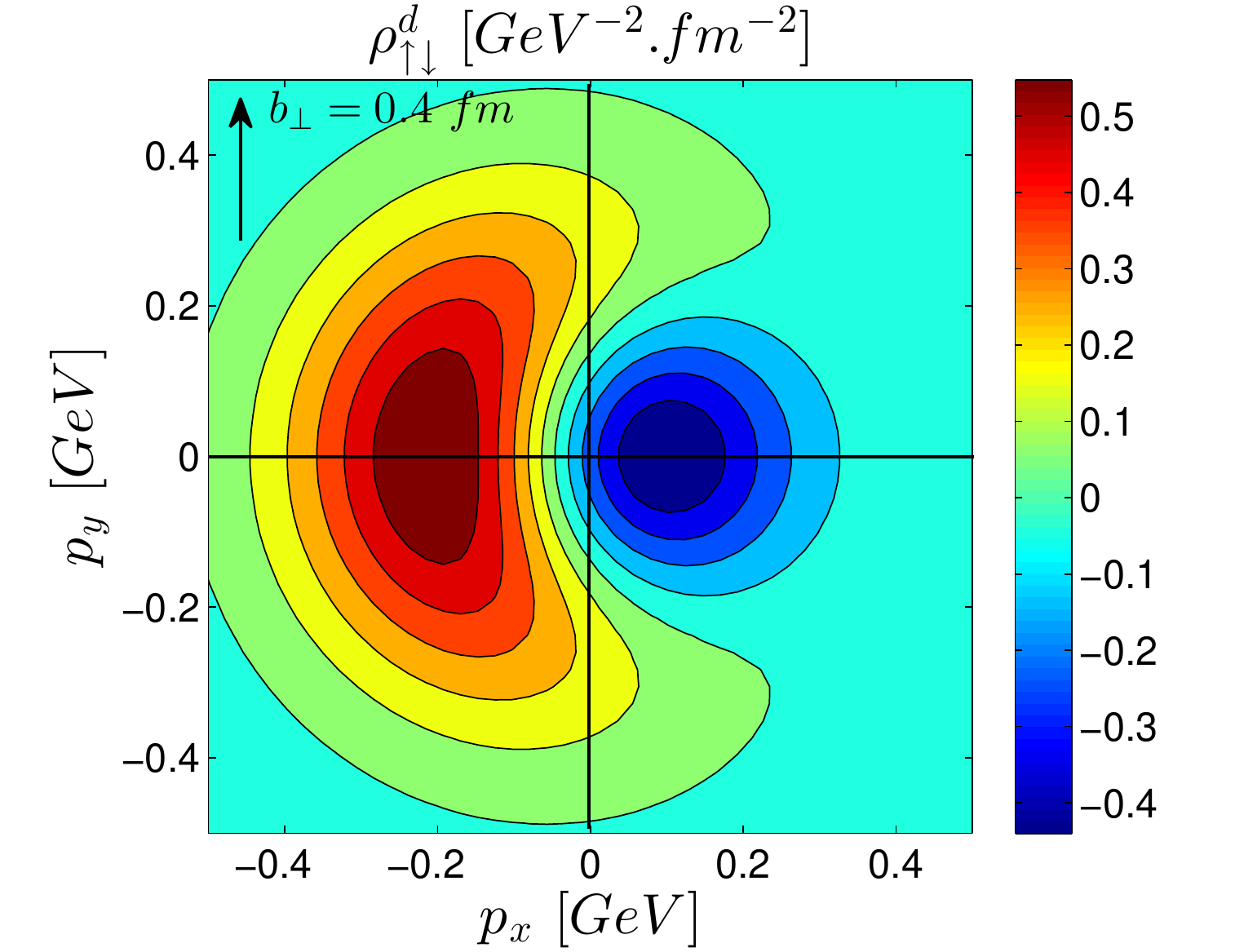}}\\
\subfigure[]{\includegraphics[width=5.cm,height=4.cm]{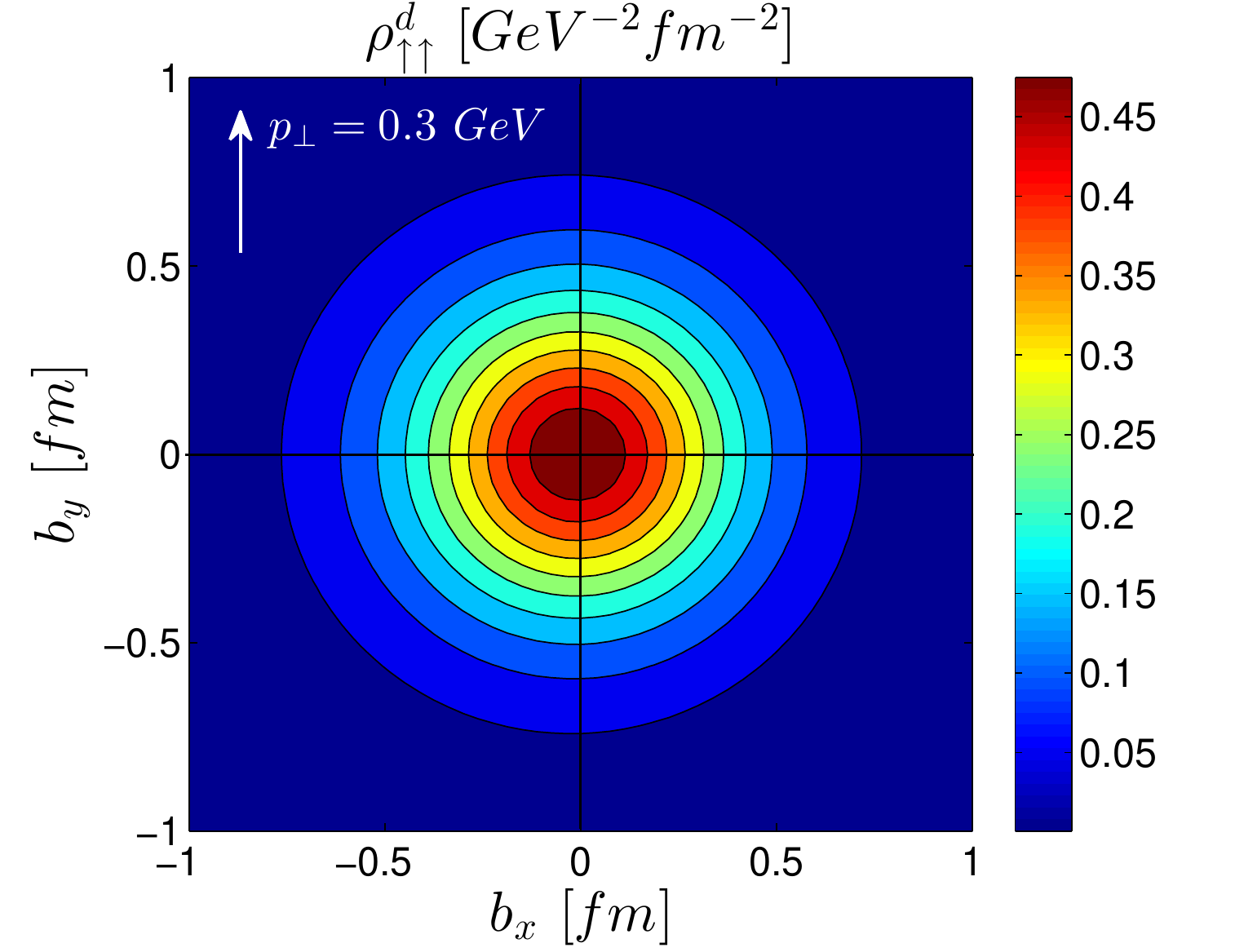}}
\subfigure[]{\includegraphics[width=5.cm,height=4.cm]{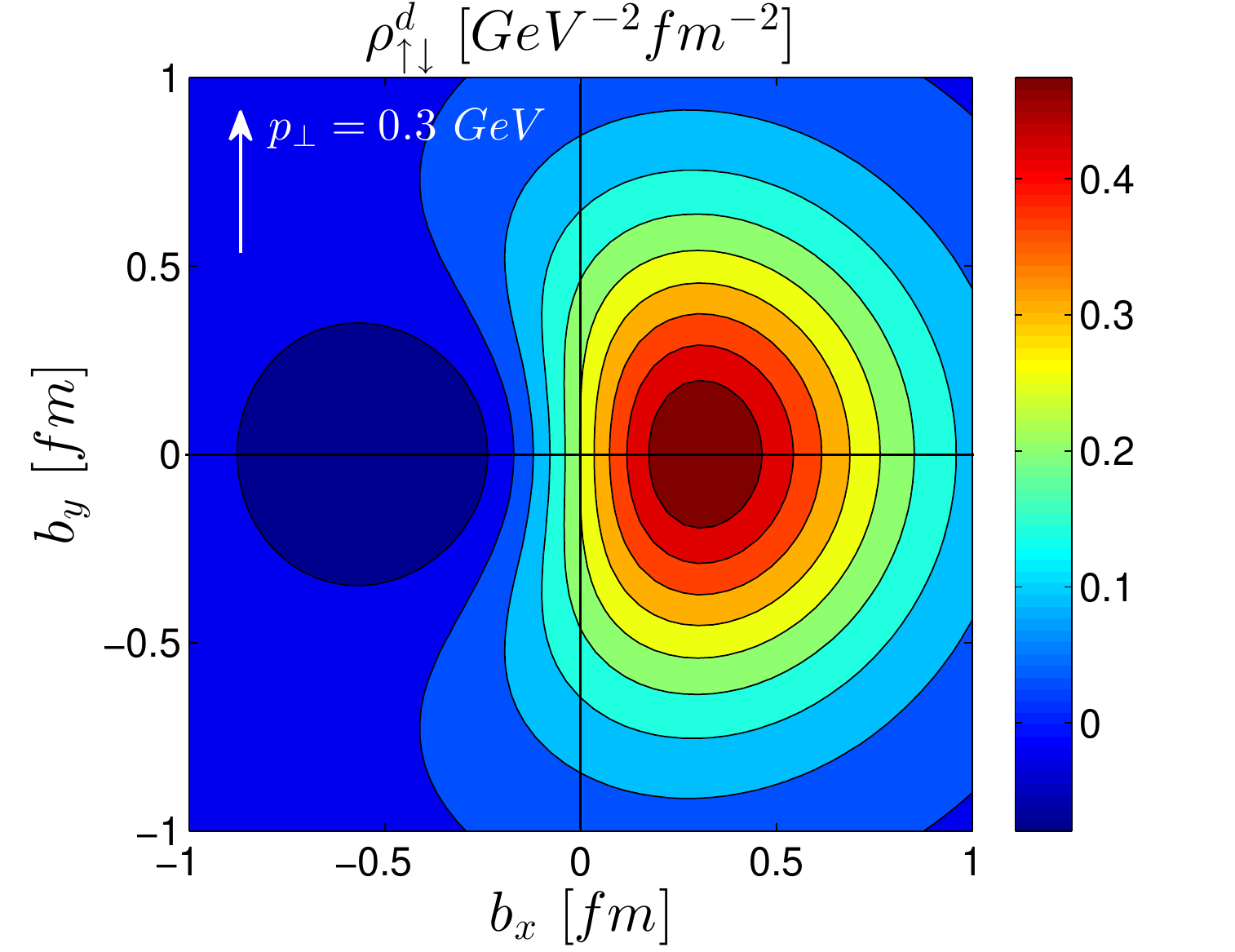}}\\
\subfigure[]{\includegraphics[width=5.cm,height=4.cm]{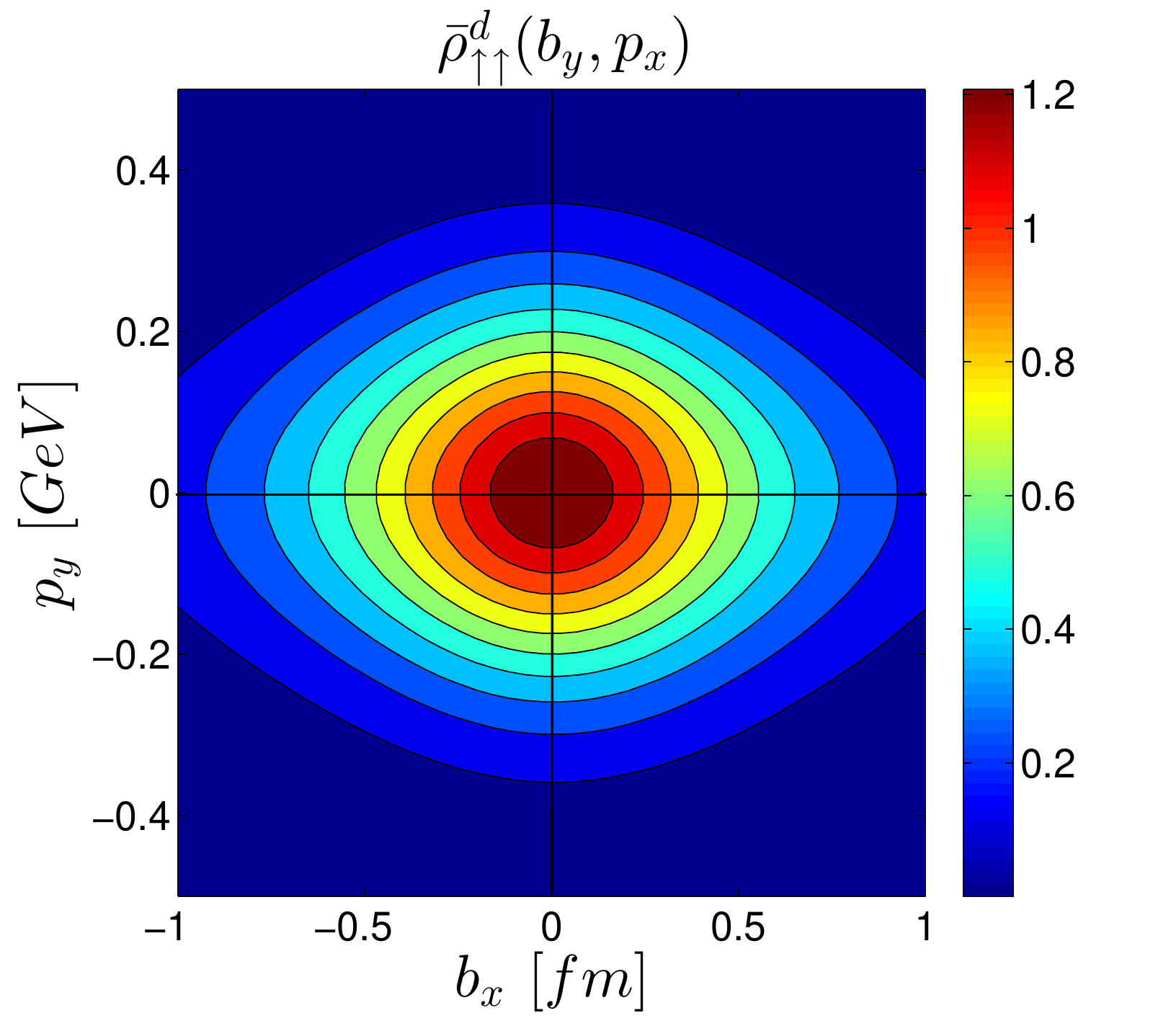}}
\subfigure[]{\includegraphics[width=5.cm,height=4.cm]{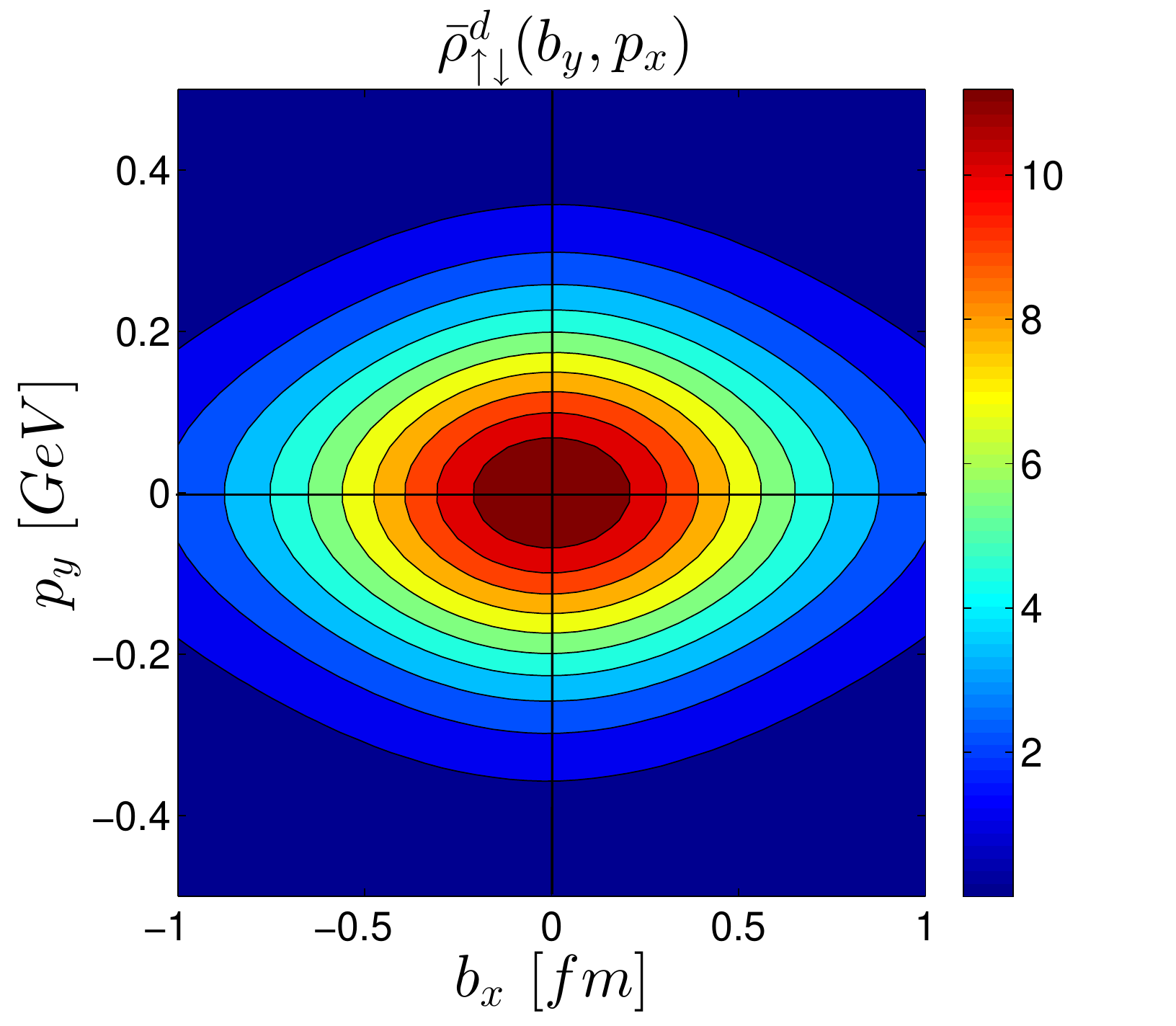}}
\caption{\label{fig_Lamlam_d}The distributions $\rho_{\Lambda\lambda}$ are shown in the transverse momentum plane, transverse coordinate plane and mixed plane for $d$  quarks. The distributions in the mixed plane are given in $GeV^0 fm^0$.}
\end{figure}
\begin{figure}[htb]
\centering
\subfigure[]{\includegraphics[width=5.cm,height=4.cm]{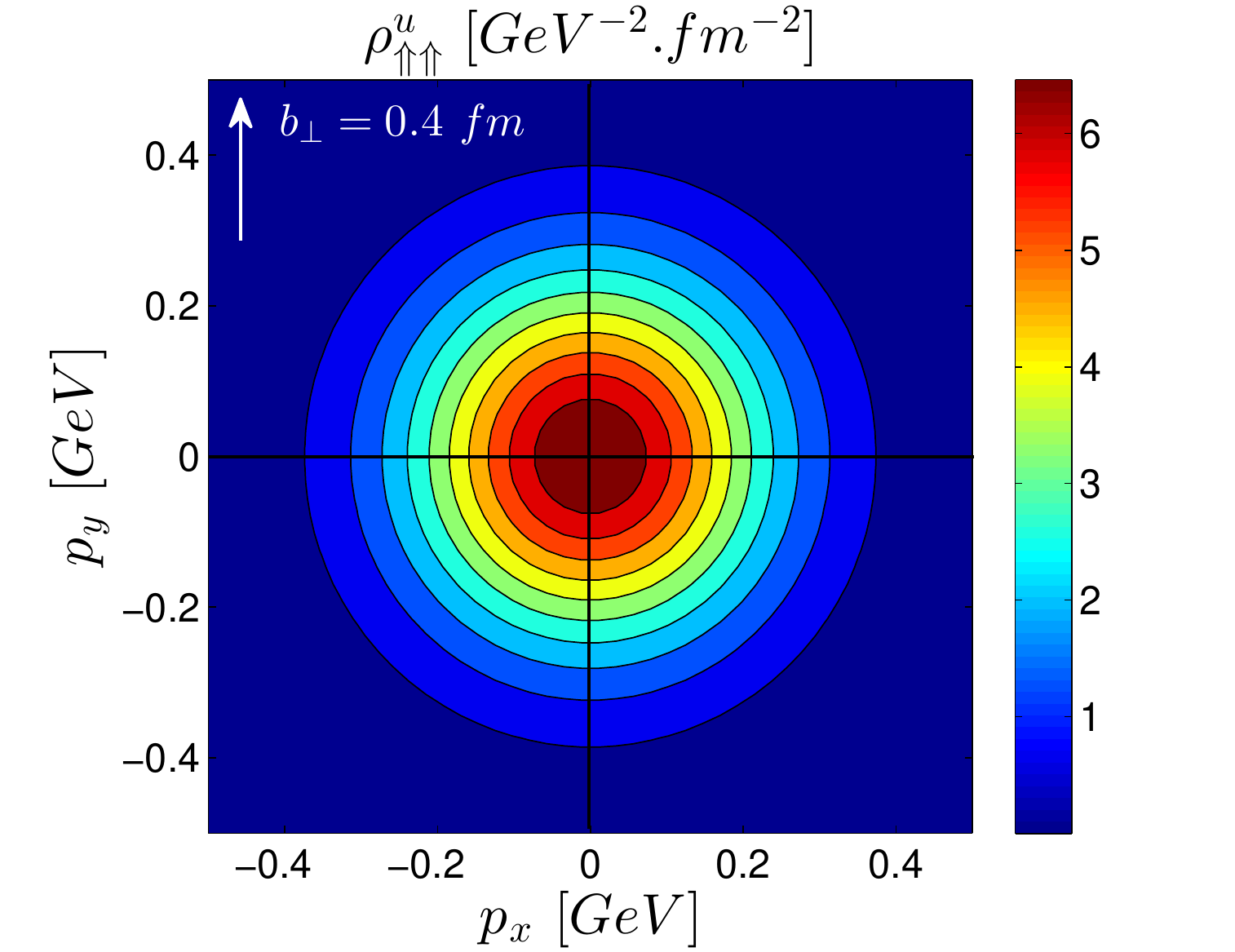}}
\subfigure[]{\includegraphics[width=5.cm,height=4.cm]{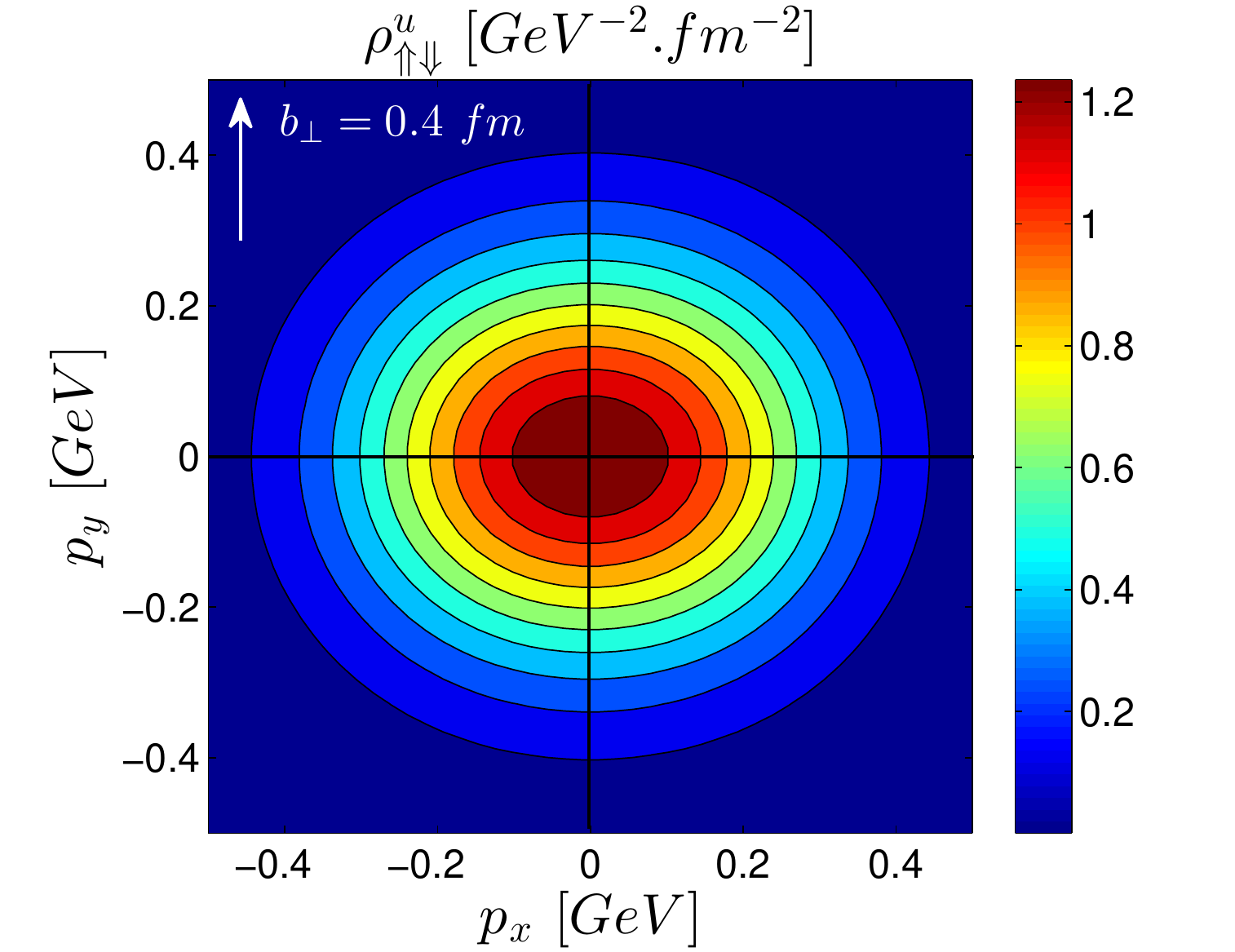}}\\
\subfigure[]{\includegraphics[width=5.cm,height=4.cm]{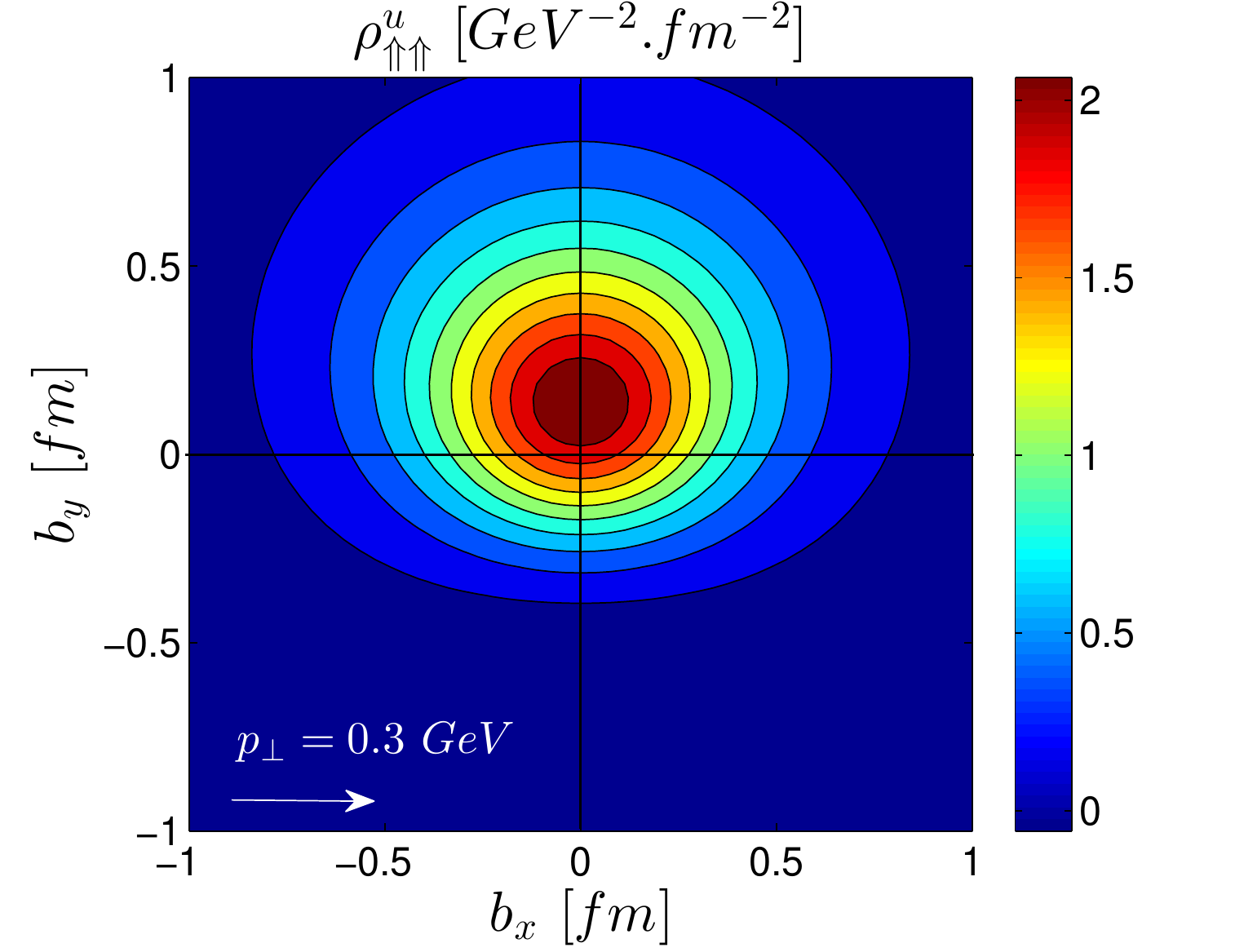}}
\subfigure[]{\includegraphics[width=5.cm,height=4.cm]{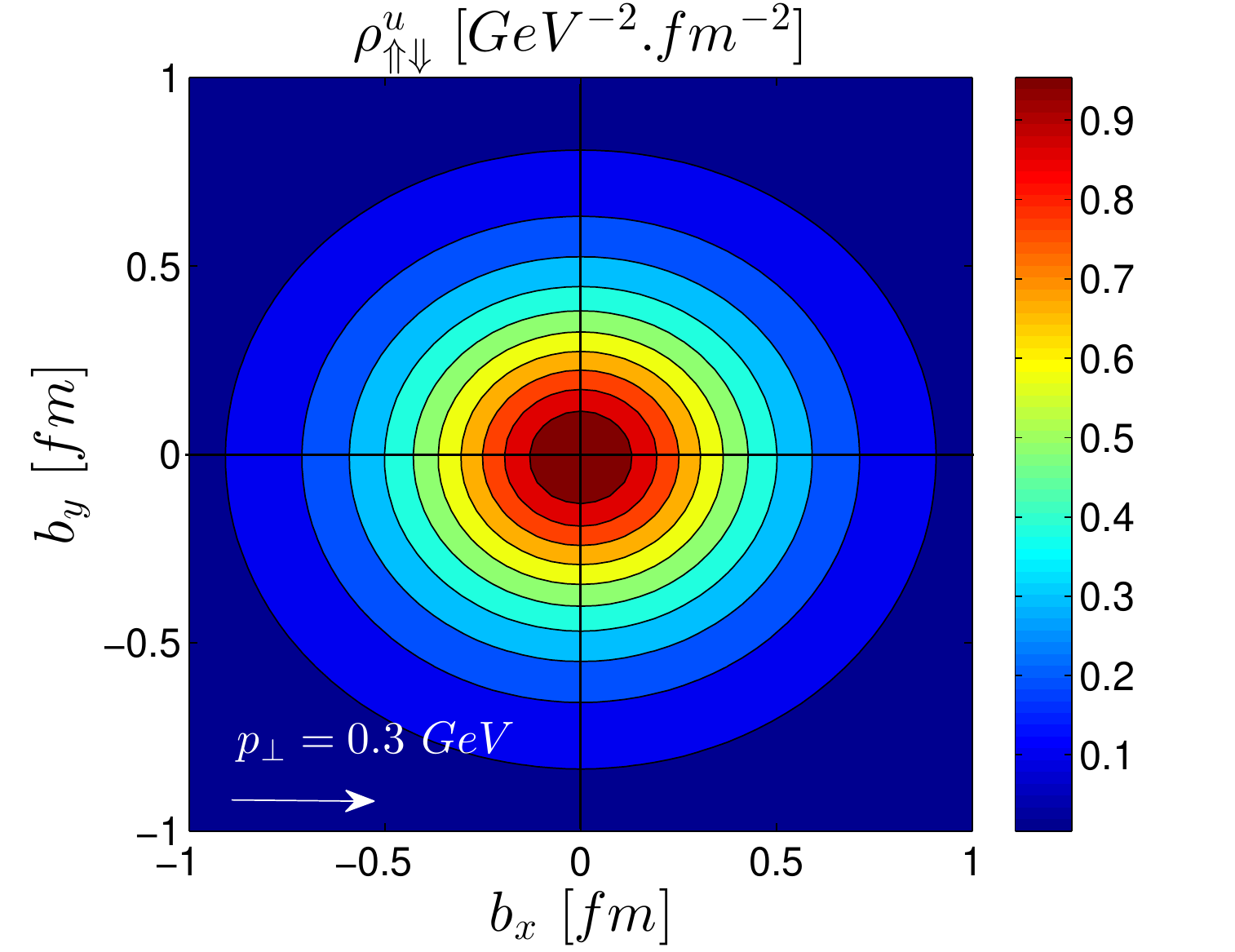}}\\
\subfigure[]{\includegraphics[width=5.cm,height=4.cm]{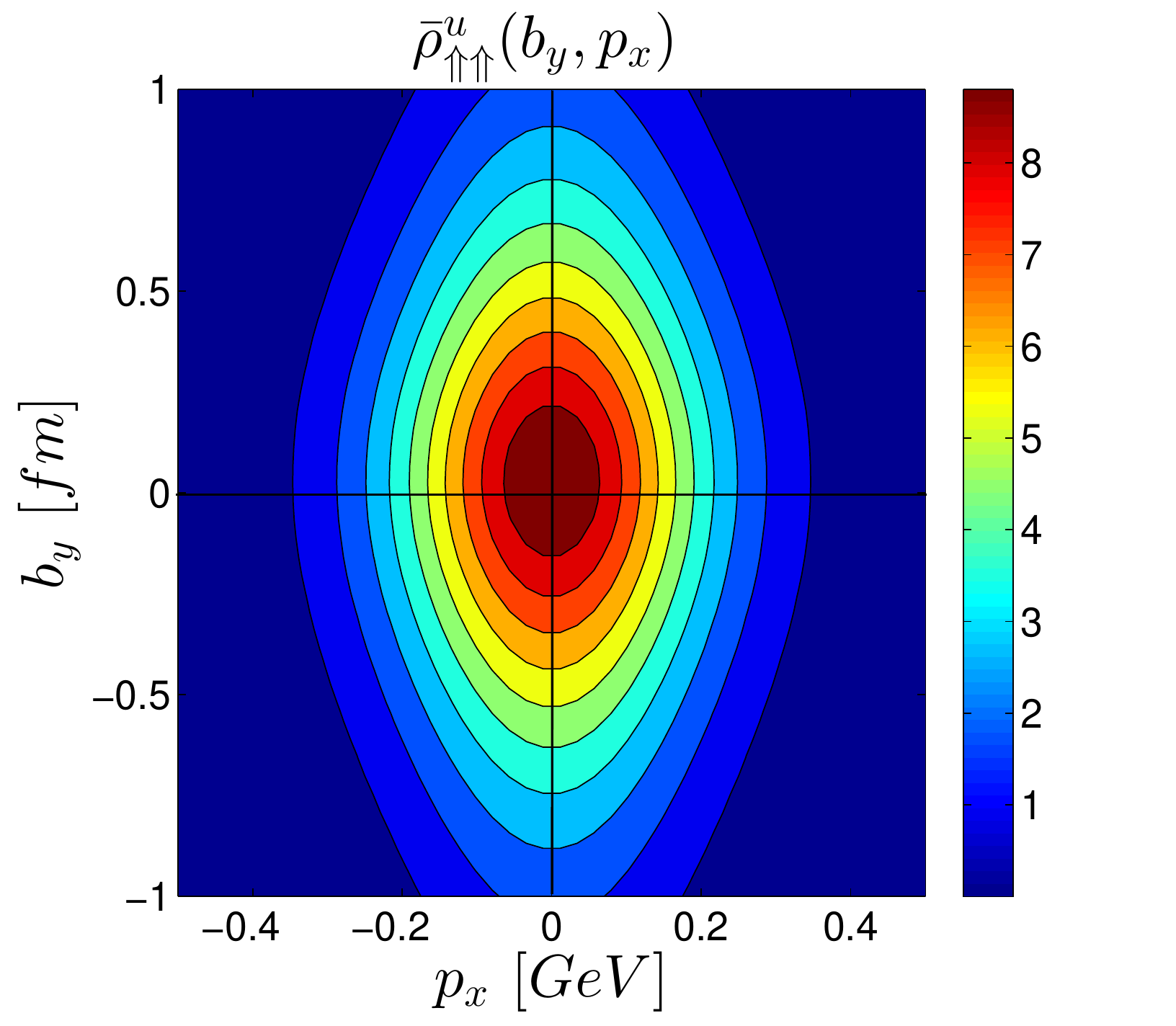}}
\subfigure[]{\includegraphics[width=5.cm,height=4.cm]{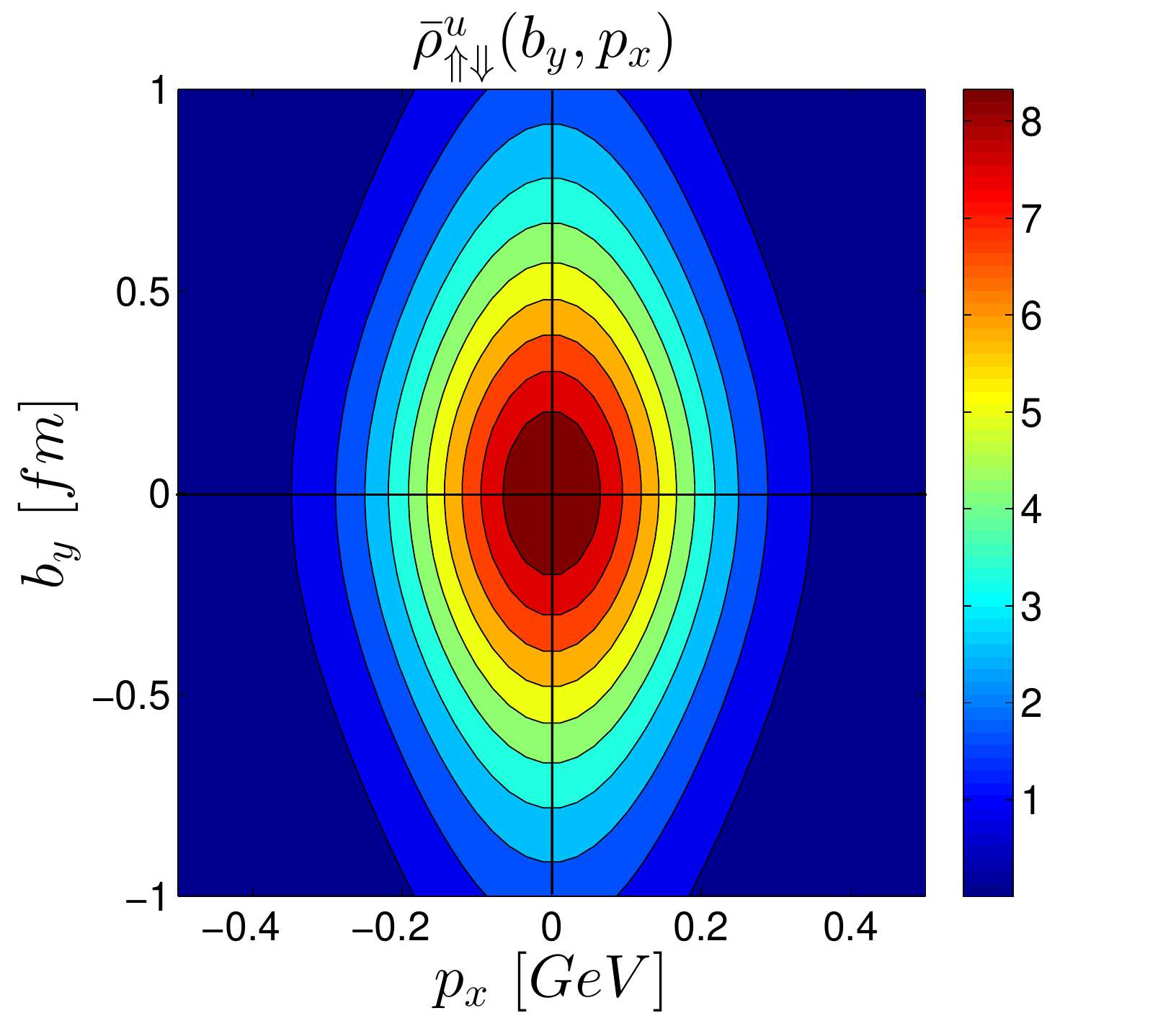}}
\caption{\label{fig_LamlamT_u}$\rho_{\Lambda_T\lambda_T}$ in transverse momentum plane, transverse impact parameter plane and mixed plane for $u$ quark(Eq.(\ref{rho_Lamlam_T})). $\Lambda_T=\Uparrow, \Downarrow $ and $\lambda_T=\Uparrow, \Downarrow$ represent the transverse polarization along x-axis for proton and quarks  respectively. The mixed plane distributions are given in $GeV^0 fm^0$. }
\end{figure}
\begin{figure}[htb]
\centering
\subfigure[]{\includegraphics[width=5.cm,height=4.cm]{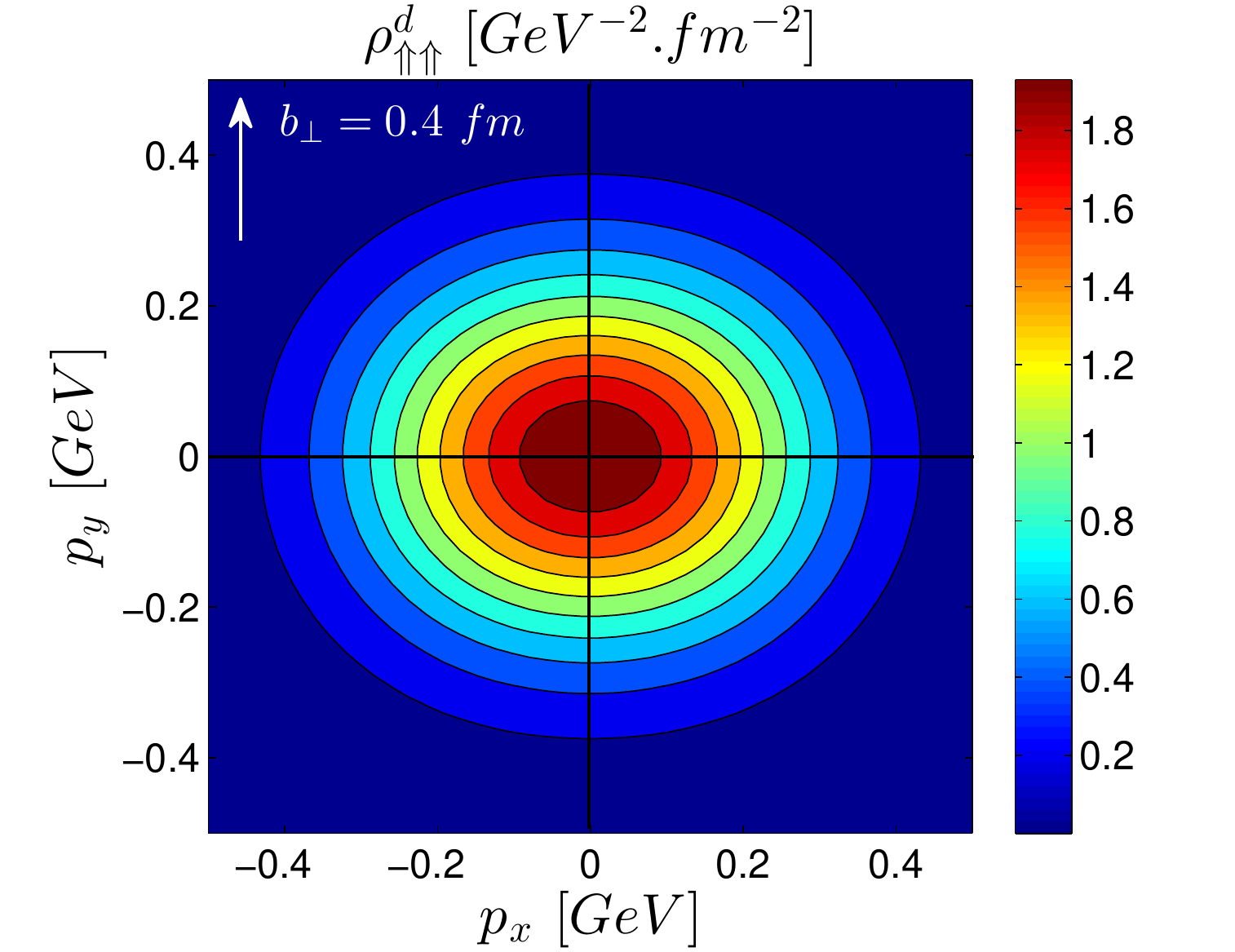}}
\subfigure[]{\includegraphics[width=5.cm,height=4.cm]{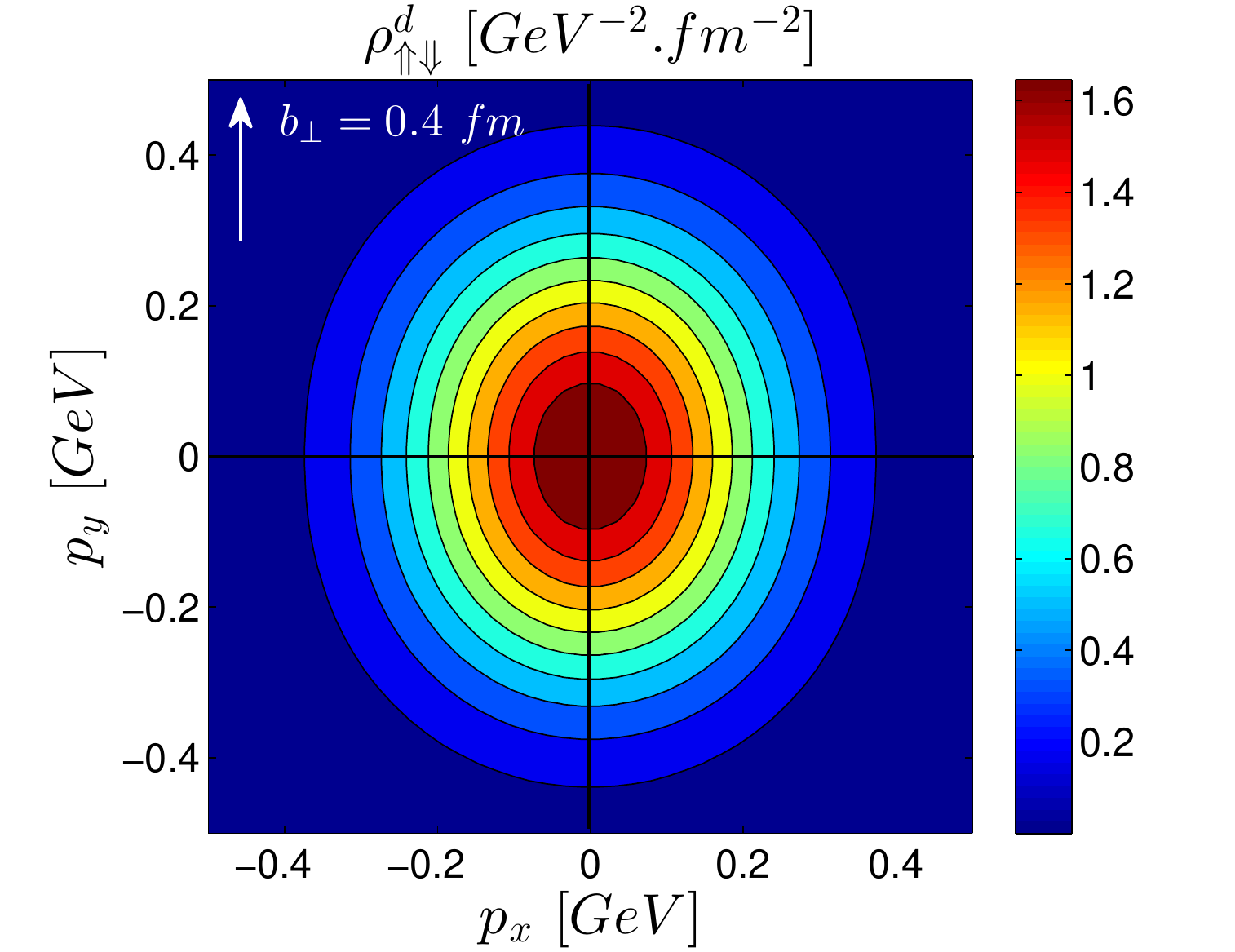}}\\
\subfigure[]{\includegraphics[width=5.cm,height=4.cm]{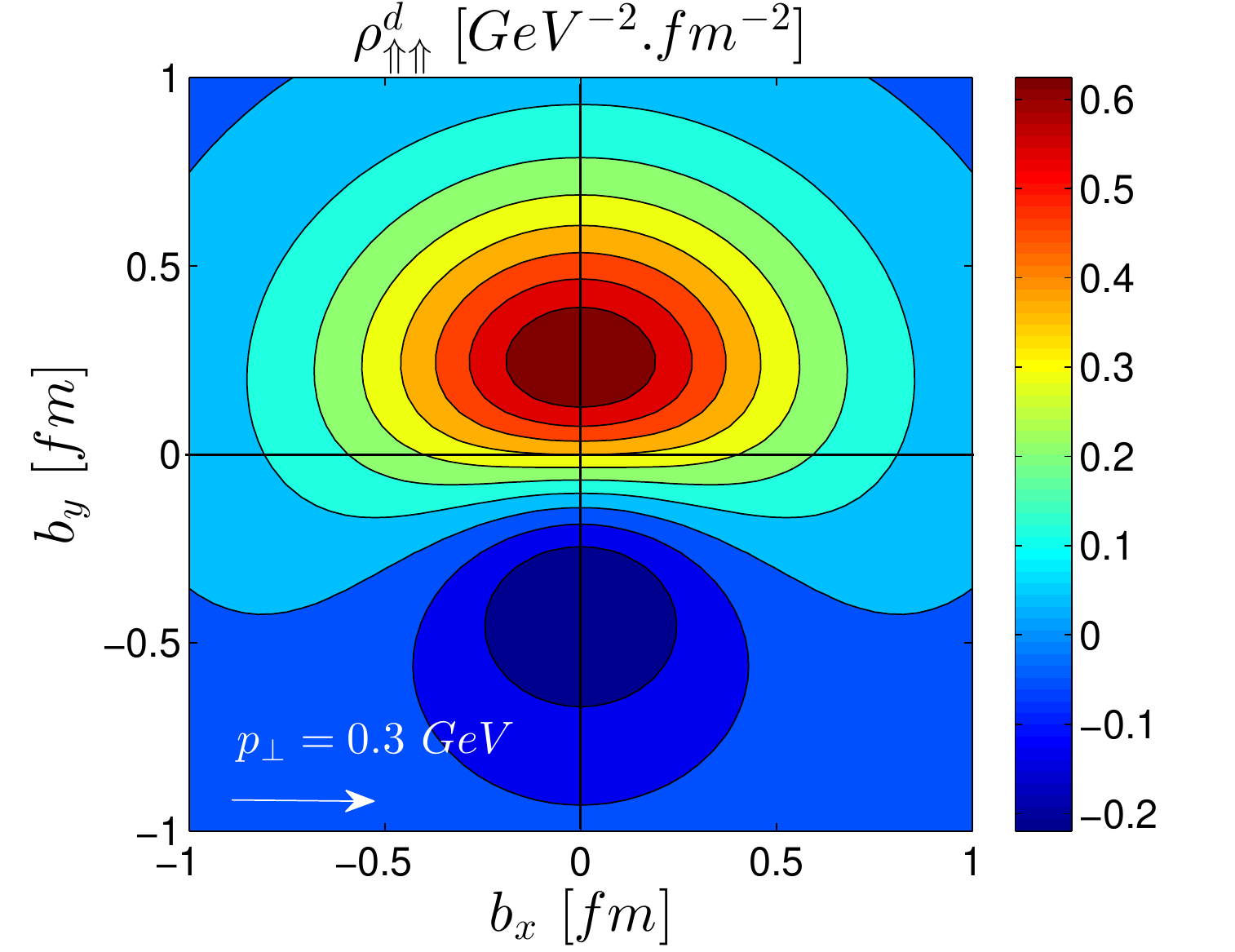}}
\subfigure[]{\includegraphics[width=5.cm,height=4.cm]{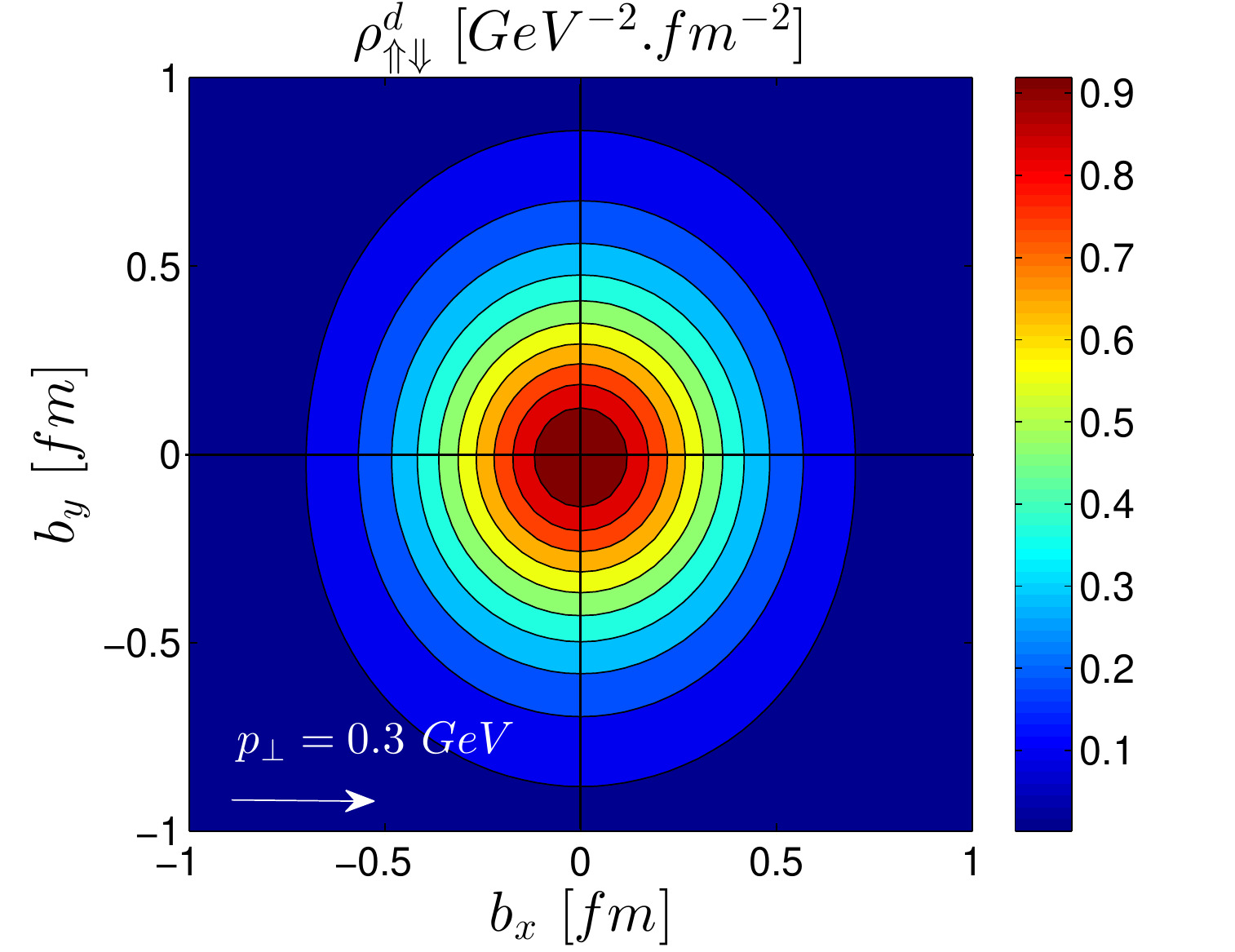}}\\
\subfigure[]{\includegraphics[width=5.cm,height=4.cm]{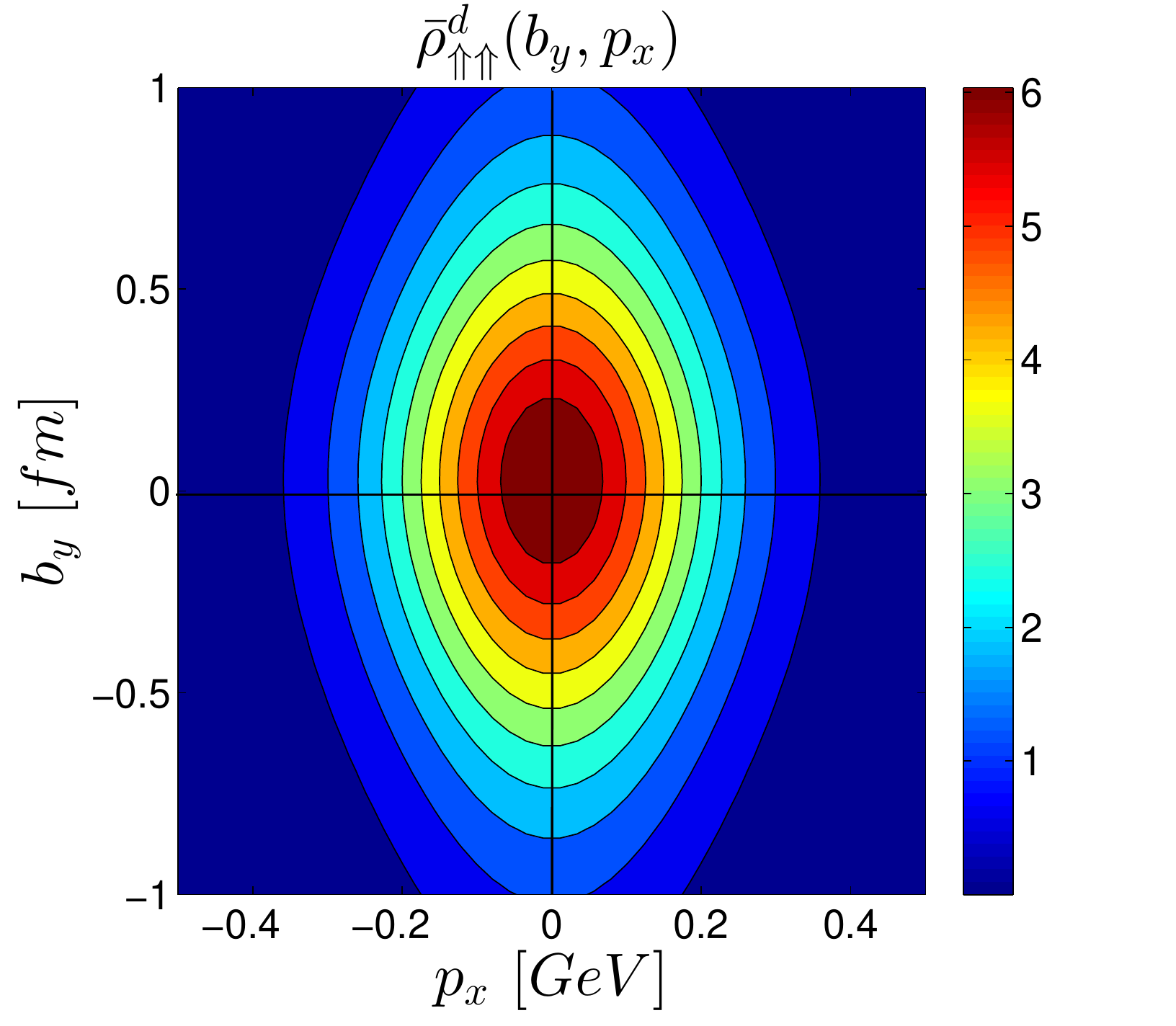}}
\subfigure[]{\includegraphics[width=5.cm,height=4.cm]{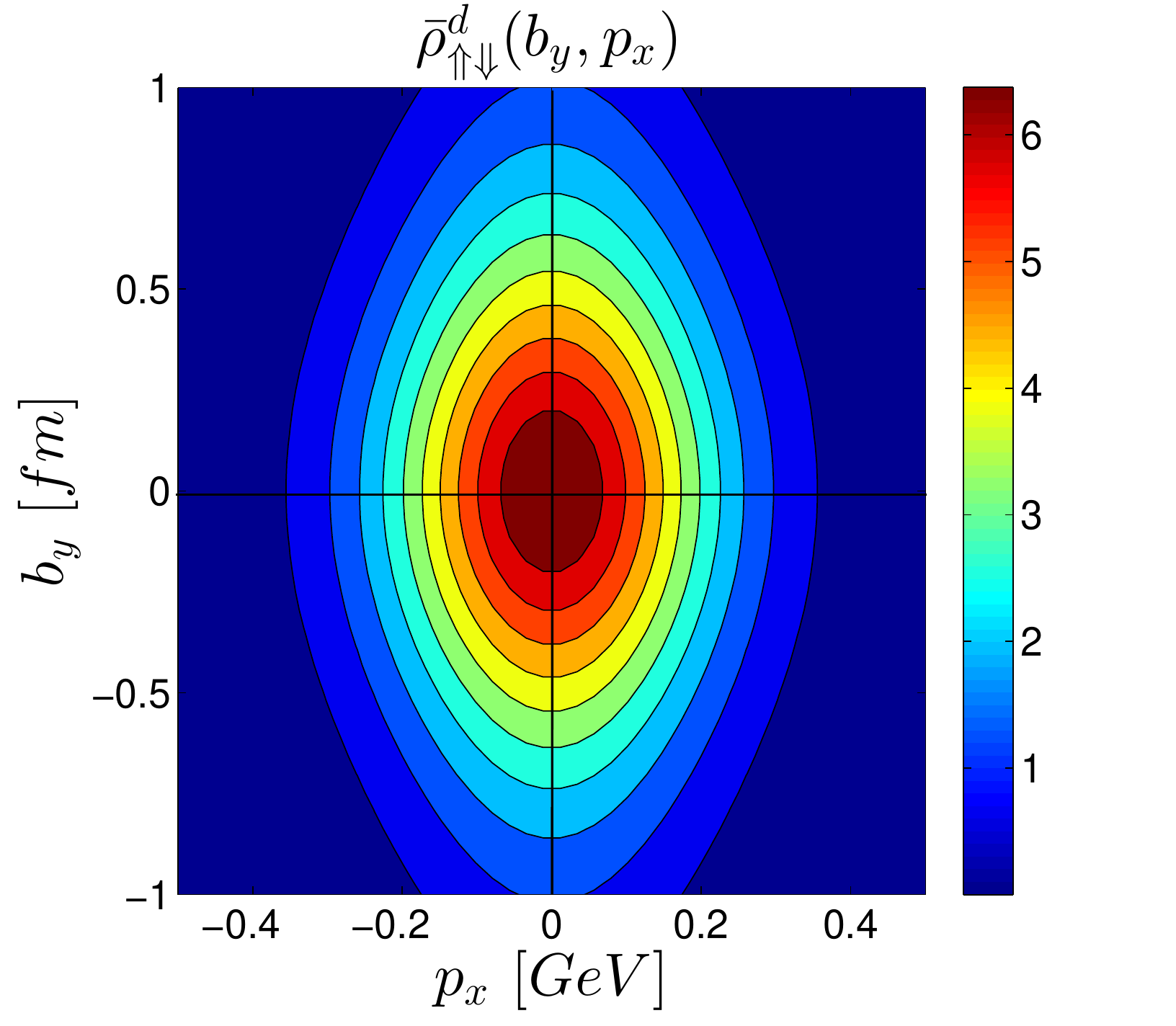}}
\caption{\label{fig_LamlamT_d}$\rho_{\Lambda_T\lambda_T}$ in transverse momentum plane, transverse impact parameter plane and mixed plane for $d$  quark(Eq.(\ref{rho_Lamlam_T})). $\Lambda_T=\Uparrow, \Downarrow $ and $\lambda_T=\Uparrow, \Downarrow$ represent the transverse polarization along x-axis for proton and quarks  respectively. The distributions in the mixed plane are given in $GeV^0 fm^0$.}
\end{figure}
%-----------------------------
\subsection{Spin-spin correlations}\label{spincor}
%-----------------------
The longitudinal Wigner distributions $\rho^\nu_{\Lambda\lambda}(\bfb,\bfp)$, with the polarization of proton $\Lambda=\uparrow$ and quark polarization $\lambda=\uparrow,\downarrow$( Eq.(\ref{rho_Lamlam})) are shown in Fig.\ref{fig_Lamlam_u} and Fig.\ref{fig_Lamlam_d} for $u$ and $d$  quarks respectively. One can observe that the distributions
$\rho^\nu_{\Lambda\lambda}(\bfb,\bfp)$ in transverse momentum plane as well as in transverse impact parameter plane look circularly symmetric for $\Lambda = \lambda $ whereas for $\Lambda \neq \lambda $ the distributions get distorted along $p_x$ or $b_x$ for both $u$ and $d$ quarks. Since the polarity of $\rho^\nu_{UL}$ is opposite to $\rho^\nu_{LU}$ and the magnitudes of the distributions are more or less same, thus in Eq.(\ref{rho_Lamlam}), the contributions from $\rho_{LU}$ and $\rho_{UL}$ get almost canceled for $\Lambda=\lambda$ and the only dominating contributions coming from $\rho_{UU}$ and $\rho_{LL}$ which are circularly symmetric in both planes. Again for $\Lambda \neq \lambda $ the contributions from $\rho_{LU}$ and $\rho_{UL}$ add up and causes the distortion. Note that the distortions from the circular symmetry in transverse momentum plane and in transverse impact parameter plane are in opposite direction to each other. Here we have shown the distributions for $\Lambda=\uparrow$; the other possible spin combinations in transverse momentum plane and in transverse impact parameter plane can be found from $ \rho^\nu_{\downarrow\lambda^\prime}(\bfb,p_x,p_y)=\rho^\nu_{\uparrow\lambda}(\bfb,-p_x,p_y)$, and $\rho^\nu_{\downarrow\lambda^\prime}(b_x,b_y,\bfp)=\rho^\nu_{\uparrow\lambda}(-b_x,b_y,\bfp)$ respectively, where  $\lambda^\prime\neq\lambda$. The mixed transverse densities $\tilde{\rho}^\nu_{\Lambda\lambda}(b_x,p_y)$ are shown in Fig.\ref{fig_Lamlam_u}(c),(f) for $u$ quark and in Fig.\ref{fig_Lamlam_d}(c),\ref{fig_Lamlam_d}(f) for $d$ quark. $\tilde{\rho}^\nu_{\Lambda\lambda}(b_x,p_y)$ exhibits the similar axially symmetric nature of $\tilde\rho_{UU}(b_x,p_y)$ with lower magnitude for $\Lambda = \lambda $. This is because of the other contribution, $\tilde\rho_{LL}(b_x,p_y)$ which is opposite to $\tilde\rho_{UU}(b_x,p_y)$. For $\Lambda \neq \lambda $, although there are additional quadrupole contributions from $\tilde\rho_{UL}$ and $\tilde\rho_{LU}$ in $\tilde\rho_{\uparrow\downarrow}(b_x,p_y)$, the contributions from $\rho_{UU}$ and $\rho_{LL}$ are very large compared to the quadrupole contributions and thus $\tilde\rho_{\uparrow\downarrow}(b_x,p_y)$ effectively show the similar behavior of $\tilde\rho_{UU}$ with larger magnitude for both $u$ and $d$ quark.

 We observe that  the quark OAM tends to be aligned with proton spin and anti-aligned to the quark spin for both $u$ and $d$ quarks. The difference in correlation strength between  quark OAM-proton spin correlation and quark OAM-spin correlation is very small(see Fig.\ref{fig_UL} and Fig.\ref{fig_LU}). Therefore, if the quark spin is parallel to the proton spin, i,e. $\Lambda=\uparrow, \lambda=\uparrow $ the contributions of $\rho_{UL}$ and $\rho_{LU}$ interfere destructively resulting the circular symmetry for $u$  and $d$ quarks, see Fig. \ref{fig_Lamlam_u}(a,b) and \ref{fig_Lamlam_d}(a,b).
If the quark spin is anti-parallel to the proton spin, i,e. $\Lambda=\uparrow, \lambda= \downarrow$ the contributions of $\rho_{UL}$ and $\rho_{LU}$ interfere constructively resulting a dipolar distribution for $u$  and $d$ quarks, see Fig \ref{fig_Lamlam_u}(d,e) and \ref{fig_Lamlam_d}(d,e).

From Eq.(\ref{rho_Lamlam_T}), we plot the transverse wigner distribution $\rho_{\Lambda_T\lambda_T}(\bfb,\bfp)$ in Fig.\ref{fig_LamlamT_u} for $u$ quark and Fig.\ref{fig_LamlamT_d} for $d$  quark to understand the transverse spin-spin correlations. The Fig.\ref{fig_LamlamT_u} represents the distribution of a $u$ quark with transverse polarization $\lambda_T=\Uparrow,\Downarrow$ (along x-axis) in a proton with transverse polarization $\Lambda_T=\Uparrow$ along x-axis. 
In the transverse momentum plane, we see an elliptical distribution for both the quarks(Fig.\ref{fig_LamlamT_u}(a,d) and \ref{fig_LamlamT_d}(a,d)) because the distortions $\rho_{TU}, \rho_{UT}$ are circularly symmetric  and $\rho_{TT}$ is elliptically symmetric. In the transverse impact parameter plane we observed significant deviation comes from the dipolar nature of the distortions $\rho^\nu_{UT}$ and $\rho^\nu_{TU}$. For $\Lambda_T\lambda_T=\Uparrow\Uparrow$, they interfere constructively and causes a large deviation as sheen in Fig.\ref{fig_LamlamT_u}(b),\ref{fig_LamlamT_d}(b). We also observed that the distributions change axis with the flip of transverse polarization of quarks.

%%%%%%%%%%
 \section{GTMDs and their evolution}\label{sec_GTMD}
%%%%%%%%%%%%%
The generalized transverse momentum dependent distributions can be extracted from the different Wigner distributions as shown in Eqs.(\ref{rhoUU_F}-\ref{rhoTTperp_H}). The GTMDs reduce to the TMDs and GPDs at certain kinematical limits. The $F_{1,4}$ and $G_{1,1}$ contribute to the spin-OAM correlation as discussed in Sec.\ref{oam}.  $F_{1,4}$ and OAM in MIT bag model has been calculated in \cite{MIT}. There are altogether 11 non zero GTMDs at the leading twist in this  model. In this model, comparing Eq.(\ref{rhoUU_F}-\ref{rhoTTperp_H}) with Eq.(\ref{rhoUU_nu}-\ref{rhoTTp_nu}), the explicit form of the GTMDs are   
\be 
F^\nu_{1,1}(x,\Dp^2,\bfp^2) &=& N^\nu_{UU}\frac{1}{16\pi^3}\bigg[|A^{\nu}_1(x)|^2 +
\bigg(\bfp^2 - \frac{\Dp^2}{4}(1-x)^2\bigg)\frac{1}{M^2x^2}|A^{\nu}_2(x)|^2 \bigg]\exp\big[-2\tilde{a}(x)\tilde{\textbf{p}}^2_\perp \big],\nonumber\\ \label{F11} \\
F^\nu_{1,2}(x,\Dp^2,\bfp^2) &=& 0,\label{F12} \\
F^\nu_{1,3}(x,\Dp^2,\bfp^2) &=&\frac{1}{2}F^\nu_{1,1}(x,\Dp^2,\bfp^2)+N^\nu_{TU}\frac{1}{16\pi^3}\bigg[\frac{(1-x)}{x}A^{\nu}_1(x)A^{\nu}_2(x)\bigg] \exp\big[-2\tilde{a}(x)\tilde{\textbf{p}}^2_\perp \big],\label{F13}\\
F^\nu_{1,4}(x,\Dp^2,\bfp^2) &=& -N^\nu_{LU}\frac{1}{16\pi^3}\bigg[\frac{(1-x)}{x^2}|A^{\nu}_2(x)|^2\bigg] \exp\big[-2\tilde{a}(x)\tilde{\textbf{p}}^2_\perp \big],\label{F14}\\
G^\nu_{1,1}(x,\Dp^2,\bfp^2) &=& N^\nu_{UL}\frac{1}{16\pi^3}\bigg[\frac{(1-x)}{x^2}|A^{\nu}_2(x)|^2\bigg] \exp\big[-2\tilde{a}(x)\tilde{\textbf{p}}^2_\perp \big] ,\label{G11}\\
G^\nu_{1,2}(x,\Dp^2,\bfp^2) &=& N^\nu_{TL}\frac{1}{16\pi^3}\bigg[\frac{2}{x}A^{\nu}_1(x)A^{\nu}_2(x)\bigg] \exp\big[-2\tilde{a}(x)\tilde{\textbf{p}}^2_\perp \big] ,\label{G12}\\
G^\nu_{1,3}(x,\Dp^2,\bfp^2) &=&0,\\
G^\nu_{1,4}(x,\Dp^2,\bfp^2) &=& N^\nu_{LL}\frac{1}{16\pi^3}\bigg[|A^{\nu}_1(x)|^2 -
\bigg(\bfp^2 - \frac{\Dp^2}{4}(1-x)^2\bigg)\frac{1}{M^2x^2}|A^{\nu}_2(x)|^2 \bigg]\exp\big[-2\tilde{a}(x)\tilde{\textbf{p}}^2_\perp \big],\label{G14}\nonumber\\
\ee
\be
H^\nu_{1,1}(x,\Dp^2,\bfp^2) &=&0,\\
H^\nu_{1,2}(x,\Dp^2,\bfp^2) &=&N^\nu_{UT}\frac{1}{16\pi^3}\bigg[\frac{(1-x)}{x}A^{\nu}_1(x)A^{\nu}_2(x)\bigg] \exp\big[-2\tilde{a}(x)\tilde{\textbf{p}}^2_\perp \big]\\
H^\nu_{1,3}(x,\Dp^2,\bfp^2) &=& N^\nu_{TT} \frac{1}{16\pi^3}\bigg[|A^{\nu}_1(x)|^2 +
\bigg(\bfp^2 - \frac{\Dp^2}{4}(1-x)^2\bigg)\frac{1}{M^2x^2}|A^{\nu}_2(x)|^2 \bigg]\nonumber\\
&& \hspace{2.6cm}\times \exp\big[-2\tilde{a}(x)\tilde{\textbf{p}}^2_\perp \big]-\frac{1}{2M^2}\Dp^2 H^\nu_{1,2}(x,\Dp^2,\bfp^2),\label{H13}\\
H^\nu_{1,4}(x,\Dp^2,\bfp^2) &=& N^{\perp\nu}_{TT} \frac{1}{16\pi^3}\bigg[\frac{2}{x^2}|A^{\nu}_2|^2\bigg] \exp\big[-2\tilde{a}(x)\tilde{\textbf{p}}^2_\perp \big],\label{H14}\\
H^\nu_{1,5}(x,\Dp^2,\bfp^2) &=&0,\\
H^\nu_{1,6}(x,\Dp^2,\bfp^2) &=& N^\nu_{TT} \frac{1}{16\pi^3}\bigg[\frac{(1-x)^2}{2x^2}|A^{\nu}_2|^2\bigg] \exp\big[-2\tilde{a}(x)\tilde{\textbf{p}}^2_\perp \big]+\frac{1}{2}H^\nu_{1,2}(x,\Dp^2,\bfp^2),\label{H16}\\
H^\nu_{1,7}(x,\Dp^2,\bfp^2) &=& N^\nu_{LT}\frac{1}{16\pi^3}\bigg[\frac{1}{x}A^{\nu}_1(x)A^{\nu}_2(x)\bigg] \exp\big[-2\tilde{a}(x)\tilde{\textbf{p}}^2_\perp \big], \label{H17}\\
H^\nu_{1,8}(x,\Dp^2,\bfp^2) &=&0.
\ee
 The normalization constants $N^\nu_{\Lambda\lambda}$ are 
\be 
N^\nu_{UU}&=&\bigg(C^2_SN^2_S+C^2_V\big(\frac{1}{3}N^2_0+\frac{2}{3}N^2_1\big)\bigg)^\nu, \quad \quad N^\nu_{LT} =\bigg(-C^2_SN^2_S-C^2_V\big(\frac{1}{3}N^2_0-\frac{2}{3}N^2_1\big)\bigg)^\nu,\nonumber\\
N^\nu_{UL}&=&\bigg(-C^2_SN^2_S-C^2_V\big(\frac{1}{3}N^2_0+\frac{2}{3}N^2_1\big)\bigg)^\nu, \quad \quad N^\nu_{TU}=\bigg(C^2_SN^2_S+C^2_V \frac{1}{3}N^2_0\bigg)^\nu,\nonumber\\
N^\nu_{UT}&=&\bigg(C^2_SN^2_S+C^2_V\big(\frac{1}{3}N^2_0+\frac{2}{3}N^2_1\big)\bigg)^\nu, \quad \quad N^\nu_{TL}=\bigg(C^2_SN^2_S-C^2_V\frac{1}{3}N^2_0\bigg)^\nu,\\
N^\nu_{LU}&=&\bigg(C^2_SN^2_S+C^2_V\big(\frac{1}{3}N^2_0-\frac{2}{3}N^2_1\big)\bigg)^\nu, \quad \quad N^\nu_{TT}=\bigg(C^2_SN^2_S-C^2_V \frac{1}{3}N^2_0\bigg)^\nu,\nonumber\\
N^\nu_{LL}&=&\bigg(C^2_SN^2_S+C^2_V(\frac{1}{3}N^2_0-\frac{2}{3}N^2_1\big)\bigg)^\nu, \quad \quad
N^{\perp\nu}_{TT}=\bigg(-C^2_SN^2_S+C^2_V\frac{1}{3}N^2_0\bigg)^\nu,\nonumber
\ee
where $N_S=0$ for $d$  quark. 
\begin{figure}[htb]
\centering
\subfigure[]{\includegraphics[width=4.cm,clip]{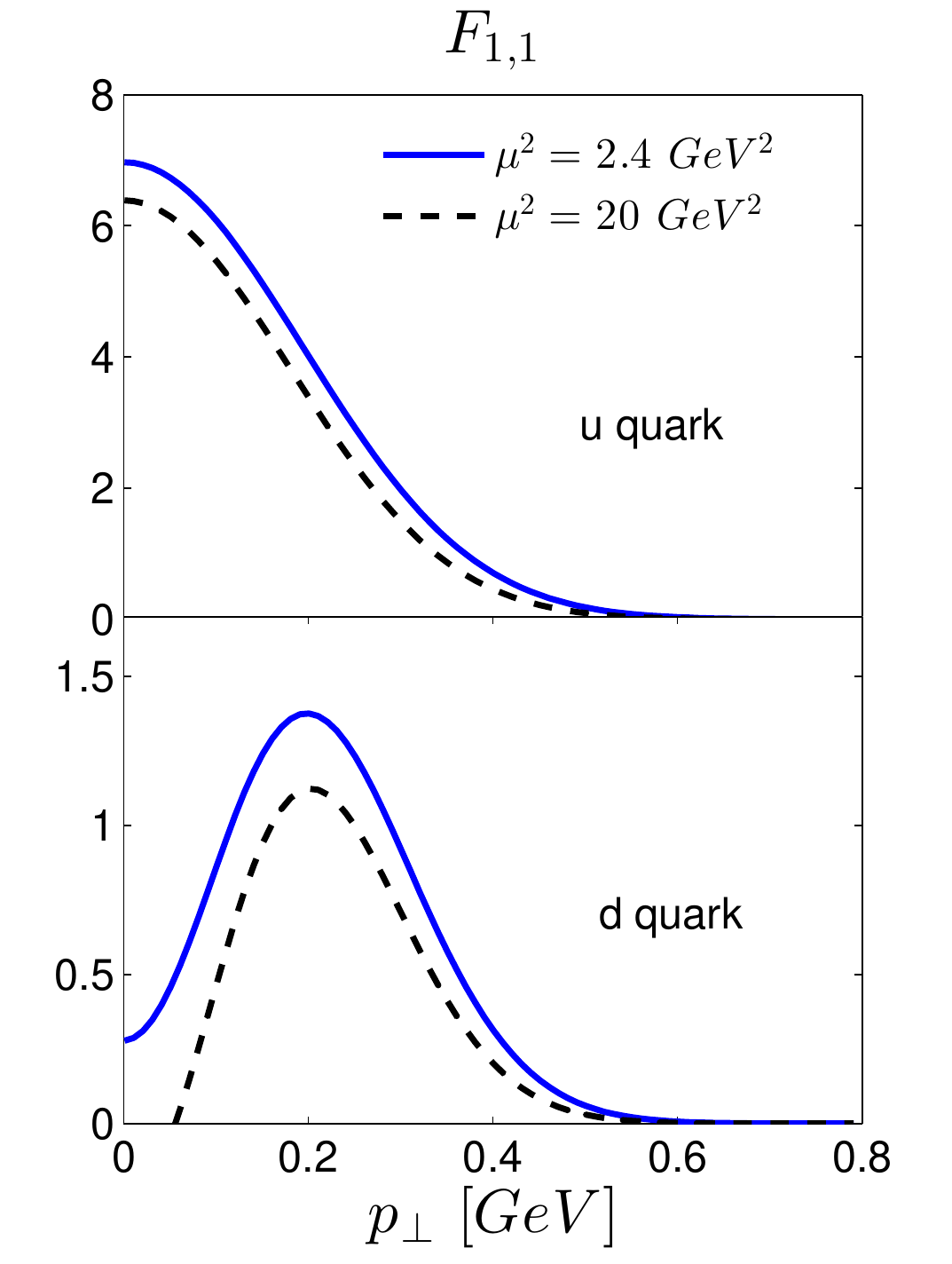}}
\subfigure[]{\includegraphics[width=4.cm,clip]{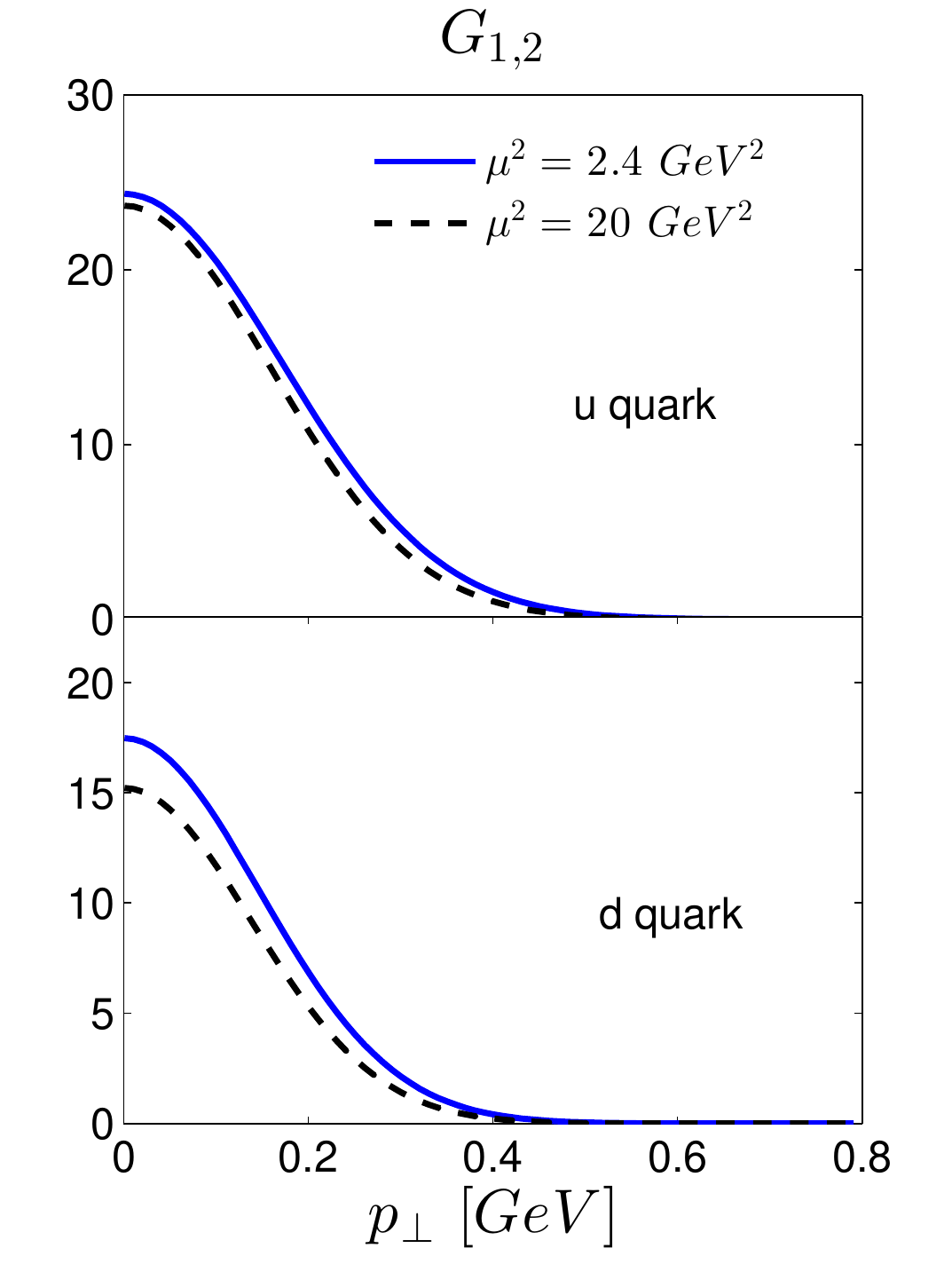}}
\subfigure[]{\includegraphics[width=4.cm,clip]{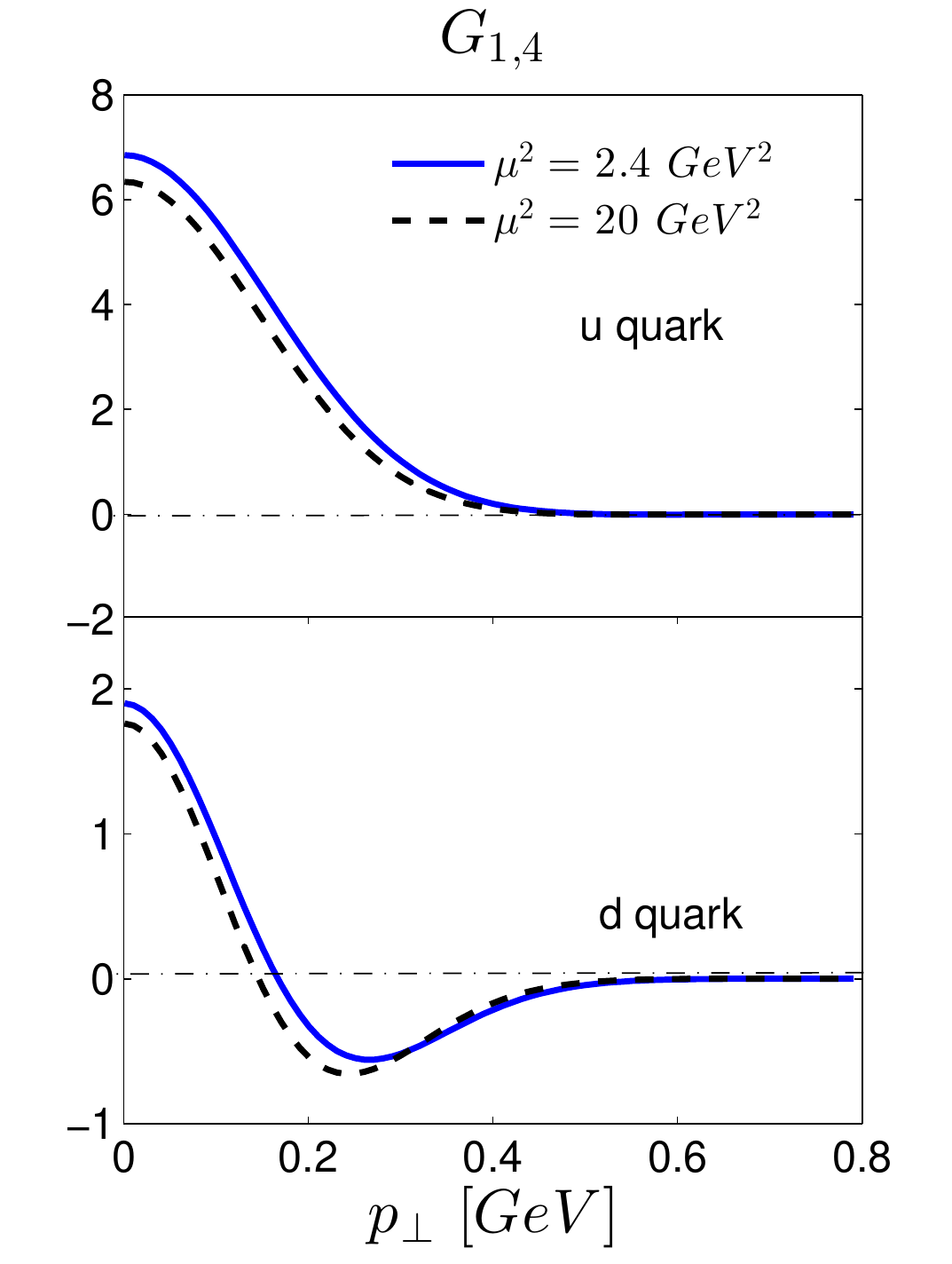}}\\
\subfigure[]{\includegraphics[width=4.cm,clip]{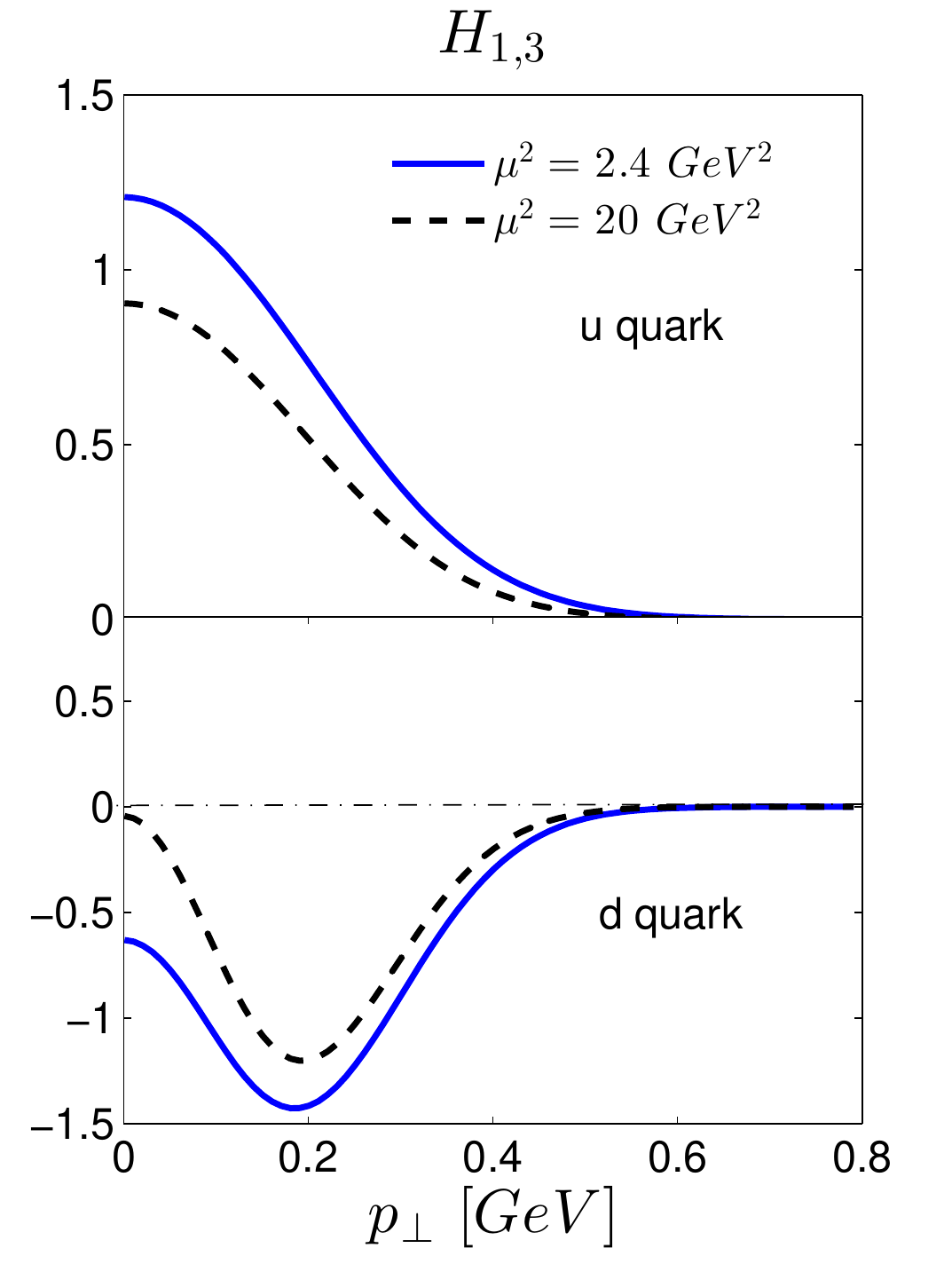}}
\subfigure[]{\includegraphics[width=4.cm,clip]{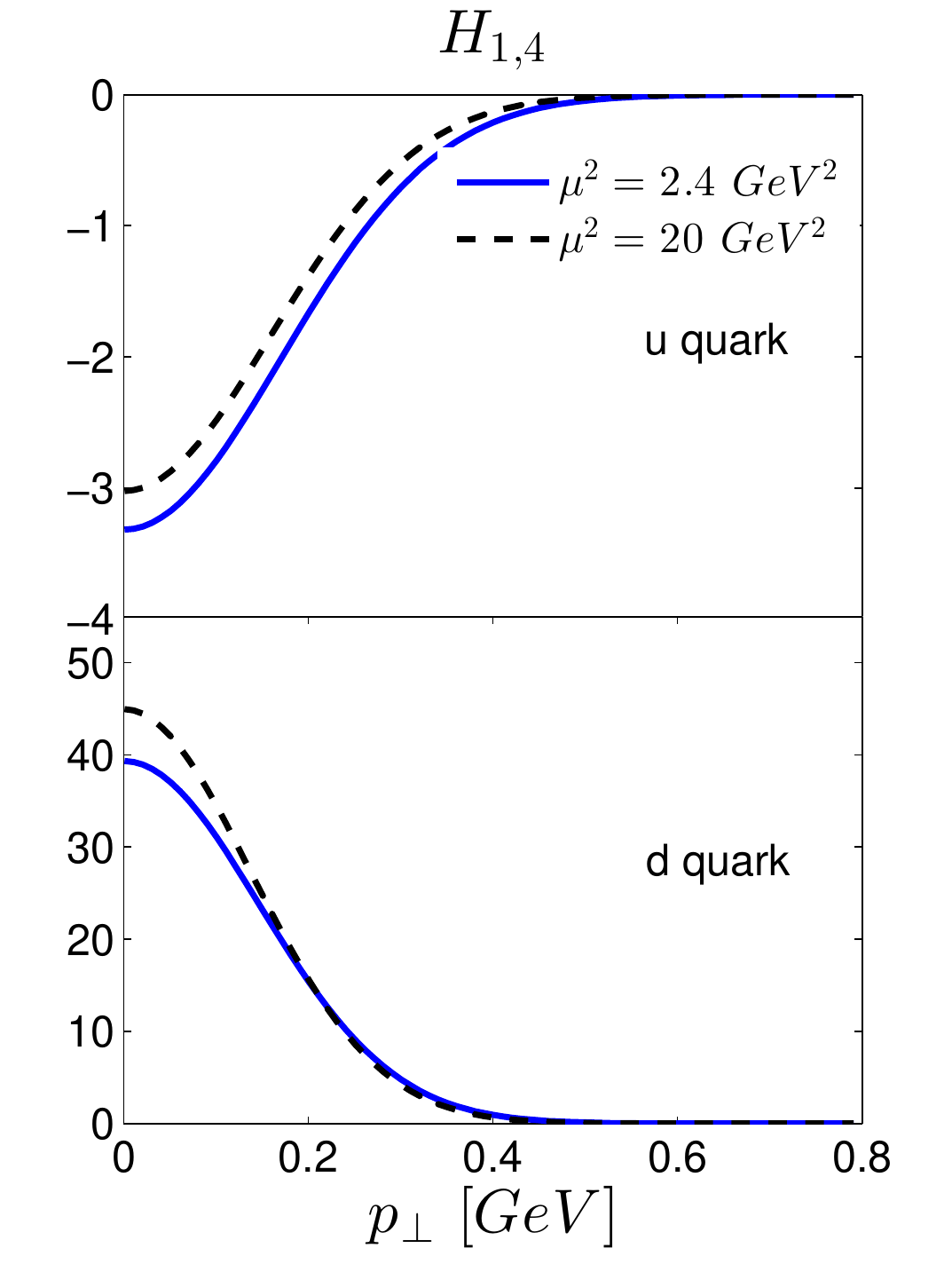}}
\subfigure[]{\includegraphics[width=4.cm,clip]{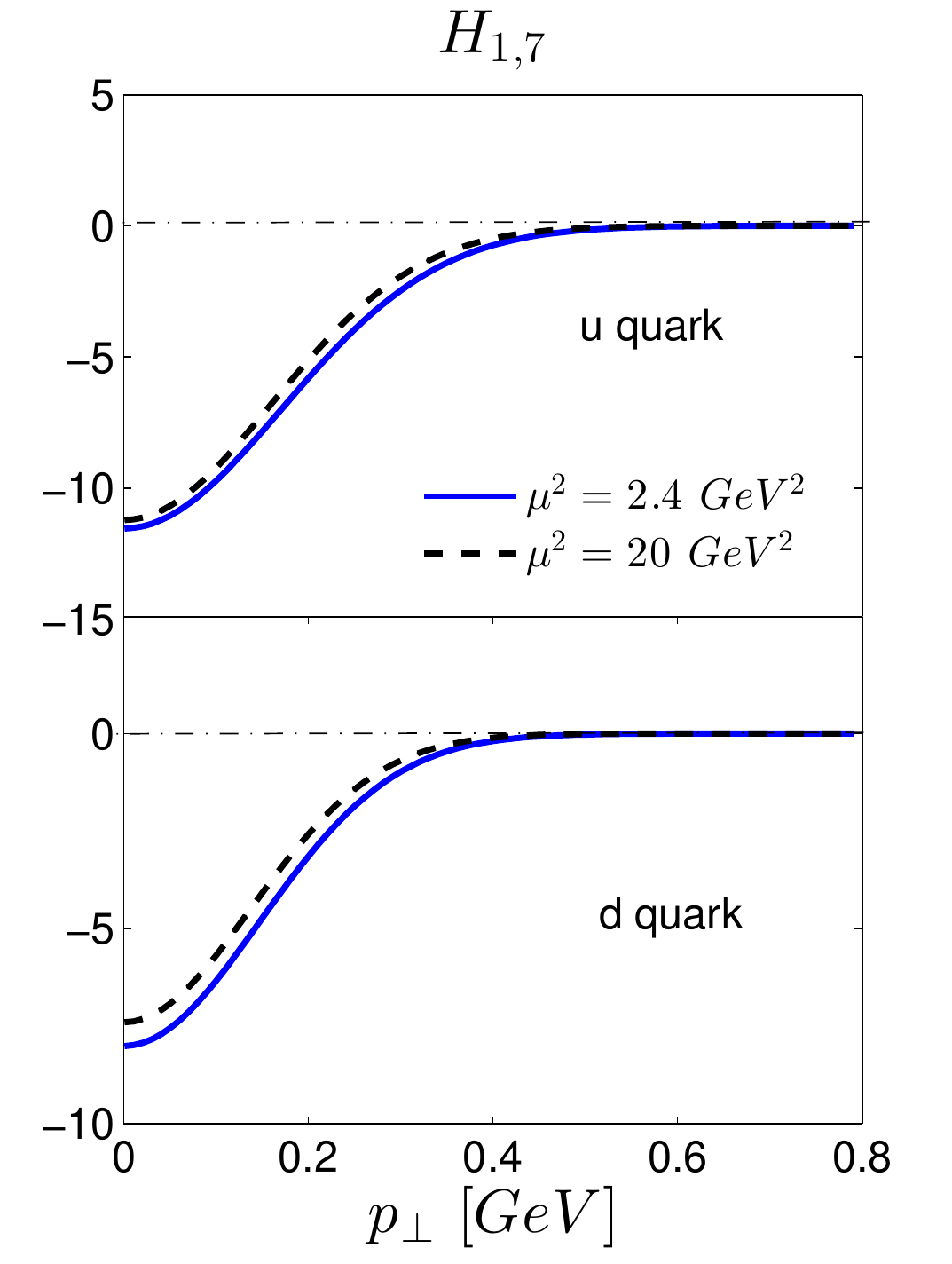}}
\caption{\label{fig_GTMD} Scale evolution of GTMDs with $x=0.3$ and $\Delta^2_\perp=0.1~GeV^2$ for u and d quarks. Where continuous lines are at $\mu^2=2.4~ GeV^2$(average $\mu^2$ value of the HERMES experiment) and doted lines are at $\mu^2=20~ GeV^2$(highest bin average $\mu^2$ value of the COMPASS experiment).}
\end{figure}

%Satisfying the hermiticity condition, in the TMD-limit there are only 8 GTMDs(6 T-even and 2 T-odd) survive at the leading twist. 
%They are  
%\be 
% F_{1,1}(x,0,\bfp^2,0,0)&=&f_1(x,\bfp^2), \\
%F_{1,2}(x,0,\bfp^2,0,0)&=&-f^\perp_{1T}(x,\bfp^2),\\
%G_{1,4}(x,0,\bfp^2,0,0)&=&g_{1L}(x,\bfp^2)\\
%G_{1,2}(x,0,\bfp^2,0,0)&=&g_{1T}(x,\bfp^2)\\
%H_{1,1}(x,0,\bfp^2,0,0)&=&-h^\perp_1(x,\bfp^2)\\
%H_{1,7}(x,0,\bfp^2,0,0)&=&h^\perp_{1L}(x,\bfp^2)\\
%H_{1,3}(x,0,\bfp^2,0,0)&=&h_{1T}(x,\bfp^2)\\
%H_{1,4}(x,0,\bfp^2,0,0)&=&h^\perp_{1T}(x,\bfp^2)
%\ee

 The scale evolution of the GTMDs are modeled considering the evolution of the parameters that reproduces the correct scale evolution of the pdfs\cite{MC}. Where the LFWFs are defined at the initial scale $\mu_0=0.313$ GeV and the hard scale evolution of the the distributions are modeled by making the parameters in the distribution scale dependent. The scale evolution of the parameters are determined by the DGLAP evolution of the PDFs:
\be 
a_i^\nu(\mu)&=&a_i^\nu(\mu_0) + A^\nu_{i}(\mu), \label{a_im}\\
b_i^\nu(\mu)&=&b_i^\nu(\mu_0) - B^\nu_{i}(\mu)\frac{4C_F}{\beta_0}\ln\bigg(\frac{\alpha_s(\mu^2)}{\alpha_s(\mu_0^2)}\bigg),\label{b_im}\\
\delta^\nu(\mu)&=& \exp\bigg[\delta^\nu_1\bigg(\ln(\mu^2/\mu_0^2)\bigg)^{\delta^\nu_2}\bigg],\label{DL}
\ee
where the $a_i^\nu(\mu_0)$ and $b_i^\nu(\mu_0)$ are the parameters at $\mu=\mu_0$. The parameter $\delta^\nu$ becomes unity at $\mu_0$ for both $u$ and $d$ quarks.
The scale dependent parts $A^\nu_{i}(\mu)$ and $B^\nu_{i}(\mu)$  evolve as 
\be 
P^\nu_{i}(\mu)&=&\alpha^\nu_{P,i} ~\mu^{2\beta^\nu_{P,i}}\bigg[\ln\bigg(\frac{\mu^2}{\mu_0^2}\bigg)\bigg]^{\gamma^\nu_{P,i}}\bigg|_{i=1,2} ,\label{Pi_evolu}
\ee
where the subscript $P$ in the right hand side of the above equation stands for $P=A,B$ corresponding to $P^\nu_{i}(\mu)=A^\nu_{i}(\mu), B^\nu_{i}(\mu)$ respectively. The detail of the scheme and the values are parameters are given in \cite{MC}. 

 Our model predictions for the GTMDs $F^\nu_{1,1}, G^\nu_{1,2}, G^\nu_{1,4}, H^\nu_{1,3}, H^\nu_{1,4}$ and $ H^\nu_{1,7}$ are shown in Fig.\ref{fig_GTMD} at the scale $\mu^2=2.4~ GeV^2$ which is the average $\mu^2$ value of the HERMES experiment, and at $\mu^2=20~ GeV^2$ which is the highest bin average $\mu^2$ value of the COMPASS experiment. At the TMD limit, the GTMDs $F^\nu_{1,1}, G^\nu_{1,2}, G^\nu_{1,4}, H^\nu_{1,3}, H^\nu_{1,4}$ and $ H^\nu_{1,7}$ give the leading twist TMDs $f^\nu_1, g^\nu_{1T},g^\nu_{1L}, h^\nu_{1T}, h^{\perp\nu}_{1T}$ and $ h^{\perp \nu}_{1L} $ respectively. We plot the GTMDs for $x=0.3$ and $\Dp^2=0.1~GeV^2$. We notice that the GTMD $G_{1,4}$ for d quark approaches towards negative at higher scales. This causes a negative axial charge for d quark as found experimentally. The scale evolution of GTMDs is considered in \cite{Eche} and shown to be the same as for TMDs.
%Experimental data for GTMDs and Wigner distribution are needed to compare our model estimations.
 
%++++++++++++++++++++++++++
\section{ Inequalities}\label{inequals}
%++++++++++++++++++++++++
It is interesting to express the transverse GTMDs in terms of the unpolarized and longitudinal GTMDs at the leading twist.  Some inequality relations for GTMDs with $\bfp ^2> \frac{\Dp^2}{4}(1-x)^2$ found  in this model are 
\be 
|H^\nu_{1,3}(x,\Dp^2,\bfp^2) + \frac{\bfp^2}{2M^2}  H^\nu_{1,4}(x,\Dp^2,\bfp^2)| & < & \frac{1}{2} |F^\nu_{1,1}(x,\Dp^2,\bfp^2)+G^\nu_{1,4}(x,\Dp^2,\bfp^2)|, \label{soff_GTMD} \\
|F^\nu_{1,1}(x,\Dp^2,\bfp^2)| & > & | H^\nu_{1,3}(x,\Dp^2,\bfp^2) + \frac{\bfp^2}{2M^2}  H^\nu_{1,4}(x,\Dp^2,\bfp^2)|,\\
|F^\nu_{1,1}(x,\Dp^2,\bfp^2)| & > & |H^\nu_{1,3}(x,\Dp^2,\bfp^2)|,\\
F^\nu_{1,1}(x,\Dp^2,\bfp^2) & > & 0,\\
F^\nu_{1,1}(x,\Dp^2,\bfp^2) & > & G^\nu_{1,4}(x,\Dp^2,\bfp^2),\\
\frac{\bfp^2}{2M^2} |H^\nu_{1,4}(x,\Dp^2,\bfp^2)| & < & \frac{1}{2} |F^\nu_{1,1}(x,\Dp^2,\bfp^2)+G^\nu_{1,4}(x,\Dp^2,\bfp^2)|.
\ee
 Eq.(\ref{soff_GTMD}) represents the Soffer bound\cite{soffer} for GTMDs. 
We observe that, at the TMD limit i,e. at $\Dp=0$, the above relations reduce to the relations discussed in \cite{MC_rel} for light front quark-scalar-diquark model.

 We can also find some inequalities for Wigner distributions  given by 
\be 
\rho^\nu _{UU}(\bfb,\bfp,x)& > & 0, \\
\rho^\nu _{UU}(\bfb,\bfp,x)& > & \rho^\nu _{LL}(\bfb,\bfp,x),\\
%\rho^\nu _{UU}(\bfb,\bfp,x)& > & \rho^\nu _{LL}(\bfb,\bfp,x)\\
\rho^\nu _{TT}(\bfb,\bfp,x) & < & \frac{1}{2}\bigg[\rho^\nu _{UU}(\bfb,\bfp,x)+\rho^\nu _{LL}(\bfb,\bfp,x)\bigg].\label{sofferbound}
\ee
%for $\bfp ^2> \frac{\Dp^2}{4}(1-x)^2$.
The Eq.(\ref{sofferbound}) can be regarded as a generalized soffer bound for the Wigner distributions. It will be interesting to check if  other models also satisfy similar inequalities.
 
%%%%%%%%%%%%%%%

%%%%%%%%%%%%%%
\section{Conclusions}\label{con}
%%%%%%%%%%%%%%%%%%%%
We calculated the Wigner distributions of quarks in a nucleon using a diquark model. The light-front wave functions are modeled using ADS/QCD prediction. We took both scalar and vector diquarks\cite{MC}.
 We have presented results of the Wigner distributions in transverse position and momentum space as well as mixed position and momentum space for unpolarized, longitudinally polarized and transversely polarized quark and proton  and compared with other model predictions.
 We have noted a few inequalities among $\rho_{UU}, \rho_{LL}$ and $\rho_{TT}$ in this model. It will be interesting to check if such inequalities are present in other models, particularly in models with gluonic degrees of freedom.  The  scale evolutions of the parton distribution functions are modeled by making the parameters scale dependent in accord with DGLAP equation.  We have used the same evolution of the parameters in our calculation  for the GTMDs.  Relations of the Wigner distributions and GTMDs with the quark orbital angular momentum and spin-spin correlations are discussed.

%%%%%%%
We thank Oleg Teryaev for many useful discussions.
%%%%%%%%%%%%%%%%

\end{document}